\documentclass[sts]{imsart}

\RequirePackage{amsthm,amsmath,amsfonts,amssymb}
\RequirePackage[numbers,sort&compress]{natbib}
\RequirePackage[colorlinks,citecolor=blue,urlcolor=blue]{hyperref}
\RequirePackage{graphicx}
\usepackage{dsfont}
\usepackage{multirow}
\usepackage{booktabs}
\usepackage{pifont}
\newcommand{\xmark}{\ding{55}}%

\usepackage{cleveref}
\startlocaldefs
\theoremstyle{plain}
\newtheorem{definition}{Definition}
\newtheorem{assumption}{Assumption}
\newtheorem{theorem}{Theorem}
\newtheorem{proposition}{Proposition}
\newtheorem{corollary}{Corollary}
\newtheorem{lemma}{Lemma}

\newcommand{\modif}[1]{{\color{black} #1}}

\usepackage{pdflscape} % landscape

\usepackage[%
    font={small,sf},
    labelfont=bf,
    format=hang,    
    format=plain,
    margin=0pt,
]{caption}

\usepackage{wrapfig}
\usepackage{subcaption}
\usepackage{bbm} % indicatrix
\usepackage[table,xcdraw, dvipsnames]{xcolor}
 \usepackage[framemethod=TikZ]{mdframed}
\newcounter{takeaway}[section]

\ifstrempty{#1}%
{\mdfsetup{%
    frametitle={%
        \tikz[baseline=(current bounding box.east),outer sep=0pt]
        \node[anchor=east,rectangle,fill=blue!20]
        {\strut Takeaways};}
    }%
}{\mdfsetup{%
    frametitle={%
        \tikz[baseline=(current bounding box.east),outer sep=0pt]
        \node[anchor=east,rectangle,fill=blue!20]
        {\strut Takeaways};}%
    }%
}%
\mdfsetup{%
    innertopmargin=10pt,linecolor=blue!20,%
    linewidth=2pt,topline=true,%
    frametitleaboveskip=\dimexpr-\ht\strutbox\relax%
}

\usepackage[list=true]{subcaption}

\usepackage{snaptodo}
\snaptodoset{block rise=2em}
\snaptodoset{margin block/.style={font=\tiny}}

\newcommand{\sout}[1]{}
\newcommand{\indep}{\perp \!\!\! \perp}

\usepackage{comment}

\usepackage{natbib}
\usepackage{here}

\usepackage{graphicx}
\usepackage{tikz}
\usetikzlibrary{backgrounds}
\usetikzlibrary{arrows,shapes}
\usetikzlibrary{tikzmark}
\usetikzlibrary{calc}

\usepackage{amssymb}
\usepackage{mathtools, nccmath}

\usepackage{xspace}

\usepackage{array}
\usepackage{ragged2e}
\newcolumntype{P}[1]{>{\RaggedRight\hspace{0pt}}p{#1}}
\newcolumntype{X}[1]{>{\RaggedRight\hspace*{0pt}}p{#1}}

\usepackage{tcolorbox}

\usetikzlibrary{arrows,shapes,positioning,shadows,trees,mindmap}
\usepackage[edges]{forest}
\usetikzlibrary{arrows.meta}
\colorlet{linecol}{black!75}
\usepackage{xkcdcolors} % xkcd colors

\usepackage{tikz}
\usetikzlibrary{backgrounds}
\usetikzlibrary{arrows,shapes}
\usetikzlibrary{tikzmark}
\usetikzlibrary{calc}
\newcommand{\highlight}[2]{\colorbox{#1!17}{$\displaystyle #2$}}
\colorlet{mhpurple}{Plum!80}

\renewcommand{\highlight}[2]{\colorbox{#1!17}{#2}}

\renewcommand{\eqref}[1]{eq.\,\ref{#1}}

\endlocaldefs

\date{\today}

\begin{document}

\begin{frontmatter}
\title{Risk ratio, odds ratio, risk difference...  Which causal measure is easier to generalize?}
%\title{A sample article title with some additional note\thanksref{T1}}
\runtitle{Which causal measure is easier to generalize?}
%\thankstext{T1}{A sample of additional note to the title.}

\begin{aug}
\author[A]{\fnms{B\'{e}n\'{e}dicte}~\snm{Colnet}\ead[label=e1]{}},
\author[B]{\fnms{Julie}~\snm{Josse}\ead[label=e2]{julie.josse@inria.fr}},
\author[C]{\fnms{Ga\"{e}l}~\snm{Varoquaux}\ead[label=e3]{gael.varoquaux@inria.fr}}
\and
\author[D]{\fnms{Erwan}~\snm{Scornet}\ead[label=e4]{erwan.scornet@polytechnique.edu}}

%%%%%%%%%%%%%%%%%%%%%%%%%%%%%%%%%%%%%%%%%%%%%%
%% Addresses                                %%
%%%%%%%%%%%%%%%%%%%%%%%%%%%%%%%%%%%%%%%%%%%%%%
\address[A]{B\'en\'edicte Colnet. Soda project-team,  Premedical project-team, INRIA.}
\address[B]{Julie Josse. Premedical project team, INRIA, University of Montpellier, France.\printead[presep={\ }]{e2}.}
\address[C]{Ga\"{e}l Varoquaux, Soda project-team, INRIA Saclay, France.\printead[presep={\ }]{e3}.}
\address[D]{Erwan Scornet. LPSM, Sorbonne Universit\' e, Paris, France.\printead[presep={\ }]{e4}.}
\end{aug}

\begin{abstract}
There are many measures to report so-called treatment or causal effects: absolute difference, ratio, odds ratio, number needed to treat, and so on. The choice of a measure, e.g. absolute versus relative, is often debated because it leads to different impressions of the benefit or risk of a treatment. Besides, different causal measures may lead to various  treatment effect heterogeneity: some input variables may have an influence on some causal measures and no effect at all on others. In addition some measures -- but not all -- have appealing properties such as collapsibility, matching the intuition of a population summary. 
In this paper, we first review common causal measures and their pros and cons typically brought forward. Doing so, we clarify the notions of collapsibility and treatment effect heterogeneity, unifying existing definitions. 
Then, we show that for any causal measures there exists a discriminative model such that the conditional average treatment effect (CATE) captures the treatment effect.
However, only the risk difference has its CATE and ATE (average treatment effect) disentangled from the baseline, regardless of the outcome type (continuous or binary). 
As our primary goal is the generalization of causal measures, we show that different sets of covariates are needed to generalize an effect to a  target population depending on \textit{(i)} the causal measure of interest,
and \textit{(ii)} the identification method chosen, that is generalizing either conditional outcome or local effects. 
\end{abstract}


\end{frontmatter}

\maketitle
%\gv{Proposition de titre: "Summary measures of causal effects are not equal when facing population shifts", pour éviter le mot "metrics" qui a beaucoup de sens différents. Ou, pour un titre plus different "Causal effects with population shifts: summary measures are not equal", "Risk Ratio, odds ratio, risk difference... Which summary measures of causal effects when facing population shifts?" pour un titre plus catchy, voir même "Risk Ratio, odds ratio, risk difference... Which summary measures of causal effects for generalization?"}

%\tableofcontents

\section{The age-old question of how to report effects}

From the physician to the patient, the term \textit{effect} of a drug on an outcome usually appears very spontaneously, within a casual discussion or in scientific documents. 
Overall, everyone agrees that for a binary treatment an effect is a comparison between two states: treated or not.
But there are various ways to report the average effect 
of a treatment.
For example, the scale on which we choose to quantify the effect of a treatment may be absolute  \citep[e.g. the number of migraine days per month is expected to diminishes by 0.8 taking Rimegepant, see][]{Edvinsson2021Migraine} or relative (e.g. the probability of having a thrombosis is expected to be multiplied by 3.8 when taking oral contraceptives \citep{Vandenbroucke1994thrombosis}).
Choosing one measure or the other has several consequences. 
First, it conveys a different impression of the same data to an external reader. \cite{naylor1992measured, Forrow1992HowResultsAreSummarized} both showed that physicians's likelihood to treat patient -- following their impression of therapeutic effect -- is impacted by the scale chosen to report clinical effect. 
Such subjective impressions may be even more prominent in newspapers, where most effects are presented in relative rather than absolute terms, creating a heightened sense of sensationalism \citep{moynihan2000coverage}. 
Second, the heterogeneity of the treatment effect  -- i.e. how the treatment effect changes from one  sub-population to another  -- 
depends on the chosen causal measure \citep[see p.199 in][]{Rothman2011bookEpidemiologyIntrod}. 
The choice of the measure to report an effect is still actively discussed \citep{Spiegelman2017Modeling, Spiegelman2017letSubject, baker2018new, Changyong2019RelationsAmongThreePop, Doi2020callToChangePractice, doi2022TimeToDoAway, xiao2021odds, xiao2022IsORPortable, Huitfeldt2021ShallWe,Lapointe2022FromMathToMeaning, liu2022rejoinder}. 
Publications on the topic come with many diverging opinions and guidelines (see Appendix~\ref{appendix:different-point-of-views} for quotes).
Yet, the question of the measure (or metric) of interest is not new. For example, as \cite{Sheps1958ShallWe} wrote in the \textit{New England Journal of Medicine} \citep[see also][]{Huitfeldt2021ShallWe}:
\begin{quote}
    `` We wish to decide whether we shall count the failures or the successes and whether we shall make relative or absolute comparisons ".
\end{quote}

Beyond conveyed impressions and captured heterogeneity, different causal measures lead to different generalization properties towards populations \citep{huitfeldt2018choice}.
The problem of generalizability (or portability) encompasses a range of different scenarios, and refers to the ability of carrying over findings to a broader population, beyond the study sample. 
Generalizability of trials' findings is crucial as, most often, clinicians use causal effects from published trials to estimate the expected response to treatments for a specific patient based on his/her baseline risks, and therefore to choose the best treatment. 
In this work, we show that some effect measures are less sensitive than others to population's differences
between the study sample and the target population.\\

Section~\ref{sec:formalization-and-key-contributions} starts with a didactic clinical example to introduce the questions, the concepts, the notations, and our main results. 
Our four contributions are detailed in Section~\ref{sec:formalization-and-key-contributions} and summarized below. 
In Section~\ref{section:causal-metrics-properties}, we review, clarify, and demonstrate typical properties of causal measures, such as treatment effect homogeneity, heterogeneity, and collapsibility. 
In Section~\ref{section:generative-models}, we show that for any causal measures there exists a discriminative model such that the conditional average treatment effect captures the treatment effect. We also show that among collapsible measures, only the Risk Difference can disentangle the treatment effect from the baseline at both strata and population level (CATE and ATE), for general settings.  
On the contrary, we exhibit specific settings in which some causal measures are able to disentangle the treatment effect from the baseline. More precisely, we study a model for binary outcome inspired by the example of the Russian Roulette, in which the Risk Difference depends on the baseline, but the Survival Ratio is constant. 
Section~\ref{sec:generalize} presents the consequences on the generalizability of causal measures. We show that the Risk Difference is easier to generalize, in the sense that it requires adjustment only on the shifted treatment effect modulators introduced in Section~\ref{section:generative-models}, and not on all shifted prognostic covariates, i.e.\ variables both predictive of the outcome and with a different distribution between populations. 
Other causal measures can be generalized in some very specific settings (e.g., homogeneous treatment effect). 
Section~\ref{sec:simulations} illustrates the takeaways through simulations. \\

As this paper builds on a prolific and diverse literature, we differentiate our original contributions from previously-known results. For this purpose, all definitions, assumptions, and lemmas from prior work contain an explicit reference in their title, while those without are original contributions.

\section{Problem setting and key results}\label{sec:formalization-and-key-contributions}

\subsection{Causal effects in the potential outcomes framework}

Among the various frameworks for causal reasoning such as \cite{Pearl2000Book}, \cite{dawid2000causal}, or \cite{hernan2020whatifbook}, we use the \emph{potential outcome} framework to characterize treatment (or causal) effects.
This framework has been proposed by Neyman in 1923 \citep[English translation in][]{SplawaNeyman1990Translation}, and popularized by Donald Rubin in the 70s \citep{imbens2015causal, hernan2020whatifbook}.
It formalizes the concept of an intervention by studying two possible values $Y^{(1)}_i$ and $Y^{(0)}_i$ for the outcome of interest (say the pain level of headache) for the two different situations where the individual $i$ has been exposed to the treatment ($A_i = 1$) or not ($A_i=0$). 
We will only consider binary exposure.
The treatment has a causal effect if the potential outcomes are different, that is testing the assumption:
\begin{equation}
\label{eq:is_equal_indivi_treat_effect}
Y^{(1)}_i \stackrel{?}{=} Y^{(0)}_i.   
\end{equation}
Unfortunately, one cannot observe the two worlds for a single individual. Statistically, it can still be possible to compare the \emph{expected} values of each potential outcome $Y^{(a)}$ but it requires a population-level approach, broadening from a specific individual.
The paradigmatic example is a randomized experiment (called Randomized Controlled Trial --RCT-- in clinical research or A/B test in marketing): randomly assigning the treatment to half of the individuals enables the average comparison of the two situations.
Doing so, the previous question of interest amounts to \textit{comparing} or \textit{contrasting} two expectations:
\begin{equation}
\label{eq:is_equal_potential_expectation}
\mathbb{E} [Y^{(1)}] \stackrel{?}{=} \mathbb{E}[Y^{(0)}],
\end{equation}
where $\mathbb{E}[Y^{(a)}]$ is the expected counterfactual outcome had all individuals in the \textit{population} received the treatment level $a$. 
This quantity is defined with respect to a population: statistically, the expectation is taken on a distribution, which we denote $P_{\text{\tiny S}}$ (reflecting the \textbf{s}ource or \textbf{s}tudy sample from which evidence comes, for example a RCT). Many methodological efforts have focused on estimating the two expectations. Our focus is different: we propose theoretical guidance for choosing among different real-valued measures that allow us to compare
those two expectations at the population level, e.g. ratio, difference, or odds. What are the properties of these measures? 
How do they impact the conclusions of a study?

\subsection{Comparing two averaged situations: different treatment effect measures}\label{subsec:causal-measures-presentation}

We focus on two types of outcomes: continuous  (e.g. headache pain level) and binary (e.g. death). 
Binary outcomes are frequent in medical questions, often related to the occurrence of an event.

\subsubsection{Continuous outcome}
 For continuous outcomes,  a common measure is the absolute difference, which corresponds to the difference of means (for homogeneity of notations with the binary outcome, we denote it as the  Risk Difference - RD): 
\begin{equation*}
    \tau_{\text{\tiny RD}} := \mathbb{E}[Y^{(1)}] -  \mathbb{E}[Y^{(0)}].
\end{equation*}
A null effect corresponds to $\tau_{\text{\tiny RD}}=0$. If the outcomes are of constant sign and different from $0$, one can also consider relative measures\footnote{Allowing situations where the outcomes can be null or change sign is at risk of having undefined ratio due to $\mathbb{E}\left[Y^{(0)}\right]=0$. This is why, when considering relative measure we assume that the continuous outcome is of constant sign. Note that this is often the case in medicine. For example with blood glucose level, systolic blood pressure, etc.} such as the ratio of means (also called Risk Ratio - RR), or  relative difference of means (also called Excess Risk Ratio - ERR):
\begin{equation*}
   \tau_{\text{\tiny RR}} :=  \frac{\mathbb{E}\left[Y^{(1)}\right]}{\mathbb{E}\left[Y^{(0)}\right]}, \quad\qquad  \tau_{\text{\tiny ERR}} := \tau_{\text{\tiny RR}} -1.
\end{equation*}
A null effect now corresponds to $\tau_{\text{\tiny RR}} =1$ or $\tau_{\text{\tiny ERR}}=0$. Contrary to the difference of means which equals the mean of the differences, the ratio of means $\tau_{\text{\tiny RR}}$ is not equal to the mean of the ratios.
Note that the ranges of the three metrics are different, e.g. if $\mathbb{E}\left[Y^{(1)}\right] =200$ and $\mathbb{E}\left[Y^{(0)}\right]  = 100$, then $ \tau_{\text{\tiny RD}} = 100$, while $ \tau_{\text{\tiny RR}} = 2$ and $ \tau_{\text{\tiny ERR}} = 1$. 

\subsubsection{Binary outcome}
Due to the binary nature of the outcome, the two expectations of \eqref{eq:is_equal_potential_expectation} can now also be understood as the probability of the event to occur $ \mathbb{E}[Y^{(a)}] = \mathbb{P}[Y^{(a)} = 1]$.
As long as the phenomenon is non-deterministic in the sense that  $\mathbb{P}[Y^{(0)} = 1] \neq 0$, previous relative measures $\tau_{\text{\tiny RR}}$ and $\tau_{\text{\tiny ERR}}$ can be used for binary outcomes. Other measures, such as the Survival Ratio (SR) can be considered if $\mathbb{P}[Y^{(1)} = 1] \neq 1$: SR is nothing but a \textit{reversed} Risk Ratio (RR) where null events are counted instead of positive events. Doing so, one could also define a reversed Excess Risk Ratio (ERR), which we denote Relative Susceptibility.
The Odds Ratio (OR) is another very common measure, as it serves as a link between follow-up studies and case-control studies \citep{Greenland1987Interpretation, king2002estimating}.
Another measure called the Number Needed to Treat (NNT) has been proposed more recently \citep{Laupacis1988AnAssessmentOfClinically}: it helps the interpretation of the Risk Difference by counting how many individuals should be treated to observe one individual answering positively to the treatment. Depending on the direction of the effect, NNT can also be called Number Needed to Harm (NNH) when the events are side effects or Number of Prevented Events (NPE) when it comes to prevention. One unappealing aspect of NNT, NNH and NPE is that the null effect corresponds to an infinite value of these measures which implies that when the difference between the two treatments is not statistically significant, the confidence interval for the number needed to treat is difficult to describe \citep{altman1998confidence}.
For simplicity of the exposition, in this work, we only consider NNT \modif{\citep[see also][for a discussion]{stang2010common}}. The exact expression of the above measures are given here:  
\begin{align*}
   \tau_{\text{\tiny SR}} & :=  \frac{\mathbb{P}\left[Y^{(1)} = 0\right]}{\mathbb{P}\left[Y^{(0)} = 0\right]},   \quad  \tau_{\text{\tiny OR}}  := \frac{\frac{\mathbb{P}[Y^{(1)} = 1]}{\mathbb{P}[Y^{(1)} = 0]}}{\left(  \frac{\mathbb{P}[Y^{(0)} = 1]}{\mathbb{P}[Y^{(0)} = 0]}\right)},\\
   & \qquad \quad \quad \quad \tau_{\text{\tiny NNT}}  := \tau_{\text{\tiny RD}}^{-1}.
\end{align*}
Other measures can be found in the literature, such as the log Odds Ratio (log-OR). We recall each measure in Appendix~\ref{appendix:list-of-measures}, where Figure~\ref{fig:big-plot-with-all-metrics} illustrates the differences between measures, for different values of the expected outcomes of controls and treated. We also compute all these measures on a clinical example in Section~\ref{subsec:illustrative-example-and-key-results}.

\paragraph*{Treatment effects on subgroups}
Treatment effects can also be reported within subgroups of a population (i.e. stratified risks) to show how sub-populations react to the treatment. 
Therefore, one could also define each of the previously introduced measures on sub-populations.
For the rest of the work, we denote by $X$ a set of covariates\footnote{Those covariates are baseline or pre-treatment covariates. See \cite{VanderWeele2007FourTypes} for a detailed explanation.}. We denote by $\tau(x)$ the treatment effect on the subpopulation $X=x$ for any causal measure. 
For example $\tau_{\text{\tiny RD}}(x)$ denotes the Risk Difference on the subgroup for which $X=x$. The quantity $\tau(x)$ is often referred to as the Conditional Average Treatment Effect (CATE). \\

\paragraph*{Assumptions} Throughout this paper, and for the Average Treatment Effect and the Conditional Average Treatment Effect to be well-defined, we assume that $\mathds{E}[|Y^{(0)} ], \mathds{E}[|Y^{(1)}|] < \infty $ and $\mathds{E}[|Y^{(0)}| |X], \mathds{E}[|Y^{(1)}| |X]< \infty $. Such assumptions are satisfied as soon as the response variable is bounded.

\subsection{Key messages: from effect measures to generalization}\label{subsec:illustrative-example-and-key-results}
\subsubsection{An illustrative example}
We consider clinical data assessing the benefit of antihyperintensive therapy ($A$) against stroke ($Y$) \citep{macmahon1990blood, Cook1995NNT}. We denote by $Y=1$ a stroke, and $Y=0$ no stroke. Individuals can be categorized into two groups depending on their diastolic blood pressure: either $X=0$ corresponding to a mild baseline risk of stroke or $X=1$ corresponding to a moderate baseline risk of stroke: 
 $\mathbb{P}[ Y^{(0)} = 1 \mid X = 0]\leq\mathbb{P}[ Y^{(0)} =1\mid X =1]$. In this example, $X=1$ (resp. $X=0$) corresponds to a baseline risk of 2 events for 10 individuals (resp. 15  events for 1,000 individuals). All the measures previously introduced are computed from values reported in the original articles and presented in Table~\ref{tab:introduction-diastolic}. 
\begin{table}
    \caption{\textbf{Different treatment measures give different impressions of the phenomenon}: The outcome is stroke in 5 years ($Y=1$ denoting stroke and $Y=0$ no stroke) and stratification is done along a binary covariate $X$ (moderate $X=1$ or mild $X=0$). Each measure are computed from aggregated data taken from \cite{macmahon1990blood, Cook1995NNT}. No confidence intervals are represented as our focus is the interpretation of the measure and not statistical significance.}
    \label{tab:introduction-diastolic}
\begin{tabular}{|c|c|c|c|c|c|c|}
\hline
      & $\tau_{\text{\tiny RD}}$ & $\tau_{\text{\tiny RR}}$ &  $\tau_{\text{\tiny SR}}$ &  $\tau_{\text{\tiny NNT}}$ & $\tau_{\text{\tiny OR}}$  & $\tau_{\text{\tiny ERR}}$ \\ \hline \hline
\cellcolor[HTML]{CBCEFB}\textbf{All ($P_\text{\tiny S}$)} &   $-0.045$ &   \textbf{0.6} &  $1.05$  &  $22$   &  $0.57$  &  \textbf{-0.4}  \\ \hline \hline
\cellcolor[HTML]{ECF4FF}\textbf{X = 0} &   $-0.006$ &   \textbf{0.6} &  $1.01$  &  $167$   &  $0.60$ &  \textbf{-0.4}   \\ \hline 
\cellcolor[HTML]{ECF4FF}\textbf{X = 1} &  $-0.080$  &  \textbf{0.6}  &  $1.10$  &  $13$   &  $0.55$ &  \textbf{-0.4}  \\ \hline 
\end{tabular}
\end{table}
A Risk Ratio below 1 means that there is an inverse association, that is a decreased risk of stroke in the treated group compared with the control group. 
More precisely, the treated group has 0.6 times the risk of having a stroke outcome when compared with the non-treated group.
On this example, one can also recover that the Odds Ratio approximates the Risk Ratio in a stratum where prevalence of the outcome is low ($X=0$), but not if the prevalence is higher ($X=1$) (derivations recalled in Appendix~\ref{appendix:list-of-measures}).
The survival ratio of 1.05 captures that there is an increased chance of not having a stroke when treated compared to the control by a factor 1.05.
Note that the Survival Ratio takes really different values than the Risk Ratio: it corresponds to the Risk Ratio where labels $Y$ are swapped for occurrences and non-occurrences, illustrating that Risk Ratio is not symmetric to the choice of outcome 0 and 1 --e.g. counting the living or the dead \citep{Sheps1958ShallWe}. This lack of symmetry is usually considered as a drawback of the survival ratio and Risk Ratio compared to the odds ratio.
Indeed, the odds ratio is robust to a change of labels: swapping labels leads to changing the odds ratio $\tau_{\text{\tiny OR}}$ by its inverse $\tau_{\text{\tiny OR}}^{-1}$ (see Appendix~\ref{appendix:list-of-measures}). 

Finally, the Risk Difference translates the effect on a absolute scale: treatment reduces by $0.045$ the probability to suffer from a stroke when treated\footnote{When it comes to binary outcomes, such absolute effects are rather presented as reducing by 45 events over $1,000$ individuals.}.
The NNT is the number of patients you need to treat to prevent one additional bad outcome. Here the NNT is 22, meaning that on average, one has to treat 22 people with the drug to prevent one additional stroke.
NNT may seem simpler to interpret than a difference in probability and it enables us to quickly assess the cost (e.g., in terms of money) of a positive outcome.

\subsubsection{Contributions: how to choose a causal measure?}

\textit{This section intends to present key results in an intuitive manner. Complete mathematical definitions are given in Sections~\ref{section:causal-metrics-properties}, \ref{section:generative-models}} and \ref{sec:generalize}.

\paragraph*{Contribution 1: Properties of causal measures [Sections~\ref{section:causal-metrics-properties}]}
Different causal measures can have different properties (homogeneous/heterogeneous treatment, logic-respecting, collapsibility), which may in turn impact their interpretation. We give precise definitions of all these properties and establish relations between them. For instance, to understand the importance of collapsibility, let us dive into the following example.
If we were only provided with subgroup effects, and not the population effect ($P_\text{\tiny S}$ or \textbf{All} on Figure~\ref{tab:introduction-diastolic}), an intuitive procedure to obtain the population effect from local effects would be to average subgroups effects. More explicitly, collapsibility allows us to write
\begin{flalign}\label{eq:toy-example-collapsibility}
    &&
    {\scriptstyle \tau_{\text{\tiny RD}} =  \tikzmarknode{amp}{\highlight{ForestGreen}{\color{black} ${\scriptstyle p_\text{\tiny S}(X=1)}$ }}\cdot\tau_{\text{\tiny RD}}(X=1) +  \tikzmarknode{amp}{\highlight{ForestGreen}{\color{black} $\scriptstyle{p_\text{\tiny S}(X=0)}$  }}\cdot\tau_{\text{\tiny RD}}(X=0),}&&
\end{flalign}%
\begin{tikzpicture}[overlay,remember picture,>=stealth,nodes={align=left,inner ysep=1pt},<-]
% For "t_{j+1}"
%\path (tj1.north) ++ (-3.85,-1.8em) node[anchor=north west,color=NavyBlue!85] (tj1text){\textsf{\footnotesize property of (j+1)\textsuperscript{th} item}};
\path (amp.north) ++ (-5.3,-1.8em) node[anchor=north west,color=ForestGreen] (sotext){\textsf{\footnotesize \% individuals with $X=1$ in} ${\footnotesize P_{\text{\tiny S}}}$};
\path (amp.north) ++ (-0.8,-1.8em) node[anchor=north west,color=ForestGreen] (sotext){\textsf{\footnotesize \% individuals with  $X=0$ in} ${\footnotesize P_{\text{\tiny S}}}$};
\end{tikzpicture}

\noindent where $P_{\text{\tiny S}}$ is the source population from which the study was sampled, and $p_\text{\tiny S}(X=x)$ is the proportion of individual with $X=x$ in this population. In our example study above,  $p_\text{\tiny S}(X=0) = 0.53$ \citep{Cook1995NNT}, thus for the risk difference, the formula retrieves the population effect from the sub-group effects:
\begin{align*}
\tau_{\text{\tiny RD}} = -0.47\cdot0.08 - 0.53\cdot 0.006 = 0.0452. 
\end{align*}
When a population-effect measure can be written as a weighted average of subgroup effects with positive weights and summing to 1, it is said to be \textit{collapsible} \citep[Definition \ref{def:indirect-collapsibility}, based on][]{Huitfeldt2019collapsible}, or \textit{directly collapsible} \citep[Definition \ref{def:direct-collapsibility}, based on][]{pearl1999collapsibility, Pearl2000Book} if the weights are simply  equal to the population's proportions.
While the Risk Difference is directly collapsible, this is not true for all measures (e.g. the Number Needed to Treat is such that $0.47 \cdot 167 + 0.53 \cdot 13 = 85 \neq 22$). We precisely define collapsibility and which measures are collapsible (or not) in Section~\ref{subsec:collapsibility}, and summarized the results in Table~\ref{tab:small-summary-measures}.

\paragraph*{Contribution 2: A measure can disentangle treatment effect from baseline risk [Section~\ref{section:generative-models}]}

Table~\ref{tab:introduction-diastolic} shows that the choice of the measure gives different impressions of the heterogeneity of the effect, i.e. how much the effects measures change on different subgroups. 
Such differences can be due to different baseline risks. For example, it seems that a higher number needed to treat on the subgroup with low prevalence ($X=0$) is expected as, even without the treatment, individuals already have a low risk of stroke.
Is it possible to disentangle the baseline variation with the treatment effect in itself? 
Surprisingly, in this example, one measure is constant (or \textit{homogeneous}) over the strata $X$: the Risk Ratio. 
 We will show that among collapsible measures, only the Risk Difference can disentangle in all generality the baseline risk with the treatment effect at both strata and population level (CATE and ATE). Other causal measures are able to do so only in specific settings (e.g., homogeneous treatment effect). 
This is the case in 
the example given in Table~\ref{tab:introduction-diastolic} for the Risk Ratio. 
 For binary outcomes, we exhibit a specific model (inspired from the Russian Roulette) in which natural causal measures to consider are $(i)$ the Conditional Risk Ratio when the effect is beneficial or $(ii)$ the Conditional Survival  Ratio when the effect is detrimental.

\paragraph{Contribution 3: There exist two generalization strategies, via potential outcomes or local effects [Section~\ref{subsection:two_generalization_strategies}]}

Collapsibility may come into play when one is interested in the population effect on a target population $P_{\text{\tiny T}}$ different from the original source population $P_{\text{\tiny S}}$, \emph{e.g.} with a different proportion of individuals with diastolic pressure ($\forall x \in \{0, 1\},\, p_\text{\tiny S}(x) \neq p_\text{\tiny T}(x) $). 

In Section~\ref{sec:generalize}, we provide two different strategies to generalize causal measures via the generalization of conditional outcomes or local effects. The first approach is valid for any causal measures, whereas the second one may require fewer variables, but can be applied to collapsible measures only (see Contribution 4 below). The second strategy works as follows.
Considering the Risk Difference, the average treatment effect $\tau_{\text{\tiny RD}}^{\text{\tiny T}}$ on the target population is given by 
\begin{align}
 {\scriptstyle \tau_{\text{\tiny RD}}^{P_{\text{\tiny T}}} =   \tikzmarknode{amp}{\highlight{Bittersweet}{ ${\scriptstyle p_\text{\tiny T}(X=1)}$ }}\cdot \tau_{\text{\tiny RD}}^{P_{\text{\tiny T}}} (X=1) +  \tikzmarknode{amp}{\highlight{Bittersweet}{ ${\scriptstyle p_\text{\tiny T}(X=0)} $}}\cdot\tau_{\text{\tiny RD}}^{P_{\text{\tiny T}}}(X=0),}
\end{align}
\begin{tikzpicture}[overlay,remember picture,>=stealth,nodes={align=left,inner ysep=1pt},<-]
\path (amp.north) ++ (-5.3,-1.8em) node[anchor=north west,color=Bittersweet] (sotext){\textsf{\footnotesize \% individuals with $X=1$ in} ${\footnotesize P_{\text{\tiny T}}}$};
\path (amp.north) ++ (-0.8,-1.8em) node[anchor=north west,color=Bittersweet] (sotext){\textsf{\footnotesize \% individuals with  $X=0$ in} ${\footnotesize P_{\text{\tiny T}}}$};
\end{tikzpicture}\\[.1em]
where $\tau_{\text{\tiny RD}}^{P_{\text{\tiny T}}}(x)$ are local effects in the target population $P_{\text{\tiny T}}$. If we assume that the CATE on the source $\tau_{\text{\tiny RD}}^{P_{\text{\tiny S}}}(x)$ and target population $\tau_{\text{\tiny RD}}^{P_{\text{\tiny T}}}(x)$ are the same, we can swap them into the above equation, giving the average effect on the target population
%\newpage
\begin{flalign}\label{eq:toy-example-standardization}
    { \scriptstyle \tau_{\text{\tiny RD}}^{P_{\text{\tiny T}}} =  \tikzmarknode{amp}{\highlight{Bittersweet}{\color{black} ${\scriptstyle p_\text{\tiny T}(X=1)}$ }}\cdot\tau_{\text{\tiny RD}}^{P_{\text{\tiny S}}} (X=1) +  \tikzmarknode{amp}{\highlight{Bittersweet}{\color{black} ${\scriptstyle p_\text{\tiny T}(X=0)}$  }}\cdot\tau_{\text{\tiny RD}}^{P_{\text{\tiny S}}}(X=0).}
\end{flalign}
\begin{tikzpicture}[overlay,remember picture,>=stealth,nodes={align=left,inner ysep=1pt},<-]
\path (amp.north) ++ (-5.3,-1.8em) node[anchor=north west,color=Bittersweet] (sotext){\textsf{\footnotesize \% individuals with $X=1$ in} $P_{\text{\tiny T}}$};
\path (amp.north) ++ (-0.8,-1.8em) node[anchor=north west,color=Bittersweet] (sotext){\textsf{\footnotesize \% individuals with  $X=0$ in} $P_{\text{\tiny T}}$};
\end{tikzpicture}
\vspace{0.15cm}

Therefore, a natural procedure to generalize a collapsible causal measure to a target population is to replace the proportions $p_\text{\tiny S}(X=0)$ (resp. $p_\text{\tiny S}(X=1)$) in \eqref{eq:toy-example-collapsibility} by their counterpart $p_\text{\tiny T}(X=0)$ (resp. $p_\text{\tiny T}(X=1)$) computed on the target population. This procedure can be found under various names: \textit{standardization}, \textit{re-weighting}, \textit{recalibration} \citep{miettinen1972standardization, Rothman2000ModernEpidemiology, Bareinboim2014ExternalValidity}. We will call it \textit{generalization}, as it   follows the work initiated by \cite{stuart2011use}, which explicitly tackles the generalization of a trial with a sample of a target population. 
We show below that procedure from \eqref{eq:toy-example-standardization} is theoretically grounded, for collapsible causal measures.

\paragraph*{Contribution 4: All causal measures are not equal when facing a population shift [Section~\ref{sec:generalization}]}
Current line of works usually advocate to adjust on all prognostic covariates being shifted between the two populations. 
Using Contribution 2 and 3, we will show that the Risk Difference is likely to be more easily generalizable than other causal measures, as it requires less covariates to adjust on (only the shifted treatment effect modulators, and not all shifted prognostic covariates). Other causal measures can be generalized using an extended set of variables via generalization of the conditional outcomes. In some specific settings, e.g. when the treatment effect is homogeneous, some measures can be easily generalized as the Risk Ratio 
in Table~\ref{tab:introduction-diastolic}.

\subsection{Related work: many different viewpoints on effect measures}

\paragraph*{The choice of measure, a long debate}
The question of which treatment-effect measure is most appropriate (RR, SR, RD, OR, NNT, log-OR, etc) is age-old \citep{Sheps1958ShallWe, Greenland1987Interpretation,Laupacis1988AnAssessmentOfClinically,  Cook1995NNT, Sackett1996DownWO, davies1998can, king2002estimating, Schwartz2006ratio, Cummings2009RelativeMeritsRRAndOR}.
Health authorities advise to report both absolute and relative causal effect \citep[item 17b]{schulz2010consort}, but in practice public health publications mostly report relative risk \citep{king2012use}. 
And yet, the question is still a heated debate: in the last 5 years, numerous publications have advocated different practices \citep[see Appendix~\ref{appendix:different-point-of-views} for details]{Spiegelman2017Modeling, Spiegelman2017letSubject, lesko2018considerations, baker2018new, Changyong2019RelationsAmongThreePop, George2020WhatsTR, Doi2020callToChangePractice, doi2022TimeToDoAway, xiao2021odds, xiao2022IsORPortable, Huitfeldt2021ShallWe,Lapointe2022FromMathToMeaning}. 
Most of these works focus on the interpretation of the metrics and simple properties such as symmetry \citep{Cummings2009RelativeMeritsRRAndOR}, heterogeneity of effects \citep{Rothman2011bookEpidemiologyIntrod, VanderWeele2007FourTypes, lesko2018considerations}, or collapsibility \citep{simpson1951interpretation, Whittemore1978Collapsibility, Miettinen1981Essence, Greenland1987Interpretation, pearl1999collapsibility, Cummings2009RelativeMeritsRRAndOR, Greenland2011adjustments, Hernan2011unraveled, Sjolander2016NoteOnNoncollapsibility, Huitfeldt2019collapsible, Daniel2020MakingApple, liu2022rejoinder,Didelez2021collapsibility} --some works discuss the paradoxes induced by a lack of collapsibility without using this exact term, e.g. in oncology \citep{ding2016subgroup, liu2022correct}. We shed new light on this debate with a framing on generalization and non-parametric discriminative models of the outcome (Section~\ref{section:generative-models}).

\paragraph*{Connecting to the generalization literature}

The problem of external validity is a growing concern in clinical research \citep{rothwell2005external,Rothman1013WhyRepresentativeness,Berkowitz2018GeneralizingBlood, Deeks2022IssuesInSelection}, related to
various methodological questions \citep{cook2002experimental, pearl2011transportability}. We focus on external validity concerns due to shifted covariates between the trial's population and the target population, following the line of work initiated in \cite{imai2008misunderstandings} (see their definition of \textit{sample} effect versus \textit{population} effect), or Corollary 1 of \cite{Bareinboim2014ExternalValidity}).
Generalization by standardization 
(\eqref{eq:toy-example-standardization}, \textit{i.e.} re-weighting\footnote{It can also be seen as a change of measure, where the Radon-Nikodym derivative fully characterizes the reweighting.}  local effects) has been proposed before  in epidemiology \citep{Rothman2000ModernEpidemiology}, and in an even older line of work in the demography literature \citep{yule1934some}. 
Note that \eqref{eq:toy-example-standardization} is very close to procedure from \eqref{eq:toy-example-collapsibility} which can be linked to post-stratification \citep{imbens2011experimental, miratrix2013adjusting}. Post-stratification is used to lower variance on a randomized controlled trial and therefore has no explicit link with generalization, despite using a similar statistical procedure.
Today, almost all statistical papers dealing with generalization focus on the estimation procedures that generalizes the risk difference $\tau_{\text{\tiny RD}}$ \citep{stuart2011use, tipton2013improving, Muircheartaigh2014GeneralizingApproach, kern2016assessing, lesko2017generalizing, nguyen2018sensitivitybis, Stuart2017ChapterBook, buchanan2018generalizing, dahabreh2020extending, ackerman2020generalization} (reviewed in \cite{Degtiar2021Generalizability, colnet2021causal}), seldom mentioning other measures. 
Other works focus on the generalization of the distribution of the treated outcome $\mathbb{E}\left[ Y^{(1)}\right]$ \citep{pearl2011transportability, Bareinboim2014ExternalValidity, CinelliGeneralizing2019}. A notable exception, \cite{huitfeldt2018choice}, details which choice of variables enables the standardization procedure for binary outcomes.

\paragraph*{Building up on causal research}
By writing the outcomes as generated by a non-parametric process disentangling the baseline from the treatment effect (in the spirit of \cite{robinson1988semiparam, nie2020quasioracle, Gao2021DINA}), we extend the usual assumptions for generalization.
In particular, \cite{Bareinboim2014ExternalValidity} state that their assumptions for generalization are
``\textit{the worst case analysis where every variable may potentially be an effect-modifier}''. Our work proposes more optimistic situations, by introducing a notion of effect-modifier without parametric assumptions. This enables the description of situations where fewer covariates are required for the generalization of certain measures.
\cite{CinelliGeneralizing2019} have proposed similar ideas, assuming monotonicity of the effect (i.e. the effect being either harmful or beneficial for everyone) \underline{and} the absence of shifted treatment effect modifiers, in order to generalize $\mathbb{E}[Y^{(1)}]$. More precisely they assume that what they call \textit{probabilities of causation} $\mathbb{P}[Y^{(1)} = 0 \mid Y^{(0)} = 1]$ are invariant across populations. We relax this assumption to allow more general situations. Doing so, we also extend work from \cite{huitfeldt2018choice, Huitfeldt2019EffectHeterogeneity}, showing how those probabilities are linked with the causal measures of interest. 
Interestingly, all our derivations retrieves \cite{Sheps1958ShallWe} intuition and results when the outcome is binary (which was the only situation described by Sheps). Our work also proposes conclusions for a continuous outcome which was not treated by \cite{Sheps1958ShallWe, CinelliGeneralizing2019, Huitfeldt2021ShallWe}.

\section{Causal metrics and their properties}\label{section:causal-metrics-properties}

\textit{This section uses notations introduced in Section~\ref{sec:formalization-and-key-contributions}, in particular the potential outcomes $Y^{(0)}$, $Y^{(1)}$ (which can be either binary or continuous), the binary treatment $A$, and the covariates $X$.} \\

In this section, we ground formally concepts such as homogeneity and heterogeneity of treatment effect, but also collapsibility. Those concepts are already described in the literature, via numerous and slightly different definitions  (see   Appendix~\ref{appendix:other-formal-definitions}).  
We unify existing definitions. For clarity, all definitions, assumptions, and lemmas that do not contain an explicit reference in the title are original. 

\modif{\subsection{Definition of causal measures}}
\begin{definition}[Causal effect measures -- \cite{Pearl2000Book}]\label{def:causal-measure}
Assuming a certain joint distribution of potential outcomes $P(Y^{(0)}, Y^{(1)})$, which implies that a certain treatment $A$ of interest is considered, we denote by $\tau$ any functional of the joint distribution of potential outcomes. More precisely,

\begin{align}
\mathcal{P} (Y^{(0)}, Y^{(1)})   & \rightarrow \mathbb{R} \nonumber \\
\tau : P(Y^{(0)}, Y^{(1)})  & \mapsto \tau^{P},
\end{align}
where $\mathcal{P} (Y^{(0)}, Y^{(1)}) $ is the set of all joint distributions of $(Y^{(0)}, Y^{(1)})$.
\end{definition}
This definition is also valid for any subpopulation: for any covariate $X$, the conditional causal effect measure $\tau^{P}(X)$ is defined as a functional of $P(Y^{(0)}, Y^{(1)}\mid X)$.
This definition highlights the fact that a so-called treatment or causal effect naturally depends on \emph{the population considered}.
The notation $\tau^{P}$ highlights this dependency.
\modif{Note that such causal measures are called individual measures \citep[][]{Fay2024Causal} and are non-identifiable as they depend on the joint distribution of  potential outcomes. 
} 

\modif{
In this paper, we consider population causal measures $\tau$, that depend on the marginal distribution of the potential outcome and in particular on their expectation. More precisely, we assume throughout the paper that  there exists a function $f: D_f \to \mathds{R}$ defined on $D_f \subset \mathds{R}^2$ verifying
\begin{align}
\tau^P & = f \left(\mathds{E}[Y^{(0)}] , \mathds{E}[Y^{(1)} ] \right),\label{eq_causal_measure_average} \\
\tau^P (x) & = f \left(\mathds{E}[Y^{(0)} \mid X=x ] , \mathds{E}[Y^{(1)} \mid X=x ] \right), \label{eq_causal_measure}
\end{align}
for all distributions $P(Y^{(0)}, Y^{(1)}|X)$ and for all $x \in \mathds{X}$ such that the above quantities exist. All causal measures presented in Section~\ref{subsec:causal-measures-presentation} satisfy \eqref{eq_causal_measure_average} and \eqref{eq_causal_measure}. For example, the function $f$ associated to the risk difference is simply $f: (z,z') \mapsto z'-z$ with $D_f = \mathds{R}^2$. }

\modif{Note that there are measures that go beyond the mean, such as the quantile treatment effect, which is defined as the difference between corresponding quantiles of the potential outcome distributions. In addition, more complex outcomes can be considered, including multivariate hierarchical outcomes \citep[][]{even2025rethinking} or fully distributional outcomes \citep[see, e.g.,][]{lin2023causal, katta2024interpretable}.
}

\begin{assumption}[Injectivity]
\label{ass:injection_def_domain}
Let $\tau$ be a causal measure and $f$ its associated function (\eqref{eq_causal_measure_average} and \eqref{eq_causal_measure}). Let, for all $z \in D_f^{(1)}$, 
\begin{align}
\begin{array}{cclc}
     g_z :  & D_f^{(2)}(z) & \to & \mathds{R} \\
     &  z' & \mapsto & f(z,z'),
\end{array}
\end{align} 
where $D_f^{(1)} = \{z_1, \exists z' \in \mathds{R} \textrm{ such that } (z_1,z') \in D_f\}$ and $D_f^{(2)}(z) = \{z', (z,z') \in D_f\}$. Assume that, for all $z \in D_f^{(1)}$, $g_z$ is an injection. 
\end{assumption}
Such an assumption, stating that $g_z$ is an injection, is mild: if this was not the case, two different values of $\mathds{E}[Y^{(1)}|X]$ would lead to the same CATE for a given baseline $\mathds{E}[Y^{(0)}|X].$

\subsection{Treatment effect heterogeneity depends on the measure chosen}\label{subsec:homogeneity}

Homogeneity or heterogeneity is linked to how the effects change on population subgroups.  
If the effect amplitude or direction is different in some subgroups (not due to sampling noise as we only consider the true population's values), the treatment effect is said to be heterogeneous. 
In the literature, one can find several informal definitions of heterogeneity of a treatment effect
but formal definitions are scarce. From now on, we let $\mathbb{X}$ be the covariate space.  
\begin{definition}[Treatment effect homogeneity]\label{def:homogeneity}
A causal effect measure $\tau$ is said to be homogeneous with respect to the covariate space $\mathds{X}$, if for all $x_1, x_2 \in \mathds{X},$
\begin{equation*}
   \tau^P(x_1) = \tau^P(x_2).
\end{equation*}
\end{definition}

\begin{definition}[Treatment effect heterogeneity - \cite{VanderWeele2007FourTypes}] \label{def:heterogeneity}
A causal effect measure $\tau$ is said to be heterogeneous with respect to the covariate space $\mathds{X}$, if there exist $x_1,x_2 \in \mathds{X}$ such that $\tau^P(x_1) \neq \tau^P(x_2).$
\end{definition}
Heterogeneity and homogeneity are properties defined with respect to \textit{(i)}  a covariate space $\mathds{X}$ (or equivalently covariates $X$) and \textit{(ii)} a measure. Claiming \textit{hetereogeneity or homogeneity of a treatment effect} should always be completed by the information about the considered covariates and the  measure under study. For instance in the illustrative example from Table~\ref{tab:introduction-diastolic}, the treatment effect on the Risk Difference scale is heterogeneous with respect to the baseline diastolic blood pressure level $X$, while the treatment effect on the Risk Ratio scale is homogeneous with respect to $X$. 
In Section~\ref{sec:generalization}, we will show that, under some proper assumptions, a homogeneous treatment effect is easily generalizable (Theorem~\ref{thm_homogeneous_independence_generalization}).

\subsection{Not all measures are collapsible}\label{subsec:collapsibility}

\paragraph*{Intuition}
Collapsibility is intuitively linked to heterogeneity.
Indeed, to investigate for heterogeneity, one looks up the treatment effect on subgroups of the population. 
Collapsibility is the opposite process, where local information is aggregated to obtain a global information (i.e. on a population). 
One might expect the global effect on a population to be an average of the subgroups effects, with weights corresponding to proportions of each subgroup in the target population of interest as in \eqref{eq:toy-example-collapsibility}.
Counter-intuitively, this procedure is valid only for certain causal effect measures.
For example, if the treatment effect is reported as an Odds Ratio, it is possible to find bewildering situations, such as that of the synthetic example detailed on Table~\ref{tab:odds-ratio-simpson}. 
In this example, the Odds Ratio is measured on the overall population (Table~\ref{tab:odds-ratio-simpson} (a)) and on the two subpopulations if female ($X=1$) or not ($X=0$) (Table~\ref{tab:odds-ratio-simpson} (b)).
Here, the drug's effect (on the OR scale) is found almost equal on both males (0.166) and females (0.167); however the average effect on the overall population appears weaker (0.26). 
The Odds Ratio value in the overall population is not even \emph{between} Odds Ratios of sub-populations. The situation mimics a randomized controlled trial conducted with exact population proportions and with $X$ being a covariate, so the phenomenon observed is not an effect of counfounding.\\

\begin{table}
    \centering
    \caption{\textbf{Non-collapsibility of the odds ratio on a toy example}: The tables below represent the exact proportion of an hypothetical population, considering two treatment level $A \in \{0, 1\}$ and a binary outcome. The proportion are as if a randomized controlled trial was conducted on this population. This population can be stratified in two strata: woman ($X=1$) or not ($X=0$). The odds ratio can be measured on (a) the overall population, or on (b) each of the sub-population, namely $X=0$ or $X=1$. Surprisingly, on each sub-population the odds ratios are similar, but on the overall population the odds ratio is almost two times bigger than on each sub-population. This example is largely inspired from \cite{Greenland1987Interpretation}, but several similar examples can be found elsewhere, for example in \cite{hernan2020whatifbook} (see their Fine point 4.3) or in \cite{pearl1999collapsibility} (see their Table 1). Another didactic example is provided in \cite{Daniel2020MakingApple} (see their Figure 1), with a geometrical argument.}
    \begin{subtable}{\linewidth}
        \centering
        \caption{Overall population, $\tau_{\text{\tiny OR}}  \approx 0.26$}

        \begin{tabular}{c|cc|}
        \cline{2-3}
                          & \multicolumn{1}{c|}{Y=0} & Y=1 \\ \hline
\multicolumn{1}{|c|}{A=1} & 1005                      & 95 \\ \cline{1-1}
\multicolumn{1}{|c|}{A=0} & 1074                       & 26  \\ \hline
        \end{tabular}

    \end{subtable}%
    
    \begin{subtable}{\linewidth}
        \caption{$\tau_{\text{\tiny OR}\mid X=1} \approx 0.167$ and $\tau_{\text{\tiny OR}\mid X=0} \approx 0.166$}
        \centering
\begin{tabular}{c|cc|lc|cc|}
\cline{2-3} \cline{6-7}
\textbf{X= 1}             & \multicolumn{1}{c|}{Y=0} & Y=1 &                       & \textbf{X=0} & \multicolumn{1}{c|}{Y=0} & Y=1 \\ \cline{1-3} \cline{5-7} 
\multicolumn{1}{|c|}{A=1} &              40          & 60  & \multicolumn{1}{l|}{} & A=1          & 965                       &  35 \\ \cline{1-1} \cline{5-5}
\multicolumn{1}{|c|}{A=0} & 80                       & 20  & \multicolumn{1}{l|}{} & A=0          & 994                      & 6  \\ \cline{1-3} \cline{5-7} 
\end{tabular}

    \end{subtable}% 
    \label{tab:odds-ratio-simpson}
\end{table}

This apparent paradox is due to what is called the non-collapsibility\footnote{This definition and phenomenon has been observed long ago by Simpson. See also the \cite{Hernan2011unraveled} for a discussion of Simpson's original paper with modern statistical framework. Note that \cite{Pearl2000Book} (page 176) mentions that collapsibility has been discussed earlier, for example by Pearson in 1899.} of the Odds Ratio. The fact that the average effect on a population could not be written as a weighted sum of effects on sub-populations is somehow going against the ``\textit{implicit assumptions that drive our causal intuitions}'' (\cite{Pearl2000Book}, page 180).
Non-collapsibility can also be understood through the non-linearity of a function linking the baseline (control) and response functions, see \ref{sec:appendixORnoncollapsible}.
On the contrary, an effect measure is said to be collapsible when the population effect measure can be expressed as a weighted average of the stratum-specific measures.
Note that non-collapsibility and confounding are two different concepts, as explained in several papers e.g. in \cite{pearl1999collapsibility}\footnote{``\textit{ 
the two concepts are distinct: confounding may occur with or without noncollapsibility and noncollapsibility may occur with or without confounding.}"}.
\paragraph*{Formalizing the problem} In various formal definitions found in the literature (see Section~\ref{appendix:other-formal-definitions}), collapsibility relates to the possibility of writing the marginal effect as a weighted sum of conditional effects on each subgroups. Yet two definitions coexist, depending on whether weights are forced to be equal to the proportion of individuals in each subgroup or not. We outline various definitions and their links below.

\begin{definition}[Direct collapsibility - adapted from \cite{pearl1999collapsibility, Pearl2000Book, liu2022correct, Didelez2021collapsibility}] \label{def:direct-collapsibility}
Let $\tau$ be a measure of effect (see Definition~\ref{def:causal-measure}).
The measure $\tau$ is said to be directly collapsible with respect to a set of  covariates $X$ if, for all joint distribution $P(Y^{(0)}, Y^{(1)}, X)$, we have
\begin{equation*}
   \mathbb{E}\left[  \tau^P(X)\right] = \tau^P.
\end{equation*}
\end{definition}

This definition can be found written sligthly differently in literature, see Definition~\ref{def:strict-collaps-pearl-greenland} in Appendix~\ref{appendix-def_collapsibility}.
\begin{lemma}[Direct collapsibility of the RD -- \citep{pearl1999collapsibility}]\label{lemma:direct-collapsibility-RD}
The Risk Difference $\tau_{\text{\tiny RD}}$ is directly collapsible.
\end{lemma}
This result grounds \eqref{eq:toy-example-collapsibility} in the illustrative example.
In the literature, more flexible definitions of collapsibility can be found, keeping the intuition of the population effect being a weighted sum of effects on subpopulations, with certain constraints on the weights: weights must be positive and sum to one. 
\begin{definition}[Collapsibility - adapted from \cite{Huitfeldt2019collapsible}] \label{def:indirect-collapsibility}
Let $\tau$ be a measure of effect and $\mathds{X}$ the covariate space.  Let $\mathcal{P}(X, Y^{(0)})$ be the set of all joint distributions of $(X, Y^{(0)})$. The measure $\tau$ is said to be collapsible with respect to the covariate space $\mathds{X}$ if there exists a positive weight function
\begin{align}
\mathds{X} \times \mathcal{P}(X, Y^{(0)}) \quad & \rightarrow \mathbb{R}_+ \nonumber \\
w : (X, P(X, Y^{(0)})) & \mapsto w(X, P(X, Y^{(0)})),
\end{align}
satisfying $\mathbb{E}\left[  w(X, P(X,Y^{(0)}))\right] = 1,$
such that, for all joint distributions $P(X, Y^{(0)}, Y^{(1)})$, we have
\begin{equation*}
\mathbb{E}\left[ w(X, P(X,Y^{(0)}))\, \tau^P(X) \right] = \tau^P.
\end{equation*}
\end{definition}
Note that here weights depend on $X$ and the distribution of controls $P(X, Y^{(0)})$. 
The direct collapsibility is therefore a specific case of the more general version of collapsibility from Definition~\ref{def:indirect-collapsibility}, where $ w(X, P(X,Y^{(0)}))$ corresponds to $1$.  
Allowing the weights to depend on the joint distribution of the covariates and the two potential outcomes would lead to all measures being collapsible. Besides, if one had access to the joint distribution of $(X, Y^{(0)}, Y^{(1)})$, one could generate and generalize any causal measure. We choose to consider weights that depend on $Y^{(0)}$ (instead of $Y^{(1)}$) as accessing the distribution of $Y^{(0)}$ (control cases) may be easier in practice.
\begin{lemma}[Collapsibility of the Risk Ratio and survival ratio - extending \cite{ding2016subgroup, Huitfeldt2019collapsible, Didelez2021collapsibility}]\label{lemma:collapsibility-of-RR-SR}
The Risk Ratio and survival ratio are collapsible measures. In particular, for any covariate space $\mathds{X}$, assume that, almost surely, $0 < \mathbb{E}[Y^{(0)}|X] < 1$. Then, the conditional Risk Ratio and conditional survival ratio exist and satisfy %\jj{est-ce qu'on les numeroterait pas? tu ne les utilises pas après?}\bc{je peux dire "formula of Lemme 2" ?}
\begin{align*}
    & \mathbb{E}\left[ \tau_\text{\tiny RR}^P(X) \frac{\mathbb{E}\left[Y^{(0)} \mid X \right]}{\mathbb{E}\left[Y^{(0)}\right]}\right] = \tau_\text{\tiny RR}^P \\
     \textrm{and} ~ & \mathbb{E}\left[\tau_\text{\tiny SR}^P(X) \frac{1-\mathbb{E}\left[ Y^{(0)} \mid X \right]}{1-\mathbb{E}\left[ Y^{(0)}\right] }\right] = \tau_\text{\tiny SR}^P.
\end{align*}
\end{lemma}
Knowing the actual form of the weights will be very helpful when coming to generalization in Section~\ref{sec:generalization}.
The proof of collapsibility for the Risk Ratio, the Survival Ratio and other causal measures are established in Appendix~\ref{proof:collapsibilty}. 
%Note that  introduced in Appendix (see Section~XX) follows directly from the collapsibility of $ \tau_\text{\tiny RR}$ and $ \tau_\text{\tiny SR}$.
Note that results of Lemma~\ref{lemma:collapsibility-of-RR-SR} are already presented in \cite{Huitfeldt2019collapsible} or \cite{ding2016subgroup} (see their Equation 2.3) 
%also recall these results 
but only for a binary outcome and categorical  covariate\footnote{Note that this result can be found under slightly different forms such as in \cite{huitfeldt2018choice, Didelez2021collapsibility}, with a categorical $X$ and using Bayes formula, $\tau_\text{\tiny RR} = \sum_x \tau_\text{\tiny RR}(x)\, \mathbb{E}\left[X=x \mid Y^{(0)} = 1 \right]$.}.   
Thus, Lemma~\ref{lemma:collapsibility-of-RR-SR} extends their results for any covariate space $\mathds{X}$  (including categorical and continuous variables) and any type of outcome $Y$ (continuous or binary).

\begin{lemma}[Non-collapsibility of the OR, log-OR, and NNT, based on \cite{Daniel2020MakingApple}]\label{lemma:non-collapsibility}   
The odds ratio $\tau_{\text{\tiny OR}}$,  log odds ratio $\tau_{\text{\tiny log-OR}}$, and Number Needed to Treat $\tau_{\text{\tiny NNT}}$ are non-collapsible measures.
\end{lemma}
 The proof is in Appendix~\ref{proof:collapsibilty}. 
  While the non-collapsibi-lity of the odds ratio and the log Odds Ratio have been reported multiple times \citep[see, e.g.,][and the example from Table~\ref{tab:odds-ratio-simpson}]{Daniel2020MakingApple}, we have not found references stating results about the NNT. 
  When considering the OR, the marginal effect $\tau$ can be smaller or bigger than the range of local effects $\tau(x)$. 
 Accordingly, \cite{liu2022correct} introduces the term \textit{logic} for such characteristic.

\begin{definition}[Logic-respecting measure -- \cite{liu2022correct}]\label{def:logic-respecting}
A measure $\tau$ is said to be logic-respecting if, for any covariate  space $\mathds{X}$ and any distribution $P(X, Y^{(0)}, Y^{(1)})$, 
\begin{equation*}
    \tau^P \in \left[\min_{x \in \mathds{X}}(\tau^P(x)), \;\; \max_{x\in \mathds{X}}(\tau^P(x)) \right].
\end{equation*}
\end{definition}
\begin{lemma}[All collapsible measures are logic-respecting, but not the opposite]\label{lemma:logic-respecting-measures} Several properties can be noted:
\begin{itemize}
    \item[\textit{(i)}] All collapsible measures are logic-respecting measures.
    \item[\textit{(ii)}] The Number Needed to Treat is a logic-respecting measure.
    \item[\textit{(iii)}] The OR and the log-OR are not logic-repecting measures.
\end{itemize}
\end{lemma}

The proof is in Appendix~\ref{proof:logic-respecting}. While the NNT is not collapsible, this measure does not show the same paradoxical behavior as the OR  (see Table~\ref{tab:odds-ratio-simpson}). This is due to the fact that the NNT results from a monotonic transformation of the RD, which is collapsible (see Lemma \ref{lemma:annexe_monotinic_causal_measure} in Appendix~\ref{app:supplementary_lemma}). 
The numerous mentions of paradoxes with the OR are probably more driven by its non logic-respecting property than by its non-collapsibility. 
This also probably explains why some definitions of collapsibility proposed in the literature do not explicitly separate the notions of collapsibility and logic-respecting  as they do not detail how weights are defined (see for example Definitions~\ref{def:collapsibility-huitfeldt} or \ref{def:collapsibility-didelez} in Appendix). All properties of this section are summarized in Table \ref{tab:small-summary-measures}.

 {\footnotesize 
 \begin{table}
 \begin{minipage}{0.5\textwidth}
     \caption{\textbf{Causal measures and their properties}: highlighting the properties of collapsibility (Definition~\ref{def:indirect-collapsibility}) and logic respecting (Definition~\ref{def:logic-respecting}). An exhaustive table is available in Appendix (see Table~\ref{tab:list-measures-with-all-properties}).}
     \label{tab:small-summary-measures}
     \begin{tabular}{|
>{\columncolor[HTML]{ECF4FF}}c |
>{\columncolor[HTML]{E0FAE0}}c |
>{\columncolor[HTML]{E0FAE0}}c |}
\hline
\cellcolor[HTML]{CBCEFB}\textbf{Measure} & \cellcolor[HTML]{CBCEFB}\textbf{Collapsible} & \cellcolor[HTML]{CBCEFB}\textbf{Logic-respecting} \\ \hline
Risk Difference (RD)                                       & Yes                                          & Yes                                               \\
Number Neeeded to Treat (NNT)                                      & \cellcolor[HTML]{FCF1F1}No                                           & Yes                                               \\
Risk Ratio (RR)                                       & Yes                                          & Yes                                               \\
Survival Ratio (SR)                                       & Yes                                          & Yes                                               \\
Odds Ratio (OR)                                       & \cellcolor[HTML]{FCF1F1}No                   & \cellcolor[HTML]{FCF1F1}No                        \\ \hline
\end{tabular}
\end{minipage}
 \end{table}
 }

\section{Disentangling the treatment effect from the baseline}\label{section:generative-models}

We now propose to reverse the thinking: rather than starting from a given metric, we propose to reason from generic non-parametric discriminative models (for continuous and binary outcomes). Such models allow us to distinguish covariates that affect only baseline level from those that modulate treatment effects. We make us of this distinction in Section~\ref{sec:generalize}  to determine which measures are easier to generalize.

\subsection{One discriminative model per causal measure}

Using the binary nature of $A$, it is possible to decompose the response $Y$ in two parts: baseline level and modification induced by the treatment. Such decompositions are generic and do not rely on any parametric assumptions.

\begin{lemma}
\label{lem_generative_models}
Let $\tau$ be a causal measure defined in \eqref{eq_causal_measure} satisfying Assumption~\ref{ass:injection_def_domain}. Then, for all distributions $P(Y^{(0)}, Y^{(1)} | X)$, there exist two unique functions $b,m : \mathds{X} \to \mathds{R}$ such that, for all $x \in \mathds{X}$ such that 
\begin{align}
\label{eq_def_domain_lemma}
\left(\mathds{E}[Y^{(0)} \mid X=x ] , \mathds{E}[Y^{(1)} \mid X=x ] \right) \in D_f,
\end{align}
we have
\begin{align}
    \mathds{E}[Y^{(0)} | X =x ] & = b(x), \nonumber \\
   \textrm{and} \quad  \mathds{E}[Y^{(1)} | X=x ] & = g_{b(x)}^{-1} (m(x)). \label{eq_gen_model_lemma}
\end{align}
Under the model defined in~\eqref{eq_gen_model_lemma}, for all $x$ satisfying \eqref{eq_def_domain_lemma}, 
\begin{align}
    \tau^P(x) = m(x).
\end{align}
\end{lemma}
The proof can be found in Appendix~\ref{app_subsection_proof_Lemma1_genmodels}. Lemma~\ref{lem_generative_models} shows that for any causal measure, there exists an appropriate discriminative model such that, under this model, the conditional causal measure captures the treatment effect, \modif{defined by the function $m$. In particular, for any given causal measure, we can simulate linear treatment effects, by choosing a linear function $m$, then choosing a baseline $b$ and finally generating the conditional expectations of the potential outcomes as in \Cref{lem_generative_models}. The discriminative model of \Cref{lem_generative_models} for the Risk Difference is presented in  \Cref{lemma:working-model-continuous-Y} below. Such a model is often used for data generation  \citep[see][]{imai2013estimating, athey2019generalized, nie2021quasi, kunzel2019metalearners} even if directly modelling the conditional expectation of the potential outcomes is also possible \citep[see, e.g., scenarios 2, 3 in][]{kunzel2019metalearners}.
}

\begin{corollary}
\label{lemma:working-model-continuous-Y}
Consider the Risk Difference. In the framework of Lemma~\ref{lem_generative_models}, we have $b(X) = \mathds{E}[Y^{(0)}|X]$. Besides, $g_z(z') = z' - z$ and $g_z^{-1}(z') = z' + z,$ which leads to
\begin{align}
   \mathds{E}[Y^{(1)}|X] = m(X) + b(X).
\end{align}
Such a model can also be written as, for all $a\in \{0,1\}$, 
\begin{align}
    \mathds{E}[Y^{(a)}|X]  = b(X) + a m(X). \label{corollary_decomposition_linear}
\end{align}
Besides, we have $\tau_{\text{\tiny RD}}^P(X) = m(X)$, 
 \begin{align*}
     & \tau_{\text{\tiny RD}}^P= \mathbb{E}\left[ m(X)\right],\qquad 
     \tau_{\text{\tiny RR}}^P = 1 + \frac{\mathbb{E}\left[m(X) \right]}{\mathbb{E}\left[b(X) \right]},\\
     & \quad \quad \quad  \text{and}\quad 
     \tau_{\text{\tiny ERR}}^P = \frac{\mathbb{E}\left[m(X) \right]}{\mathbb{E}\left[b(X) \right]}.
 \end{align*}
\end{corollary}

\begin{figure}[!h]
\includegraphics[width=0.4\textwidth]{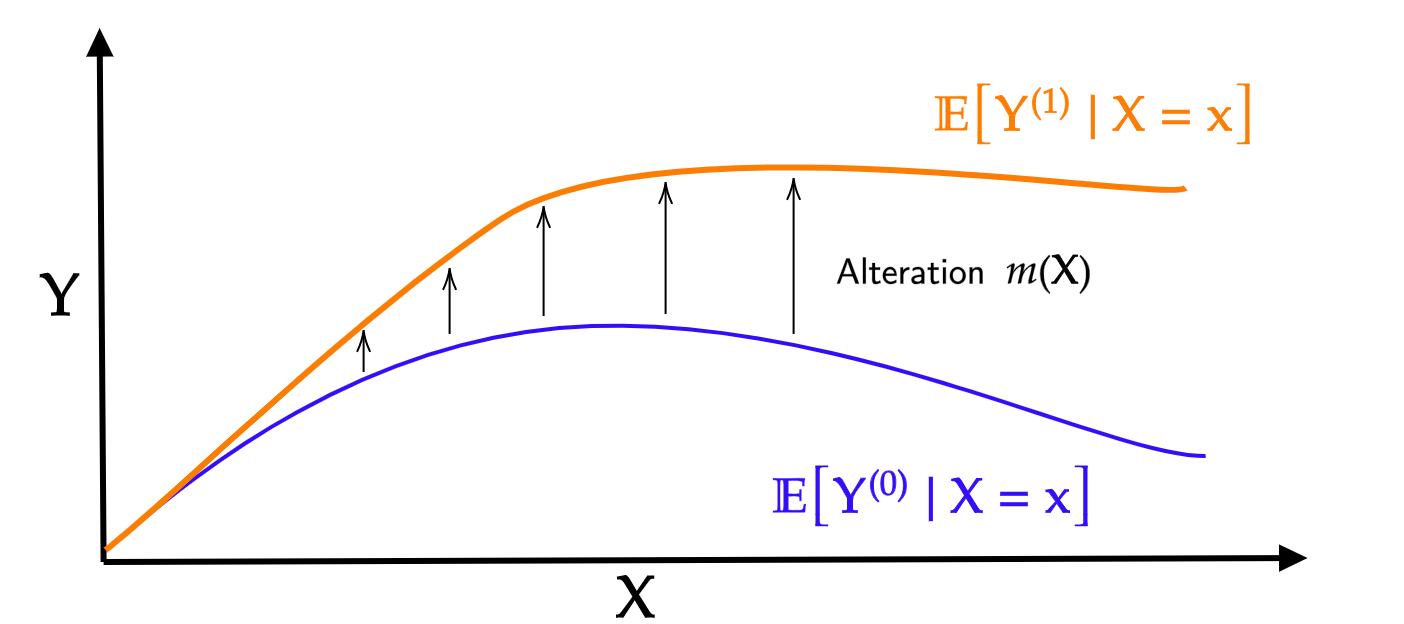}
\caption{\textbf{Intuition behind \eqref{corollary_decomposition_linear}}: This illustration highlights that, for a given set of covariates $X$, one can assume that there exist two functions accounting for the expected outcome value for any individual with baseline characteristics $X$. Then, it is possible to denote $m(X)$ as the alteration or \textbf{m}odification of the \textbf{b}aseline $b(X):=  \mathbb{E}[Y^{(0)} \mid X]$ response.}
	\label{fig:alteration.png}
 \end{figure}

The formula in \eqref{corollary_decomposition_linear} is related to the Robinson \citep{robinson1988semiparam}  decomposition  \citep[see also][for a completely linear model]{angrist2008mostlyharmless}.
This model allows to interpret the difference between the distributions of treated and control groups as the alteration $m(X)$ of a baseline model $b(X)$  by the treatment. The function $b$ corresponds to the \textbf{b}aseline, and $m$ to the \textbf{m}odifying function due to treatment. Figure~\ref{fig:alteration.png} gives the intuition backing \Cref{lemma:working-model-continuous-Y}.

Corollary~\ref{lemma:working-model-continuous-Y} also illustrates how the relative measures $ \tau_{\text{\tiny RR}}$ and $ \tau_{\text{\tiny ERR}}$ depend on both the effect $m(X)$ \textbf{and}  the baseline $b(X)$. On the contrary, $ \tau_{\text{\tiny RD}}^P$ and $ \tau_{\text{\tiny RD}}^P(X)$ are independent of the baseline $b(X)$. \\

Based on \Cref{lem_generative_models}, 
one can make explicit the discriminative model associated to any causal measure. In particular, the Conditional Odds Ratio equals  the treatment effect in the logistic model (see Section~\ref{appendix:usual-point-of-view}). 
Below, we detail the discriminative model associated to the Risk Ratio. 

\begin{corollary}
\label{lemma:working-model-continuous-Y-RR}
Consider the Risk Ratio. In the framework of Lemma~\ref{lem_generative_models}, we have $b(X) = \mathds{E}[Y^{(0)}|X]$. Besides, $g_z(z') = z'/z$ and $g_z^{-1}(z') = z z',$ which leads to
\begin{align}
   \mathds{E}[Y^{(1)}|X] = b(X) m(X).
\end{align}
Such a model can also be written as, for all $a\in \{0,1\}$, 
\begin{align}
    \mathds{E}[Y^{(a)}|X]  = b(X) (m(X))^a.
\end{align}
Consequently, we have
 \begin{equation}
     \tau_{\text{\tiny RR}}^P(X) = m(X) \quad \textrm{and} \quad \tau_{\text{\tiny RR}}^P= \frac{\mathbb{E}[Y^{(1)}]}{\mathbb{E}[Y^{(0)}]} = \frac{\mathbb{E}\left[  b(X) m(X)\right]}{\mathbb{E}\left[  b(X)\right]}.
     %,\qquad 
     %\tau_{\text{\tiny RD}} = \mathbb{E}[b(X) (m(X)-1)],\qquad \text{and}\quad 
     %\tau_{\text{\tiny ERR}} = \frac{\mathbb{E}\left[  b(X) (m(X)-1) \right]}{\mathbb{E}\left[  b(X)\right]}.
 \end{equation}
\end{corollary}

Under the discriminative model associated with the Risk Ratio (as stated in Corrolary~\ref{lemma:working-model-continuous-Y-RR}), the conditional Risk Ratio captures the treatment effect, but the Risk Ratio computed on the overall population depends on the baseline: the Risk Ratio is unable to disentangle the treatment effect from the baseline both at a strata level and at the population level (CATE and ATE).

\subsection{Only the Risk Difference can disentangle the treatment effect from the baseline}
\modif{\Cref{lemma:working-model-continuous-Y} and \Cref{lemma:working-model-continuous-Y-RR} suggest different behaviors of causal measures: the RD is able to disentangle the baseline and the treatment effect at both local and global level (CATE and ATE), contrary to the RR, only able to do so at the local level. We want to investigate if other measures than RD are able to disentangle baseline and treatment effect. To this aim, we need to introduce a formal definition of disentanglement, with respect to a collection of possible joint distributions $P(X, Y^{(0)}, Y^{(1)})$. In the sequel, for any collection $\mathcal{P}$ of distributions $P(X, Y^{(0)}, Y^{(1)})$, we let  
\begin{align}
 \mathcal{P}(Y^{(0)}|X) = \{ P(Y^{(0)}|X): P \in \mathcal{P}\}   
\end{align}
be the collection of all baseline distributions. Besides, for any causal measure $\tau$, we also let 
\begin{align}
\mathcal{P}(\tau(\cdot)) = \{ \tau^P(\cdot): P \in \mathcal{P}\}    
\end{align}
be the set of all possible CATE. 
\begin{definition}[Disentanglement of a causal measure $\tau$ on a collection $\mathcal{P}$]
\label{def:distanglement}
Let $\tau$ be a causal measure. Let $\mathcal{P}$ be a collection of distributions $P(X, Y^{(0)}, Y^{(1)})$. We say that $\tau$ has its CATE and ATE disentangled from the baseline on the collection of distribution $\mathcal{P}$ if, for all functions $m \in \mathcal{P}(\tau(\cdot))$, the two following statements hold: 
\begin{align*}
    \left\lbrace P(Y^{(0)}|X): P \in \mathcal{P} \textrm{ s.t. } \tau^P(\cdot) = m(\cdot)\right\rbrace = \mathcal{P}(Y^{(0)}|X)
    \end{align*} 
and, for all $P \in \mathcal{P}$ satisfying $\tau^P(\cdot) = m(\cdot)$, there exists a constant $C_{m, P(X)}$ which depends only on $m$ and $P(X)$, such that $\tau^P = C_{m, P(X)}$.
\end{definition}
While \Cref{def:distanglement} appears technical, its meaning is rather simple: a causal measure has its CATE and ATE disentangled from the baseline on a collection of distributions if $(i)$ specifying a specific form for the treatment effect (via the function $m$) does not restrict the set of possible baseline distributions and $(ii)$ if for any given form of the treatment effect, the ATE depends only on $m$ and the covariate distribution. This corresponds respectively to the first and second statement of \Cref{def:distanglement}. The collection of distributions $\mathcal{P}$ represents all possible distributions of the conditional outcomes and the covariate for a given problem. For generic settings as described in \Cref{lem_generative_models}, \Cref{lemma:working-model-continuous-Y} or \Cref{lemma:working-model-continuous-Y-RR}, since we did not specify any form for $m$ or $b$, the collection $\mathcal{P}$ would naturally be the set $\mathcal{P}_{all}(X,Y^{(0)}, Y^{(1)})$, defined as the set of all joint distributions $P(X,Y^{(0)}, Y^{(1)})$. In order to better understand this notion of disentanglement, let us consider two specific settings. For any $ S \subset \{1,  \hdots, d\}$, we let $X_S$  be the subvector of $X$ composed of components indexed by $S$.}

\modif{ 
\begin{lemma}
\label{lem:disentanglement_homogeneous_treatment}
Consider a collapsible causal measure $\tau$. 
\begin{itemize}
    \item (homogeneous treatment effect) Let $\mathcal{P}_{all}(X, Y^{(0)})$ be the set of all joint distributions $P(X, Y^{(0)})$. Let $m \in \mathds{R}$ and let 
    \begin{align}
        \mathcal{P} & = \Big\lbrace P(X, Y^{(0)}, Y^{(1)}): \nonumber \\
        & P(X, Y^{(0)}) \in \mathcal{P}_{all}(X, Y^{(0)}) \textrm{ and } \tau^P(\cdot) = m \Big\rbrace. \nonumber
    \end{align}
    Then the causal measure $\tau$ has its CATE and ATE disentangled from the baseline on $\mathcal{P}$ and, for all $P \in \mathcal{P}$,  $\tau^P = m$. 
    \item (independence between baseline and treatment effect) Assume that the collapsibility weights of $\tau$ depends only on the baseline distribution $Y^{(0)}|X$. Let $ S \subset \{1,  \hdots, d\}$ and   
    \begin{align}
        \mathcal{P} & = \Big\lbrace P(X, Y^{(0)}, Y^{(1)}) \in \mathcal{P}_{all}(X,Y^{(0)}, Y^{(1)}): \nonumber \\
        & \qquad X_S \indep X_{S^c},   Y^{(0)}|X = Y^{(0)}|X_S, \nonumber \\
        & \qquad \tau^P(X) = \tau^P(X_{S^c}) \Big\rbrace. \nonumber
    \end{align}
    Then the causal measure $\tau$ has its CATE and ATE disentangled from the baseline on $\mathcal{P}$ and, for all $P \in \mathcal{P}$, $\tau^{P} =   \mathds{E}[ \tau^{P}(X)]$.
\end{itemize}
\end{lemma}
According to \Cref{lem:disentanglement_homogeneous_treatment}, a collapsible causal measure disentangles the treatment effect from the baseline on the collection of distributions that correspond to homogeneous treatment effects. To put it differently, if we are in a favorable setting (favorable collection of distributions) in which the causal measure is homogeneous, then the causal measure disentangles the treatment effect from the baseline (on this setting). The same conclusion holds for causal measures whose collapsibility  weights depend only on the conditional distribution $Y^{(0)}|X$ (this is the case for the RR) and for settings in which the baseline distributions $Y^{(0)}|X$ are independent of treatment effect distribution. For example, this is the case for the Risk Ratio, when the baseline and the treatment effect depend respectively on $X_S$ and $X_{S^c}$ with $X_S \indep X_{S^c}$. }

\modif{
Disentanglement is particularly interesting for generalization as generalizing the treatment effect of a causal measure with such a property to a target population would not require  estimating the baseline.
According to \Cref{lem:disentanglement_homogeneous_treatment}, we see that disentanglement on specific settings (collection of distributions) is always possible. However, when generalizing an effect to a target population, we typically do not have any information on the target distribution (especially the distribution of the conditional outcomes). Thus, we want to analyze which causal measure is able to disentangle its CATE and ATE from the baseline on the collection of all possible distributions. 
Unfortunately, among all collapsible measures, only linear causal measures are able to do so, as proved below.}
\modif{
\begin{theorem}
\label{th_onlyRD_separates_baseline_tteffect}
Let $\tau$ be an injective  collapsible causal measure (see \Cref{def:indirect-collapsibility} and \Cref{ass:injection_def_domain}) defined in \eqref{eq_causal_measure}. If the causal measure $\tau$ is able to disentangle its CATE and ATE from the baseline on the collection $\mathcal{P}_{all}(X,Y^{(0)}, Y^{(1)})$, then there exist $a,b,c \in \mathds{R}$ such that, for all distributions $P(X, Y^{(0)}, Y^{(1)}) \in \mathcal{P}_{all}(X,Y^{(0)}, Y^{(1)})$,
\begin{align}
    \tau^P(X) = a \mathds{E}[Y^{(1)}|X] + b \mathds{E}[Y^{(0)}|X] + c.
\end{align}
\end{theorem}
The proof is postponed to Appendix~\ref{app_subsection_proof_thm1_genmodels}. 
Theorem~\ref{th_onlyRD_separates_baseline_tteffect} shows that up to renormalization, the Risk Difference is the only causal measure capable of disentangling the treatment effect from the baseline on the collection of all joint distributions. The strength of this result comes from the fact that the definition of disentanglement is more restrictive when we consider a large collection of distributions. Whereas any causal measure satisfies this definition for restricted collection (homogeneous effect, see \Cref{lem:disentanglement_homogeneous_treatment}), only the Risk Difference disentangles its CATE and ATE from the baseline on the whole collection of joint distributions $P(X, Y^{(0)}, Y^{(1)})$. Although restrictive, such an assumption mimics the practical situation in which one has no information on the shape of the baseline or on the treatment effect. We will show in \Cref{sec:generalization} that, due to its disentangling ability, generalizing the Risk Difference may be possible based on a restricted set of covariates. }

\modif{The notion of disentanglement (or independence) between the baseline function and the treatment effect function (CATE) has also been discussed in \citet{richardson2017modeling}. They state that the two cannot be independent due to constraints on the range of the potential outcomes, which corresponds to the following discussion on bounded outcomes. Their work focuses on the conditional quantities (baseline function and CATE) and not on the ATE which is central in our work, in order to understand how generalization can be obtained without estimating the baseline function \citep[see also][for a discussion about the independence]{wang2022homogeneity}. }

\modif{
\paragraph*{Case of bounded outcomes}
Let us consider a specific setting in which the potential outcomes are bounded, that is, almost surely,  
\begin{align*}
& c_1(X) = \min \left(\mathds{E}\left[Y^{(0)}|X \right], \mathds{E}\left[Y^{(1)}|X  \right]\right) > 0, \\
 & c_2(X) = \max \left(\mathds{E}\left[Y^{(0)}|X\right], \mathds{E}\left[Y^{(1)}|X \right] \right) < \infty.  
\end{align*} 
%Hence, almost surely, $c_1(X) < \mathds{E}\left[Y^{(0)}|X\right], \mathds{E}\left[Y^{(1)}|X\right] < c_2.$ 
Since $\tau_{\text{\tiny RD}}^P(X) = \mathds{E}\left[Y^{(1)}|X\right] - \mathds{E}\left[Y^{(0)}|X\right]$, we must have
\begin{align}
c_1(X) - \mathds{E}\left[Y^{(0)}|X\right]    \leq \tau_{\text{\tiny RD}}^P(X) \leq c_2(X) - \mathds{E}\left[Y^{(0)}|X\right]. \label{eq_baseline_condition}
\end{align}
Consider the collection $\mathcal{P}_{bounded}$ of all possible distributions whose conditional expectations of potential outcomes are bounded. Then the Risk Difference is not able to disentangle the treatment effect from the baseline on $\mathcal{P}_{bounded}$. Indeed, fixing the CATE  $\tau_{\text{\tiny RD}}^P(\cdot)$ put constraints on the baseline distribution, and thus the first statement in \Cref{def:distanglement} does not hold. 
These constraints are more stringent as the CATE takes extreme values (i.e., close to $c_1(x) - c_2(x)$ or $c_2(x) - c_1(x)$), which requires the baseline to be close to $c_1(x)$ or $c_2(x)$. We adapted \Cref{def:distanglement} to the case of bounded outcomes, and extended \Cref{th_onlyRD_separates_baseline_tteffect}
to such a definition, to prove that the Risk Difference is the only causal measure capable of disentangling the treatment effect from the baseline in bounded settings (see \Cref{th_onlyRD_separates_baseline_tteffect_bounded_outcome} in \Cref{sec:app:th2_bounded_outcomes} and 
\ref{app_subsection_proof_thm2_genmodels}).}

\modif{
In a binary setting, potential outcomes naturally belong to $[0,1]$ and the expected potential outcomes turn into 
\begin{align*}
    & \mathds{E}[Y^{(0)} | X] = \mathds{P}[Y^{(0)} = 1 | X], \\
    \textrm{and} \quad &   \mathds{E}[Y^{(1)} | X] = \mathds{P}[Y^{(1)} = 1 | X].
\end{align*}
In this context, Theorem~\ref{th_onlyRD_separates_baseline_tteffect_bounded_outcome} proves that the only causal measure able to disentangle the treatment effect (CATE and ATE)  from the baseline on a generic collection of distributions is the Risk Difference. However, in specific binary settings, other causal measures may allow us to retrieve information on the underlying causal process. This is the subject of the next section.
}

\subsection{A specific binary outcome model: the Russian Roulette}
\label{subsec:russian_roulette}

\subsubsection{Intuition of the entanglement model}
\label{sec:binary-outcomes-intrication}

\begin{figure}[H]
	\caption{\textbf{Illustration of the properties of the Risk Difference} under the russian roulette example. With low baseline risk, the Risk Difference can capture the effect of the roulette 1/6 whereas with high baseline the effects tend to 0.}
	\label{fig:RouletteRusse}
    \includegraphics[width=0.35\textwidth]{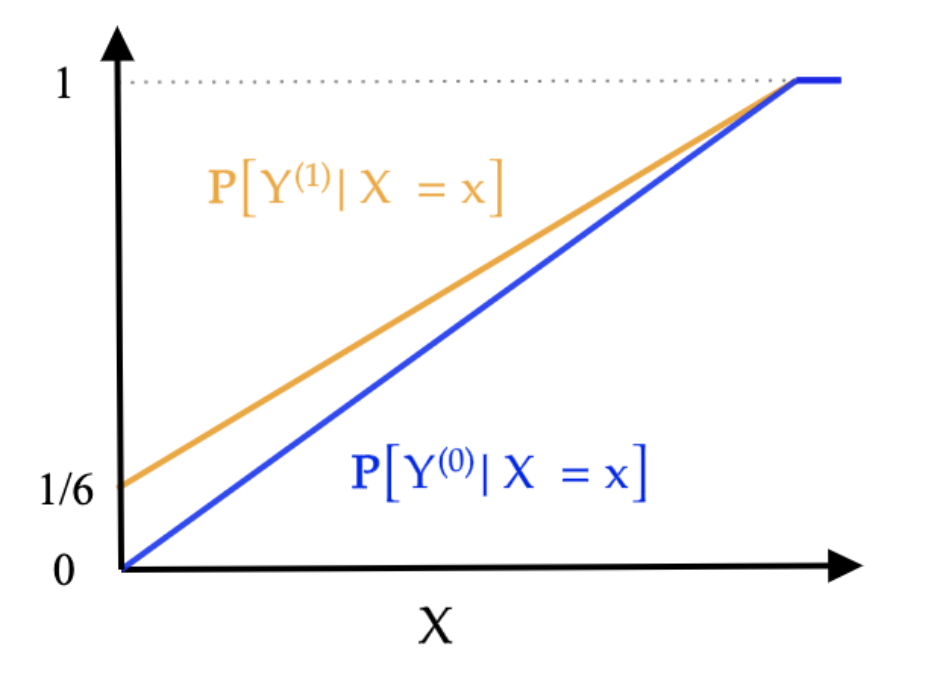}
\end{figure}

We borrow the intuitive example of the Russian Roulette from \cite{Huitfeldt2019LessWrong}, further used by \cite{CinelliGeneralizing2019}.
When playing the Russian Roulette, everyone has the same probability of $1/6$ to die each time they play.
We know this because of the intrinsic mechanism of the Russian Roulette. 
Now, assume that we have not access to this information. 
Consider a hypothetical randomized trial to estimate the effect of the Russian Roulette: a random set of individuals is forced to play Russian Roulette, and the others just wait.
For logistic reasons, the experiment is done on a certain time frame, i.e. we collect the outcome 28 days after the ``treatment'' administration, mimicking a typical clinical outcome defined as mortality after 28 days of hospitalization.
During this time frame, individuals can die from other reasons, such as diseases or poor health conditions.
For an individual with characteristics $x$, denoting $b(x)$ his/her probability to die without the Russian Roulette, and counting a death as $Y=1$ and survival $Y=0$, one has:
\begin{align}\label{eq:intuition-of-intrication}
    \mathbb{P}[ Y^{(a)} = 1 \mid X = x] &= b(x) + a\,\underbrace{\textcolor{RoyalBlue}{\left(1-b\left(x \right)\right)}}_\textrm{Entanglement}\, \frac{1}{6}.
\end{align}
This equation simply states the fact that each individual $X=x$ has a certain probability to die $b(x)$ by default. When getting treatment, an individual can also die from Russian Roulette if affected in the treated group $a=1$, but only if not dead otherwise.
In this equation, one can explicitly observe that the effect 
(measured via the Risk Difference) is naturally \textit{entangled} with the baseline. 
As a consequence, the treatment effect $m$ in the discriminative model associated to the Risk Difference cannot be assumed to be independent of the baseline $b(x)$, as $m(x) = (1-b(x))/6$. In particular, 
\begin{equation*}
    \tau_{\text{\tiny RD}}^P = \frac{1}{6}\left(1 - \mathbb{E}\left[b(x) \right]  \right), \quad \text{and}\quad  \lim\limits_{\mathbb{E}[b(x)] \rightarrow 1} \tau_{\text{\tiny RD}} = 0.
\end{equation*}
\modif{
In this situation, the first statement of \Cref{def:distanglement} does not hold for the collection 
\begin{align*}
\mathcal{P} = \big\lbrace & P(X, Y^{(0)}, Y^{(1)}): \\ & \mathds{E}[Y^{(1)}|X] - \mathds{E}[Y^{(0)}|X] 
= (1 - \mathds{E}[Y^{(0)}|X])/6 \big\rbrace    
\end{align*}
of distributions corresponding to the Russian Roulette setting: fixing the CATE of the Risk Difference restricts the choice of the baseline (to a unique element). Thus Theorem~\ref{th_onlyRD_separates_baseline_tteffect_bounded_outcome} does not apply. This is  illustrated in Figure \ref{fig:RouletteRusse}. In a population with a high baseline, the measured effect vanishes along the risk difference scale.
}

In other words, when considering the RD, the effect of the treatment can only be observed on people that would not have died otherwise. 
This could seem a bit odd, as the Russian Roulette example contains the idea of an \textit{homogeneous} treatment effect, that should not vary over different populations. Still, one measure, the survival ratio, shows an interesting property,
\begin{equation*}
     \tau_{\text{\tiny SR}}^P = 1 - \frac{\mathbb{E}\left[\left( 1- b\left( X\right) \right) \frac{1}{6} \right]}{\mathbb{E}\left[\left( 1- b\left( X\right) \right) \right]} = \frac{5}{6}.
\end{equation*}
The Survival Ratio thus captures the idea of homogeneity: no matter the baseline risk, the Russian Roulette acts in the same way for everyone, as noticed by \cite{Huitfeldt2019LessWrong}. Appendix~\ref{appendix:more-details-on-the-intrication-model} gives more details about the origin of this example.

\subsubsection{Formal analysis}
Equation~\ref{eq:intuition-of-intrication} only describes harmful situations while we may be interested in modelling positive or deleterious effects of the treatment. 
In addition, we want a model able to encode situations where there is heterogeneity of the treatment effect (e.g. Russian Roulette can have a higher impact on stressed out people  because the prospect of playing would create cardiac arrests). Or on a more concrete example: the seat belts could be protective for taller individuals but less protective (or even deleterious) for smaller individuals because of the design. 

\begin{lemma}[Entanglement Model]\label{lemma:intrication_model}%\gv{AMHA on retire "intrication" :)}
Considering a binary outcome $Y$, assume that 
\begin{equation*}
     \forall x \in \mathds{X},\, \forall a \in \{0,1\},\quad 0 < p_a(x) < 1,
 \end{equation*}
where $p_a(x)  :=  \mathbb{P}\left[Y^{(a)} = 1 \mid X=x\right]$. Introducing
\begin{equation*}
    m_g(x):= \mathbb{P}[ Y^{(1)} = 0 \mid Y^{(0)} = 1, X = x]
\end{equation*}
and 
\begin{equation*} 
    m_b(x):= \mathbb{P}[ Y^{(1)} = 1 \mid Y^{(0)} = 0, X = x],
\end{equation*}
allows us to write
\begin{align*}
     & \mathbb{P}[ Y^{(a)} = 1 \mid X = x]  \\
     = & ~b(x)+  a\, \big( \left( 1-b\left(x\right) \right) m_b\left(x\right) -  b\left(x\right)m_g\left(x\right) \big),
\end{align*}
with $b(x):=p_0(x)$.
\end{lemma}
Proof is available in Appendix~\ref{proof:intrication_model}.
Usually $Y=1$ denotes death or deleterious events, therefore the subscripts $b$ (resp. $g$) stands for \textit{bad} (resp. \textit{good}) events. $m_b$ (resp. $m_g$) corresponds to  the probability that a person who was previously not destined (resp. destined) to experience the outcome, does (resp. does not) experience the outcome in response to treatment. They represent the outcome switch depending on the position at baseline\footnote{Such parameters can be found to be close to the “\textit{counterfactual outcome state transition}” (COST) in \cite{huitfeldt2018choice}. For example $m_b$ would correspond to the quantity denoted by $1-H$. Also note that the intrication model also allows to apprehend what has been done by \cite{CinelliGeneralizing2019}, where the quantity they introduce being $PS_{01} := \mathbb{P}\left[ Y^{(1)} = 1 \mid Y^{(0)} = 0\right]$ corresponds to $m_b$. While their work mostly rely on the formalism of selection diagram, they define $PS_{01}$ (and therefore $m_b$) as the probability of fatal treatment among those who would survive had they not been assigned to for treatment. And conversely, $PS_{10} := \mathbb{P}\left[ Y^{(1)} = 0 \mid Y^{(0)} = 1\right]$ (corresponding to $m_g$) stands for the probability that the treatment is sufficient to save a person who would die if defined. As far as we understand, in both of these works these probabilities are not taken conditionally to $X$.}.
The expressions of classical causal measures are established in Lemma~\ref{lemma:expression-of-causal-quantities-under-generative-model-binary-outcome-intrication-model} (see Appendix~\ref{proof:expression-of-causal-quantities-under-generative-model-binary-outcome-intrication-model}). Such expressions are difficult to interpret in the general case where both $m_b$ and $m_g$ are non-zero. In fact, in such a situation $m_b(X)$ and $m_g(X)$ are not identifiable 
\citep{Pearl2000Book, huitfeldt2018choice}. However, since we are mainly interested in the total effect, one could consider the discriminative model of the risk difference, 
\begin{equation*}
    \mathbb{P}\left[ Y^{(a)} = 1 \mid X = x \right] = b(x)\,+  a\, \tau(x),
\end{equation*}
where $\tau(x):= \left( 1-b(x) \right) m_b(x)- b(x)m_g(x)$, 
with the limitations described in Section~\ref{sec:binary-outcomes-intrication}. 
Thus, we consider the case of monotonous effects.

\subsubsection{Notion of monotonous effect}
We introduce the assumption of monotonous effects, where either $\forall x,$ $ m_b(x) = 0$ or $\forall x, m_g(x) = 0$ \citep{huitfeldt2018choice, CinelliGeneralizing2019}, corresponding to scenarios where the treatment is only beneficial or deleterious\footnote{In particular, the Russian Roulette corresponds to a situation where $\forall x, m_g(x) = 0$ (Russian Roulette makes no good).}, but cannot be both. If the treatment is always beneficial (i.e. $\forall x, m_b(x) = 0$) then the probability $p_1(x)$ (see Lemma~\ref{lemma:intrication_model}) is lower than the baseline. Respectively, if the treatment is always deleterious  (i.e. $\forall x, m_g(x) = 0$) then the probability $p_1(x)$ is higher than the baseline.
This can be summarized as follows, 

\begin{align}
    & \mathbb{P}\left[ Y^{(a)} = 1 \mid X = x\right] \\
    = & ~b(x)\, \underbrace{+ \, a\,  \left( 1-b\left(x\right) \right) m_b\left(x\right)}_{\nearrow} \, \underbrace{- \,a\, b\left(x\right)m_g\left(x\right) }_{\searrow} , \label{eq_monotonous_effect_deletere_beneficial}
\end{align}
where arrows indicate whether each term of the equation is increasing or decreasing the probability of occurrences.
Equation~\eqref{eq_monotonous_effect_deletere_beneficial} highlights that the entanglement is not the same depending on the nature of the treatment (deleterious or not). A beneficial effect ($m_b(x)=0$) is more visible on a high baseline population ($b(x)$ close to 1). On the opposite, a deleterious effect ($m_g(x)=0$) is visible only on the population with low baseline  ($1-b(x)$ close to 1). 
In other words, an effect increasing the probability of occurences acts only on individuals on which occurences has not already happened yet. 

\begin{lemma}[Risk Ratio and Survival Ratio under a monotonous effect]\label{lemma:monotonous-effect}
Ensuring conditions of Lemma~\ref{lemma:intrication_model},

\begin{itemize}
    \item Assuming that the treatment is beneficial (i.e. $\forall x,$ $ m_b(x) = 0$), then

    \begin{align*}
    & \tau_{\text{\tiny RR}}^P(X)  = 1  -  m_g(X) \\
     \textrm{and} \quad 
     & \tau_{\text{\tiny RR}}^P = 1  - \frac{\mathbb{E}\left[  b(X) m_g(X)\right]}{\mathbb{E}\left[ b(X)\right]}. 
 \end{align*}

 \item  Assuming that the treatment is harmful (i.e. $\forall x,$ $ m_g(x) = 0$), then

 \begin{align*}
 & \tau_{\text{\tiny SR}}^P(X) = 1  -   m_b(X),\\
  \textrm{and} \quad 
     & \tau_{\text{\tiny SR}}^P = 1  - \frac{\mathbb{E}\left[ \left( 1-b(X) \right) m_b(X)\right]}{\mathbb{E}\left[ 1-b(X)\right]}.
 \end{align*}
\end{itemize}

\end{lemma}

 These results formalize what has been proposed several times in the literature, for example by \cite{Sheps1958ShallWe}, and later by \cite{huitfeldt2018choice, Huitfeldt2021ShallWe}, or with what has been called the \textit{Generalised Relative Risk Reduction} \citep{baker2018new}. In particular, Sheps finishes her paper with the following quote
 \begin{quote}
   ``A beneficial or harmful effect may be estimated from the proportions of persons affected. The absolute measure does not provide a measure of this sort. The choice of an appropriate measure resolves itself largely into the choice of an appropriate base or denominator for a relative comparisons. [$\dots$] the appropriate denominator consists of the number of persons who could have been affected by the factor in question".
 \end{quote}
This recommendation is consistent with Lemma~\ref{lemma:monotonous-effect}. In other words, the sign of the effect dictates on which labels the relative comparison should be made i.e., dividing by $\mathbb{P}\left[Y^{(0)}=1\right]$ or $\mathbb{P}\left[Y^{(0)}=0\right]$, to obtain a CATE disentangled from the baseline. While the CATE of SR (resp. RR) is interpretable, as it allows us to retrieve a deleterious (resp. beneficial) local effect, the corresponding ATE is not able to disentangle the baseline from the causal effect. This is consistent with the result of Theorem~\ref{th_onlyRD_separates_baseline_tteffect_bounded_outcome} that states that only the Risk Difference can disantangle baseline and treatment effetcs at both CATE and ATE levels.

\section{Generalization}
\label{sec:generalize}

\begin{center}
\begin{figure}[!h]
\begin{center}
    \includegraphics[width= 0.5\textwidth]{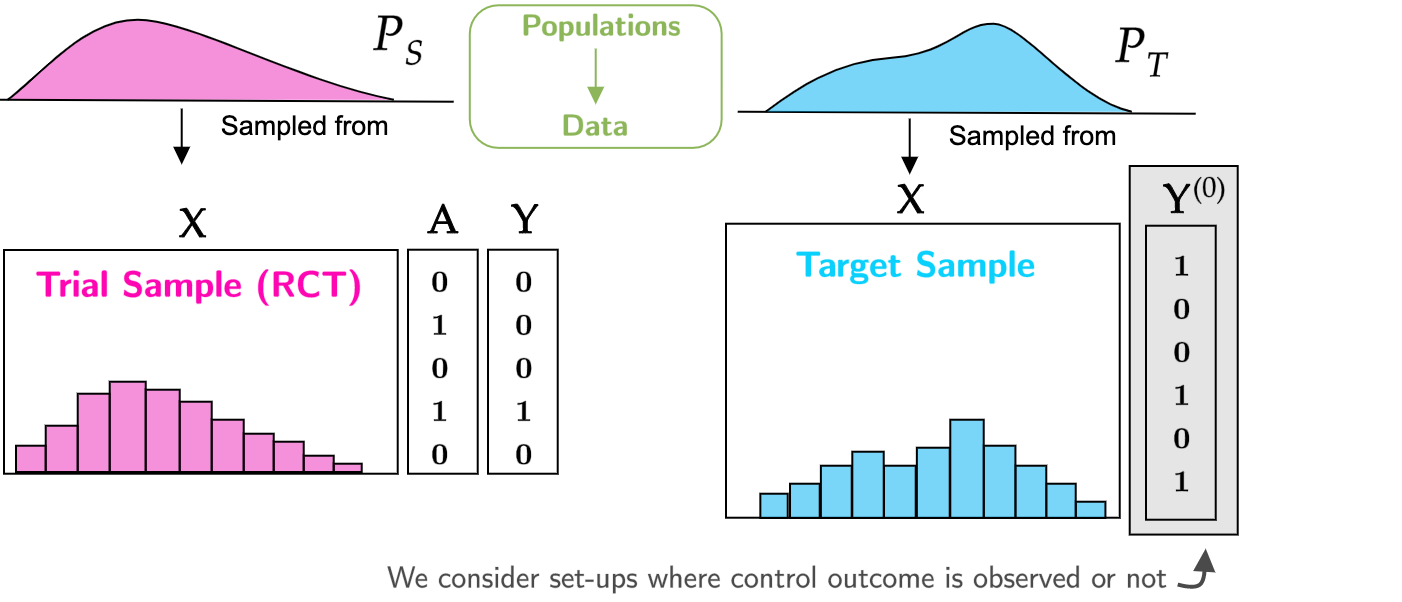}
\end{center}
    \caption{\textbf{Generalization in practice}: We typically consider a situation where the treatment effect is estimated from a Randomized Controlled Trial (RCT) where individuals are sampled from a population $P_{\text{\tiny S}}$. When willing to extend these findings to $P_{\text{\tiny T}}$, we assume to have access to a representative sample of the patients of interest, with information on their covariates $P_{\text{\tiny T}}(X)$, and also \underline{maybe} the outcome under no treatment $P_{\text{\tiny T}}(X, Y^{(0)})$.}
    \label{fig:observed-data}
\end{figure}    
\end{center}

As highlighted in Section~\ref{sec:formalization-and-key-contributions}, an RCT conducted in a population $P_{\text{\tiny S}}$  allows for the estimation of a treatment effect $\tau^{P_{\text{\tiny S}}}$ on this population. What would the result be if the individuals in the trial were rather sampled from a population $P_{\text{\tiny T}}$ with different covariates distribution? 
This question is linked to external validity, and more precisely to a sub-problem of external validity being \textit{generalizability} or \textit{transportability}. We say that findings from a trial sampled from $P_{\text{\tiny S}}$ can be generalized to $P_{\text{\tiny T}}$ when $\tau^{P_{\text{\tiny T}}}$ can be estimated without running a trial on $P_{\text{\tiny T}}$, but only using data from the RCT and baseline information on the target population $P_{\text{\tiny T}}$ (the covariates $X$, and sometimes also the control outcome $Y^{(0)}$), as summarized on Figure~\ref{fig:observed-data}. 

\subsection{Two different strategies for generalizability}
\label{subsection:two_generalization_strategies}

There exist two identification strategies, generalizing \textit{(i)} the conditional outcomes or \textit{(ii)} the local effect measure itself, leading to different assumptions required for generalizing.
For both strategies, we consider the settings where information gathered on the source population covers at least the support of the target population. 

\begin{assumption}[Overlap or positivity]\label{a:overlap}
The support of the target population is included in the source population:
$\operatorname{supp}(P_{\text{\tiny T}}) \subset \operatorname{supp}(P_{\text{\tiny S}})$.
\end{assumption}

This assumption\footnote{Note that Assumption~\ref{a:overlap} can be phrased as ``the measure $P_{\text{\tiny T}}$ is absolutely continuous with respect to $P_{\text{\tiny S}}$''.} is the counterpart of the the so-called positivity or overlap assumption in observational studies. It means that all members of the target population have positive probability of being selected into the source population. In the specific case of generalization, such a common assumption is violated when the source population is a randomized controlled trial conducted on a restricted population (for e.g. because of strict eligibility criteria for safety reasons) compared to the target/whole population. Still, in practice, it is possible to restrict the support of the target population to the source population. This would allow generalizing an effect from the source population to the target population, answering the following question “what would the effect be on the target population if the same eligibility criteria were used?”. \\

The first approach aims at generalizing conditional expectations $\mathbb{E}_{\text{\tiny S}}[ Y^{(a)} \mid X]$ of the potential outcomes to the target population. Such a strategy is valid only under the following assumption.

\begin{assumption}[Transportability or S-ignorability or Exchangeability between populations]\label{a:transportability-wide}
For all $x \in \operatorname{supp}(P_{\text{\tiny T}}) \, \cap \, \operatorname{supp}(P_{\text{\tiny S}})$, for all $a \in \{0,1 \}$, $$\mathbb{E}_{\text{\tiny S}}[ Y^{(a)} \mid X = x ] = \mathbb{E}_{\text{\tiny T}}[ Y^{(a)} \mid X = x].$$
\end{assumption}
This assumption\footnote{This assumption is also commonly found expressed as $Y^{(0)}, Y^{(1)} \indep  I \mid X$, where $I$ is an indicator of the population membership \citep{stuart2011use, pearl2015findings, lesko2017generalizing}. Such assumptions can also be expressed using selection diagram \citep{pearl2011transportability}. 
} boils down to: $X$ contains all covariates that are \textit{both} shifted between the two populations $P_{\text{\tiny S}}$ and $P_{\text{\tiny T}}$ \textit{and}  prognostic of the outcome.  
Assumption~\ref{a:transportability-wide} enables the identification of $\tau^{P_{\text{\tiny T}}}$ using information from $P_{\text{\tiny S}}(X,Y^{(0)}, Y^{(1)})$ and only the covariate distribution $P_{\text{\tiny T}}(X)$ in the target population, as shown in the following Proposition (see Appendix~\ref{proof:generalizability-section-3} for the proof). \modif{In the case where the outcome is available in the target population, \cite{robertson2021center} advocate for tests to determine the plausibility of Assumption~\ref{a:transportability-wide}. 
}

\begin{proposition}[Generalizing conditional outcomes]\label{proposition:generalization-density}
Consider two distributions $P_{\text{\tiny S}}$ and $P_{\text{\tiny T}}$ satisfying Assumptions~\ref{a:overlap} and \ref{a:transportability-wide}. Then, the conditional outcomes are generalizable, that is for all $a \in  \{0,1\}$,
\begin{align*}
     \mathbb{E}_{\text{\tiny T}}[ Y^{(a)}] &=  \mathbb{E}_{\text{\tiny T}}\left[ \mathbb{E}_{\text{\tiny S}}[ Y^{(a)} \mid X]  \right] && \text{G-formula} \\
    &= \mathbb{E}_{\text{\tiny S}}\left[ \frac{p_{\text{\tiny T}}(X)}{p_{\text{\tiny S}}(X)}\mathbb{E}_{\text{\tiny S}}[ Y^{(a)} \mid X]  \right] %&& \text{Re-weighting.}
\end{align*}
where $p_{\text{\tiny T}}(X)/p_{\text{\tiny S}}(X)$ corresponds to the ratio of covariate densities in the
source and target populations.
\end{proposition}

The first formula in Proposition~\ref{proposition:generalization-density} suggests a strategy to generalize the potential outcomes: first, one can compute of $\mu_{a, \text{\tiny S}}(x) = \mathbb{E}_{\text{\tiny S}}\left[ Y^{(a)} \mid X =x \right]$ using the source distribution $P_{\text{\tiny S}}(X,Y^{(0)}, Y^{(1)})$, then one can compute $\mathds{E}_{\text{\tiny T}} [\mu_{a, \text{\tiny S}}(X)]$ using the covariate target distribution $P_{\text{\tiny T}}(X)$. Any causal measure $\tau$ satisfying Equation~\eqref{eq_causal_measure} can be generalized on the target distribution using this strategy.

When the causal measure is collapsible, rather than using a conditional outcome model, the second approach relies on the local effects $\tau^{P_{\text{\tiny S}}}(x)$ to get the target population's effect $\tau^{P_{\text{\tiny T}}}$, such as in Equation~\ref{eq:toy-example-standardization}. Importantly,  Assumption~\ref{a:transportability-wide} can then be relaxed into a new, less restrictive, Assumption~\ref{a:transportability}.

\begin{assumption}[Transportability of the treatment effect]\label{a:transportability}
For all $x \in \operatorname{supp}(P_{\text{\tiny T}}) \cap \operatorname{supp}(P_{\text{\tiny S}})$, $$\quad \tau^{P_{\text{\tiny S}}}(x) = \tau^{P_{\text{\tiny T}}}(x).$$
\end{assumption}
Here, \Cref{a:transportability}\footnote{This assumption is also commonly found expressed as $Y^{(0)}- Y^{(1)} \indep I \mid X$ when it comes to the generalization of the risk difference ($I$ being an indicator of the population membership). 
%\es{I like the way you express things mathematically and then interpret them qualitatively. Both are very valuable to readers!}
Note that the transportability assumptions conveys the idea of some homogeneity assumption (close to the spirit of Definition~\ref{def:homogeneity}). This is highlighted by \cite{huitfeldt2018choice} who refer to Assumptions~\ref{a:transportability-wide} and \ref{a:transportability} as ``\textit{different homogeneity conditions to operationalize standardization}".} can be phrased as: $X$ contains all covariates that are \textit{both} shifted between the two populations $P_{\text{\tiny R}}$ and $P_{\text{\tiny T}}$ \textit{and}  treatment effect modulators.

\begin{proposition}[Generalizing local effects]\label{prop:generalization-of-local-effects}
Consider two distributions $P_{\text{\tiny S}}$ and $P_{\text{\tiny T}}$ and a collapsible causal measure $\tau$ satisfying Assumptions~\ref{a:overlap} and \ref{a:transportability}. Then, $\tau$ is generalizable to the target population via the formula 
\begin{align*}
    \tau^{P_{\text{\tiny T}}}  &= \mathbb{E}_{\text{\tiny T}}\left[ w(X, P_{\text{\tiny T}}(X, Y^{(0)})) \tau^{P_{\text{\tiny S}}}(X)  \right] \\%&& \text{G-formula}\\
    &= \mathbb{E}_{\text{\tiny S}}\left[ \frac{p_{\text{\tiny T}}(X)}{p_{\text{\tiny S}}(X)} \, w(X, P_{\text{\tiny T}}(X, Y^{(0)})) \, \tau^{P_{\text{\tiny S}}}(X)  \right] && \text{Re-weighting.}
\end{align*}
where $p_{\text{\tiny T}}(X)/p_{\text{\tiny S}}(X)$ corresponds to the ratio of covariate densities in the source and target populations, and $w(X,$ $ P_{\text{\tiny T}}(X, Y^{(0)}))$ corresponds to the collapsibility weights (see Definition~\ref{def:indirect-collapsibility}). 
\end{proposition}
The proof is postponed to Appendix~\ref{proof:generalizability-section-3}. The first formula in Proposition~\ref{prop:generalization-of-local-effects} leads to the following generalization strategy: the quantity $\tau^{P_{\text{\tiny S}}}(X) = \mathds{E}_{\text{\tiny S}}[Y^{(1)} - Y^{(0)} | X]$ can be computed using the source distribution $P_{\text{\tiny S}}(X,Y^{(0)}, Y^{(1)})$ and both the collapsibility weights and the expectation in the first formula of Proposition~\ref{prop:generalization-of-local-effects} can be computed using the target distribution $P_{\text{\tiny T}}(X,Y^{(0)})$. 
Note that the second formula suggests the classical re-weighting estimation strategy also called IPSW (Inverse Propensity of Sampling Weighting, see  \cite{colnet2021causal} for a review on the Risk Difference).

The two above strategies rely on two different assumptions. However,
it is very important to note that Assumption~\ref{a:transportability} is lighter than Assumption~\ref{a:transportability-wide} as highlighted in \cite{nguyen2018sensitivitybis, huitfeldt2018re, colnet2022sensitivity}.
\modif{To see this, consider the following  example, in which the source and the target distributions correspond to populations of two different hospitals. Assume that, in the source population 
\begin{align}
    \mathds{E}_S[Y^{(a)}|X] = b(X) + a m(X),
\end{align}
whereas in the target population
\begin{align}
        \mathds{E}_T[Y^{(a)}|X] = \gamma + b(X) + a m(X).
\end{align}
The constant $\gamma$ modifies the baseline of the target population and may result from a target population more likely than the source population to undergo undesirable events, due to exogeneous variables not included in the covariates $X$. In this case,  \Cref{a:transportability} holds but \Cref{a:transportability-wide} does not.}
As a consequence,
using local effects -- Proposition~\ref{prop:generalization-of-local-effects} -- as opposed to conditional outcomes --Proposition~\ref{proposition:generalization-density} -- may allow generalizing collapsible causal measure \emph{with fewer covariates}, as detailed in the next section.

\subsection{Are some measures easier to generalize than others?}
\label{sec:generalization}

Section~\ref{subsection:two_generalization_strategies} exposes two transportability assumptions depending on which conditional quantity from the source population is generalized: the conditional outcome (Assumption~\ref{a:transportability-wide}) or the local effect (Assumption~\ref{a:transportability}).
While Assumption~\ref{a:transportability-wide} requires that all covariates being prognostic and shifted in the two populations have been observed, Assumption~\ref{a:transportability} involves all covariates modulating treatment effect and shifted. In this section, we analyze precisely which strategy can be used for a given causal measure, and what is the required set of covariates for such a strategy. 
We start by specifying which variables are shifted between the two populations.  

\begin{assumption}[Shifted covariates set]\label{def:shidted-covariates}
We assume that only some components of $X$ are shifted between the source and the target population. More precisely, we denote by $\textrm{Sh} \subset \{1, \hdots, d \}$ the set of indices corresponding to the components of $X$ that are shifted between the source and the target population, that is, for all integrable functions $f: \mathds{X} \to \mathbb{R}$, for all $x \in Supp(P_{\text{\tiny T}})$, 
\begin{align*}
    \mathbb{E}_{\text{\tiny S}}[ f(X) | X_{Sh} = x_{Sh}] =  \mathbb{E}_{\text{\tiny T}}[ f(X) | X_{Sh} = x_{Sh}],
\end{align*}
and the complementary set of covariates $X_{Sh^c}$ is independent of $X_{Sh}$.
\end{assumption}
\modif{\Cref{def:shidted-covariates} states that the information of the shifted covariates $X_{Sh}$ is enough to transport any functional of the covariates. The last condition, the independence between $X_{Sh^c}$ and $ X_{Sh}$, may seem overly restrictive. However, without this assumption, the set of shifted covariate may not be unique. Indeed, if there was some dependence between $X_{Sh^c}$ and $ X_{Sh}$, some changes in $ X_{Sh}$ between $P_{\text{\tiny S}}$ and $P_{\text{\tiny T}}$ would result in changes in $X_{Sh^c}$, thus some components of  $X_{Sh^c}$ would also be shifted.  Our analysis is based on the fact that there exist variables that are not shifted (either conditionally or unconditionally), therefore requiring the last statement of \Cref{def:shidted-covariates}.}

To formalize which covariates are implied in local treatment effect, we introduce notations to distinguish covariates status, either intervening on the baseline level or modulating the effect. 

\begin{assumption}[Two types of covariates]\label{def:two-kind-covariates}
Let $\tau$ be a causal measure and let $b: \mathds{X} \to \mathbb{R}$ and $m: \mathds{X} \to \mathbb{R}$ be the function describing the associated model (see Lemma~\ref{lem_generative_models}). 
We assume that the function $b$ depends only on $X_B$, a subset of covariates indexed by $B \subset \{1, \hdots, d \}$. Similarly, we assume that the function $m$ depends only on $X_M$, a subset of covariates indexed by $M \subset \{1, \hdots, d \}$.
\end{assumption}

The baseline $b$ and the treatment effect $m$ are assumed to depend on certain sets of variables (Assumption~\ref{def:two-kind-covariates}). Determining such sets is an active area of research \citep{hines2022variable, benard2023variable} and falls beyond the scope of this paper. Instead, we fix the sets $X_B$ and $X_M$, and the set of shifted covariates between the source and target population, and analyze which covariates are required for generalization, depending on the considered strategy (generalizing potential outcomes or local effects)\footnote{Note that the size of $X_B$, $X_M$ and $X_{Sh}$ completely depends on the data distribution and the causal measure: if the causal measure does not allow disentangling the treatment effect from the baseline at a strata level then $X_M = X_B$, whereas if all variables are shifted then $X_{Sh} = X$.}.

Generalizing conditional outcomes requires to have access to all shifted prognostic covariates.

\begin{theorem}\label{theorem:all-covariates}
    Consider an injective causal measure (Assumption~\ref{ass:injection_def_domain}). For all distributions $P_{\text{\tiny S}}(X, Y^{(0)}, Y^{(1)})$ and $P_{\text{\tiny T}}(X, Y^{(0)}, Y^{(1)})$ satisfying Assumption~\ref{a:overlap} (overlap assumption) and Assumption~\ref{a:transportability-wide}, 
    generalization of the conditional outcomes is possible if one has access to all shifted covariates involved in the baseline and the treatment effect, that is $X_{(B\cup M)\cap Sh}$.
\end{theorem}

\begin{figure}
    \centering
    \includegraphics[width=0.4\textwidth]{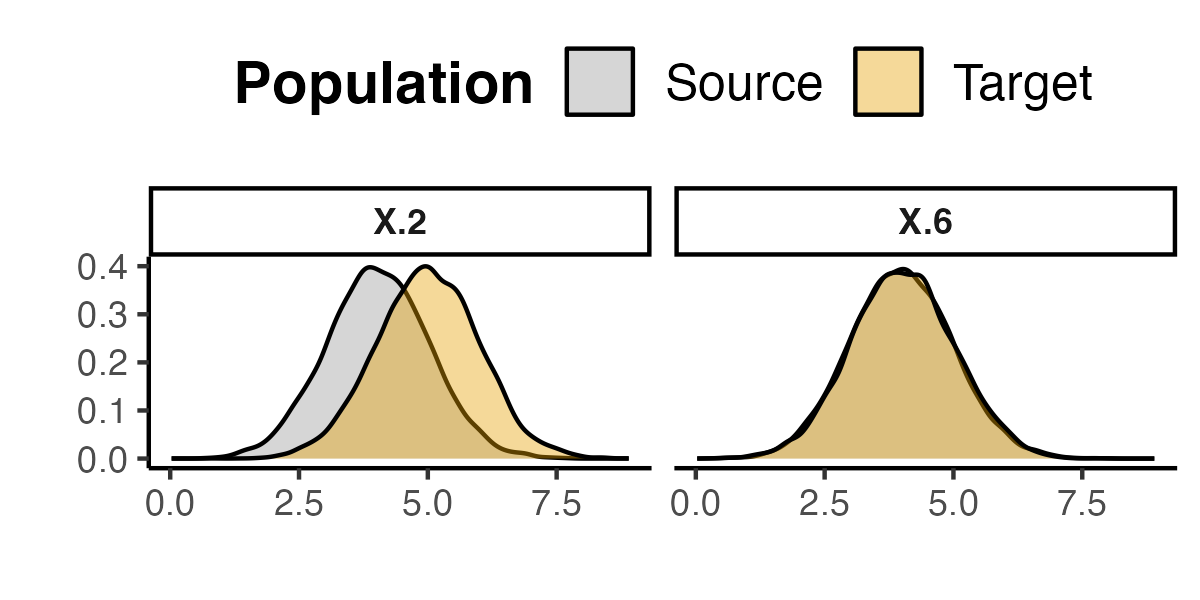}
    \caption{$2 \in \textrm{Shift}$, and $6 \not\in \textrm{Shift}$.}
    \label{fig:illustration-shifted}
\end{figure}

The proof can be found in Appendix~\ref{subsubsec:proof_generalization_conditional_outcomes}. To illustrate what are the different covariate sets, we introduce the data generative model of the simulations (see Section~\ref{sec:simulations}), where we assume that six covariates are prognostic and that data are generated as
\begin{equation}\label{eq:simulation-continuous-generative-model}
    Y=b\left(X_1, X_2, X_3, X_4, X_5, X_6\right) + A\, m\left(X_1, X_2, X_5\right) + \varepsilon.
\end{equation}
 Doing so, $B=(1,2,3,4,5,6)$, and $M = (1,2,5)$. In addition, the two populations are constructed such that $X_1, \dots X_4$ are shifted covariates, but not $X_5, X_6$. Figure~\ref{fig:illustration-shifted} illustrates what shifted and non-shifted means.  Theorem~\ref{theorem:all-covariates} states that generalization of the conditional outcomes is possible when observing  $X_1, \dots X_4$.
 
Having access to all shifted prognostic covariates in the \underline{two} data samples seems challenging (and maybe too optimistic). This situation could explain all the numerous recent research works about sensitivity analysis when necessary covariates are not observed or partially observed when generalizing \citep{nguyen2018sensitivitybis, nie2021covariate, colnet2022sensitivity}. In such a context, generalizing local effects (instead of conditional outcomes) is a promising strategy, as it may require less covariates, as shown in Theorem~\ref{theorem:restricted-set-for-Y-continuous-RD} below. 

\begin{theorem}
\label{theorem:restricted-set-for-Y-continuous-RD}
 Consider the Risk Difference $\tau_{\text{\tiny RD}}$. For all distributions $P_{\text{\tiny S}}(X, Y^{(0)}, Y^{(1)})$ and $P_{\text{\tiny T}}(X, Y^{(0)}, Y^{(1)})$ satisfying Assumption~\ref{a:overlap} (overlap assumption) and Assumption~\ref{a:transportability}, observing  all shifted treatment effect modifiers $X_{M \cap Sh}$ is sufficient for generalizing $\tau_{\text{\tiny RD}}$.
\end{theorem}
The proof can be found in Appendix~\ref{sec:proof_theorem_generalizing_local_effects}. 
Theorem~\ref{theorem:restricted-set-for-Y-continuous-RD} shows that the Risk Difference can be generalized via the local effect strategy with fewer covariates and under a weaker assumption compared to Theorem~\ref{theorem:all-covariates}. 
Back to \eqref{eq:simulation-continuous-generative-model}, one would require only $X_1$ and $X_2$ to generalize the Risk Difference with the local effects strategy, as $X_5$ is not shifted.

\modif{Now consider a nonlinear injective collapsible causal measure $\tau$, which verifies by definition of collapsibility and \Cref{a:transportability}:
\begin{align}
\tau^{P_{\text{\tiny T}}}  &= \mathbb{E}_{\text{\tiny T}}\left[ w(X, P_{\text{\tiny T}}(X, Y^{(0)})) \tau^{P_{\text{\tiny S}}}(X)  \right],
\end{align}
for all functions $\tau^{P_{\text{\tiny S}}}(X)$. Assume furthermore that \Cref{theorem:restricted-set-for-Y-continuous-RD} holds for this measure, and consider settings for which $X_B \cap X_M = \emptyset$. Since $X_{M \cap Sh}$ is sufficient for generalizing $\tau$, $\tau$ is independent of $X_B$. Thus, the collapsibility weights are independent of $X_B$. Based on the proof of \Cref{th_onlyRD_separates_baseline_tteffect} (from \eqref{eq_proof_th1_for_th4} to the end of the proof), one can show that $\tau$ is indeed linear, which contradicts our first assumption. Thus, \Cref{theorem:restricted-set-for-Y-continuous-RD} highlights the particular status of the RD compared to other nonlinear  collapsible causal measures.}

There are two specific situations in which all collapsible causal measures can be generalized: in presence of a homogeneous effect or when the baseline and the treatment effect are independent. \Cref{thm_homogeneous_independence_generalization} is equivalent to \Cref{lem:disentanglement_homogeneous_treatment} in the generalization framework.
\begin{lemma}
\label{thm_homogeneous_independence_generalization}
Let $\tau$ be a collapsible causal measure. 
\begin{itemize}
    \item (homogeneous treatment effect) For all distributions $P_{\text{\tiny S}}(X, Y^{(0)}, Y^{(1)})$ and $P_{\text{\tiny T}}(X, Y^{(0)}, Y^{(1)})$ satisfying Assumption~\ref{a:overlap} (overlap assumption), Assumption~\ref{a:transportability} and such that there exists $C \in \mathds{R}$ satisfying, for all $x \in  Supp(P_{\text{\tiny T}})$, $\tau^{P_{\text{\tiny S}}}(x) = C$, we have
    \begin{align}
    \tau^{P_{\text{\tiny T}}} = \tau^{P_{\text{\tiny S}}} = C. 
    \end{align}

    \item (independence between treatment effect and collapsibility weights) For all distributions $P_{\text{\tiny S}}(X,$ $ Y^{(0)}, Y^{(1)})$ and $P_{\text{\tiny T}}(X, Y^{(0)}, Y^{(1)})$ satisfying Assumption~\ref{a:overlap} (overlap assumption), Assumption~\ref{a:transportability} and such that $\tau(X)$ is independent of the collapsibility weights $w(X, P(X, Y^{(0)}))$ (Definition~\ref{def:indirect-collapsibility}), we have, 
    \begin{align}
        \tau^{P_{\text{\tiny T}}} =  \tau^{P_{\text{\tiny S}}} = \mathds{E}\left[ \tau^{P_{\text{\tiny S}}}(X) \right].
    \end{align}
\end{itemize}
\end{lemma}

\section{Illustration through simulations}\label{sec:simulations}

We use synthetic simulations to illustrate Theorems~\ref{theorem:all-covariates} and \ref{theorem:restricted-set-for-Y-continuous-RD}, that is different covariates sets are required to identify the target population effect depending on \textit{(i)} the causal measure of interest   and \textit{(ii)} the generalization method.
All implementations details, as well as the estimation strategies are provided in Appendix~\ref{appendix:additional_simulations}. Other experiments studying the impact of missing covariates or misspecified models are also presented in Appendix~\ref{appendix:additional_simulations} (see Figure~\ref{fig:simulations-continuous-Y-missing-X1} and Figure~\ref{fig:simulations-continuous-Y-mispe}). 
The code to reproduce the simulations is available on \href{https://github.com/BenedicteColnet/ratio-versus-difference}{github} (see repository \texttt{BenedicteColnet/ratio-versus-difference}).
 \begin{figure}[h!]
    \centering
    \includegraphics[width=0.5\textwidth]{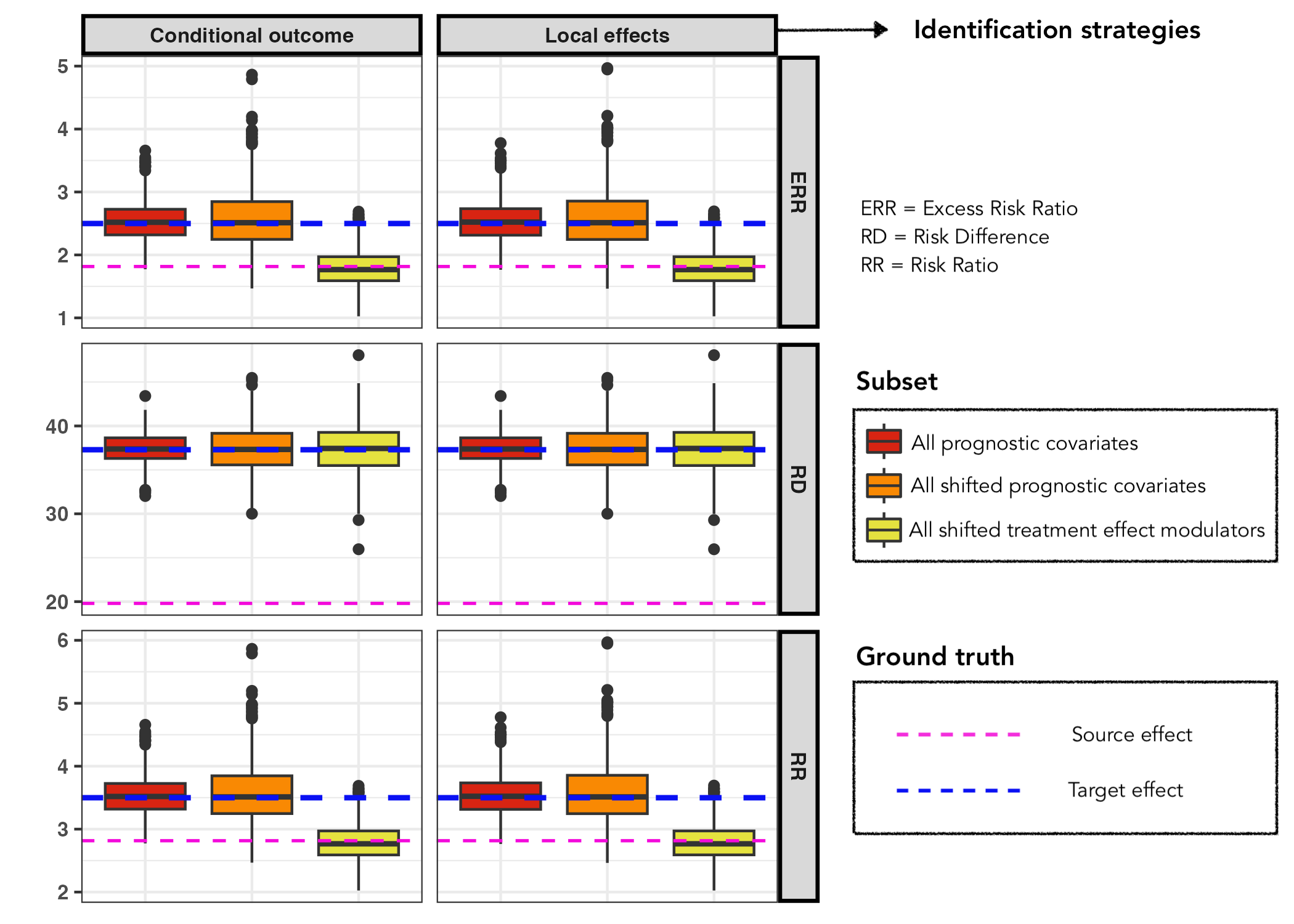}
    \caption{\textbf{Results of the simulations for a continuous outcomes}: where the generative model corresponds to \eqref{eq:simulation-continuous-generative-model}, and where $b(.)$ and $m(.)$ are linear functions (more details are given in Appendix, see \eqref{eq:Ymodel-simulation-continuous}). Column 1 corresponds to generalizing conditional outcome (Proposition~\ref{proposition:generalization-density}), column 2 corresponds to generalizing local effect (Proposition~\ref{prop:generalization-of-local-effects}). For these two approaches we use different covariates set, with \textcolor{Goldenrod}{\textbf{shifted treatment effect modulators}} ($X_1$, $X_2$), \textcolor{RedOrange}{\textbf{shifted prognostic covariates}} ($X_1$, $X_2$, $X_3$, and $X_4$), and all \textcolor{Mahogany}{\textbf{prognostic covariates}}  ($X_1$, $X_2$, $X_3$, $X_4$, $X_4$ and $X_6$). According to Theorems~\ref{theorem:all-covariates} and \ref{theorem:restricted-set-for-Y-continuous-RD}, only the Risk Difference can be generalized with a restricted covariates set. Simulations are performed with $1000$ repetitions, a source sample size of $500$ and target sample size of $1,000$. Estimation is performed with plug-in g-formula modeling all responses with an OLS approach as detailed in Section~\ref{appendix:continuous-estimation-steps}.}
    \label{fig:simulations-continuous-Y}
\end{figure}

\subsection{Continuous outcome}
We propose a situation where the continuous outcome is generated from six prognostic covariates $X_1, \dots X_6$ as detailed in \eqref{eq:simulation-continuous-generative-model}. More precisely, $B=\{1,2,3,4,5,6\}$, and $M = \{1,2,5\}$, while only covariates $X_1, X_2, X_3, X_4$ are shifted between $P_{\text{\tiny S}}$ and $P_{\text{\tiny T}}$. For this simulation, both $b(.)$ and $m(.)$ are linear functions of the covariates (see Section~\ref{appendix:continuous-generative-model}), so that estimation with an OLS procedure is well-specified. 
Figure~\ref{fig:simulations-continuous-Y} presents results, where the \textcolor{magenta}{\textbf{pink}} dashed line represents the source causal effect and the \modif{\textbf{blue}} dashed line represents the target causal effect. 
As expected for the outcome generalization strategy (Theorem~\ref{theorem:all-covariates}), all causal measure can be generalized using all prognostic and shifted covariates (orange boxplots).
Note that adding all prognostic covariates (red boxplots) leads to more precision, in accordance with what is proposed in \cite{colnet2022reweighting} for the risk difference\footnote{This is similar to adding an instrument or an outcome-related covariate in an adjustement set when estimating causal effect from a single observational data set \citep{brookhart2006variable}.}.
According to Theorem~\ref{theorem:restricted-set-for-Y-continuous-RD}, the Risk Difference $\tau_{\text{\tiny RD}}$ can be generalized via the local effect strategy using  less covariates, namely the shifted treatment effect modulators, $X_1$ and $X_2$ (yellow boxplot). We observe that all other causal measures require access to all shifted prognostic covariates in both strategies in order to retrieve the target effect.

\subsection{Binary outcome}

We enrich the example of the Russian Roulette assuming that the effect of the Russian Roulette itself is modulated by covariates.
This gloomy example is, of course, completely fictitious and is used for better understanding.
We adapt the discriminative model of \eqref{eq:intuition-of-intrication} into
\begin{align}\label{eq:simulation-binary-generative-model}
    & \mathbb{P}\left[ Y^{(a)} = 1 \mid X = x \right] \\
    & = b(X_1, X_2, X_3) + a\,\left(1-b\left(X_1, X_2, X_3\right)\right)\,m_b(X_2, X_3), \nonumber 
\end{align}
where $X_1=\texttt{lifestyle}$, $X_2=\texttt{stress}$, and $X_3=\texttt{gender}$, a situation where individuals' baseline risk of death depends on their lifestyle, stress, and gender. 
\begin{figure}
    \centering
    \includegraphics[width=0.5\textwidth]{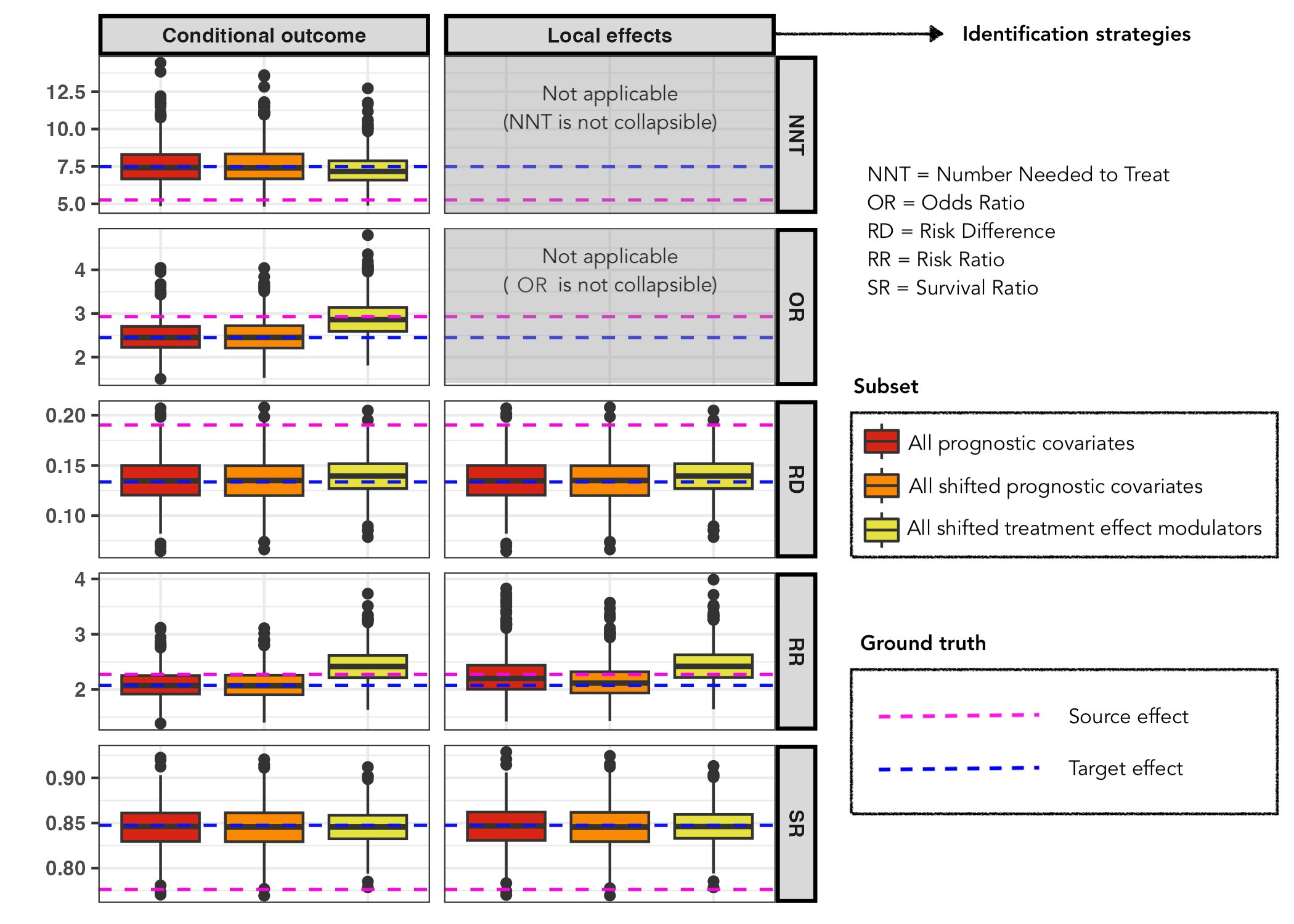}
    \caption{\textbf{Simulation with binary outcome $Y$}: for a monotonous and deleterious effect. 
    Adjusting on \textcolor{RedOrange}{\textbf{shifted prognostic covariates}} (\texttt{stress} and \texttt{lifestyle}), or with all \textcolor{Mahogany}{\textbf{prognostic covariates}} (\texttt{stress, lifestyle,} and \texttt{gender}) enables generalization of all causal measures by generalization of the conditional outcome or re-weighting of local effect if possible (only for collapsible measures, namely RR, SR, and RD). 
    On this simulation, estimation is done with IPSW estimator, source (resp. target) sample being of size $n=5\,000$ (resp. $m=20,000$), with $1000$ repetitions.} 
    \label{fig:simulations-binary-Y-russian-roulette}
\end{figure}
We assume that the effect of the Russian Roulette can be modulated by stress (imagine individuals having a heart attack as soon as the gun is approaching their head) and gender (the executioner being more merciful when facing a women). 
We further assume that \texttt{gender} is the only covariate with no shift between the two populations. In particular, we suppose that $P_{\text{\tiny S}}$ is composed of more people with a good lifestyle but are very stressed, while in $P_{\text{\tiny T}}$ individuals have a poor lifestyle but a low stress.  
Details on the generative model are provided in Appendix (see Section~\ref{appendix:simulation-binary-data-generative-model}).

Results are shown in Figure~\ref{fig:simulations-binary-Y-russian-roulette}. Note that on the simulation both the NNT and the OR cannot be generalized via the local effect strategy, as these measures are not collapsible.
As expected for the outcome generalization strategy (Theorem~\ref{theorem:all-covariates}), all causal measure can be generalized using all prognostic and shifted covariates (orange boxplots). Note that this appears to hold also for the local effect strategy.

\section{Conclusion}

The choice of a population-level measure of treatment effect has been much debated. Indeed, all causal measures do not share the same properties, which may lead to different interpretation of the treatment effect. In particular, we show that collapsibility is a very important property, as it allows computing the average treatment effect via a reweighting of local effects on substrata. 
Among population causal collapsible measures, only the Risk Difference is able to disentangle the treatment effect from the baseline at both a strata (CATE) and population (ATE) level (see Theorem~\ref{th_onlyRD_separates_baseline_tteffect}). This generic result holds for both continuous and binary outcomes, but only for a restricted range of baseline functions in the case of bounded outcomes (see Theorem~\ref{th_onlyRD_separates_baseline_tteffect_bounded_outcome}). \modif{Besides, in binary settings, the CATE of the Survival Ratio (resp. the Risk Ratio) has a specific interpretation for harmful (resp. beneficial) treatment effects, even if its ATE is not disentangled from the baseline for  all distributions.}  
Our analysis of the different properties of causal measures leads us to establish two different strategies for generalization, based on the potential outcomes (Proposition~\ref{proposition:generalization-density}) or the local effect (Proposition~\ref{prop:generalization-of-local-effects}). The first approach can be applied to any causal measure but requires a stronger assumption. The local effect strategy can be applied to collapsible causal measures only but requires a less stringent assumption, and potentially fewer covariates than the first approach. In particular, we show that all shifted prognostic variables are needed to generalize the potential outcomes (Theorem~\ref{theorem:all-covariates}), while only shifted treatment effect modifiers are needed to generalize the Risk Difference via the local effect procedure (Theorem~\ref{theorem:restricted-set-for-Y-continuous-RD}). 
 However,  identifying which covariates are treatment modifiers is still an open problem despite recent advances \citep[][]{hines2022variable, benard2023variable, paillard2025measuring}.
Regardless of the outcome type (continuous or binary), the Risk Difference may require fewer variables than other causal measures to be generalized. Note that this is not always the case: if the treatment effect of the Risk Difference is entangled with the baseline (as in the Russian Roulette example), then generalizing the Risk Difference via local effect would require all shifted prognostic variables. 
Note that all other causal measures are able to separate the treatment effect from the baseline in very specific settings (e.g., homogeneous treatment effect), and are thus easily generalizable in these contexts (see  \Cref{thm_homogeneous_independence_generalization}). \modif{\Cref{tab:measures_properties} presents a comprehensive view of the properties discussed in this paper for different causal measures.}

\modif{
In this paper, we have focused on causal measures that can be expressed as functions of the expectations of both potential outcomes, for both binary and continuous outcomes. We have not addressed time-to-event outcomes, which require accounting for censoring. In this context, although the hazard ratio remains the standard measure, there is a growing body of literature advocating for the use of the restricted mean survival time (RMST), as it offers a clear causal interpretation and is collapsible unlike the hazard ratio \citep[see, e.g.,][]{dumas2025hazard, Hernan2010}. Generalizations of these approaches have been explored in \citet{Wen2025}.
}

\modif{Finally, our analysis focuses on population quantities, which notably leads us to derive identifiability formula for generalizing the average treatment effect (see \Cref{proposition:generalization-density} and \Cref{prop:generalization-of-local-effects}). This is a necessary first step to derive estimators of the average treatment effect on the target population. New problems arise from considering estimation and practical (finite-sample) setting. Indeed, it is unlikely that all estimators derived from the identifiability formula in \Cref{subsection:two_generalization_strategies} have the same bias and variance. Besides, doubly-robust estimators can adapt more easily to a variety of situations \citep[see, e.g.][for RR analysis in observational studies]{boughdiri2025quantifying}. 
An interesting avenue for further research consists in studying the properties of the causal measures and their estimators when faced to overlap issues \citep[][]{Huang2025}, unobserved confounders \citep[see simulations and sensitivity analyses in][]{robins2000sensitivity, sturmer2010treatment, carnegie2016assessing, huang2023design}, missing data or noncompliance \citep{nagelkerke2000estimating, chen2025generalizing}.}

%\section*{Acknowledgments} 
%First of all we would like to thank clinician François-Xavier \textsc{Ageron} who first raised our interest on generalizability and the choice of measure. We also would like to thank fruitful discussions with our clinicians collaborators, in particular with François-Camille \textsc{Grolleau} and Raphaël \textsc{Porcher}.We thank Maxime \textsc{Fosset} and Marine \textsc{Le Morvan} who gave precious comments on the manuscript. Thank you also to Wouter \textsc{van Amsterdam} for careful proofreading. Finally, we would like to thank Anders \textsc{Huitfieldt}: his research papers (and blog articles) have been precious sources of inspiration for our work.

\section*{Fundings}
Authors are all funded by their respective employer (Inria or Sorbonne University). This work/project was partially and publicly funded through ANR (the French National Research Agency) under the ``\textit{Investissements d’avenir}" program with the reference {\scriptsize \texttt{ANR-16-IDEX-0006}}. GV acknowledges funding from Intercept-T2D, with the reference {\scriptsize \texttt{HORIZON-HLTH-2022-STAYHLTH-02-01}}.

\bibliographystyle{chicago}
\bibliography{references}

\begin{thebibliography}{}

\bibitem[\protect\citeauthoryear{Ackerman, Lesko, Siddique, Susukida, and
  Stuart}{Ackerman et~al.}{2021}]{ackerman2020generalization}
Ackerman, B., C.~R. Lesko, J.~Siddique, R.~Susukida, and E.~A. Stuart (2021).
\newblock Generalizing randomized trial findings to a target population using
  complex survey population data.
\newblock {\em Statistics in medicine\/}~{\em 40\/}(5), 1101--1120.

\bibitem[\protect\citeauthoryear{Altman}{Altman}{1998}]{altman1998confidence}
Altman, D.~G. (1998).
\newblock Confidence intervals for the number needed to treat.
\newblock {\em Bmj\/}~{\em 317\/}(7168), 1309--1312.

\bibitem[\protect\citeauthoryear{Angrist and Pischke}{Angrist and
  Pischke}{2008}]{angrist2008mostlyharmless}
Angrist, J.~D. and J.-S. Pischke (2008, December).
\newblock {\em Mostly Harmless Econometrics: An Empiricist's Companion}.
\newblock Princeton University Press.

\bibitem[\protect\citeauthoryear{Athey, Tibshirani, and Wager}{Athey
  et~al.}{2019}]{athey2019generalized}
Athey, S., J.~Tibshirani, and S.~Wager (2019).
\newblock Generalized random forests.
\newblock {\em The Annals of Statistics\/}~{\em 47\/}(2), 1148--1178.

\bibitem[\protect\citeauthoryear{Baker and Jackson}{Baker and
  Jackson}{2018}]{baker2018new}
Baker, R. and D.~Jackson (2018).
\newblock A new measure of treatment effect for random-effects meta-analysis of
  comparative binary outcome data.
\newblock {\em arXiv preprint arXiv:1806.03471\/}.

\bibitem[\protect\citeauthoryear{B{\'e}nard and Josse}{B{\'e}nard and
  Josse}{2023}]{benard2023variable}
B{\'e}nard, C. and J.~Josse (2023).
\newblock Variable importance for causal forests: breaking down the
  heterogeneity of treatment effects.
\newblock {\em arXiv preprint arXiv:2308.03369\/}.

\bibitem[\protect\citeauthoryear{Berkowitz, Sussman, Jonas, and Basu}{Berkowitz
  et~al.}{2018}]{Berkowitz2018GeneralizingBlood}
Berkowitz, S.~A., J.~B. Sussman, D.~E. Jonas, and S.~Basu (2018).
\newblock Generalizing intensive blood pressure treatment to adults with
  diabetes mellitus.
\newblock {\em Journal of the American College of Cardiology\/}~{\em 72\/}(11),
  1214--1223.
\newblock SPECIAL FOCUS ISSUE: BLOOD PRESSURE.

\bibitem[\protect\citeauthoryear{Boughdiri, Josse, and Scornet}{Boughdiri
  et~al.}{2025}]{boughdiri2025quantifying}
Boughdiri, A., J.~Josse, and E.~Scornet (2025).
\newblock Quantifying treatment effects: Estimating risk ratios via
  observational studies.
\newblock In {\em Forty-second International Conference on Machine Learning}.

\bibitem[\protect\citeauthoryear{Brookhart, Schneeweiss, Rothman, Glynn, Avorn,
  and St{\"u}rmer}{Brookhart et~al.}{2006}]{brookhart2006variable}
Brookhart, M.~A., S.~Schneeweiss, K.~J. Rothman, R.~J. Glynn, J.~Avorn, and
  T.~St{\"u}rmer (2006).
\newblock Variable selection for propensity score models.
\newblock {\em American journal of epidemiology\/}~{\em 163\/}(12), 1149--1156.

\bibitem[\protect\citeauthoryear{Buchanan, Hudgens, Cole, Mollan, Sax, Daar,
  Adimora, Eron, and Mugavero}{Buchanan
  et~al.}{2018}]{buchanan2018generalizing}
Buchanan, A.~L., M.~G. Hudgens, S.~R. Cole, K.~R. Mollan, P.~E. Sax, E.~S.
  Daar, A.~A. Adimora, J.~J. Eron, and M.~J. Mugavero (2018).
\newblock Generalizing evidence from randomized trials using inverse
  probability of sampling weights.
\newblock {\em Journal of the Royal Statistical Society: Series A (Statistics
  in Society)\/}~{\em 181}, 1193--1209.

\bibitem[\protect\citeauthoryear{Carnegie, Harada, and Hill}{Carnegie
  et~al.}{2016}]{carnegie2016assessing}
Carnegie, N.~B., M.~Harada, and J.~L. Hill (2016).
\newblock Assessing sensitivity to unmeasured confounding using a simulated
  potential confounder.
\newblock {\em Journal of Research on Educational Effectiveness\/}~{\em
  9\/}(3), 395--420.

\bibitem[\protect\citeauthoryear{Chen and Huang}{Chen and
  Huang}{2025}]{chen2025generalizing}
Chen, Z. and M.~Huang (2025).
\newblock Generalizing causal effects with noncompliance: Application to deep
  canvassing experiments.
\newblock {\em arXiv preprint arXiv:2506.00149\/}.

\bibitem[\protect\citeauthoryear{Cinelli and Pearl}{Cinelli and
  Pearl}{2020}]{CinelliGeneralizing2019}
Cinelli, C. and J.~Pearl (2020).
\newblock Generalizing experimental results by leveraging knowledge of
  mechanisms.
\newblock {\em European Journal of Epidemiology\/}.

\bibitem[\protect\citeauthoryear{Cole and MacMahon}{Cole and
  MacMahon}{1971}]{cole1971attributable}
Cole, P. and B.~MacMahon (1971).
\newblock Attributable risk percent in case-control studies.
\newblock {\em British journal of preventive \& social medicine\/}~{\em
  25\/}(4), 242.

\bibitem[\protect\citeauthoryear{Colnet, Josse, Varoquaux, and Scornet}{Colnet
  et~al.}{2022a}]{colnet2022sensitivity}
Colnet, B., J.~Josse, G.~Varoquaux, and E.~Scornet (2022a).
\newblock Causal effect on a target population: a sensitivity analysis to
  handle missing covariates.
\newblock {\em Journal of Causal Inference\/}~{\em 10\/}(1), 372--414.

\bibitem[\protect\citeauthoryear{Colnet, Josse, Varoquaux, and Scornet}{Colnet
  et~al.}{2022b}]{colnet2022reweighting}
Colnet, B., J.~Josse, G.~Varoquaux, and E.~Scornet (2022b).
\newblock Reweighting the rct for generalization: finite sample analysis and
  variable selection.
\newblock {\em arXiv preprint arXiv:2208.07614\/}.

\bibitem[\protect\citeauthoryear{Colnet, Mayer, Chen, Dieng, Li, Varoquaux,
  Vert, Josse, and Yang}{Colnet et~al.}{2024}]{colnet2021causal}
Colnet, B., I.~Mayer, G.~Chen, A.~Dieng, R.~Li, G.~Varoquaux, J.-P. Vert,
  J.~Josse, and S.~Yang (2024).
\newblock Causal inference methods for combining randomized trials and
  observational studies: a review.
\newblock {\em Statistical science\/}~{\em 39\/}(1), 165--191.

\bibitem[\protect\citeauthoryear{Cook and Sackett}{Cook and
  Sackett}{1995}]{Cook1995NNT}
Cook, R.~J. and D.~L. Sackett (1995).
\newblock The number needed to treat: a clinically useful measure of treatment
  effect.
\newblock {\em Bmj\/}~{\em 310\/}(6977), 452--454.

\bibitem[\protect\citeauthoryear{Cook, Campbell, and Shadish}{Cook
  et~al.}{2002}]{cook2002experimental}
Cook, T.~D., D.~T. Campbell, and W.~Shadish (2002).
\newblock {\em Experimental and quasi-experimental designs for generalized
  causal inference}.
\newblock Houghton Mifflin Boston, MA.

\bibitem[\protect\citeauthoryear{Cornfield et~al.}{Cornfield
  et~al.}{1951}]{cornfield1951method}
Cornfield, J. et~al. (1951).
\newblock A method of estimating comparative rates from clinical data;
  applications to cancer of the lung, breast, and cervix.
\newblock {\em Journal of the National Cancer Institute\/}, 1269--1275.

\bibitem[\protect\citeauthoryear{Cummings}{Cummings}{2009}]{Cummings2009RelativeMeritsRRAndOR}
Cummings, P. (2009, 05).
\newblock {The Relative Merits of Risk Ratios and Odds Ratios}.
\newblock {\em Archives of Pediatrics \& Adolescent Medicine\/}~{\em 163\/}(5),
  438--445.

\bibitem[\protect\citeauthoryear{Dahabreh, Robertson, Steingrimsson, Stuart,
  and Hern{\'a}n}{Dahabreh et~al.}{2020}]{dahabreh2020extending}
Dahabreh, I.~J., S.~E. Robertson, J.~A. Steingrimsson, E.~A. Stuart, and M.~A.
  Hern{\'a}n (2020).
\newblock Extending inferences from a randomized trial to a new target
  population.
\newblock {\em Statistics in Medicine\/}~{\em 39\/}(14), 1999--2014.

\bibitem[\protect\citeauthoryear{Daniel, Zhang, and Farewell}{Daniel
  et~al.}{2020}]{Daniel2020MakingApple}
Daniel, R., J.~Zhang, and D.~Farewell (2020, 12).
\newblock Making apples from oranges: Comparing noncollapsible effect
  estimators and their standard errors after adjustment for different covariate
  sets.
\newblock {\em Biometrical Journal\/}~{\em 63}.

\bibitem[\protect\citeauthoryear{Davies, Crombie, and Tavakoli}{Davies
  et~al.}{1998}]{davies1998can}
Davies, H. T.~O., I.~K. Crombie, and M.~Tavakoli (1998).
\newblock When can odds ratios mislead?
\newblock {\em Bmj\/}~{\em 316\/}(7136), 989--991.

\bibitem[\protect\citeauthoryear{Dawid}{Dawid}{2000}]{dawid2000causal}
Dawid, A.~P. (2000).
\newblock Causal inference without counterfactuals.
\newblock {\em Journal of the American statistical Association\/}~{\em
  95\/}(450), 407--424.

\bibitem[\protect\citeauthoryear{Deeks}{Deeks}{2002}]{Deeks2022IssuesInSelection}
Deeks, J. (2002, June).
\newblock Issues in the selection of a summary statistic in meta-analysis of
  clinical trials with binary outcomes.
\newblock {\em Statistics in Medicine\/}~{\em 21\/}(11), 1575--1600.

\bibitem[\protect\citeauthoryear{Degtiar and Rose}{Degtiar and
  Rose}{2023}]{Degtiar2021Generalizability}
Degtiar, I. and S.~Rose (2023).
\newblock A review of generalizability and transportability.
\newblock {\em Annual Review of Statistics and Its Application\/}~{\em 10},
  501--524.

\bibitem[\protect\citeauthoryear{Didelez and Stensrud}{Didelez and
  Stensrud}{2022}]{Didelez2021collapsibility}
Didelez, V. and M.~J. Stensrud (2022).
\newblock On the logic of collapsibility for causal effect measures.
\newblock {\em Biometrical Journal\/}~{\em 64\/}(2), 235--242.

\bibitem[\protect\citeauthoryear{Ding, Lin, and Hsu}{Ding
  et~al.}{2016}]{ding2016subgroup}
Ding, Y., H.-M. Lin, and J.~C. Hsu (2016).
\newblock Subgroup mixable inference on treatment efficacy in mixture
  populations, with an application to time-to-event outcomes.
\newblock {\em Statistics in medicine\/}~{\em 35\/}(10), 1580--1594.

\bibitem[\protect\citeauthoryear{Doi, Furuya-Kanamori, Xu, Lin, Chivese, and
  Thalib}{Doi et~al.}{2020}]{Doi2020callToChangePractice}
Doi, S., L.~Furuya-Kanamori, C.~Xu, L.~Lin, T.~Chivese, and L.~Thalib (2020,
  11).
\newblock Questionable utility of the relative risk in clinical research: A
  call for change to practice.
\newblock {\em Journal of Clinical Epidemiology\/}.

\bibitem[\protect\citeauthoryear{Doi, Furuya-Kanamori, Xu, Chivese, Lin, Musa,
  Hindy, Thalib, and Harrell~Jr}{Doi et~al.}{2022}]{doi2022TimeToDoAway}
Doi, S.~A., L.~Furuya-Kanamori, C.~Xu, T.~Chivese, L.~Lin, O.~A. Musa,
  G.~Hindy, L.~Thalib, and F.~E. Harrell~Jr (2022).
\newblock The odds ratio is “portable” across baseline risk but not the
  relative risk: Time to do away with the log link in binomial regression.
\newblock {\em Journal of Clinical Epidemiology\/}~{\em 142}, 288--293.

\bibitem[\protect\citeauthoryear{Doi, Furuya-Kanamori, Xu, Lin, Chivese, and
  Thalib}{Doi et~al.}{2022}]{Suhail2022CallForChange}
Doi, S.~A., L.~Furuya-Kanamori, C.~Xu, L.~Lin, T.~Chivese, and L.~Thalib
  (2022).
\newblock Controversy and debate: Questionable utility of the relative risk in
  clinical research: Paper 1: A call for change to practice.
\newblock {\em Journal of Clinical Epidemiology\/}~{\em 142}, 271--279.

\bibitem[\protect\citeauthoryear{Dumas and Stensrud}{Dumas and
  Stensrud}{2025}]{dumas2025hazard}
Dumas, E. and M.~J. Stensrud (2025).
\newblock How hazard ratios can mislead and why it matters in practice.
\newblock {\em European Journal of Epidemiology\/}, 1--7.

\bibitem[\protect\citeauthoryear{Edvinsson}{Edvinsson}{2021}]{Edvinsson2021Migraine}
Edvinsson, L. (2021).
\newblock Oral rimegepant for migraine prevention.
\newblock {\em The Lancet\/}~{\em 397\/}(10268), 4--5.

\bibitem[\protect\citeauthoryear{Even and Josse}{Even and
  Josse}{2025}]{even2025rethinking}
Even, M. and J.~Josse (2025).
\newblock Rethinking the win ratio: A causal framework for hierarchical outcome
  analysis.
\newblock {\em arXiv preprint arXiv:2501.16933\/}.

\bibitem[\protect\citeauthoryear{Fay and Li}{Fay and Li}{2024}]{Fay2024Causal}
Fay, M.~P. and F.~Li (2024, October).
\newblock Causal interpretation of the hazard ratio in randomized clinical
  trials.
\newblock {\em Clinical Trials\/}~{\em 21\/}(5), 623--635.
\newblock Epub 2024 Apr 28.

\bibitem[\protect\citeauthoryear{Feng, Wang, and Hongyue}{Feng
  et~al.}{2019}]{Changyong2019RelationsAmongThreePop}
Feng, C., B.~Wang, and W.~Hongyue (2019, 07).
\newblock The relations among three popular indices of risks.
\newblock {\em Statistics in Medicine\/}~{\em 38}.

\bibitem[\protect\citeauthoryear{Forrow, Taylor, and Arnold}{Forrow
  et~al.}{1992}]{Forrow1992HowResultsAreSummarized}
Forrow, L., W.~C. Taylor, and R.~M. Arnold (1992).
\newblock Absolutely relative: How research results are summarized can affect
  treatment decisions.
\newblock {\em The American Journal of Medicine\/}~{\em 92\/}(2), 121--124.

\bibitem[\protect\citeauthoryear{Gao and Hastie}{Gao and
  Hastie}{2021}]{Gao2021DINA}
Gao, Z. and T.~Hastie (2021).
\newblock Estimating heterogeneous treatment effects for general responses.

\bibitem[\protect\citeauthoryear{Gatsonis and Sally}{Gatsonis and
  Sally}{2017}]{Stuart2017ChapterBook}
Gatsonis, C. and M.~C. Sally (2017).
\newblock {\em Methods in Comparative Effectiveness Research}, pp.\  177--199.
\newblock Chapman \& Hall.

\bibitem[\protect\citeauthoryear{George, Stead, and Ganti}{George
  et~al.}{2020}]{George2020WhatsTR}
George, A., T.~S. Stead, and L.~Ganti (2020).
\newblock What’s the risk: Differentiating risk ratios, odds ratios, and
  hazard ratios?
\newblock {\em Cureus\/}~{\em 12}.

\bibitem[\protect\citeauthoryear{Greenland}{Greenland}{1987}]{Greenland1987Interpretation}
Greenland, S. (1987, 05).
\newblock {Interpretation and choice of effect measures in epidemiologic
  analysies}.
\newblock {\em American Journal of Epidemiology\/}~{\em 125\/}(5), 761--768.

\bibitem[\protect\citeauthoryear{Greenland and Pearl}{Greenland and
  Pearl}{2011}]{Greenland2011adjustments}
Greenland, S. and J.~Pearl (2011, 12).
\newblock Adjustments and their consequences—collapsibility analysis using
  graphical models.
\newblock {\em International Statistical Review / Revue Internationale de
  Statistique\/}~{\em 79}.

\bibitem[\protect\citeauthoryear{Greenland, Robbins, and Pearl}{Greenland
  et~al.}{1999}]{pearl1999collapsibility}
Greenland, S., J.~M. Robbins, and J.~Pearl (1999, 01).
\newblock Confounding and collapsibility in causal inference.
\newblock {\em Statistical Science\/}~{\em 14}, 29--46.

\bibitem[\protect\citeauthoryear{Guyatt, Rennie, Meade, and Cook}{Guyatt
  et~al.}{2015}]{Cook2014UserGuide}
Guyatt, G., D.~Rennie, M.~O. Meade, and D.~J. Cook (2015).
\newblock {\em Users' Guides to the Medical Literature : A Manual for
  Evidence-Based Clinical Practice.}
\newblock New York: McGraw-Hill Education.

\bibitem[\protect\citeauthoryear{Hern\`an, Clayton, and Keiding}{Hern\`an
  et~al.}{2011}]{Hernan2011unraveled}
Hern\`an, M., D.~Clayton, and N.~Keiding (2011, 03).
\newblock The simpson's paradox unraveled.
\newblock {\em International journal of epidemiology\/}~{\em 40}, 780--5.

\bibitem[\protect\citeauthoryear{Hern\`an and Robins}{Hern\`an and
  Robins}{2020}]{hernan2020whatifbook}
Hern\`an, M. and J.~Robins (2020).
\newblock {\em Causal Inference: What If.}

\bibitem[\protect\citeauthoryear{Hernán}{Hernán}{2010}]{Hernan2010}
Hernán, M.~A. (2010).
\newblock The hazards of hazard ratios.
\newblock {\em Epidemiology\/}~{\em 21\/}(1), 13--15.

\bibitem[\protect\citeauthoryear{Hines, Diaz-Ordaz, and Vansteelandt}{Hines
  et~al.}{2022}]{hines2022variable}
Hines, O., K.~Diaz-Ordaz, and S.~Vansteelandt (2022).
\newblock Variable importance measures for heterogeneous causal effects.
\newblock {\em arXiv preprint arXiv:2204.06030\/}.

\bibitem[\protect\citeauthoryear{Huang}{Huang}{2025}]{Huang2025}
Huang, M. (2025, Mar).
\newblock Overlap violations in external validity: Application to ugandan cash
  transfer programs.
\newblock {\em Annals of Applied Statistics\/}~{\em 19\/}(1), 351--370.

\bibitem[\protect\citeauthoryear{Huang, Soriano, and Pimentel}{Huang
  et~al.}{2023}]{huang2023design}
Huang, M., D.~Soriano, and S.~D. Pimentel (2023).
\newblock Design sensitivity and its implications for weighted observational
  studies.
\newblock {\em arXiv preprint arXiv:2307.00093\/}.

\bibitem[\protect\citeauthoryear{Huitfeldt}{Huitfeldt}{2019}]{Huitfeldt2019LessWrong}
Huitfeldt, A. (2019).
\newblock Effect heterogeneity and external validity in medicine.
\newblock {\em Available in: https://www. lesswrong.
  com/posts/wwbrvumMWhDfeo652\/}.

\bibitem[\protect\citeauthoryear{Huitfeldt, Fox, Murray, Hr{\'o}bjartsson, and
  Daniel}{Huitfeldt et~al.}{2021}]{Huitfeldt2021ShallWe}
Huitfeldt, A., M.~P. Fox, E.~J. Murray, A.~Hr{\'o}bjartsson, and R.~M. Daniel
  (2021).
\newblock Shall we count the living or the dead?
\newblock {\em arXiv preprint arXiv:2106.06316\/}.

\bibitem[\protect\citeauthoryear{Huitfeldt, Goldstein, and Swanson}{Huitfeldt
  et~al.}{2018}]{huitfeldt2018choice}
Huitfeldt, A., A.~Goldstein, and S.~A. Swanson (2018).
\newblock The choice of effect measure for binary outcomes: Introducing
  counterfactual outcome state transition parameters.
\newblock {\em Epidemiologic methods\/}~{\em 7\/}(1).

\bibitem[\protect\citeauthoryear{Huitfeldt, Stensrud, and Suzuki}{Huitfeldt
  et~al.}{2019}]{Huitfeldt2019collapsible}
Huitfeldt, A., M.~Stensrud, and E.~Suzuki (2019, 01).
\newblock On the collapsibility of measures of effect in the counterfactual
  causal framework.
\newblock {\em Emerging Themes in Epidemiology\/}~{\em 16}.

\bibitem[\protect\citeauthoryear{Huitfeldt and Stensrud}{Huitfeldt and
  Stensrud}{2018}]{huitfeldt2018re}
Huitfeldt, A. and M.~J. Stensrud (2018).
\newblock Re: generalizing study results: a potential outcomes perspective.
\newblock {\em Epidemiology\/}~{\em 29\/}(2), e13--e14.

\bibitem[\protect\citeauthoryear{Huitfeldt, Swanson, Stensrud, and
  Suzuki}{Huitfeldt et~al.}{2019}]{Huitfeldt2019EffectHeterogeneity}
Huitfeldt, A., S.~Swanson, M.~Stensrud, and E.~Suzuki (2019, 12).
\newblock Effect heterogeneity and variable selection for standardizing causal
  effects to a target population.
\newblock {\em European Journal of Epidemiology\/}~{\em 34}.

\bibitem[\protect\citeauthoryear{Imai, King, and Stuart}{Imai
  et~al.}{2008}]{imai2008misunderstandings}
Imai, K., G.~King, and E.~A. Stuart (2008).
\newblock Misunderstandings between experimentalists and observationalists
  about causal inference.
\newblock {\em Journal of the royal statistical society: series A (statistics
  in society)\/}~{\em 171\/}(2), 481--502.

\bibitem[\protect\citeauthoryear{Imai and Ratkovic}{Imai and
  Ratkovic}{2013}]{imai2013estimating}
Imai, K. and M.~Ratkovic (2013).
\newblock Estimating treatment effect heterogeneity in randomized program
  evaluation.
\newblock {\em The Annals of Applied Statistics\/}, 443--470.

\bibitem[\protect\citeauthoryear{Imbens}{Imbens}{2011}]{imbens2011experimental}
Imbens, G.~W. (2011).
\newblock Experimental design for unit and cluster randomid trials.
\newblock {\em International Initiative for Impact Evaluation Paper\/}.

\bibitem[\protect\citeauthoryear{Imbens and Rubin}{Imbens and
  Rubin}{2015}]{imbens2015causal}
Imbens, G.~W. and D.~B. Rubin (2015).
\newblock {\em {Causal Inference in Statistics, Social, and Biomedical
  Sciences}}.
\newblock Cambridge UK: Cambridge University Press.

\bibitem[\protect\citeauthoryear{Jim{\'e}nez, Guallar, and
  Martin-Moreno}{Jim{\'e}nez et~al.}{1997}]{Jimnez1997GraphicalDisplayUseful}
Jim{\'e}nez, F.~J., E.~Guallar, and J.~M. Martin-Moreno (1997).
\newblock A graphical display useful for meta-analysis.
\newblock {\em European Journal of Public Health\/}~{\em 7}, 101--105.

\bibitem[\protect\citeauthoryear{Katta, Parikh, Rudin, and Volfovsky}{Katta
  et~al.}{2024}]{katta2024interpretable}
Katta, S., H.~Parikh, C.~Rudin, and A.~Volfovsky (2024).
\newblock Interpretable causal inference for analyzing wearable, sensor, and
  distributional data.
\newblock In {\em International Conference on Artificial Intelligence and
  Statistics}, pp.\  3340--3348. PMLR.

\bibitem[\protect\citeauthoryear{Kern, Stuart, Hill, and Green}{Kern
  et~al.}{2016}]{kern2016assessing}
Kern, H.~L., E.~A. Stuart, J.~Hill, and D.~P. Green (2016).
\newblock Assessing methods for generalizing experimental impact estimates to
  target populations.
\newblock {\em Journal of research on educational effectiveness\/}~{\em
  9\/}(1), 103--127.

\bibitem[\protect\citeauthoryear{King and Zeng}{King and
  Zeng}{2002}]{king2002estimating}
King, G. and L.~Zeng (2002).
\newblock Estimating risk and rate levels, ratios and differences in
  case-control studies.
\newblock {\em Statistics in medicine\/}~{\em 21\/}(10), 1409--1427.

\bibitem[\protect\citeauthoryear{King, Harper, and Young}{King
  et~al.}{2012}]{king2012use}
King, N.~B., S.~Harper, and M.~E. Young (2012).
\newblock Use of relative and absolute effect measures in reporting health
  inequalities: structured review.
\newblock {\em Bmj\/}~{\em 345}.

\bibitem[\protect\citeauthoryear{K{\"u}nzel, Sekhon, Bickel, and Yu}{K{\"u}nzel
  et~al.}{2019}]{kunzel2019metalearners}
K{\"u}nzel, S.~R., J.~S. Sekhon, P.~J. Bickel, and B.~Yu (2019).
\newblock Metalearners for estimating heterogeneous treatment effects using
  machine learning.
\newblock {\em Proceedings of the national academy of sciences\/}~{\em
  116\/}(10), 4156--4165.

\bibitem[\protect\citeauthoryear{L'abb{\'e}, Detsky, and O'rourke}{L'abb{\'e}
  et~al.}{1987}]{Labbe1987MetaAnalysis}
L'abb{\'e}, K.~A., A.~S. Detsky, and K.~O'rourke (1987).
\newblock Meta-analysis in clinical research.
\newblock {\em Annals of internal medicine\/}~{\em 107 2}, 224--33.

\bibitem[\protect\citeauthoryear{Lapointe-Shaw, Babe, Austin, Costa, and
  Jones}{Lapointe-Shaw et~al.}{2022}]{Lapointe2022FromMathToMeaning}
Lapointe-Shaw, L., G.~Babe, P.~C. Austin, A.~P. Costa, and A.~Jones (2022,
  Nov.).
\newblock Reporting risk: from math to meaning.
\newblock {\em Canadian Journal of General Internal Medicine\/}~{\em 17\/}(4),
  59–66.

\bibitem[\protect\citeauthoryear{Laupacis, Sackett, and Roberts}{Laupacis
  et~al.}{1988}]{Laupacis1988AnAssessmentOfClinically}
Laupacis, A., D.~L. Sackett, and R.~S. Roberts (1988).
\newblock An assessment of clinically useful measures of the consequences of
  treatment.
\newblock {\em New England Journal of Medicine\/}~{\em 318\/}(26), 1728--1733.
\newblock PMID: 3374545.

\bibitem[\protect\citeauthoryear{Lesko, Buchanan, Westreich, Edwards, Hudgens,
  and Cole}{Lesko et~al.}{2017}]{lesko2017generalizing}
Lesko, C.~R., A.~L. Buchanan, D.~Westreich, J.~K. Edwards, M.~G. Hudgens, and
  S.~R. Cole (2017).
\newblock Generalizing study results: a potential outcomes perspective.
\newblock {\em Epidemiology (Cambridge, Mass.)\/}~{\em 28\/}(4), 553.

\bibitem[\protect\citeauthoryear{Lesko, Henderson, and Varadhan}{Lesko
  et~al.}{2018}]{lesko2018considerations}
Lesko, C.~R., N.~C. Henderson, and R.~Varadhan (2018).
\newblock Considerations when assessing heterogeneity of treatment effect in
  patient-centered outcomes research.
\newblock {\em Journal of clinical epidemiology\/}~{\em 100}, 22--31.

\bibitem[\protect\citeauthoryear{Lin, Kong, and Wang}{Lin
  et~al.}{2023}]{lin2023causal}
Lin, Z., D.~Kong, and L.~Wang (2023).
\newblock Causal inference on distribution functions.
\newblock {\em Journal of the Royal Statistical Society Series B: Statistical
  Methodology\/}~{\em 85\/}(2), 378--398.

\bibitem[\protect\citeauthoryear{Liu, Wang, Tian, and Hsu}{Liu
  et~al.}{2022}]{liu2022rejoinder}
Liu, Y., B.~Wang, H.~Tian, and J.~C. Hsu (2022).
\newblock Rejoinder for discussions on correct and logical causal inference for
  binary and time-to-event outcomes in randomized controlled trials.
\newblock {\em Biometrical Journal\/}~{\em 64\/}(2), 246--255.

\bibitem[\protect\citeauthoryear{Liu, Wang, Yang, Hui, Xu, Kil, and Hsu}{Liu
  et~al.}{2022}]{liu2022correct}
Liu, Y., B.~Wang, M.~Yang, J.~Hui, H.~Xu, S.~Kil, and J.~C. Hsu (2022).
\newblock Correct and logical causal inference for binary and time-to-event
  outcomes in randomized controlled trials.
\newblock {\em Biometrical Journal\/}~{\em 64\/}(2), 198--224.

\bibitem[\protect\citeauthoryear{MacMahon, Peto, Collins, Godwin, Cutler,
  Sorlie, Abbott, Neaton, Dyer, and Stamler}{MacMahon
  et~al.}{1990}]{macmahon1990blood}
MacMahon, S., R.~Peto, R.~Collins, J.~Godwin, J.~Cutler, P.~Sorlie, R.~Abbott,
  J.~Neaton, A.~Dyer, and J.~Stamler (1990).
\newblock Blood pressure, stroke, and coronary heart disease: part 1, prolonged
  differences in blood pressure: prospective observational studies corrected
  for the regression dilution bias.
\newblock {\em The Lancet\/}~{\em 335\/}(8692), 765--774.

\bibitem[\protect\citeauthoryear{Miettinen}{Miettinen}{1972}]{miettinen1972standardization}
Miettinen, O.~S. (1972).
\newblock Standardization of risk ratios.
\newblock {\em American Journal of Epidemiology\/}~{\em 96\/}(6), 383--388.

\bibitem[\protect\citeauthoryear{Miettinen and Cook}{Miettinen and
  Cook}{1981}]{Miettinen1981Essence}
Miettinen, O.~S. and E.~F. Cook (1981, 10).
\newblock {Counfounding: Essence and Detection}.
\newblock {\em American Journal of Epidemiology\/}~{\em 114\/}(4), 593--603.

\bibitem[\protect\citeauthoryear{Mills}{Mills}{1999}]{Mills1999PillScare}
Mills, A. (1999, 11).
\newblock {Clinical implications. Avoiding problems in clinical practice after
  the pill scare}.
\newblock {\em Human Reproduction Update\/}~{\em 5\/}(6), 639--653.

\bibitem[\protect\citeauthoryear{Miratrix, Sekhon, and Yu}{Miratrix
  et~al.}{2013}]{miratrix2013adjusting}
Miratrix, L.~W., J.~S. Sekhon, and B.~Yu (2013).
\newblock {Adjusting treatment effect estimates by post-stratification in
  randomized experiments}.
\newblock {\em Journal of the Royal Statistical Society Series B\/}~{\em 75},
  369{\textendash}396.

\bibitem[\protect\citeauthoryear{Moynihan, Bero, Ross-Degnan, Henry, Lee,
  Watkins, Mah, and Soumerai}{Moynihan et~al.}{2000}]{moynihan2000coverage}
Moynihan, R., L.~Bero, D.~Ross-Degnan, D.~Henry, K.~Lee, J.~Watkins, C.~Mah,
  and S.~B. Soumerai (2000).
\newblock Coverage by the news media of the benefits and risks of medications.
\newblock {\em New England journal of medicine\/}~{\em 342\/}(22), 1645--1650.

\bibitem[\protect\citeauthoryear{Nagelkerke, Fidler, Bernsen, and
  Borgdorff}{Nagelkerke et~al.}{2000}]{nagelkerke2000estimating}
Nagelkerke, N., V.~Fidler, R.~Bernsen, and M.~Borgdorff (2000).
\newblock Estimating treatment effects in randomized clinical trials in the
  presence of non-compliance.
\newblock {\em Statistics in medicine\/}~{\em 19\/}(14), 1849--1864.

\bibitem[\protect\citeauthoryear{Naylor, Chen, and Strauss}{Naylor
  et~al.}{1992}]{naylor1992measured}
Naylor, C.~D., E.~Chen, and B.~Strauss (1992).
\newblock Measured enthusiasm: does the method of reporting trial results alter
  perceptions of therapeutic effectiveness?
\newblock {\em Annals of Internal Medicine\/}~{\em 117\/}(11), 916--921.

\bibitem[\protect\citeauthoryear{Neuhaus~S and Jewell}{Neuhaus~S and
  Jewell}{1993}]{Neuhaus1993GeometricApproach}
Neuhaus~S, J.~M. and N.~P. Jewell (1993, 12).
\newblock {A geometric approach to assess bias due to omitted covariates in
  generalized linear models}.
\newblock {\em Biometrika\/}~{\em 80\/}(4), 807--815.

\bibitem[\protect\citeauthoryear{Nguyen, Ackerman, Schmid, Cole, and
  Stuart}{Nguyen et~al.}{2018}]{nguyen2018sensitivitybis}
Nguyen, T., B.~Ackerman, I.~Schmid, S.~Cole, and E.~Stuart (2018, 12).
\newblock Sensitivity analyses for effect modifiers not observed in the target
  population when generalizing treatment effects from a randomized controlled
  trial: Assumptions, models, effect scales, data scenarios, and implementation
  details.
\newblock {\em PLOS ONE\/}~{\em 13}, e0208795.

\bibitem[\protect\citeauthoryear{Nie, Imbens, and Wager}{Nie
  et~al.}{2021}]{nie2021covariate}
Nie, X., G.~Imbens, and S.~Wager (2021).
\newblock Covariate balancing sensitivity analysis for extrapolating randomized
  trials across locations.
\newblock {\em arXiv preprint arXiv:2112.04723\/}.

\bibitem[\protect\citeauthoryear{Nie and Wager}{Nie and
  Wager}{2020}]{nie2020quasioracle}
Nie, X. and S.~Wager (2020, 09).
\newblock Quasi-oracle estimation of heterogeneous treatment effects.
\newblock {\em Biometrika\/}~{\em 108}.

\bibitem[\protect\citeauthoryear{Nie and Wager}{Nie and
  Wager}{2021}]{nie2021quasi}
Nie, X. and S.~Wager (2021).
\newblock Quasi-oracle estimation of heterogeneous treatment effects.
\newblock {\em Biometrika\/}~{\em 108\/}(2), 299--319.

\bibitem[\protect\citeauthoryear{Nuovo, Melnikow, and Chang}{Nuovo
  et~al.}{2002}]{Nuovo2002ReportingNNT}
Nuovo, J., J.~Melnikow, and D.~Chang (2002, 06).
\newblock {Reporting Number Needed to Treat and Absolute Risk Reduction in
  Randomized Controlled Trials}.
\newblock {\em JAMA\/}~{\em 287\/}(21), 2813--2814.

\bibitem[\protect\citeauthoryear{O'Muircheartaigh and Hedges}{O'Muircheartaigh
  and Hedges}{2013}]{Muircheartaigh2014GeneralizingApproach}
O'Muircheartaigh, C. and L.~Hedges (2013, 11).
\newblock Generalizing from unrepresentative experiments: A stratified
  propensity score approach.
\newblock {\em Journal of the Royal Statistical Society: Series C (Applied
  Statistics)\/}~{\em 63}.

\bibitem[\protect\citeauthoryear{Paillard, LOBO, Kolodyazhniy, Thirion, and
  Engemann}{Paillard et~al.}{2025}]{paillard2025measuring}
Paillard, J., A.~D.~R. LOBO, V.~Kolodyazhniy, B.~Thirion, and D.-A. Engemann
  (2025).
\newblock Measuring variable importance in heterogeneous treatment effects with
  confidence.
\newblock In {\em Forty-second International Conference on Machine Learning}.

\bibitem[\protect\citeauthoryear{Pearl}{Pearl}{2000}]{Pearl2000Book}
Pearl, J. (2000).
\newblock {\em Causality: {Models}, Reasoning, and Inference}.
\newblock Cambridge University Press.

\bibitem[\protect\citeauthoryear{Pearl}{Pearl}{2015}]{pearl2015findings}
Pearl, J. (2015).
\newblock Generalizing experimental findings.
\newblock {\em Journal of Causal Inference\/}~{\em 3\/}(2), 259--266.

\bibitem[\protect\citeauthoryear{Pearl and Bareinboim}{Pearl and
  Bareinboim}{2011}]{pearl2011transportability}
Pearl, J. and E.~Bareinboim (2011).
\newblock Transportability of causal and statistical relations: A formal
  approach.
\newblock In {\em Proceedings of the Twenty-Fifth AAAI Conference on Artificial
  Intelligence}, AAAI'11, pp.\  247–254. AAAI Press.

\bibitem[\protect\citeauthoryear{Pearl and Bareinboim}{Pearl and
  Bareinboim}{2014}]{Bareinboim2014ExternalValidity}
Pearl, J. and E.~Bareinboim (2014).
\newblock {External Validity: From Do-Calculus to Transportability Across
  Populations}.
\newblock {\em Statistical Science\/}~{\em 29\/}(4), 579 -- 595.

\bibitem[\protect\citeauthoryear{Richardson, Robins, and Wang}{Richardson
  et~al.}{2017}]{richardson2017modeling}
Richardson, T.~S., J.~M. Robins, and L.~Wang (2017).
\newblock On modeling and estimation for the relative risk and risk difference.
\newblock {\em Journal of the American Statistical Association\/}~{\em
  112\/}(519), 1121--1130.

\bibitem[\protect\citeauthoryear{Robertson, Steingrimsson, Joyce, Stuart, and
  Dahabreh}{Robertson et~al.}{2021}]{robertson2021center}
Robertson, S.~E., J.~A. Steingrimsson, N.~R. Joyce, E.~A. Stuart, and I.~J.
  Dahabreh (2021).
\newblock Center-specific causal inference with multicenter trials:
  reinterpreting trial evidence in the context of each participating center.
\newblock {\em arXiv preprint arXiv:2104.05905\/}.

\bibitem[\protect\citeauthoryear{Robins, Rotnitzky, and Scharfstein}{Robins
  et~al.}{2000}]{robins2000sensitivity}
Robins, J.~M., A.~Rotnitzky, and D.~O. Scharfstein (2000).
\newblock Sensitivity analysis for selection bias and unmeasured confounding in
  missing data and causal inference models.
\newblock In {\em Statistical models in epidemiology, the environment, and
  clinical trials}, pp.\  1--94. Springer.

\bibitem[\protect\citeauthoryear{Robinson}{Robinson}{1988}]{robinson1988semiparam}
Robinson, P. (1988).
\newblock Root- n-consistent semiparametric regression.
\newblock {\em Econometrica\/}~{\em 56\/}(4), 931--54.

\bibitem[\protect\citeauthoryear{Rothman}{Rothman}{2011}]{Rothman2011bookEpidemiologyIntrod}
Rothman, K.~J. (2011).
\newblock {\em Epidemiology: an introduction\/} (2 ed.).
\newblock Oxford University Press.

\bibitem[\protect\citeauthoryear{Rothman, Gallacher, and Hatch}{Rothman
  et~al.}{2013}]{Rothman1013WhyRepresentativeness}
Rothman, K.~J., J.~E. Gallacher, and E.~E. Hatch (2013, 08).
\newblock {Why representativeness should be avoided}.
\newblock {\em International Journal of Epidemiology\/}~{\em 42\/}(4),
  1012--1014.

\bibitem[\protect\citeauthoryear{Rothman and Greenland}{Rothman and
  Greenland}{2000}]{Rothman2000ModernEpidemiology}
Rothman, K.~J. and S.~Greenland (2000).
\newblock {\em Modern Epidemiology\/} (2 ed.).
\newblock Lippincott Williams and Wilkins.

\bibitem[\protect\citeauthoryear{Rothwell}{Rothwell}{2005}]{rothwell2005external}
Rothwell, P.~M. (2005).
\newblock External validity of randomised controlled trials: ``to whom do the
  results of this trial apply?''.
\newblock {\em The Lancet\/}~{\em 365}, 82--93.

\bibitem[\protect\citeauthoryear{Sackett, Deeks, and Altman}{Sackett
  et~al.}{1996}]{Sackett1996DownWO}
Sackett, D.~L., J.~J. Deeks, and D.~G. Altman (1996).
\newblock Down with odds ratios!
\newblock {\em Evidence Based Medicine\/}~{\em 1}, 164 -- 166.

\bibitem[\protect\citeauthoryear{Schulz, Altman, Moher, and Group*}{Schulz
  et~al.}{2010}]{schulz2010consort}
Schulz, K.~F., D.~G. Altman, D.~Moher, and C.~Group* (2010).
\newblock Consort 2010 statement: updated guidelines for reporting parallel
  group randomized trials.
\newblock {\em Annals of internal medicine\/}~{\em 152\/}(11), 726--732.

\bibitem[\protect\citeauthoryear{Schwartz, Woloshin, Dvorin, and
  Welch}{Schwartz et~al.}{2006}]{Schwartz2006ratio}
Schwartz, L.~M., S.~Woloshin, E.~L. Dvorin, and H.~G. Welch (2006, December).
\newblock Ratio measures in leading medical journals: structured review of
  accessibility of underlying absolute risks.
\newblock {\em BMJ (Clinical research ed.)\/}~{\em 333\/}(7581), 1248.

\bibitem[\protect\citeauthoryear{Sheps}{Sheps}{1958}]{Sheps1958ShallWe}
Sheps, M.~C. (1958).
\newblock Shall we count the living or the dead?
\newblock {\em New England Journal of Medicine\/}~{\em 259\/}(25), 1210--1214.
\newblock PMID: 13622912.

\bibitem[\protect\citeauthoryear{Simpson}{Simpson}{1951}]{simpson1951interpretation}
Simpson, E.~H. (1951).
\newblock The interpretation of interaction in contingency tables.
\newblock {\em Journal of the Royal Statistical Society: Series B
  (Methodological)\/}~{\em 13\/}(2), 238--241.

\bibitem[\protect\citeauthoryear{Sjölander, Dahlqwist, and
  Zetterqvist}{Sjölander et~al.}{2016}]{Sjolander2016NoteOnNoncollapsibility}
Sjölander, A., E.~Dahlqwist, and J.~Zetterqvist (2016, May).
\newblock A note on the noncollapsibility of rate differences and rate ratios.
\newblock {\em Epidemiology (Cambridge, Mass.)\/}~{\em 27\/}(3), 356—359.

\bibitem[\protect\citeauthoryear{Spiegelman, Khudyakov, Wang, and
  Vanderweele}{Spiegelman et~al.}{2017}]{Spiegelman2017letSubject}
Spiegelman, D., P.~Khudyakov, M.~Wang, and T.~Vanderweele (2017, 11).
\newblock Evaluating public health interventions: 7. let the subject matter
  choose the effect measure: Ratio, difference, or something else entirely.
\newblock {\em American journal of public health\/}~{\em 108}, e1--e4.

\bibitem[\protect\citeauthoryear{Spiegelman and VanderWeele}{Spiegelman and
  VanderWeele}{2017}]{Spiegelman2017Modeling}
Spiegelman, D. and T.~J. VanderWeele (2017).
\newblock Evaluating public health interventions: 6. modeling ratios or
  differences? let the data tell us.
\newblock {\em American Journal of Public Health\/}~{\em 107}, 1087--1091.

\bibitem[\protect\citeauthoryear{Splawa-Neyman, Dabrowska, and
  Speed}{Splawa-Neyman et~al.}{1990}]{SplawaNeyman1990Translation}
Splawa-Neyman, J., D.~M. Dabrowska, and T.~P. Speed (1990).
\newblock {On the Application of Probability Theory to Agricultural
  Experiments. Essay on Principles. Section 9}.
\newblock {\em Statistical Science\/}~{\em 5\/}(4), 465 -- 472.

\bibitem[\protect\citeauthoryear{Stang, Poole, and Bender}{Stang
  et~al.}{2010}]{stang2010common}
Stang, A., C.~Poole, and R.~Bender (2010).
\newblock Common problems related to the use of number needed to treat.
\newblock {\em Journal of clinical epidemiology\/}~{\em 63\/}(8), 820--825.

\bibitem[\protect\citeauthoryear{Stuart, Cole, Bradshaw, and Leaf}{Stuart
  et~al.}{2011}]{stuart2011use}
Stuart, E.~A., S.~R. Cole, C.~P. Bradshaw, and P.~J. Leaf (2011).
\newblock The use of propensity scores to assess the generalizability of
  results from randomized trials.
\newblock {\em Journal of the Royal Statistical Society: Series A (Statistics
  in Society)\/}~{\em 174}, 369--386.

\bibitem[\protect\citeauthoryear{St{\"u}rmer, Rothman, Avorn, and
  Glynn}{St{\"u}rmer et~al.}{2010}]{sturmer2010treatment}
St{\"u}rmer, T., K.~J. Rothman, J.~Avorn, and R.~J. Glynn (2010).
\newblock Treatment effects in the presence of unmeasured confounding: dealing
  with observations in the tails of the propensity score distribution—a
  simulation study.
\newblock {\em American journal of epidemiology\/}~{\em 172\/}(7), 843--854.

\bibitem[\protect\citeauthoryear{Tipton}{Tipton}{2013}]{tipton2013improving}
Tipton, E. (2013).
\newblock Improving generalizations from experiments using propensity score
  subclassification: Assumptions, properties, and contexts.
\newblock {\em Journal of Educational and Behavioral Statistics\/}~{\em 38},
  239--266.

\bibitem[\protect\citeauthoryear{Turrini and Bourgain}{Turrini and
  Bourgain}{2021}]{Bourgain2021Appraising}
Turrini, M. and C.~Bourgain (2021, August).
\newblock {Appraising screening, making risk in/visible. The medical debate
  over Non-Rare Thrombophilia (NRT) testing before prescribing the pill}.
\newblock {\em {Sociology of Health and Illness}\/}~{\em 43\/}(7), 1627--1642.

\bibitem[\protect\citeauthoryear{Vandenbroucke, Koster, Rosendaal, Briët,
  Reitsma, and Bertina}{Vandenbroucke
  et~al.}{1994}]{Vandenbroucke1994thrombosis}
Vandenbroucke, J., T.~Koster, F.~Rosendaal, E.~Briët, P.~Reitsma, and
  R.~Bertina (1994).
\newblock Increased risk of venous thrombosis in oral-contraceptive users who
  are carriers of factor v leiden mutation.
\newblock {\em The Lancet\/}~{\em 344\/}(8935), 1453--1457.
\newblock Originally published as Volume 2, Issue 8935.

\bibitem[\protect\citeauthoryear{VanderWeele and Robins}{VanderWeele and
  Robins}{2007}]{VanderWeele2007FourTypes}
VanderWeele, T.~J. and J.~M. Robins (2007, September).
\newblock Four types of effect modification: a classification based on directed
  acyclic graphs.
\newblock {\em Epidemiology (Cambridge, Mass.)\/}~{\em 18\/}(5), 561—568.

\bibitem[\protect\citeauthoryear{Wager and Athey}{Wager and
  Athey}{2018}]{wager2018estimation}
Wager, S. and S.~Athey (2018).
\newblock Estimation and inference of heterogeneous treatment effects using
  random forests.
\newblock {\em Journal of the American Statistical Association\/}~{\em
  113\/}(523), 1228--1242.

\bibitem[\protect\citeauthoryear{Wang}{Wang}{2022}]{wang2022homogeneity}
Wang, L. (2022).
\newblock On the homogeneity of measures for binary associations.
\newblock {\em arXiv preprint arXiv:2210.05179\/}.

\bibitem[\protect\citeauthoryear{Wen, Steingrimsson, Robertson, and
  Dahabreh}{Wen et~al.}{2025}]{Wen2025}
Wen, L., J.~A. Steingrimsson, S.~E. Robertson, and I.~J. Dahabreh (2025, Jul).
\newblock Multi-source analyses of average treatment effects with failure time
  outcomes.
\newblock {\em Lifetime Data Analysis\/}.
\newblock Online ahead of print.

\bibitem[\protect\citeauthoryear{Whittemore}{Whittemore}{1978}]{Whittemore1978Collapsibility}
Whittemore, A.~S. (1978).
\newblock Collapsibility of multidimensional contingency tables.
\newblock {\em Journal of the Royal Statistical Society. Series B
  (Methodological)\/}~{\em 40\/}(3), 328--340.

\bibitem[\protect\citeauthoryear{Xiao, Chen, Cole, MacLehose, Richardson, and
  Chu}{Xiao et~al.}{2022}]{xiao2022IsORPortable}
Xiao, M., Y.~Chen, S.~Cole, R.~MacLehose, D.~Richardson, and H.~Chu (2022).
\newblock Is or “portable” in meta-analysis? time to consider bivariate
  generalized linear mixed model.
\newblock {\em Journal of clinical epidemiology\/}~{\em 142}, 280.

\bibitem[\protect\citeauthoryear{Xiao, Chu, Cole, Chen, MacLehose, Richardson,
  and Greenland}{Xiao et~al.}{2021}]{xiao2021odds}
Xiao, M., H.~Chu, S.~Cole, Y.~Chen, R.~MacLehose, D.~Richardson, and
  S.~Greenland (2021).
\newblock Odds ratios are far from "portable": A call to use realistic models
  for effect variation in meta-analysis.

\bibitem[\protect\citeauthoryear{Yadlowsky, Pellegrini, Lionetto, Braune, and
  Tian}{Yadlowsky et~al.}{2021}]{yadlowsky2021estimation}
Yadlowsky, S., F.~Pellegrini, F.~Lionetto, S.~Braune, and L.~Tian (2021).
\newblock Estimation and validation of ratio-based conditional average
  treatment effects using observational data.
\newblock {\em Journal of the American Statistical Association\/}~{\em
  116\/}(533), 335--352.

\bibitem[\protect\citeauthoryear{Yule}{Yule}{1934}]{yule1934some}
Yule, G.~U. (1934).
\newblock On some points relating to vital statistics, more especially
  statistics of occupational mortality.
\newblock {\em Journal of the Royal Statistical Society\/}~{\em 97\/}(1),
  1--84.

\end{thebibliography}

\appendix

\newpage

\onecolumn

\begin{center}
    {\Large \textbf{Supplementary Materials}}
\end{center}

\section{Treatment effect measures}\label{appendix:list-of-measures}

\textit{This section completes Section~\ref{sec:formalization-and-key-contributions} (and more specially Section~\ref{subsec:causal-measures-presentation}) by exposing the different treatment (or causal) effect measures.}

\subsection{About the definition of causal measures}

Recall that all causal measures used in this paper or present in applied medical work can be defined as follows. 
\begin{definition}[Causal effect measures -- \cite{Pearl2000Book}]
Assuming a certain joint distribution of potential outcomes $P(Y^{(0)}, Y^{(1)})$, which implies that a certain treatment $A$ of interest is considered, we denote by $\tau^P$ any functional of the joint distribution of potential outcomes. More precisely,

\begin{align}
\mathcal{P} \quad & \rightarrow \mathbb{R} \\
P(Y^{(0)}, Y^{(1)}) & \mapsto \tau^{P}
\end{align}
\end{definition}

This definition is also valid for any subpopulation, as for any covariate $X$, $\tau^{P}(X)$ is defined as a functional of $P(Y^{(0)}, Y^{(1)}\mid X)$. This definition is the one used in this article.

\textit{Why do we say that those measures are causal?} Note that the same definition could have been made on the distribution $P(A,Y)$, comparing expectation on two distributions: $P(Y \mid A=1)$ and $P(Y \mid A=0)$. For example, within the statistical community, the odds ratio is often known as the strength of the association between two events, $A=1$ and $A=0$ and therefore defined as:

\begin{align*}
    OR := \frac{P(Y = 1 \mid A=1)}{P(Y = 0 \mid A=1)} \cdot \frac{P(Y = 0 \mid A=0)}{P(Y = 1 \mid A=0)}.
\end{align*}

In such a situation, the OR measure would be an associational measure and not a causal measure, except if there is no confounding in the distribution considered (for e.g. in the case of a Randomized Controlled Trial). To avoid discussion about confounding, in this paper we never consider distribution such as $Y \mid A, X$ or $Y \mid A$. We rather consider $Y^{(a)}  \mid X$. For any new reader discovering the potential outcomes framework, we refer to the first chapters of \cite{imbens2015causal} for a clear and complete exposition of this notations inherited from Neyman. Note that \cite{Didelez2021collapsibility} make the same distinction when discussing collapsibility questions.

\begin{figure}[!htbp]
\centering
\begin{subfigure}{.4\linewidth}
\includegraphics[width=\linewidth]{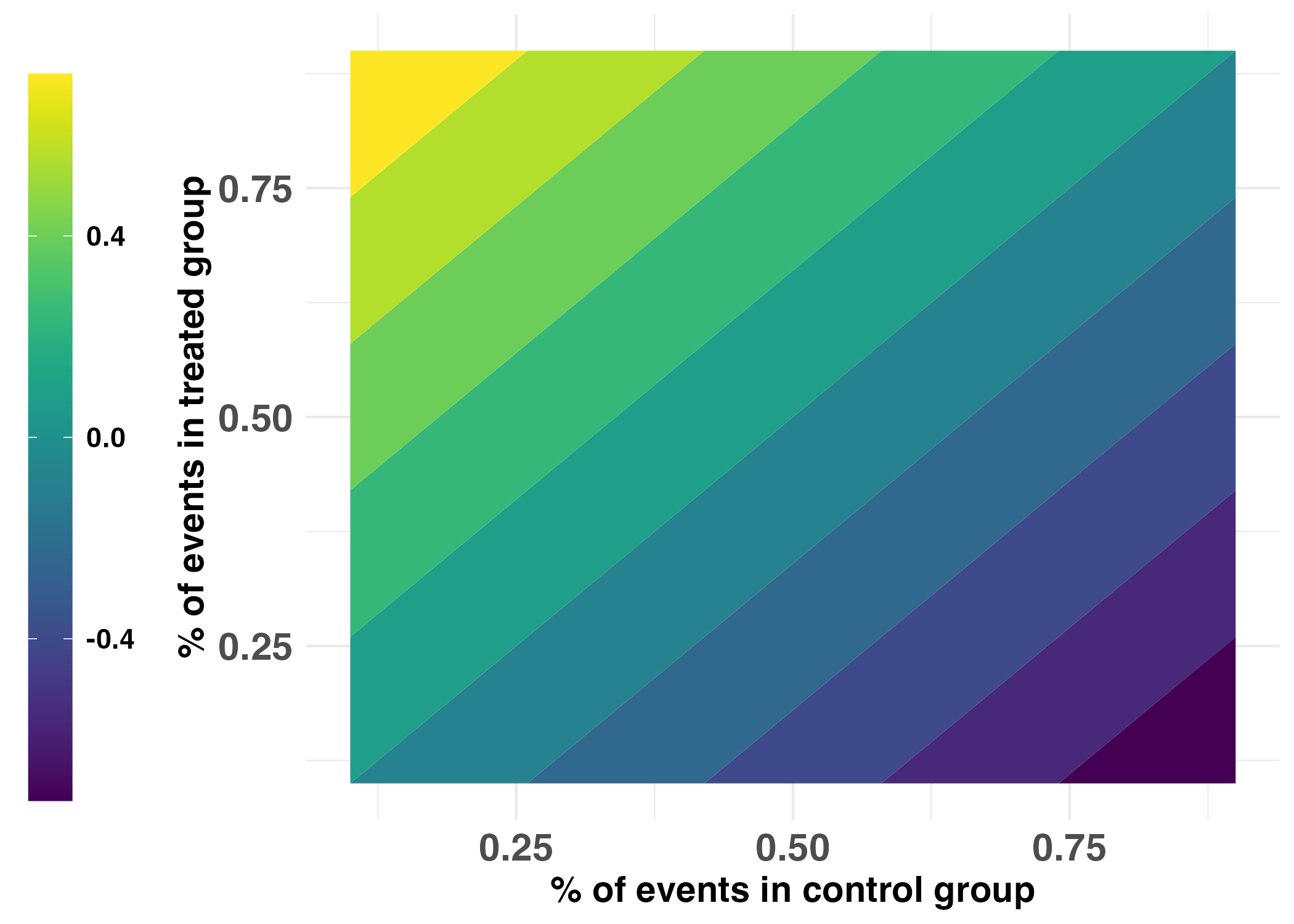}
\caption{\textbf{Risk Difference (RD)}}\label{fig:RD}
\end{subfigure}%
\hspace{1cm}
\begin{subfigure}{.4\textwidth}% glorified minipage
\includegraphics[width=\linewidth]{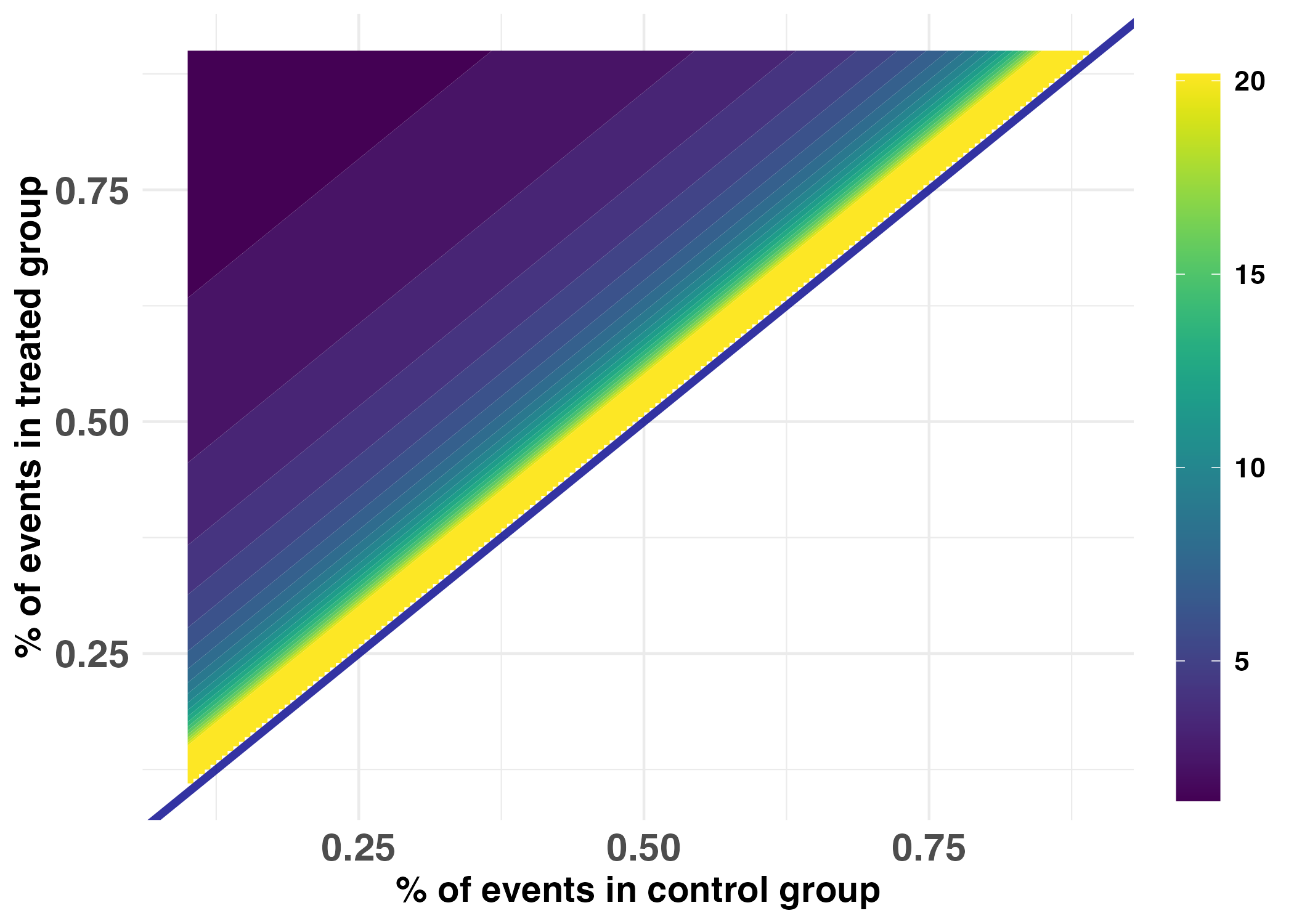}
\caption{\textbf{Number Needed to Treat (NNT)}}\label{fig:NNT}
\end{subfigure}%
\\
\vspace{0.5cm}
\begin{subfigure}{.4\linewidth}
\includegraphics[width=\linewidth]{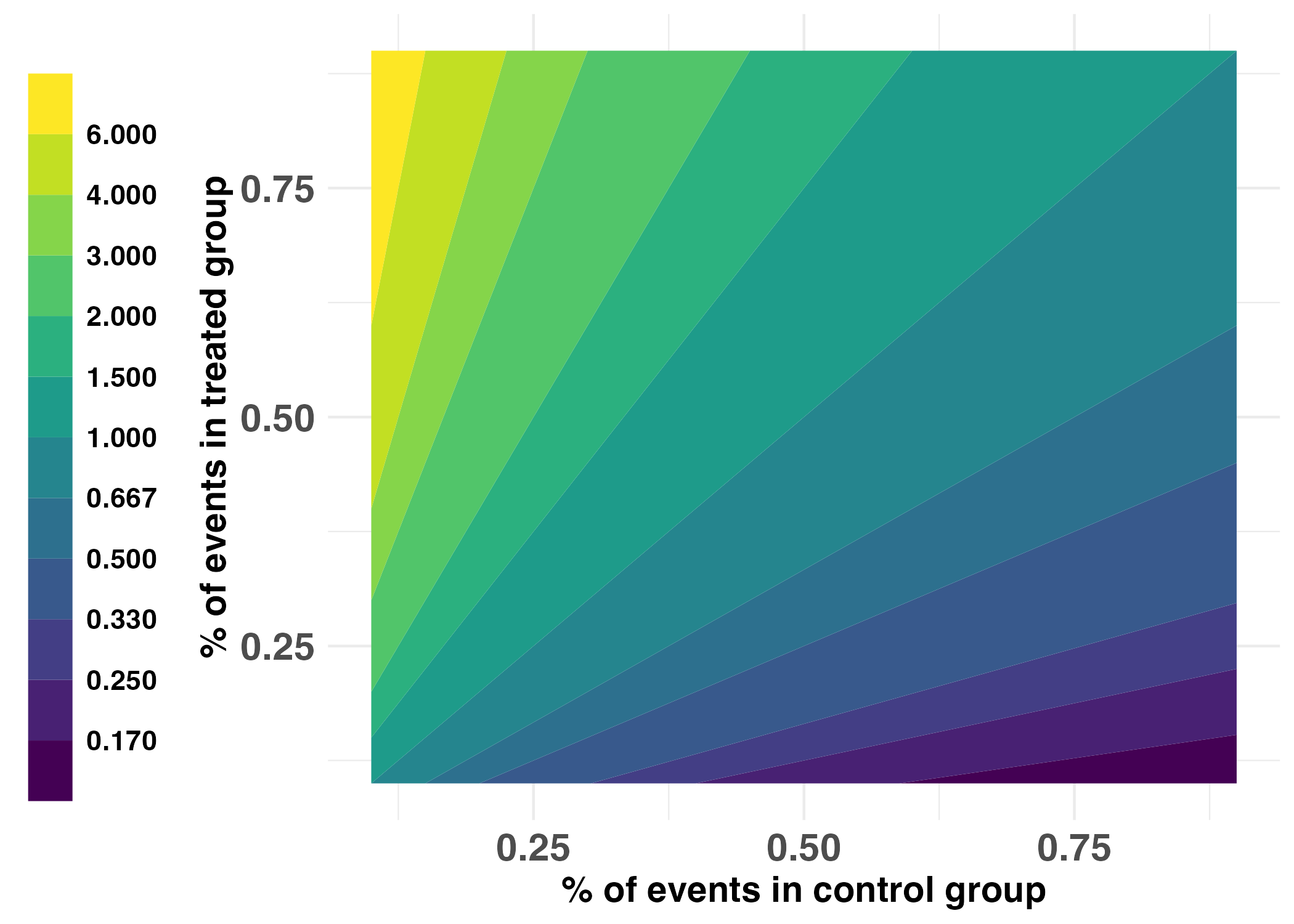}
\caption{\textbf{Risk Ratio (RR)}}\label{fig:RR}
\end{subfigure}%
\hspace{1cm}
\begin{subfigure}{.4\linewidth}% glorified minipage
\includegraphics[width=\linewidth]{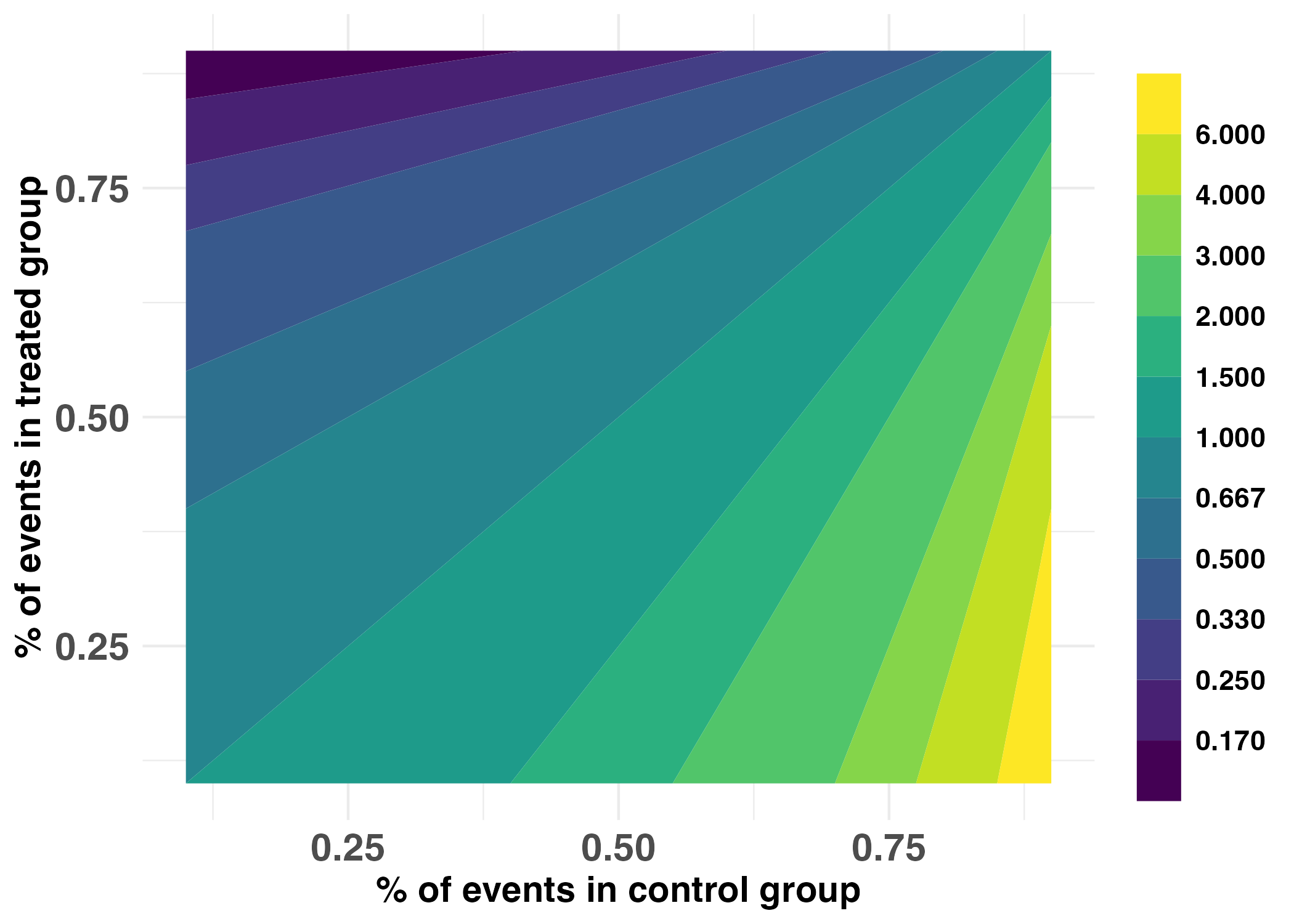}
\caption{\textbf{Survival Ratio (SR)}}\label{fig:SR}
\end{subfigure}%
\\
\vspace{0.5cm}
\begin{subfigure}{.4\linewidth}
\includegraphics[width=\linewidth]{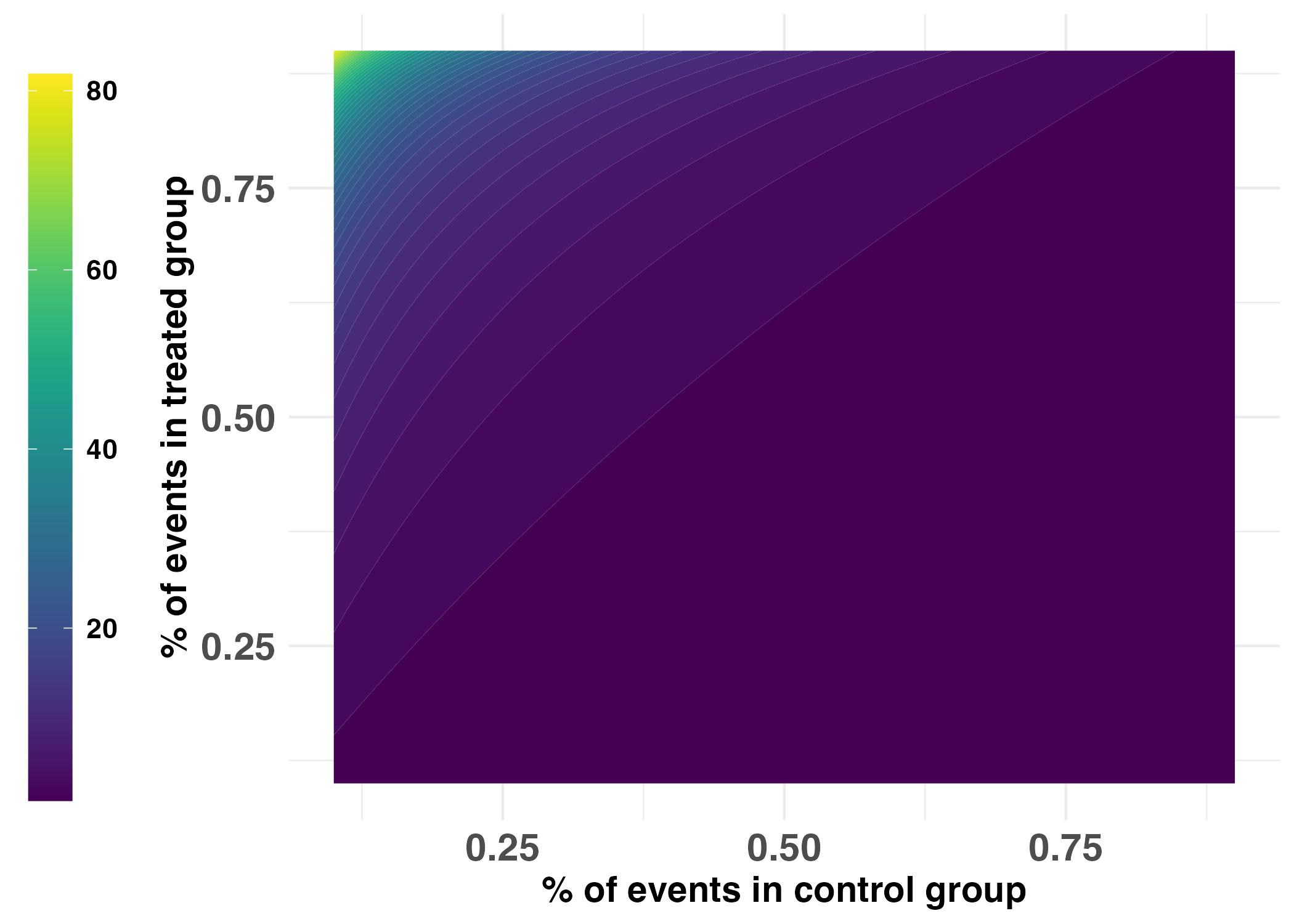}
\caption{\textbf{Odds Ratio (OR)}}\label{fig:OR}
\end{subfigure}%
\hspace{1cm}
\begin{subfigure}{.4\linewidth}% glorified minipage
\includegraphics[width=\linewidth]{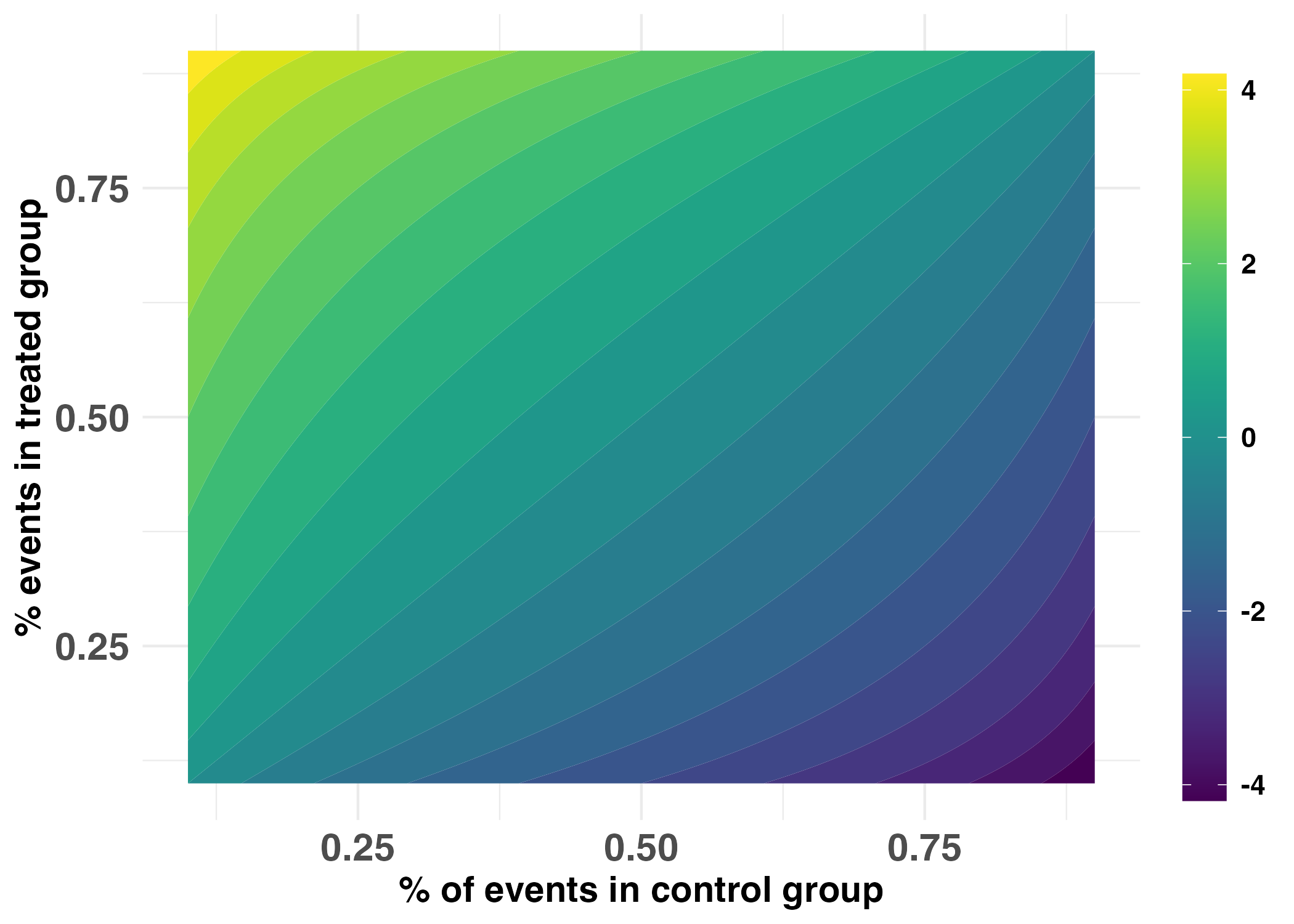}
\caption{\textbf{Log Odds Ratio (log-OR)}}\label{fig:logOR}
\end{subfigure}%
\\
\vspace{1cm}
\begin{minipage}{0.49\textwidth}

\caption{\textbf{Plots of the ranges of the different metrics as a function of the proportion of events in control group}, namely $\mathbb{E}[ Y^{(0)}]$ (x-axis), and of the proportion of events in treated group, namely $\mathbb{E}[ Y^{(1)}]$ (y-axis). See Subfigure~\ref{fig:how-to-read}. As both the colors and the different scale illustrate, the ranges of the effect considerably differ with the metric chosen. Similar plots can be found under the name "\textit{L'Abbé plots}" \citep{Labbe1987MetaAnalysis, Jimnez1997GraphicalDisplayUseful,Deeks2022IssuesInSelection} in research works related to meta-analysis.} %Note that for the NNT (Figure~\ref{fig:NNT}) we only represented the quarter of the plot when an event encodes death as usually done in the medical field, with threshold at 20 for readability of the scale. 
\label{fig:big-plot-with-all-metrics}
\end{minipage}
\begin{minipage}{0.49\textwidth}
  \begin{flushright}
    \begin{subfigure}{1.1\linewidth}
\centering
\includegraphics[width=\textwidth]{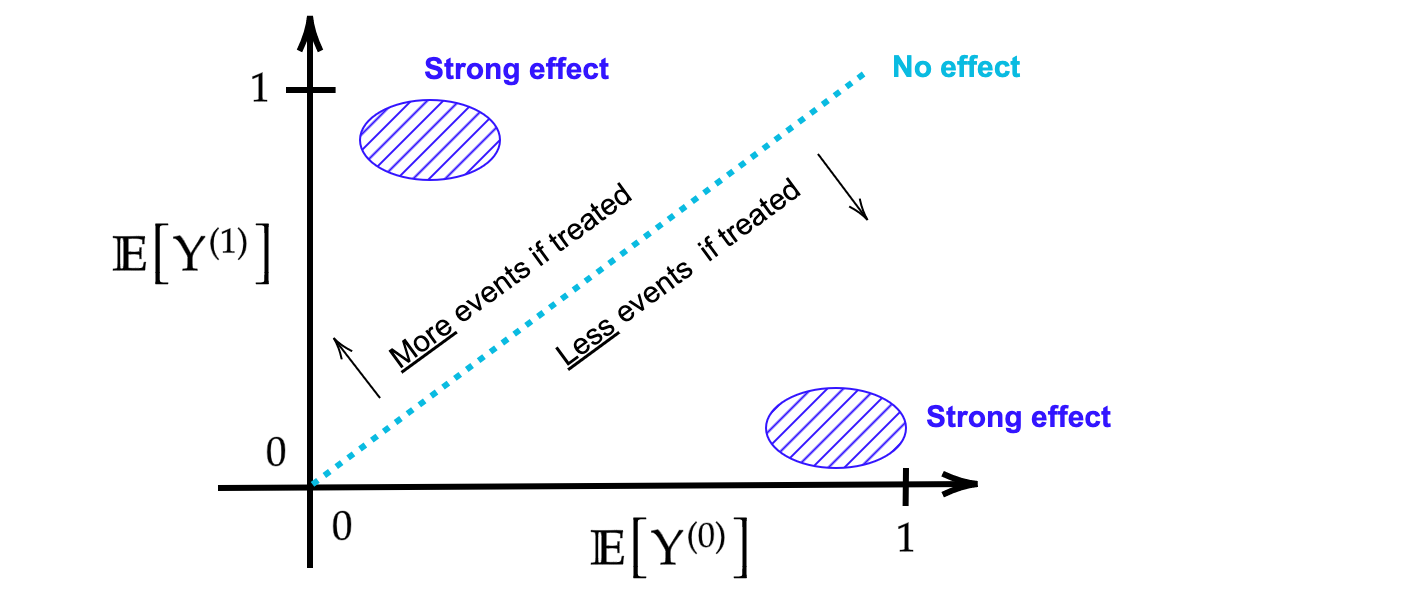}
\caption{\textbf{Legend}}\label{fig:how-to-read}
\end{subfigure}%
  \end{flushright}
\end{minipage}
\end{figure}

\subsection{Common treatment effect measures}

As highlighted by Definition~\ref{def:causal-measure}, many measures could be proposed. Here we detail common measures found in applied works and propose an illustration for the case of binary outcomes (Figure~\ref{fig:big-plot-with-all-metrics}). 
Most of the time, the distinction is made on whether or not the measure is an absolute or a relative effect.

\subsubsection{Absolute measures}

\begin{definition}[Risk Difference (RD)]
The risk difference is a causal effect measure defined as the difference of the expectations (also called risks),

\begin{equation*}
    \tau_{\text{\tiny RD}} = \mathbb{E}[Y^{(1)}] - \mathbb{E}[Y^{(0)}].
\end{equation*}

%Note that the risk difference can also be written, $\tau_{\text{\tiny RD}} = \mathbb{E}[ Y^{(1)} - Y^{(0)}] = \mathbb{P}[Y^{(1)} = 1] - \mathbb{P}[Y^{(0)} = 1]$.
\end{definition}
RD is also named Absolute Risk Reduction (ARR), Absolute Effect (AE), Absolute Difference (AD), or Excess Risk (ER).

\begin{definition}[Number Needed to Treat (NNT)]\label{def:nnt}

The number needed to treat (NNT) is a causal effect measure defined as the average number of individuals or observations who need to be treated to prevent one additional outcome,

\begin{equation*}
  \tau_{\text{\tiny NNT}}  = \frac{1}{\mathbb{E}[Y^{(1)}= 1] - \mathbb{E}[Y^{(0)}=1]}
\end{equation*}

\end{definition}
 The Number Needed to Treat (NNT) has been proposed as a measure rather recently \citep{Laupacis1988AnAssessmentOfClinically}. A harmful treatment is usually called the Number Needed to Harm (NNH) and made positive.

\subsubsection{Relative measures}

\begin{definition}[Risk Ratio]
The Risk Ratio is a causal effect measure defined as the ratio of the expectations,

\begin{equation*}
     \tau_{\text{\tiny RR}} = \frac{\mathbb{E}[Y^{(1)}]}{\mathbb{E}[Y^{(0)}]} 
\end{equation*}

\end{definition}

The Risk Ratio (RR) is also named Relative Risk (RR), Relative Response (RR), or Incidence Proportion Ratio (IPR)

\begin{definition}[Survival Ratio]
The survival ratio is a causal effect measure defined as the Risk Ratio were labels are swapped,

\begin{equation*}
     \tau_{\text{\tiny SR}} = \frac{1-\mathbb{E}[Y^{(1)}]}{1-\mathbb{E}[Y^{(0)}]} 
\end{equation*}

\end{definition}

It is possible to introduce a measure that captures both the Risk Difference, but normalized by the baseline.

\begin{definition}[Excess relative risk (ERR)]

\begin{equation*}
  \tau_{\text{\tiny ERR}}  = \frac{\mathbb{E}[Y^{(1)}] - \mathbb{E}[Y^{(0)}]}{\mathbb{E}[Y^{(0)}]}
\end{equation*}

\end{definition}

The Excess relative risk (ERR) has been proposed by \cite{cole1971attributable}. Note that,

\begin{equation*}
    \tau_{\text{\tiny ERR}}   = \tau_{\text{\tiny RR}} - 1 .
\end{equation*}

\begin{definition}[Relative Susceptibility (RS)]
\begin{equation*}
    \tau_{\text{\tiny RS}}  := \frac{\mathbb{E}[Y^{(1)}] - \mathbb{E}[Y^{(0)}]}{1-\mathbb{E}\left[Y^{(0)}\right]}.
\end{equation*}
\end{definition}

Note that,

\begin{equation*}
      \tau_{\text{\tiny RS}} =  1 - \tau_{\text{\tiny SR}}.
\end{equation*}

Finally, another measure is often used based on odds. Odds are a way of representing probability in particular for betting. For example a throw with a die will produce a one with odds 1:5. The odds is the ratio of the probability that the event occurs to the probability it does not.

\begin{definition}[Odds Ratio (OR)]
The odds ratio is a causal effect measure defined as the ratio of the odds of the treated and control groups,
\begin{equation*}
     \tau_{\text{\tiny OR}} := \frac{\mathbb{P}[Y^{(1)} = 1]}{1-\mathbb{P}[Y^{(1)} = 1]}\, \left(  \frac{\mathbb{P}[Y^{(0)} = 1]}{1-\mathbb{P}[Y^{(0)} = 1]}\right)^{-1}.
\end{equation*}

\end{definition}
Odds Ratio (OR) is sometimes named Marginal Causal Odds Ratio (MCOR). This is by opposition to a conditional Odds Ratio, being defined as,

\begin{equation*}
     \tau_{\text{\tiny OR}}(X) := \frac{\mathbb{E}[Y^{(1)} = 1 \mid X = x]}{1-\mathbb{E}[Y^{(1)} = 1\mid X = x]}\, \left(  \frac{\mathbb{E}[Y^{(0)} = 1\mid X = x]}{1- \mathbb{E}[Y^{(0)} = 1\mid X = x]}\right)^{-1},
\end{equation*}

often used due to its homogeneity when considering a logistic discriminative model of the outcome (see Section~\ref{proof:non-collapsibility-OR} for a detailed proof).
The OR is known to approximate the RR at low baseline (see for example the illustrative example of Table~\ref{tab:introduction-diastolic}).

\begin{proof}
{\footnotesize {\color{Blue} 
    $
   \mathbb{P}[Y^{(1)} = 1] \le \mathbb{P}[Y^{(0)} = 1] \ll 1 \implies \tau_{\text{\tiny OR}}  = \frac{\mathbb{P}[Y^{(1)} = 1]}{1-\mathbb{P}[Y^{(1)} = 1]}\cdot  \frac{1-\mathbb{P}[Y^{(0)} = 1]}{\mathbb{P}[Y^{(0)} = 1]} \approx\frac{\mathbb{P}[Y^{(1)} = 1]}{1}\cdot \frac{1}{\mathbb{P}[Y^{(0)} = 1]} =  \tau_{\text{\tiny RR}}$. }}
\end{proof}

These derivations can be found as late as in the 50's in case-control studies about lung cancer \citep{cornfield1951method}.
Also note that,

\begin{equation*}
    \tau_{\text{\tiny OR}} = \tau_{\text{\tiny RR}} \cdot \tau_{\text{\tiny SR}}^{-1} .
\end{equation*}

\begin{proof}

{\footnotesize {\color{Blue} 
    \begin{align*}
           \tau_{\text{\tiny OR}}  &= \frac{\mathbb{P}[Y^{(1)} = 1]}{1-\mathbb{P}[Y^{(1)} = 1]}\, \left(  \frac{\mathbb{P}[Y^{(0)} = 1]}{1-\mathbb{P}[Y^{(0)} = 1]}\right)^{-1} \\
           &= \frac{\mathbb{P}[Y^{(1)} = 1]}{\mathbb{P}[Y^{(1)} = 0]}\, \left(  \frac{\mathbb{P}[Y^{(0)} = 1]}{\mathbb{P}[Y^{(0)} = 0]}\right)^{-1}\\
           &=  \frac{\mathbb{P}[Y^{(1)} = 1]}{\mathbb{P}[Y^{(1)} = 0]}\,  \frac{\mathbb{P}[Y^{(0)} = 0]}{\mathbb{P}[Y^{(0)} = 1]}\\
           &= \frac{\mathbb{P}[Y^{(1)} = 1]}{\mathbb{P}[Y^{(0)} = 1]}\,  \frac{\mathbb{P}[Y^{(0)} = 0]}{\mathbb{P}[Y^{(1)} = 0]} \\
           &=  \tau_{\text{\tiny RR}} \cdot \tau_{\text{\tiny SR}}^{-1} 
    \end{align*}
    }}
\end{proof}

One can observe on Figure~\ref{fig:big-plot-with-all-metrics} (see subplots Figures~\ref{fig:OR} and \ref{fig:logOR}) the range on which the OR varies depends on the direction of the effect. Therefore, the OR is often presented encapsulated in a logarithm.
\begin{definition}[Log Odds Ratio (log-OR)]
    \begin{equation*}
        \tau_{\text{\tiny log-OR}}  := \operatorname{log}\left( \frac{\mathbb{P}[Y^{(1)} = 1]}{\mathbb{P}[Y^{(1)} = 0]}\right) - \operatorname{log}\left( \frac{\mathbb{P}[Y^{(0)} = 1]}{\mathbb{P}[Y^{(0)} = 0]}\right)
    \end{equation*}
\end{definition}

\section{Definitions found in the literature}\label{appendix:other-formal-definitions}

This section completes Section~\ref{section:causal-metrics-properties} with formalization of homogeneity of effects, heterogeneity of effects, and collapsibility we have found in the literature. Doing so, we highlight that definitions can be more or less formal, and therefore can lead to different apprehension of phenomenons, in particular collapsibility. 
%The purpose of the definitions given in Section~\ref{section:causal-metrics-properties} is to provide  intuitions behind all previously proposed definitions, while uniformizing them.

\subsection{Effect modification}

\textit{This section supports definitions proposed in Section~\ref{subsec:homogeneity}.}

Note that effect modification or heterogeneity is mentioned in many places, but not always clearly defined. This is highlighted by the following quote:

\begin{quote}
    We searched the National Library of Medicine Books, National Library of Medicine Catalog, Current Index to Statistics database, ISI web of science, and websites of 25 major regulatory agencies and organizations for papers and guidelines on study design, analysis and interpretation of treatment effect heterogeneity. Because there is not standard terminology for this topic, a structured search strategy was not sensitive nor specific and we found many resources through “snowball” searching, that is, reviewing citations in, and citations of, key methodological and policy papers. -- \citep{lesko2018considerations}
\end{quote}

\subsubsection{Definitions of heterogeneity of effect or effect modification found in the literature}

\begin{definition}[\cite{Rothman2011bookEpidemiologyIntrod}, page 51]\label{def-heterogeneity-Greenland}
    Suppose we divide our cohort into two or more distinct categories, or strata. In each stratum, we can construct an effect measure of our choosing. These stratum-specific effect measures may or may not equal on another. Rarely would we have any reason to suppose that they do equal one another.
    If indeed they are not equal, we say that the effect measure is \textit{heterogeneous} or \textit{modified} across strata.
    If they are equal we say that the measure is \textit{homogeneous}, \textit{constant}, or \textit{uniform} across strata.
    A major point about effect-measure modification is that, if effects are present, it will usually be the case that only one or none of the effect measures will be uniform across strata.
\end{definition}

\begin{definition}[\cite{VanderWeele2007FourTypes}]\label{def-VanderWeele2007FourTypes}
We say that a variable $Q$ is a treatment effect modifier for the causal risk difference of $A$ on $Y$ if $Q$ is not affected by $A$ and if there exist two levels of $A$, $a_0$ and $a_1$, such that $\mathbb{E}\left[ Y^{(a_1)}   \mid Q = q\right]-\mathbb{E}\left[ Y^{(a_0)}   \mid Q = q\right]$ is not constant in $q$.
\end{definition}

\subsubsection{Effect heterogeneity depends on the chosen scale: an illustration}

\begin{figure}[H]
    \centering
    \includegraphics[width=0.8\textwidth]{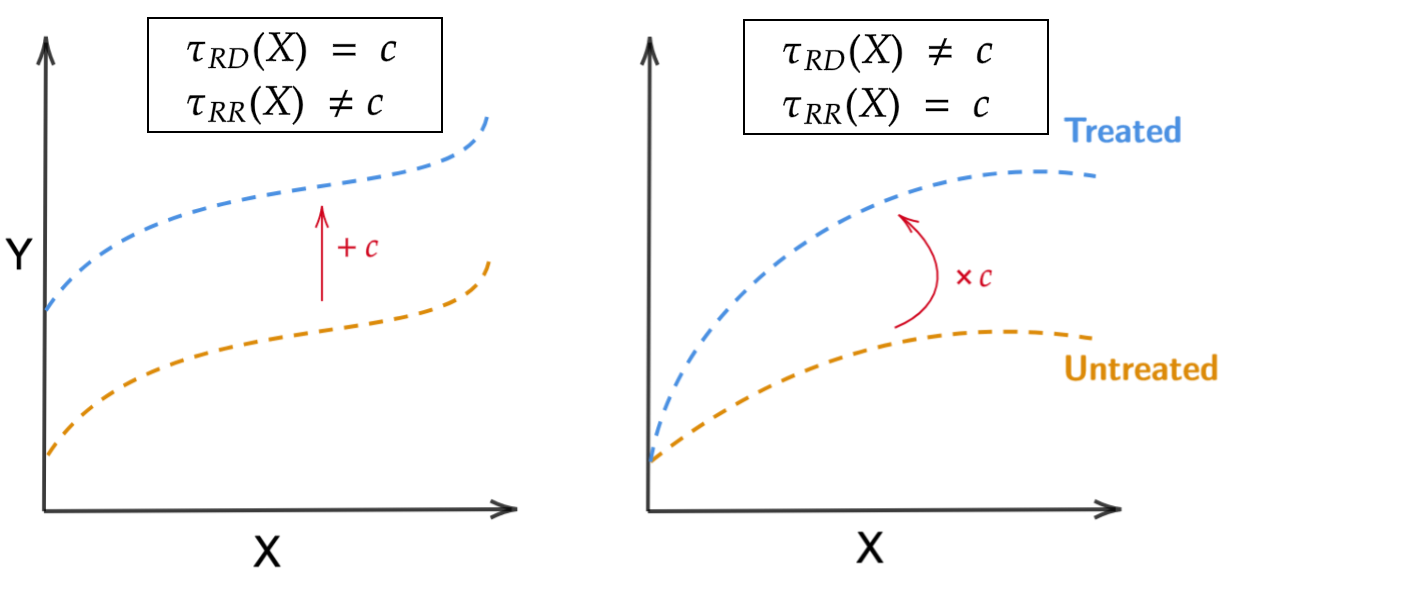}
    \caption{\textbf{Heterogeneity of a treatment effect depends on the scale}: Illustrative schematics where the data discriminative model on the left leads to a constant treatment effect on the absolute scale (RD) when conditioning on $X$, while on the data discriminative model on the right leads to an homogeneous treatment effect on the relative scale (RR). In both of the situations, homogeneity of treatment effect of one scale (RR or RD) leads to heterogeneity on the other scale. Note that a similar schematic is presented in \cite{Rothman2011bookEpidemiologyIntrod} (see their Figure 11–1, p. 199)}
    \label{fig:hetero-schematic}
\end{figure}

A treatment effect heterogeneity depends on the causal measure $\tau$ chosen (the scale). 
This idea is well-known in epidemiology \citep{Rothman2011bookEpidemiologyIntrod, lesko2018considerations}.
To be convinced by such phenomenon, the drawing in Figure~\ref{fig:hetero-schematic} illustrates what could be two data discriminative models leading to two different homogeneity and heterogeneity patterns.

\subsection{Different definitions of collapsibility in the literature}
\label{appendix-def_collapsibility}

\textit{This section supports definitions proposed in Section~\ref{subsec:collapsibility}.}
\subsubsection{Unformal definitions}
We have found many unformal definitions in the literature, such as:

\begin{quote}
    In a single study with a non-confounding stratification variable, if the stratum-specific effects are homogenous, then they are expected to be the same as the crude effect, a desirable property known as collapsibility of an effect measure.  -- \citep{xiao2022IsORPortable}
\end{quote}

\begin{quote}
    RR but, not OR, have a mathematical property called collapsibility; this means the size of the Risk Ratio will not change if adjustment is made for a variable that is not a confounder. -- \citep{Cummings2009RelativeMeritsRRAndOR}
\end{quote}

and

\begin{quote}
    Collapsibility means that in the absence of confounding, a weighted average of stratum-specific ratios (e.g., using Mantel-Haenszel methods) will equal the ratio from a single 2 by 2 table of the pooled (collapsed) counts from the stratum-specific tables. This means that a crude (unadjusted) ratio will not change if we adjust for a variable that is not a confounder.  -- \citep{Cummings2009RelativeMeritsRRAndOR}
\end{quote}

\subsubsection{Formal definitions}
\begin{definition}[Strict collapsibility \cite{pearl1999collapsibility}]\label{def:strict-collaps-pearl-greenland}
We say a measure of association between $ Y^{(0)}$ and $ Y^{(1)}$ is strictly collapsible accross $X$ if it is constant accross the strata (subtables) and this constant value equals the value obtained from the marginal table.
\end{definition}
Similar definition as Definition~\ref{def:strict-collaps-pearl-greenland} have been proposed in \cite{liu2022correct, Didelez2021collapsibility}.
\begin{definition}[\cite{Pearl2000Book}]\label{def:collapsibility-pearl}
Let $\tau \left(P\left( Y^{(0)}, Y^{(1)} \right) \right)$ be any functional that measures the association between $Y^{(0)}$ and $Y^{(1)}$ in the joint distribution  $P\left( Y^{(0)}, Y^{(1)} \right)$. We say that $\tau$ is collapsible on a variable $V$ if

\begin{align*}
    \mathbb{E}\left[ \tau \left(P\left( Y^{(0)}, Y^{(1)} \mid V \right)  \right) \right] = \tau \left(P\left( Y^{(0)}, Y^{(1)} \right) \right)
\end{align*}
\end{definition}

Note that in his book, Judea Pearl rather present the definition of collapsibility with respect two any two covariates, not necessarily potential outcomes. Indeed, collapsibility is a statistical concept at first. As in this work we are explicitely concerned with causal metrics, this definition has been written here with potential outcomes.

\begin{definition}
[\cite{Huitfeldt2019collapsible}]\label{def:collapsibility-huitfeldt}
Let $\tau \left(P\left( Y^{(0)}, Y^{(1)} \right) \right)$ be any function of the parameters $Y^{(0)}$ and $Y^{(1)}$ in the joint distribution $P\left( Y^{(0)}, Y^{(1)} \right)$. We say that $\tau$ is collapsible on a variable $V$ with weights $w_v$ if,

\begin{align*}
    \frac{\sum_v w_v \tau \left(P\left( Y^{(0)}, Y^{(1)} \right) \mid V = v \right)}{\sum_v w_v} &= \tau \left(P\left( Y^{(0)}, Y^{(1)} \right) \right)
\end{align*}
 \end{definition}

\begin{definition}[\cite{Didelez2021collapsibility}]\label{def:collapsibility-didelez}
Let $\tau=\tau \left(P\left( Y^{(0)}, Y^{(1)} \right) \right)$ be a measure of association between $Y^{(0)}$ and $Y^{(1)}$; that is, $\tau$ is a functional of the joint distribution $P\left( Y^{(0)}, Y^{(1)} \right)$. Let $\tau_x= \tau(Y,A\mid X=x)$  be a measure of conditional association between $Y$ and $A$ given $X=x$; that is, $\tau_x$ is a functional of the conditional distribution $P(Y,A \mid X = x)$. The measure $\tau$ is called \textit{collapsible over $X$}, if $\tau$ is a weighted average of $\tau_x$ for $x \in \mathds{X}$. Strict collapsibility demands that $\tau=\tau_x$.
\end{definition}
%Note that under this definition, collapsibility seems to be defined over a certain covariate $X$. Doing the odds-ratio could be defined as a strictly collapsible as soon as the outcome is generated from a parametric logistic model such as in \eqref{eq:typical-model-used-binary-Y} (this is explicitely given as an example in their work). Note that in our work we define collapsibility differently, such that OR can not be considered as a collapsible measure.

\section{Proofs}

In this section we detail all the derivations needed to understand the results of this article. 

\subsection{Collapsibility}\label{proof:collapsibilty}

\textit{Note that not all proofs are novel work. Collapsibility results have been reported multiple times and in multiple ways as explained in the main paper. For clarity we still recall them. We indicate when the proofs are not novel or when similar proofs exist elsewhere. When we indicate nothing, this means that we have not found those results in other published work.}

\subsubsection{Proof of Lemma~\ref{lemma:direct-collapsibility-RD}}

\textit{N.B: The proof for the direct collapsibility of the RD \underline{is not} a novel contribution.}

\begin{proof}

\begin{align*}
    \tau_{\text{\tiny RD}}^P&= \mathbb{E}\left[Y^{(1)} - Y^{(0)}  \right] && \text{By definition} \\
&= \mathbb{E}\left[  \mathbb{E}\left[Y^{(1)} - Y^{(0)} \mid X  \right]  \right] && \text{Law of total expectation} \\
&= \mathbb{E}\left[ \tau_{\text{\tiny RD}}^P\left(X \right) \right].
\end{align*}

\end{proof}

\paragraph*{Remark} To observe the phenomenon as weighting, one can also write this last quantity as an integral.

\begin{align*}
   \mathbb{E}\left[  \mathbb{E}\left[Y^{(1)} - Y^{(0)} \mid X  \right]  \right] 
&=  \int_{\mathds{X}}  \mathbb{E}\left[Y^{(1)} - Y^{(0)} \mid X  \right] f(x)\,dx  &&\text{Re-writing} \\
&= \int_{\mathds{X}}  \tau_{\text{\tiny RD}}^P\left( x \right)f(x)\,dx.
\end{align*}

Here, one can observe that weights are the density of $x$ in the population. Most of the time \citep{pearl2011transportability, huitfeldt2018choice, Didelez2021collapsibility} express such quantity on categorical covariates $X$, therefore using a sum.

\subsubsection{Proof of Lemma~\ref{lemma:collapsibility-of-RR-SR}}

\textit{N.B: The proof for the collapsibility of the RR and SR are extensions of \cite{Huitfeldt2019collapsible}.}

\paragraph*{General comment}
In this subsection we detail the proof for collapsibility of the RR, and SR. 
Before detailing the proof, we want to highlight why the RR (and SR) is not directly collapsible.

\begin{align*}
    \tau_{\text{\tiny RR}}^P &= \frac{\mathbb{E}\left[ Y^{(1)}\right]}{\mathbb{E}\left[ Y^{(0)}\right]}  \\
    &= \frac{\mathbb{E}\left[ \mathbb{E}\left[ Y^{(1)} \mid X  \right]\right]}{\mathbb{E}\left[ \mathbb{E}\left[ Y^{(0)} \mid X  \right]\right]}  \\
    & \neq \mathbb{E}\left[ \frac{\mathbb{E}\left[ Y^{(1)} \mid X\right]}{\mathbb{E}\left[ Y^{(0)} \mid X\right]}\right],
\end{align*}
%\es{ici, on a envie d'être un peu plus précis : donner des cas dans lesquels ce n'est pas vrai}
in all generality. For example, assuming that $\mathbb{E}\left[ Y^{(0)} \mid X\right]$ and $\mathbb{E}\left[ Y^{(1)} \mid X\right]$ are independent, we have 
\begin{align*}
\mathbb{E}\left[ \frac{\mathbb{E}\left[ Y^{(1)} \mid X\right]}{\mathbb{E}\left[ Y^{(0)} \mid X\right]}\right] & = \mathbb{E}[Y^{(1)}]\mathbb{E}\left[ \frac{1}{\mathbb{E}\left[ Y^{(0)} \mid X\right]}\right] > \frac{\mathbb{E}[Y^{(1)}]}{\mathbb{E}[Y^{(0)}]} = \tau_{\text{\tiny RR}}^P,
\end{align*}
by Jensen inequality, assuming additionally that $\mathbb{E}\left[ Y^{(0)} \mid X\right] > 0$. 

\paragraph*{Risk Ratio (RR)}

\begin{proof}

\begin{align*}
    \tau_{\text{\tiny RR}}^P &= \frac{\mathbb{E}\left[ Y^{(1)}\right]}{\mathbb{E}\left[ Y^{(0)}\right]}  && \text{By definition of the RR}\\
    &= \frac{\mathbb{E}\left[ \mathbb{E}\left[ Y^{(1)} \mid X \right]\right]}{\mathbb{E}\left[ Y^{(0)}\right]}  && \text{Law of total expectation used on $\mathbb{E}\left[ Y^{(1)}\right]$} \\
    &= \frac{\mathbb{E}\left[\frac{ \mathbb{E}\left[ Y^{(1)} \mid X \right]}{ \mathbb{E}\left[ Y^{(0)} \mid X \right]} \mathbb{E}\left[ Y^{(0)} \mid X \right]\right]}{\mathbb{E}\left[ Y^{(0)}\right]}  && \text{$\mathbb{E}\left[ Y^{(0)} \mid X \right] \neq 0$ ~almost surely} \\
    &= \mathbb{E}\left[\frac{ \mathbb{E}\left[ Y^{(1)} \mid X \right]}{ \mathbb{E}\left[ Y^{(0)} \mid X \right]} \frac{\mathbb{E}\left[ Y^{(0)} \mid X \right]}{\mathbb{E}\left[ Y^{(0)}\right] }\right] && \text{$\mathbb{E}\left[ Y^{(0)}\right]$ is a constant} \\
    &= \mathbb{E}\left[\tau_\text{\tiny RR}^P(X) \frac{\mathbb{E}\left[ Y^{(0)} \mid X \right]}{\mathbb{E}\left[ Y^{(0)}\right] }\right]. && \text{$\frac{ \mathbb{E}\left[ Y^{(1)} \mid X \right]}{ \mathbb{E}\left[ Y^{(0)} \mid X \right]}:= \tau_\text{\tiny RR}^P(X)$ }
\end{align*}

\end{proof}

\paragraph*{Survival Ratio (SR)}

\begin{proof}
\begin{align*}
    \tau_{\text{\tiny SR}}^P &= \frac{1-\mathbb{E}\left[ Y^{(1)}\right]}{1-\mathbb{E}\left[ Y^{(0)}\right]}  && \text{By definition of the SR}\\
    &= \frac{1-\mathbb{E}\left[ \mathbb{E}\left[ Y^{(1)} \mid X \right]\right]}{1-\mathbb{E}\left[ Y^{(0)}\right]}  && \text{Law of total expectation} \\
    &= \frac{\mathbb{E}\left[\frac{1- \mathbb{E}\left[ Y^{(1)} \mid X \right]}{ 1-\mathbb{E}\left[ Y^{(0)} \mid X \right]} \left( 1-\mathbb{E}\left[ Y^{(0)} \mid X \right] \right)\right]}{1-\mathbb{E}\left[ Y^{(0)}\right]}  && \text{$1-\mathbb{E}\left[ Y^{(0)} \mid X \right] \neq 0$~ almost surely} \\
    &= \mathbb{E}\left[\tau_\text{\tiny SR}^P(X) \frac{1-\mathbb{E}\left[ Y^{(0)} \mid X \right]}{1-\mathbb{E}\left[ Y^{(0)}\right] }\right] && \text{$1-\mathbb{E}\left[ Y^{(0)}\right]$ is a constant}
    \end{align*}

\end{proof}

The Excess Risk Ratio (ERR) (resp. Risk Susceptibility) collapsibility are proven using the same derivations than RR (resp. SR).

\paragraph*{Excess Risk Ratio (ERR)}

\begin{proof}
\begin{align*}
    \tau_{\text{\tiny ERR}}^P & =\frac{ \mathbb{E}\left[ Y^{(1)} - Y^{(0)} \right]}{ \mathbb{E}\left[  Y^{(0)} \right]} \\
    &= \frac{ \mathbb{E}\left[  \mathbb{E}\left[  Y^{(1)} - Y^{(0)} \mid X  \right]\right]}{ \mathbb{E}\left[  Y^{(0)} \right]} \\
    &=  \mathbb{E}\left[  \frac{\mathbb{E}\left[  Y^{(1)} - Y^{(0)} \mid X  \right]}{ \mathbb{E}\left[  Y^{(0)} \right]}\right]\\
    &= \mathbb{E}\left[  \frac{\mathbb{E}\left[  Y^{(1)} - Y^{(0)} \mid X  \right]}{ \mathbb{E}\left[  Y^{(0)} \right]} \frac{\mathbb{E}\left[  Y^{(0)} \mid X \right]}{\mathbb{E}\left[  Y^{(0)} \mid X \right]}  \right]\\
    &= \mathbb{E}\left[   \tau_{\text{\tiny ERR}}^P(X) \frac{\mathbb{E}\left[  Y^{(0)} \mid X \right]}{ \mathbb{E}\left[  Y^{(0)} \right]} \right]
\end{align*}
\end{proof}

\paragraph*{Risk Susceptibility (RS)}

\begin{proof}
\begin{align*}
    \tau_{\text{\tiny RS}}^P & = \frac{ \mathbb{E}\left[ Y^{(1)} - Y^{(0)} \right]}{ 1- \mathbb{E}\left[  Y^{(0)} \right]} \\
    &=  \frac{ \mathbb{E}\left[\mathbb{E}\left[ Y^{(1)} - Y^{(0)} \mid X \right] \right]}{ 1- \mathbb{E}\left[  Y^{(0)} \right]} \\
    &=  \mathbb{E}\left[ \frac{ \mathbb{E}\left[ Y^{(1)} - Y^{(0)} \mid X \right]}{1- \mathbb{E}\left[  Y^{(0)} \right]}  \right] \\
    &=  \mathbb{E}\left[ \frac{ \mathbb{E}\left[ Y^{(1)} - Y^{(0)} \mid X \right]}{1- \mathbb{E}\left[  Y^{(0)} \right]} \frac{1- \mathbb{E}\left[  Y^{(0)} \mid X \right]}{1-\mathbb{E}\left[  Y^{(0)} \mid X \right]}  \right] \\
    &= \mathbb{E}\left[\tau_{\text{\tiny RS}}^P(X) \frac{1- \mathbb{E}\left[  Y^{(0)} \mid X \right]}{1- \mathbb{E}\left[  Y^{(0)}\right]} \right]
\end{align*}
\end{proof}

\subsubsection{Proof of Lemma~\ref{lemma:non-collapsibility}: Non-collapsibility of the OR, log-OR, and NNT}\label{proof:non-collapsibility-OR}

\textbf{Odds Ratio (OR).}
According to the first point of Lemma~\ref{lemma:logic-respecting-measures}, all collapsible measure are logic-respecting. However, according to the third point of Lemma~\ref{lemma:logic-respecting-measures}, OR is not logic-respecting. Therefore OR is not collapsible. 

\medskip 

\textbf{Log Odds Ratio (log-OR).}
The same reasoning as above holds for the log Odds Ratio. 

\medskip 
%\textit{N.B: the proof for the non collapsibility of the OR and log-OR \underline{is not} a novel contribution}\\

%See proof showing that OR is not logic-respecting in Section~\ref{proof:logic-respecting}, which leads to non-collapsibility of the OR. If a measure is not logic-respecting, then the population's effect can not be written as a weighted sum of local effects with positive weights. \\

\textbf{Number Needed to Treat (NNT).} \\

\begin{proof}
Recall that 
\begin{align}
\tau_{\text{\tiny  NNT}}^P = \frac{1}{\mathbb{E}[Y^{(1)}] - \mathbb{E}[Y^{(0)}]} \quad \textrm{and} \quad \tau_{\text{\tiny  NNT}}^P(X) = \frac{1}{\mathbb{E}[Y^{(1)}| X ] - \mathbb{E}[Y^{(0)} | X]}.
\end{align}

Assume that the NNT causal measure is collapsible, that is there exist weights $w(X, P(X,Y^{(0)}))$ such that for all distributions $P(X, Y^{(0)}, Y^{(1)})$ we have
\begin{align}
\mathbb{E}\left[ w(X, P(X,Y^{(0)}))\, \tau_{\text{\tiny  NNT}}^P(X) \right] = \tau_{\text{\tiny  NNT}}^P,\qquad \text{with }   w \ge 0,\, \text{and} \quad  \mathbb{E}\left[  w(X, P(X,Y^{(0)}))\right] = 1. \label{eq_proof_NNT_def_collapsible}
\end{align}

Note that 
\begin{align}
\tau_{\text{\tiny  NNT}}^P = \frac{1}{\mathbb{E}\left[ \frac{1}{\tau_{\text{\tiny  NNT}}^P(X)}\right]},
\end{align}
which, combined with the previous equation, leads to 
\begin{align}
\mathbb{E}\left[ w(X, P(X,Y^{(0)}))\, \tau_{\text{\tiny  NNT}}^P(X) \right] = \frac{1}{\mathbb{E}\left[ \frac{1}{\tau_{\text{\tiny  NNT}}^P(X)}\right]}.
\end{align}
Assuming that $\tau_{\text{\tiny  NNT}}^P(X) \geq 0$, by Jensen inequality, we have 
\begin{align}
\mathbb{E}\left[ w(X, P(X,Y^{(0)}))\, \tau_{\text{\tiny  NNT}}^P(X) \right] & \leq \mathbb{E}\left[ \tau_{\text{\tiny  NNT}}^P(X)\right]\\
\mathbb{E}\left[ \left( w(X, P(X,Y^{(0)})) - 1 \right) \, \tau_{\text{\tiny  NNT}}^P(X) \right] & \leq 0. \label{eq_proof_NNT1}
\end{align}
Fix $\varepsilon >0$. Assume now that there exists a measurable set $B \subset \mathcal{X}$ with positive measure, such that for all $x \in B$, $w(X, P(X,Y^{(0)})) > 1+ \varepsilon$. By choosing the distribution of $Y^{(1)}$ such that $\mathbb{E}[Y^{(1)} | X]$ is arbitrary close to $\mathbb{E}[Y^{(0)} | X]$ on $B$, one has that $\tau_{\text{\tiny  NNT}}^P(X)$ is arbitrary large, so that $\left( w(X, P(X,Y^{(0)})) - 1 \right) \, \tau_{\text{\tiny  NNT}}^P(X) $ is arbitrary large on $B$, which contradicts  \eqref{eq_proof_NNT1}. This proves that $w(X, P(X,Y^{(0)})) \leq 1$ almost surely. Since $\mathbb{E}[w(X, P(X,Y^{(0)}))] =1$, this implies that almost surely $w(X, P(X,Y^{(0)})) = 1$. Thus, one should have 
\begin{align}
\mathbb{E}\left[ \tau_{\text{\tiny  NNT}}^P(X) \right] = \frac{1}{\mathbb{E}\left[ \frac{1}{\tau_{\text{\tiny  NNT}}^P(X)}\right]}, 
\end{align}
which, according to Jensen inequality, holds only if $\tau_{\text{\tiny  NNT}}^P(X)$ is constant. Thus the Number Needed to Treat satisfies the collapsibility equation \eqref{eq_proof_NNT_def_collapsible} only in the specific case of homogeneous treatment effect. \\
This proves that the NNT is not collapsible. 

%\begin{align*}
%      \tau_{\text{\tiny NNT}}&= \frac{1}{ \mathbb{E}\left[ Y^{(1)} - Y^{(0)} \right]}  \\
%      &=  \frac{1}{ \mathbb{E}\left[ \mathbb{E}\left[ Y^{(1)} - Y^{(0)} \mid X \right] \right]}\\ 
 %     &= \mathbb{E}\left[ \frac{1}{\tau_{\text{\tiny NNT}}(X)}   \right]^{-1}.
%\end{align*}
 
\end{proof}

\subsection{Proof of Lemma~\ref{lemma:annexe_monotinic_causal_measure} }\label{app:supplementary_lemma}

\begin{lemma}
\label{lemma:annexe_monotinic_causal_measure}
Let $\tau_1 $ be any collapsible causal measure defined by Equation~\eqref{eq_causal_measure}, that is 
\begin{align}
\tau_1^P (x) = f \left(\mathds{E}[Y^{(0)} \mid X=x ] , \mathds{E}[Y^{(1)} \mid X=x ] \right),
\end{align}
and 
\begin{align}
\tau_1^P = f \left(\mathds{E}[Y^{(0)}  ] , \mathds{E}[Y^{(1)} ] \right),
\end{align}
Consider $\tau_2$ another causal measure, such that, there exists $h$ satisfying 
\begin{align}
\tau_2^P(x) = h(\tau_1(x)) \textrm{ and } \tau_2 = h(\tau_1). 
\end{align}
If $h$ is bijective and monotonic, then $\tau_2$ is logic-respecting. 
\end{lemma}

\begin{proof}[Proof of Lemma~\ref{lemma:annexe_monotinic_causal_measure}]

Since $\tau_1$ is collapsible, we know that, for all distributions of $(X, Y^{(0)}, Y^{(1)})$, 
\begin{align}
\tau_1^P = \mathds{E}[\tau_1^P(X) w(X, P(X,Y^{(0)}))].
\end{align}
Since $\tau_2 = h(\tau_1)$, we obtain
\begin{align}
\tau_2^P = h \left( \mathds{E}[h^{-1} (\tau_2^P(X)) w(X, P(X,Y^{(0)}))] \right).
\end{align}
Assume that $h$ is increasing, then $h^{-1}$ is increasing and 
\begin{align}
   h^{-1} \left( \min_x \tau_2^P(x) \right) \leq h^{-1} (\tau_2^P(X)) \leq h^{-1} \left( \max_x \tau_2^P(x) \right),
\end{align}
which implies, since $h$ is increasing, 
\begin{align}
    \min_x \tau_2^P(x) \leq h \left( \mathds{E}[h^{-1} (\tau_2^P(X)) w(X, P(X,Y^{(0)}))] \right) \leq \max_x \tau_2^P(x),
\end{align}
and thus, 
\begin{align}
\min_x \tau_2^P(x) \leq  \tau_2^P \leq \max_x \tau_2^P(x).   
\end{align}
Consequently, the causal measure $\tau_2$ is logic-respecting. The same reasoning holds for a decreasing function $h$. 

\end{proof}

\subsection{Proof of Lemma~\ref{lemma:logic-respecting-measures} (about logic-respecting measures)}\label{proof:logic-respecting}

\subsubsection{All collapsible measures are logic respecting}
\begin{proof}
    We recall from Definition~\ref{def:indirect-collapsibility} that a measure $\tau$ is said to be collapsible (directly or not), if there exist positive weights $w(X, P(X,Y^{(0)}))$ verifying  $\mathbb{E}\left[w(X, P(X,Y^{(0)}))\right] = 1$, such that
   
    \begin{equation*}
        \tau^P = \mathbb{E}\left[ w(X, P(X,Y^{(0)}))\tau^P(X)  \right].
    \end{equation*}

    Then,

    \begin{align*}
         \tau^P &\le  \mathbb{E}\left[w(X, P(X,Y^{(0)})) \max_x \left( \tau^P(X) \right)  \right]\\
         \tau^P &\le  \mathbb{E}\left[w(X, P(X,Y^{(0)})) \right] \max_x \left( \tau^P(x) \right)  \\
         \tau^P &\le\max_x \left( \tau^P(x) \right)
    \end{align*}

    using the properties of the weights. Similarly, one can show that,

  \begin{align*}
        \mathbb{E}\left[w(X, P(X,Y^{(0)})) \min_x \left( \tau^P(x) \right)  \right] &\le  \tau^P.
    \end{align*}

    This proves that $\tau$ is logic-respecting, according to Definition~\ref{def:logic-respecting}.
    
\end{proof}

\subsubsection{Number Needed to Treat is a logic-respecting measure}

\begin{proof}

   First, note that,
\begin{align*}
      \tau_{\text{\tiny NNT}}^P&= \frac{1}{ \mathbb{E}\left[ Y^{(1)} - Y^{(0)} \right]}  \\
      &=  \frac{1}{ \mathbb{E}\left[ \mathbb{E}\left[ Y^{(1)} - Y^{(0)} \mid X \right] \right]} && \text{Law of total expectation}\\ 
      &= \mathbb{E}\left[ \frac{1}{\tau_{\text{\tiny NNT}}^P(X)}   \right]^{-1} . && \text{$\tau_{\text{\tiny NNT}}^P(X):= 1/\mathbb{E}\left[ Y^{(1)} - Y^{(0)} \mid X \right]$}
\end{align*}

By definition, $\min_x \left(\tau_{\text{\tiny NNT}}^P(x) \right) \le \tau_{\text{\tiny NNT}}^P(X)$ almost surely, such that taking the inverse and the expectation leads to

\begin{equation*}
  \mathbb{E}\left[\frac{1}{\tau_{\text{\tiny NNT}}^P(X) }  \right] \le  \mathbb{E}\left[\frac{1}{\min_x \left(\tau_{\text{\tiny NNT}}^P(x) \right)} \right] = \frac{1}{\min_x \left(\tau_{\text{\tiny NNT}}^P(x) \right)}, 
\end{equation*}
%as $\min_x \left(\tau_{\text{\tiny NNT}}(X) \right)$ is a constant. 
which implies
%Taking again the inverse of $ \mathbb{E}\left[\frac{1}{\tau_{\text{\tiny NNT}}(X) }  \right] $ allows to recover the marginal $\tau_{\text{\tiny NNT}}$, so that
\begin{equation*}
  \min_x \left(\tau_{\text{\tiny NNT}}^P(x) \right)\le  \tau_{\text{\tiny NNT}}^P.
\end{equation*}
The exact same reasoning leads to 
\begin{equation*}
  \tau_{\text{\tiny NNT}}^P \le \max_x \left(\tau_{\text{\tiny NNT}}^P(x) \right).
\end{equation*}
Consequently, 
\begin{equation*}
  \min_x \left(\tau_{\text{\tiny NNT}}^P(x) \right)\le  \tau_{\text{\tiny NNT}}^P \le \max_x \left(\tau_{\text{\tiny NNT}}^P(x) \right),
\end{equation*}
which concludes the proof.
\end{proof}

\subsubsection{OR and log-OR are not logic-respecting}
\label{sec:appendixORnoncollapsible}
%\es{je mettrais la preuve ici et ferai référence au Lemme 4 et à ce résultat dans la partie précédente}

Proving that the OR is not logic-respecting can be done with a counter-example as in Table~\ref{tab:odds-ratio-simpson}. 
 Previous works propose to understand non-collapsibility through the non-linearity of a function linking the baseline (control) and response functions. This link function is named the \textit{characteristic collapsibility function} (CCF) and have been proposed by \cite{Neuhaus1993GeometricApproach} and is nicely recalled in \cite{Daniel2020MakingApple} (see their Appendix 1A). This proof relies on Jensen inequality. The proof we recall here is largely inspired from these works, but written within the formalism of our paper.

%On the contrary, some measures are not collapsible. In particular for the odds ratio, this can be viewed through the example from Table~\ref{tab:odds-ratio-simpson}. It is possible to understand non-collapsibility through the non-linearity of a function linking the baseline (control) and response functions. This link function is named the \textit{characteristic collapsibility function} (CCF) and have been proposed by \cite{Neuhaus1993GeometricApproach} and is nicely recalled in \cite[Appendix 1A]{Daniel2020MakingApple}. \es{of what ?}\bc{Erwan: j'ai mis ici}

\begin{proof}
    
Assume a discriminative model such as

\begin{equation}\label{eq:proof-gen-model-non-collapsible}
    \operatorname{logit}\left( \mathbb{P}(Y^{(a)} =1 \mid X, A=a) \right) = b(X) + a\,m,
\end{equation}

where $b(X)$ can be any function of the vector $X$ to $\mathbb{R}$, and where $m$ is a non-null constant. Without loss of generality, one can further assume that $m>0$. Under such model, on has a property on the conditional log-OR or OR, being that:

\begin{equation}\label{eq_proof_local_ORx}
    \tau_{\text{\tiny log-OR}}^P(X) =  \operatorname{log}\left( \frac{\mathbb{P}(Y^{(1)} =1 \mid X)}{1- \mathbb{P}(Y^{(1)} =1 \mid X)} \cdot \left ( \frac{\mathbb{P}(Y^{(0)} =1 \mid X)}{1- \mathbb{P}(Y^{(0)} =1 \mid X)} \right)^{-1} \right) =  b(X) + m - b(X) =  m,
\end{equation}

or similarly that 

\begin{equation*}
     \tau_{\text{\tiny OR}}^P(X) = e^{b(X) + m}\cdot e^{-b(X)} =  e^{m}.
\end{equation*}

In other words, for any $x$ the OR $\tau_{\text{\tiny OR}}^P(x)$ (resp. log-OR) is the same and equal to $e^{m}$ (resp. $m$).\\

Now, we propose to go from this conditional causal measure to the marginal measure. When looking for the marginal OR, one can first estimate $\mathbb{P}(Y^{(1)}=1)$ and $\mathbb{P}(Y^{(0)}=1)$, and then compute the OR. To do so, we propose to rewrite $ \mathbb{P}(Y^{(1)}=1 \mid X)$ as a function of $ \mathbb{P}(Y^{(0)}=1 \mid X)$. From \eqref{eq:proof-gen-model-non-collapsible} one has,

\begin{equation*}
   \operatorname{logit}\left( \mathbb{P}(Y^{(0)}=1 \mid X) \right) =  b(X) ,
\end{equation*}
so that 
\begin{align}\label{eq:proof-non-collapsible}
 \operatorname{logit}\left( \mathbb{P}(Y^{(1)}=1 \mid X) \right)  &=  \operatorname{logit}\left( \mathbb{P}(Y^{(0)}=1 \mid X) \right) + m ,
\end{align}
which is equivalent to
\begin{align}\label{eq:proof-non-collapsible}
  \mathbb{P}(Y^{(1)}=1 \mid X) &=\operatorname{expit}\left(  \operatorname{logit}\left(   \mathbb{P}(Y^{(0)}=1 \mid X) \right) + m \right).
\end{align}

Letting, for all $z \in [0,1]$, 
\begin{align}\label{eq:proof-non-collapsible}
  f(z) &=\operatorname{expit}\left(  \operatorname{logit}\left(   z \right) + m \right),
\end{align}
we have
\begin{align}\label{eq:proof-non-collapsible}
  \mathbb{P}(Y^{(1)}=1 \mid X) &=f \left(  \mathbb{P}(Y^{(0)}=1 \mid X) \right) .
\end{align}
Note that the function $f$ is concave for positive $m$ (it is possible to derive it, but we propose an illustration on Figure~\ref{fig:schema-proof-non-collapsible} to help to be convinced). Then, using Jensen inequality, we obtain,

\begin{align*}
 \mathbb{P}(Y^{(1)}=1) 
 &=   \mathbb{E}\left[   \mathbb{P}(Y^{(1)}=1 \mid X )\right]\\
 &= \mathbb{E}\left[ f \left(  \mathbb{P}(Y^{(0)}=1 \mid X) \right) \right] \\
 %& = \mathbb{E}\left[   f(X) \right]\\
 & < f(\mathbb{E}\left[   \mathbb{P}(Y^{(0)}=1 \mid X) \right]\\
   %&= \mathbb{E}\left[  \operatorname{expit}\left(  \operatorname{logit}\left(   \mathbb{P}(Y^{(0)}=1 \mid X) \right) + m \right) \right]\\
     &= \operatorname{expit}\left(  \operatorname{logit}\left(  \mathbb{E}\left[  \mathbb{P}(Y^{(0)}=1 \mid X) \right]\right) + m \right) && \text{Jensen and $m >0 $} \\
     &= \operatorname{expit}\left(  \operatorname{logit}\left( \mathbb{P}(Y^{(0)}=1 )\right) + m \right),
\end{align*}

and because the $ \operatorname{logit}$ is a monotonous function, then,

\begin{align*}
     \operatorname{logit}\left( \mathbb{P}(Y^{(1)}=1)\right) &<  \operatorname{logit}\left( \mathbb{P}(Y^{(0)}=1 )\right) + m,
\end{align*}

so that 

\begin{align*}
     \operatorname{logit}\left( \mathbb{P}(Y^{(1)}=1)\right) - \operatorname{logit}\left( \mathbb{P}(Y^{(0)}=1 )\right) = \tau_{\text{\tiny log-OR}}^P  &< m,
\end{align*}
where $m=\tau_{\text{\tiny log-OR}}^P(x)$ (see \eqref{eq_proof_local_ORx}).
This allows to conclude that there exists a data discriminative process for which the odds ratio at the population level can not be written as a positively weighted sum of conditional odds ratio. 
%Would $m$ be strictly negative, then the function would be convex and the inequality would be in the other direction. \es{on peut enlever cette remarque je pense}

\begin{figure}
    \centering
    \includegraphics[width=0.4\textwidth]{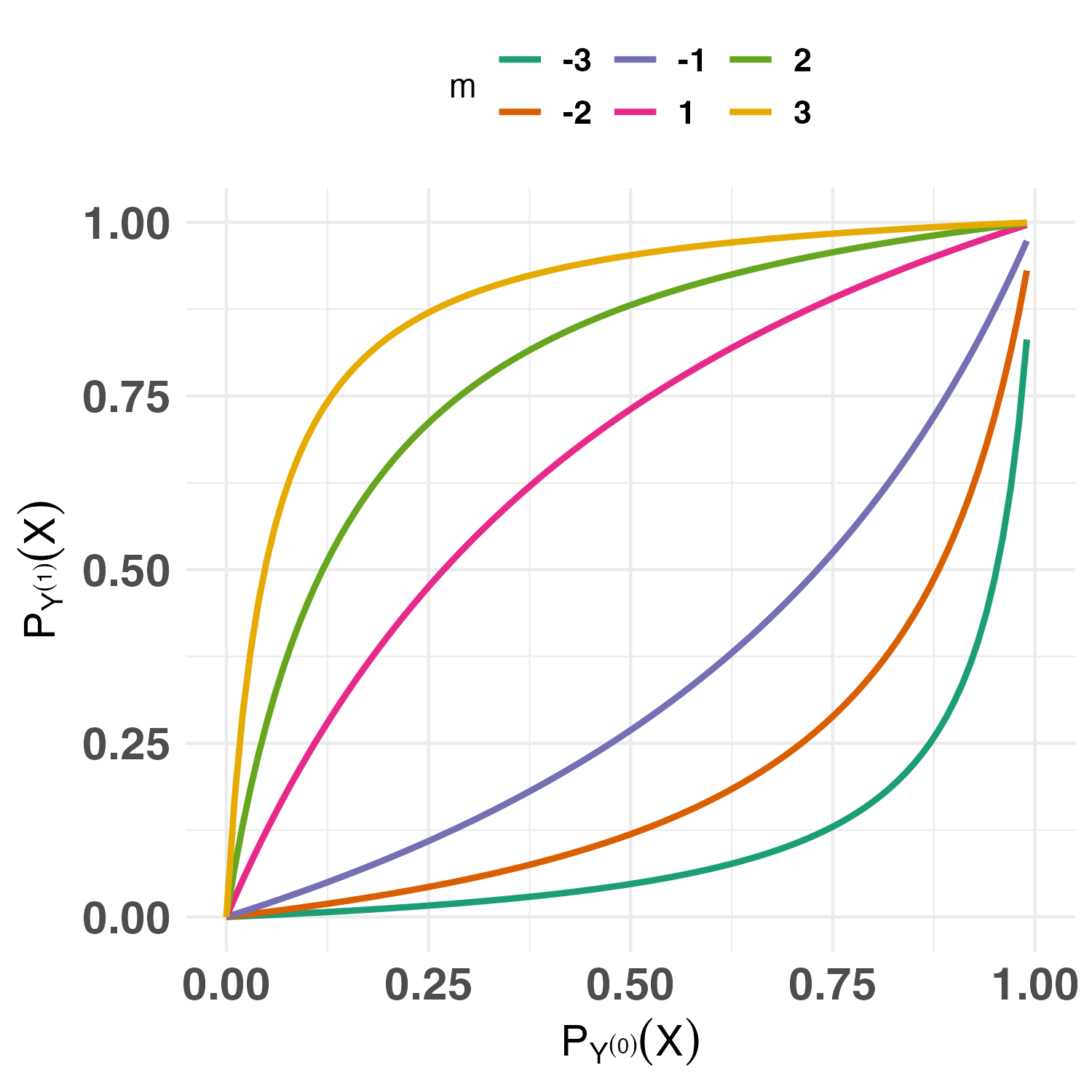}
    \caption{Implementation of the formulae from \eqref{eq:proof-non-collapsible} for different values of $m$. This illustrates the concavity of the function linking $\mathbb{P}(Y^{(0)}=1 \mid X) $ to $\mathbb{P}(Y^{(1)}=1 \mid X)$ when assuming the discriminative model of \eqref{eq:proof-gen-model-non-collapsible}. }
    \label{fig:schema-proof-non-collapsible}
\end{figure}

Note that the example provided in Table~\ref{tab:odds-ratio-simpson} is for a negative $m$, showing constant effect on the two substrata and a higher effect on the marginal population.
\end{proof}

\subsection{Proofs related to generalizability}\label{proof:generalizability-section-3}

\subsubsection{Proof of Proposition~\ref{proposition:generalization-density}}

\begin{proof}

   Consider $a\in \{0,1 \}$, then
\begin{align*}
   \mathbb{E}_{\text{\tiny T}}\left[Y^{(a)} \right] &=   \mathbb{E}_{\text{\tiny T}}\left[\mathbb{E}_{\text{\tiny T}}\left[Y^{(a)}\mid X \right]\right] && \text{Total expectation}\\
   &=  \mathbb{E}_{\text{\tiny T}}\left[\mathbb{E}_{\text{\tiny S}}\left[Y^{(a)}\mid X \right]\right] && \text{Transportability -- Assumptions~\ref{a:transportability-wide}}\\
   &= \mathbb{E}_{\text{\tiny S}}\left[ \frac{p_{\text{\tiny T}}(X)}{p_{\text{\tiny S}}(X)}  \mathbb{E}_{\text{\tiny S}}\left[Y^{(a)}\mid X \right]\right] && \text{Overlap -- Assumptions~\ref{a:overlap}} \\
\end{align*}

\end{proof}

The last step can also be written as follow:

\begin{align*}
    \mathbb{E}_{\text{\tiny T}}\left[\mathbb{E}_{\text{\tiny S}}\left[Y^{(a)}\mid X \right]\right]  &= \int \mathbb{E}_{\text{\tiny S}}\left[Y^{(a)}\mid X=x \right] p_{\text{\tiny T}}(x)\,dx && \text{By definition}\\
    &=  \int \mathbb{E}_{\text{\tiny S}}\left[Y^{(a)}\mid X=x \right]p_{\text{\tiny T}}(x) \frac{p_{\text{\tiny S}}(x)}{p_{\text{\tiny S}}(x)}\,dx && \text{Assumption~\ref{a:overlap}: $\frac{p_{\text{\tiny T}}}{p_{\text{\tiny S}}}(x)$ is defined} \\
   &= \int\mathbb{E}_{\text{\tiny S}}\left[Y^{(a)}\mid X =x\right]p_{\text{\tiny S}}(x) \frac{p_{\text{\tiny T}}(x)}{p_{\text{\tiny S}}(x)}\,dx && \text{Re-arrangement} \\
   &=\mathbb{E}_{\text{\tiny S}}\left[ \frac{p_{\text{\tiny T}}(X)}{p_{\text{\tiny S}}(X)}\mathbb{E}_{\text{\tiny S}}\left[Y^{(a)}\mid X \right]\right]
\end{align*}

\subsubsection{Proof of Proposition~\ref{prop:generalization-of-local-effects}}

\begin{proof}

If $\tau$ is collapsible, then there exists weights $w(X, P_{\text{\tiny T}}(X,Y^{(0)}))$  such that

\begin{align*}
   \tau^{P_{\text{\tiny T}}}&= \mathbb{E}_{\text{\tiny T}}\left[ w(X, P_{\text{\tiny T}}(X,Y^{(0)})) \tau^{P_{\text{\tiny T}}} (X)\right] && \text{Collapsibility}\\
   &= \mathbb{E}_{\text{\tiny T}}\left[ w(X, P_{\text{\tiny T}}(X,Y^{(0)})) \tau^{P_{\text{\tiny S}}} (X)\right] && \text{Transportability -- Assumption~\ref{a:transportability}}\\
   &= \mathbb{E}_{\text{\tiny S}}\left[ \frac{p_{\text{\tiny T}}(X)}{p_{\text{\tiny S}}(X)} w(X, P_{\text{\tiny T}}(X,Y^{(0)})) \tau^{P_{\text{\tiny S}}} (X)\right] && \text{Overlap -- Assumption~\ref{a:overlap}}.
\end{align*}
\end{proof}

\subsection{Proofs related to non-parametric discriminative models (Section~\ref{section:generative-models})}

%\textit{As we have not found elsewhere the approach of writing non-parametric models and to relate them to measures of effect,  to the best of our knowledge, all proofs in this subsection are novel.}
Proofs of Corollary~\ref{lemma:working-model-continuous-Y} and \ref{lemma:working-model-continuous-Y-RR} are straightforward and left to the reader.

\subsubsection{Proof of Lemma~\ref{lem_generative_models}}
\label{app_subsection_proof_Lemma1_genmodels}

By assumption, throughout the paper, $\mathds{E}[Y^{(0)}|X] < \infty$ and $\mathds{E}[Y^{(1)}|X] < \infty$. Thus, one can set 
\begin{align}
    b(X) = \mathds{E}[Y^{(0)}|X], \quad \textrm{and} \quad m(X) = g_{b(X)} \left( \mathds{E}[Y^{(1)}|X]\right).
\end{align}
Since, for all $b \in \mathds{R}$, the function $g_b$ is a bijection on its domain, we have 
\begin{align}
\mathds{E}[Y^{(1)}|X] = g_{b(X)}^{-1}(m(X)). 
\end{align}
With these notations, 
\begin{align}
    \tau^P(X) & = f(\mathds{E}[Y^{(0)}|X], \mathds{E}[Y^{(1)}|X]) \\
    & = g_{b(X)} (\mathds{E}[Y^{(1)}|X]) \\
    & = m(X).
\end{align}

\subsubsection{Proof of \Cref{lem:disentanglement_homogeneous_treatment}}
\label{sec:preuve_actuelle}

\paragraph*{First case (homogeneous treatment effect)} Let $m \in \mathds{R}$ and let 
    \begin{align}
        \mathcal{P} & = \Big\lbrace P(X, Y^{(0)}, Y^{(1)}): P(X, Y^{(0)}) \in \mathcal{P}_{all}(X, Y^{(0)}) \textrm{ and } \tau^P(\cdot) = m \Big\rbrace. \nonumber
    \end{align}
Then, we have $\mathcal{P}(\tau(\cdot)) = \{ x \mapsto m\}$. Thus, 
\begin{align*}
    \left\lbrace P(Y^{(0)}|X): P \in \mathcal{P} \textrm{ s.t. } \tau^P(\cdot) = m\right\rbrace = \mathcal{P}(Y^{(0)}|X).
    \end{align*}
 Besides, due to the collapsibility of $\tau$, we have
 for all $P \in \mathcal{P}$, \begin{align}
  \tau^P = \mathds{E} [w(X, P(X,Y^{(0)}) \tau^P(X)] =   m \mathds{E} [w(X, P(X,Y^{(0)}) ] = m,  
 \end{align}
 which is a constant depending only on $m$. Thus, the causal measure $\tau$ has its CATE and ATE disentangled from the baseline on the collection $\mathcal{P}$.

 \paragraph*{Second case (independence between baseline and treatment effect)}
 Let $ S \subset \{1,  \hdots, d\}$ and   
    \begin{align}
        \mathcal{P} & = \Big\lbrace P(X, Y^{(0)}, Y^{(1)}) \in \mathcal{P}_{all}(X,Y^{(0)}, Y^{(1)}): X_S \indep X_{S^c},   Y^{(0)}|X = Y^{(0)}|X_S, \tau^P(X) = \tau^P(X_{S^c}) \Big\rbrace. \nonumber
    \end{align}
Thus, as the baseline distribution and the treatment effect are set independently, we have , for all $h \in \mathcal{P}(\tau(\cdot))$
\begin{align*}
    \left\lbrace P(Y^{(0)}|X): P \in \mathcal{P} \textrm{ s.t. } \tau^P(\cdot) = h(\cdot) \right\rbrace = \mathcal{P}(Y^{(0)}|X).
    \end{align*}
 Besides, due to the collapsibility of $\tau$, we have
 for all for all $m \in \mathcal{P}(\tau(\cdot))$ and for all $P \in \mathcal{P}$ such that $\tau^P (\cdot) = m(\cdot)$, 
 \begin{align}
  \tau^P = \mathds{E} [w(X, P(X,Y^{(0)})) m(X)] = \mathds{E} [w(X_S) m(X_{S^c})]  = \mathds{E}[ m(X_{S^c})],  
 \end{align}
 which depends only on $m$ and on the distribution of $X$. Thus, the causal measure $\tau$ has its CATE and ATE disentangled from the baseline on the collection $\mathcal{P}$ and, for all $P \in \mathcal{P}$, $\tau^P = \mathds{E} [\tau^P(X_{S^c})]$

\subsubsection{Proof of Theorem~\ref{th_onlyRD_separates_baseline_tteffect}}
\label{app_subsection_proof_thm1_genmodels}

Recall that the conditional causal measure $\tau$ can be written as 
\begin{align}
    \tau^P (x) = f \left( \mathds{E}[Y^{(0)}|X=x] , \mathds{E}[Y^{(1)}|X=x] \right),
\end{align}
if $\left( \mathds{E}[Y^{(0)}|X=x] , \mathds{E}[Y^{(1)}|X=x] \right) \in D_f$. 
Besides, if $\left( \mathds{E}[Y^{(0)} ] , \mathds{E}[Y^{(1)}] \right) \in D_f$,
\begin{align}
    \tau^P = f \left( \mathds{E}[Y^{(0)} ] , \mathds{E}[Y^{(1)}] \right).
\end{align}
As the causal measure $\tau$ is assumed to be collapsible, there exist non-negative weights $w(X,$ $P(X,Y^{(0)}))$ verifying, for all distribution $(X, Y^{(0)}, Y^{(1)})$,  $\mathds{E}[w(X, P(X,Y^{(0)}))] = 1$ and 
\begin{equation}
\tau^P = \mathbb{E}\left[ w(X, P(X,Y^{(0)})) \tau^P(X)  \right].
\end{equation}
By assumption, for all functions $m \in \mathcal{P}_{all}(\tau(\cdot))$, 
\begin{align*}
    \left\lbrace P(Y^{(0)}|X): P \in \mathcal{P}_{all} \textrm{ s.t. } \tau^P(\cdot) = m(\cdot)\right\rbrace = \mathcal{P}_{all}(Y^{(0)}|X)
    \end{align*} 
and, for all $P \in \mathcal{P}_{all}$ satisfying $\tau^P(\cdot) = m(\cdot)$, there exists a constant $C_{m, P(X)}$ which depends only on $m$ and $P(X)$, such that $\tau^P = C_{m, P(X)}$. 
Thus, for all distributions $P(X,Y^{(0)})$ and for all functions $m: \mathds{X} \to f(D_f)$,
\begin{align}
    C_{m, P(X)}  = \mathbb{E}\left[ m(X)  w(X, P(X,Y^{(0)})) \right]. \label{eq_proof_th1_for_th4}
\end{align}
Note that the joint distribution $P(X,Y^{(0)})$ can be written as $P(Y^{(0)}|X) P(X)$. Since $m$ can be arbitrary chosen, and since the left-hand term does not depend on the distribution $Y^{(0)}|X$, one must have
\begin{align}
 w(X, P(X,Y^{(0)})) = w(X, P(X)).   
\end{align}
Therefore, 
\begin{equation}
\tau^P = \mathbb{E}\left[ \tau^P(X)  w(X, P(X)) \right], 
\end{equation}
For simplicity, we denote $w(X,P(X))$ by $w(X)$. Thus, for all distributions $(X, Y^{(0)}, Y^{(1)})$, 
\begin{equation}
\tau^P = \mathbb{E}\left[ \tau^P(X)  w(X) \right], 
\end{equation}
which is equivalent to 
\begin{align}
f \left( \mathds{E}[Y^{(0)} ] , \mathds{E}[Y^{(1)}] \right) = \mathds{E} \left[ f \left( \mathds{E}[Y^{(0)}|X] , \mathds{E}[Y^{(1)}|X] \right) w(X)\right]. \label{eq_proof_th_models21}
\end{align}
Now, assume that $\mathds{E}[Y^{(0)} | X] = C$ (constant baseline), and let $m(X) = \mathds{E}[Y^{(1)}|X].$ We have
\begin{align}
f \left( C , \mathds{E}[m(X)] \right) = \mathds{E} \left[ f \left( C , m(X)  \right) w(X)\right].
\end{align}

\paragraph*{First case} Assume that $f(C,C) = 0$. Let $B \subset \mathds{X}$ a borelian. Set 
\begin{align}
    m_B(x) = \left\lbrace 
    \begin{array}{cc}
       0  &  \textrm{if } x \in B^c\\
       C  & \textrm{if } x \in B
    \end{array}
    \right.
\end{align}
We have
\begin{align}
    \mathds{E}[m_B(X)] = C \mu_X(B),
\end{align}
and
\begin{align}
    \mathds{E}[f \left( C , m_B(X)  \right) w(X)] = \mathds{E}\left[ f \left( C , 0  \right) \mathds{1}_{X \in B^c} w(X) \right].
\end{align}
Hence, 
\begin{align}
f \left( C , C \mu_X(B) \right) = f(C,0) \mathds{E}\left[ w(X)  \mathds{1}_{X \in B^c} \right].
\end{align}
Since $x \mapsto f(C,x)$ is an injection, $f(C,0) \neq f(C,C) = 0$. Thus, for all borelian $B_1, B_2 \subset \mathds{X}$ such that $\mu_X(B_1) = \mu_X(B_2)$, 
\begin{align}
\frac{1}{\mu_X(B_1)} \mathds{E}\left[ w(X)  \mathds{1}_{X \in B_1} \right] = \frac{1}{\mu_X(B_2)}  \mathds{E}\left[ w(X)  \mathds{1}_{X \in B_2} \right]. \label{eq_proof_th_model11}    
\end{align}
Let $x_1, x_2 \in \mathds{X}$ and $(B_{1,n}), (B_{2,n})$ two sequences of decreasing open balls centered respectively at $x_1$ and $x_2$ such that, for all $n$,  $\mu_X(B_{1,n}) = \mu_X(B_{2,n})$. 
Letting $f$ the density of $X$, we have
\begin{align}
\frac{1}{\mu_X(B_1)} \mathds{E}\left[ w(X)  \mathds{1}_{X \in B_1} \right] 
&=     \frac{\mu(B_{1,n})}{\mu_X(B_{1,n})} \frac{1}{\mu(B_{1,n})} \int_{B_{1,n}} w(x)f(x) dx.
\end{align}
According to the Lebesgue density theorem, we have 
\begin{align}
\frac{\mu_X(B_{1,n})}{\mu(B_{1,n})} &= \frac{1}{\mu(B_{1,n})} \int_{B_{1,n}} f(x) dx \to f(x_1),
\end{align}
and
\begin{align}
\frac{1}{\mu(B_{1,n})} \int_{B_{1,n}} w(x)f(x) dx \to w(x_1) f(x_1).
\end{align}
Thus, 
\begin{align}
\frac{1}{\mu_X(B_{1,n})} \mathds{E}\left[ w(X)  \mathds{1}_{X \in B_{1,n}} \right] \to w(x_1)
\end{align}
and similarly, 
\begin{align}
\frac{1}{\mu_X(B_{2,n})} \mathds{E}\left[ w(X)  \mathds{1}_{X \in B_{2,n}} \right] \to w(x_2), 
\end{align}
which implies, according to equation~\eqref{eq_proof_th_model11}, $w(x_1) = w(x_2)$. Since $\mathds{E}[w(X)] = 1$, we obtain $w(x) = 1$ for all $x \in \mathds{X}$. 

\paragraph*{Second case} Assume that $f(C,0) = 0$. For all Borelian $B \subset \mathds{X}$,  
\begin{align}
    m_B(x) = \left\lbrace 
    \begin{array}{cc}
       0  &  \textrm{if } x \in B\\
       1  & \textrm{if } x \in B^c
    \end{array}
    \right.
\end{align}
We have
\begin{align}
    \mathds{E}[m_B(X)] =  \mu_X(B^c),
\end{align}
and
\begin{align}
    \mathds{E}[f \left( C , m_B(X)  \right) w(X)] = \mathds{E}\left[ f \left( C , 1  \right) \mathds{1}_{X \in B^c} w(X) \right].
\end{align}
Hence, 
\begin{align}
f \left( C ,  \mu_X(B^c) \right) = f(C,1) \mathds{E}\left[ w(X)  \mathds{1}_{X \in B^c} \right],
\end{align}
and the same reasoning as above applies. Since for all $x$, $w(x) = 1$, according to Equation~\eqref{eq_proof_th_models21}, we have
\begin{align}
f \left( \mathds{E}[Y^{(0)} ] , \mathds{E}[Y^{(1)}] \right) = \mathds{E} \left[ f \left( \mathds{E}[Y^{(0)}|X] , \mathds{E}[Y^{(1)}|X] \right) \right]. \label{eq_proof_theorem_models31}
\end{align}
Again, assume that $\mathds{E}[Y^{(0)}|X] = C$ and set $m(X) = \mathds{E}[Y^{(1)}|X]$. For any $a,b \in \mathds{R}$, and any $p \in [0,1]$, set 
\begin{align}
    m_{a,b,p}(x) = \left\lbrace 
    \begin{array}{cc}
       a  &  \textrm{with probability } p\\
       b  &  \textrm{with probability } 1-p\\
    \end{array}
    \right.
\end{align}
Hence, 
\begin{align}
f(C, ap + b(1-p)) = f(C,a) p + f(C,b) (1-p).
\end{align}
Thus, the function $x \mapsto f(C,x)$ is convex. By Jensen inequality, \eqref{eq_proof_theorem_models31} holds if and only if $x \mapsto f(C,x)$ is linear or $m(X)$ is degenerate. Since 
\eqref{eq_proof_theorem_models31} must hold for every distribution of $m(X)$, we deduce that, for all $C$, $x \mapsto f(C,x)$ is linear. The same reasoning can be applied by considering $x \mapsto f(x,C)$. Thus,  $x \mapsto f(x,C)$ is also linear for all $C$ and we obtain that there exist $a,b,c,d \in \mathds{R}$ such that
\begin{align}
    f \left (\mathds{E}[Y^{(0)}|X] , \mathds{E}[Y^{(1)}|X] \right) = a \mathds{E}[Y^{(1)}|X] \mathds{E}[Y^{(0)}|X] + b \mathds{E}[Y^{(1)}|X] + c \mathds{E}[Y^{(0)}|X] + d.
\end{align}
Considering $  \mathds{E}[Y^{(0)}|X] = \mathds{E}[Y^{(1)}|X] = m(X)$, we have
\begin{align}
f \left(\mathds{E}[h(X)], \mathds{E}[m(X)] \right) = a (\mathds{E}[m(X)])^2 + (b+c) \mathds{E}[m(X)] + d,
\end{align}
and
\begin{align}
\mathds{E} \left[ f \left( m(X) , m(X) \right) \right] = a \mathds{E}[m(X)^2] + (b+c) \mathds{E}[m(X)] + d,
\end{align}
which leads to, according to Equation~\eqref{eq_proof_theorem_models31}, 
\begin{align}
 a \mathds{V}[m(X)] = 0   
\end{align}
Since this must hold for every distribution of $m(X)$, we deduce that $a=0$. Finally, there exist $a,b,c \in \mathds{R}$ such that 
\begin{align}
    \tau^P (X) & = f( \mathds{E}[Y^{(0)}|X] , \mathds{E}[Y^{(1)} | X] ) \\
    & = a  \mathds{E}[Y^{(1)}|X] + b \mathds{E}[Y^{(0)}|X] + c. 
\end{align}

\subsubsection{Extension of \Cref{th_onlyRD_separates_baseline_tteffect} for bounded outcomes}
\label{sec:app:th2_bounded_outcomes}

\begin{definition}
\label{definition_domain_potential_outcomes}
We say that $\mathcal{A}: x \mapsto \mathcal{A}(x) \subset \mathds{R}$ is an admissible set of values for the potential outcomes if, for each $x \in \mathds{X}$,   
\begin{align}
    \mathds{E}[Y^{(0)}| X = x], \mathds{E}[Y^{(1)}| X = x] \in \mathcal{A}(x).
\end{align}
Given a causal measure $\tau$, and a CATE $\tau(\cdot)$, we let $\mathcal{A}_{B, \tau(\cdot)}(x)$ be the admissible set of values for the baseline, defined as
 \begin{align}
    & \mathcal{A}_{B, \tau(\cdot)}(x) \\
    & = \{ u \in \mathcal{A}(x), \exists v \in \mathcal{A}(x) \textrm{ such that } f(u,v) = h(x) \}. \nonumber 
\end{align}
\end{definition}

\begin{theorem}
\label{th_onlyRD_separates_baseline_tteffect_bounded_outcome}
Let $\mathcal{A}$ be an admissible set of values for the potential outcomes and assume that there exist $\alpha_1<\alpha_2$ such that, for all $x \in \mathds{X}$,  $(\alpha_1,\alpha_2) \subset \mathcal{A}(x)$.  Let $\tau$ be a collapsible (see Definition~\ref{def:indirect-collapsibility}) causal measure defined in Equation~\eqref{eq_causal_measure} satisfying Assumption~\ref{ass:injection_def_domain} and such that $\mathcal{A}(x)\times \mathcal{A}(x) \subset D_f$.  

Assume that for all distributions $P(X)$ of $X$, and for all functions $h: \mathds{X} \to f(D_f)$ such that $h(x) \in f(\mathcal{A}(x)\times \mathcal{A}(x))$, there exists $C_{P(X), h} \in \mathds{R}$ such that, for all distributions $Y^{(0)}|X$ satisfying 
$$\forall x \in \mathds{X}, ~\mathds{E}[Y^{(0)} | X = x] \in \mathcal{A}_{B, h}(x),$$
there exists a distribution $Y^{(1)}|X$ such that $\forall x \in \mathds{X},$ $ \mathds{E}[Y^{(1)} | X = x] \in \mathcal{A}(x) $ and
\begin{itemize}
    \item for all $x \in \mathds{X}, \tau^P(x) = h(x)$
    \item $\tau^P = C_{P(X), h}$.
\end{itemize}
Then, there exist $a,b,c \in \mathds{R}$ such that, for all distributions $P(X, Y^{(0)}, Y^{(1)})$ satisfying for all $x \in \mathds{X}$,  $\mathds{E}[Y^{(0)}| X = x], \mathds{E}[Y^{(1)}| X = x] \in \mathcal{A}(x)$, we have
\begin{align}
    \tau^P(X) = a \mathds{E}[Y^{(1)}|X] + b \mathds{E}[Y^{(0)}|X] + c.
\end{align}
\end{theorem}

\subsubsection{Proof of Theorem~\ref{th_onlyRD_separates_baseline_tteffect_bounded_outcome}}
\label{app_subsection_proof_thm2_genmodels}

Recall that, by assumption, there exist $\alpha_1, \alpha_2$ such that for all $x \in \mathds{X}$, $(\alpha_1,\alpha_2) \subset \mathcal{A}(x)$. Recall that the conditional causal measure $\tau$ can be written as 
\begin{align}
    \tau^P (x) = f \left( \mathds{E}[Y^{(0)}|X=x] , \mathds{E}[Y^{(1)}|X=x] \right),
\end{align}
if $\left( \mathds{E}[Y^{(0)}|X=x] , \mathds{E}[Y^{(1)}|X=x] \right) \in D_f$. 
Besides, if $\left( \mathds{E}[Y^{(0)} ] , \mathds{E}[Y^{(1)}] \right) \in D_f$,
\begin{align}
    \tau^P = f \left( \mathds{E}[Y^{(0)} ] , \mathds{E}[Y^{(1)}] \right).
\end{align}
As the causal measure $\tau$ is assumed to be collapsible, there exist non-negative weights $w(X,$ $P(X,Y^{(0)}))$ verifying, for all distribution $(X, Y^{(0)}, Y^{(1)})$,  $\mathds{E}[w(X, P(X,Y^{(0)}))] = 1$ and 
\begin{equation}
\tau^P = \mathbb{E}\left[ \tau^P(X)  w(X, P(X,Y^{(0)})) \right].
\end{equation}
By assumption, for all distributions $P(X)$ of $X$, and for all functions $h: \mathds{X} \to f(D_f)$ such that $h(x) \in f(\mathcal{A}(x)\times \mathcal{A}(x))$, there exists $C_{P(X), h} \in \mathds{R}$ such that, for all distributions $Y^{(0)}|X$ satisfying 
$$\forall x \in \mathds{X}, ~\mathds{E}[Y^{(0)} | X = x] \in \mathcal{A}_{B, h}(x),$$
there exists a distribution $Y^{(1)}|X$ such that $\forall x \in \mathds{X}, \mathds{E}[Y^{(1)} | X = x] \in \mathcal{A}(x) $ and
\begin{itemize}
    \item for all $x \in \mathds{X}, \tau^P(x) = h(x)$
    \item $\tau^P = C_{P(X), h}$.
\end{itemize}
Consequently, since $\tau$ is collapsible, 
\begin{align}
    C_{P(X), h}  = \mathbb{E}\left[ h(X)  w(X, P(X,Y^{(0)})) \right].
\end{align}
Assume that one can find two distributions $P_1$ and $P_2$ of $Y^{(0)}|X$ such that $w(X, P(X), P_1)$ and $w(X, P(X), P_2)$ differ, that is there exists a ball $B \subset \mathds{X}$ such that 
\begin{align}
\mathds{E}[w(X, P(X), P_1) \mathds{1}_{X \in B}] \neq \mathds{E}[w(X, P(X), P_2) \mathds{1}_{X \in B}].
\end{align}
Let $h(x) = (f(\alpha_1,\alpha_2) - f(\alpha_1, \alpha_1)) \mathds{1}_{x \in B} + f(\alpha_1, \alpha_1) \in f(\mathcal{A}(x) \times \mathcal{A}(x))$, we have
\begin{align}
 & \mathbb{E}\left[ h(X)  w(X,  P(X), P_1) \right] = \mathbb{E}\left[ h(X)  w(X,P(X), P_2) \right]\\
 \Leftrightarrow ~~& \mathbb{E}\left[ (f(\alpha_1,\alpha_2) - f(\alpha_1, \alpha_1)) \mathds{1}_{x \in B}  w(X,  P(X), P_1) \right] \\
 & \qquad \qquad \qquad = \mathbb{E}\left[ (f(\alpha_1,\alpha_2) - f(\alpha_1, \alpha_1)) \mathds{1}_{x \in B}  w(X,P(X), P_2) \right]\\
 \Leftrightarrow ~~ & \mathbb{E}\left[  \mathds{1}_{x \in B}  w(X,  P(X), P_1) \right] = \mathbb{E}\left[  \mathds{1}_{x \in B}  w(X,P(X), P_2) \right],
\end{align}
since $\mathds{E}[w(X,P(X), P_1)] = \mathds{E}[w(X,P(X), P_2)] = 1$ and by injectivity of $x \mapsto f(\alpha_1,x)$. Therefore, $w(X, P(X,Y^{(0)}))$ does not depend on the distribution $Y^{(0)}|X$ and one can write, for all distributions $(X, Y^{(0)}, Y^{(1)})$, 
\begin{equation}
\tau^P = \mathbb{E}\left[ \tau^P(X)  w(X) \right], 
\end{equation}
where $w(X) = w(X,P(X))$. Thus,  
\begin{align}
f \left( \mathds{E}[Y^{(0)} ] , \mathds{E}[Y^{(1)}] \right) = \mathds{E} \left[ f \left( \mathds{E}[Y^{(0)}|X] , \mathds{E}[Y^{(1)}|X] \right) w(X)\right]. \label{eq_proof_th_models2}
\end{align}
Now, assume that $\mathds{E}[Y^{(0)} | X] = \alpha_1$ (constant baseline), and let $h(X) = \mathds{E}[Y^{(1)}|X].$ We have
\begin{align}
f \left( \alpha_1 , \mathds{E}[h(X)] \right) = \mathds{E} \left[ f \left( \alpha_1 , h(X)  \right) w(X)\right].
\end{align}
Let $B \subset \mathds{X}$ a borelian. Set 
\begin{align}
    h_B(x) = \left\lbrace 
    \begin{array}{cc}
       \alpha_1  &  \textrm{if } x \in B \\
       \alpha_2  & \textrm{if } x \in B^c
    \end{array}
    \right.
\end{align}
We have
\begin{align}
    \mathds{E}[h_B(X)] & = \alpha_1 \mu_X(B) + \alpha_2 \mu_X(B^c)\\
    & = \alpha_2 + (\alpha_1 - \alpha_2) \mu_X(B),
\end{align}
and
\begin{align}
    & \mathds{E}[f \left( \alpha_1 , h_B(X)  \right) w(X)] \\
     = & \mathds{E}\left[ f \left( \alpha_1 , \alpha_1  \right) \mathds{1}_{X \in B } w(X) \right] 
     + \mathds{E}\left[ f \left( \alpha_1 , \alpha_2  \right) \mathds{1}_{X \in B^c } w(X) \right]\\
     = & f(\alpha_1,\alpha_2) + \left( f(\alpha_1,\alpha_1) - f(\alpha_1,\alpha_2) \right) \mathds{E}[\mathds{1}_{X \in B} w(X)].
\end{align}
Hence, 
\begin{align}
f \left( \alpha_1 , \alpha_2 + (\alpha_1 - \alpha_2) \mu_X(B) \right) = f(\alpha_1,\alpha_2) + \left( f(\alpha_1,\alpha_1) - f(\alpha_1,\alpha_2) \right) \mathds{E}[\mathds{1}_{X \in B} w(X)].
\end{align}
Since $x \mapsto f(\alpha_1,x)$ is an injection, $f(\alpha_1,\alpha_1) \neq f(\alpha_1,\alpha_2)$. Thus, the right-hand side in
\begin{align}
\mathds{E}[\mathds{1}_{X \in B} w(X)] = \frac{f \left( \alpha_1 , \alpha_2 + (\alpha_1 - \alpha_2) \mu_X(B) \right) - f(\alpha_1,\alpha_2)}{f(\alpha_1,\alpha_1) - f(\alpha_1,\alpha_2)}
\end{align}
depends only on $\mu_X(B)$. Hence, for all borelian $B_1, B_2 \subset \mathds{X}$ such that $\mu_X(B_1) = \mu_X(B_2)$, 
\begin{align}
\frac{1}{\mu_X(B_1)} \mathds{E}\left[ w(X)  \mathds{1}_{X \in B_1} \right] = \frac{1}{\mu_X(B_2)}  \mathds{E}\left[ w(X)  \mathds{1}_{X \in B_2} \right]. \label{eq_proof_th_model1}    
\end{align}
Let $x_1, x_2 \in \mathds{X}$ and $(B_{1,n}), (B_{2,n})$ two sequences of decreasing open balls centered respectively at $x_1$ and $x_2$ such that, for all $n$,  $\mu_X(B_{1,n}) = \mu_X(B_{2,n})$. 
Letting $f$ the density of $X$, we have
\begin{align}
\frac{1}{\mu_X(B_1)} \mathds{E}\left[ w(X)  \mathds{1}_{X \in B_1} \right] 
&=     \frac{\mu(B_{1,n})}{\mu_X(B_{1,n})} \frac{1}{\mu(B_{1,n})} \int_{B_{1,n}} w(x)f(x) dx.
\end{align}
According to the Lebesgue density theorem, we have 
\begin{align}
\frac{\mu_X(B_{1,n})}{\mu(B_{1,n})} &= \frac{1}{\mu(B_{1,n})} \int_{B_{1,n}} f(x) dx \to f(x_1),
\end{align}
and
\begin{align}
\frac{1}{\mu(B_{1,n})} \int_{B_{1,n}} w(x)f(x) dx \to w(x_1) f(x_1).
\end{align}
Thus, 
\begin{align}
\frac{1}{\mu_X(B_{1,n})} \mathds{E}\left[ w(X)  \mathds{1}_{X \in B_{1,n}} \right] \to w(x_1)
\end{align}
and similarly, 
\begin{align}
\frac{1}{\mu_X(B_{2,n})} \mathds{E}\left[ w(X)  \mathds{1}_{X \in B_{2,n}} \right] \to w(x_2), 
\end{align}
which implies, according to equation~\eqref{eq_proof_th_model1}, $w(x_1) = w(x_2)$. Since $\mathds{E}[w(X)] = 1$, we obtain $w(x) = 1$ for all $x \in \mathds{X}$. 
Since for all $x$, $w(x) = 1$, according to Equation~\eqref{eq_proof_th_models2}, we have
\begin{align}
f \left( \mathds{E}[Y^{(0)} ] , \mathds{E}[Y^{(1)}] \right) = \mathds{E} \left[ f \left( \mathds{E}[Y^{(0)}|X] , \mathds{E}[Y^{(1)}|X] \right) \right]. \label{eq_proof_theorem_models3}
\end{align}
Let $x \in \mathds{X}$. Let $a,b \in \mathcal{A}(x)$. Set $\mathds{E}[Y^{(0)}|X] = a$ and, for any   $p \in [0,1]$, 
\begin{align}
    \mathds{E}[Y^{(1)}|X=x] = \left\lbrace 
    \begin{array}{cc}
       a  &  \textrm{with probability } p\\
       b  &  \textrm{with probability } 1-p\\
    \end{array}
    \right.
\end{align}
Hence, 
\begin{align}
f(a, ap + b(1-p)) = f(a,a) p + f(a,b) (1-p).
\end{align}
Thus, the function $z' \mapsto f(a,z')$ is convex. By Jensen inequality, \eqref{eq_proof_theorem_models3} holds if and only if $z' \mapsto f(a,z')$ is linear or $h(X)$ is degenerate. Since 
\eqref{eq_proof_theorem_models3} must hold for every distribution of $h(X)$, we deduce that, for all $u \in \mathcal{A}(x)$, $z' \mapsto f(u,z')$ is linear. The same reasoning can be applied by considering $z \mapsto f(z,u)$. Thus,  $z \mapsto f(z,u)$ is also linear for all $u \in \mathcal{A}(x)$ and we obtain that there exist $\beta_1, \beta_2, \beta_3, \beta_4 \in \mathds{R}$ such that
\begin{align}
    f \left (\mathds{E}[Y^{(0)}|X] , \mathds{E}[Y^{(1)}|X] \right) = \beta_1 \mathds{E}[Y^{(1)}|X] \mathds{E}[Y^{(0)}|X] + \beta_2 \mathds{E}[Y^{(1)}|X] + \beta_3 \mathds{E}[Y^{(0)}|X] + \beta_4.
\end{align}
Considering $  \mathds{E}[Y^{(0)}|X] = \mathds{E}[Y^{(1)}|X] = h(X)$, we have
\begin{align}
f \left(\mathds{E}[h(X)], \mathds{E}[h(X)] \right) = \beta_1 (\mathds{E}[h(X)])^2 + (\beta_2+\beta_3) \mathds{E}[h(X)] + \beta_4,
\end{align}
and
\begin{align}
\mathds{E} \left[ f \left( h(X) , h(X) \right) \right] = \beta_1 \mathds{E}[h(X)^2] + (\beta_2+\beta_3) \mathds{E}[h(X)] + \beta_4,
\end{align}
which leads to, according to Equation~\eqref{eq_proof_theorem_models3}, 
\begin{align}
 \beta_1 \mathds{V}[h(X)] = 0   
\end{align}
Since this must hold for every distribution of $h(X)$, we deduce that $\beta_1=0$. Finally, there exist $a,b,c \in \mathds{R}$ such that 
\begin{align}
    \tau^P (X) & = f( \mathds{E}[Y^{(0)}|X] , \mathds{E}[Y^{(1)} | X] ) \\
    & = a  \mathds{E}[Y^{(1)}|X] + b \mathds{E}[Y^{(0)}|X] + c. 
\end{align}

\subsubsection{Proof of Lemma~\ref{lemma:intrication_model} (binary outcomes)}\label{proof:intrication_model}

\begin{proof}
Consider a binary outcome $Y$. We further assume that,
 \begin{equation*}
     \forall x \in \mathds{X},\, \forall a \in \{0,1\},\quad 0 < p_a(x) < 1,\quad \text{where } p_a(x)  := \mathbb{P}\left[Y^{(a)} = 1 \mid X=x\right],
 \end{equation*}
which means that the outcome is non-deterministic. Using the law of total expectation, one has

\begin{align*}
 p_1(x) &= \mathbb{P}\left[Y^{(1)} = 1 \mid X=x\right]\\
  & = \mathbb{P}\left[ Y^{(1)} = 1 \mid Y^{(0)} = 0, X = x\right] \mathbb{P}\left[ Y^{(0)} = 0\mid X = x\right] \\
  & \quad \quad + \mathbb{P}\left[ Y^{(1)} = 1 \mid Y^{(0)} = 1, X = x\right] \mathbb{P}\left[ Y^{(0)} = 1\mid X = x\right] \\
  &= \mathbb{P}\left[ Y^{(1)} = 1 \mid Y^{(0)} = 0, X = x\right] (1- p_0(x) )+ \mathbb{P}\left[ Y^{(1)} = 1 \mid Y^{(0)} = 1, X = x\right]  p_0(x) .
\end{align*}

Denoting
\begin{equation*}
    m_g(x):= \mathbb{P}\left[ Y^{(1)} = 0 \mid Y^{(0)} = 1, X = x\right] \quad \text{and} \quad m_b(x):= \mathbb{P}\left[ Y^{(1)} = 1 \mid Y^{(0)} = 0, X = x\right],
\end{equation*}
we finally obtain
\begin{align*}
 p_1(x) &=m_b(x) (1- p_0(x) ) + (1-  m_g(x))  p_0(x) \\
 &= p_0(x) +  m_b(x) (1- p_0(x) )  - p_0(x)m_g(x).
\end{align*}
Therefore, for all $a\in \{0,1\}$, 
\begin{align*}
 p_a(x)  &= p_0(x) +  a \left( m_b(x) (1- p_0(x) )  - p_0(x)m_g(x) \right).
\end{align*}
\end{proof}

\begin{figure}[H]
    \centering
    \includegraphics[width=0.8\linewidth]{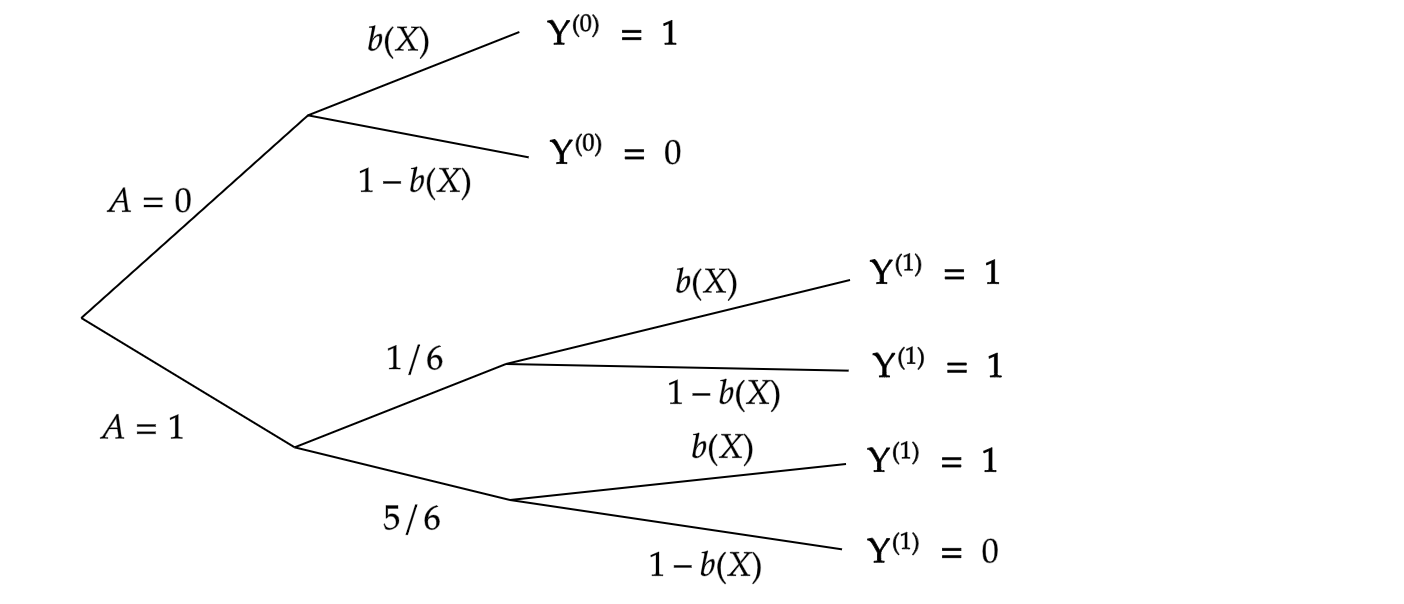}
    \caption{Illustration of the Russian Roulette problem with a probability tree}
    \label{fig:tree}
\end{figure}

Note that the rational of the proof can be captured with a probability tree. Below on Figure~\ref{fig:tree} illustrates the problem with the Russian roulette example.

\subsubsection{Proof of Lemma~\ref{lemma:expression-of-causal-quantities-under-generative-model-binary-outcome-intrication-model}}\label{proof:expression-of-causal-quantities-under-generative-model-binary-outcome-intrication-model}

\begin{lemma}[Expression of the causal measures]\label{lemma:expression-of-causal-quantities-under-generative-model-binary-outcome-intrication-model}
Ensuring conditions of Lemma~\ref{lemma:intrication_model} leads to,
\begin{align*}
  \tau_{\text{\tiny RD}}^P & = \mathbb{E}\left[ \left( 1-b\left(X\right) \right) m_b\left(X\right)\right] -  \mathbb{E}\left[ b\left(X\right)m_g\left(X\right) \right]\\
    \tau_{\text{\tiny NNT}}^P & =  \frac{1}{\mathbb{E}\left[ \left( 1-b\left(X\right) \right) m_b\left(X\right)\right] -  \mathbb{E}\left[ b\left(X\right)m_g\left(X\right) \right] }
\end{align*}
 \begin{align*}
     \tau_{\text{\tiny RR}}^P & = 1  + \frac{ \mathbb{E}\left[  \left( 1-b\left(X\right) \right) m_b\left(X\right) \right] }{ \mathbb{E}\left[ b(X)\right]} - \frac{\mathbb{E}\left[ b(X) m_g\left(X\right)\right]}{\mathbb{E}\left[ b(X)\right]}\\
         \tau_{\text{\tiny SR}}^P & = 1 - \frac{\mathbb{E}\left[ \left( 1-b\left(X\right) \right) m_b\left(X\right) \right]}{\mathbb{E}\left[ 1- b(X)\right]} + \frac{\mathbb{E}\left[ b\left(X\right)m_g\left(X\right) \right]}{\mathbb{E}\left[ 1- b(X)\right]},
 \end{align*}

 \begin{equation*}
     \tau_{\text{\tiny OR}}^P =\frac{\mathbb{E}\left[ b(X) \right]+  \mathbb{E}\left[\left( \left( 1-b\left(X\right) \right) m_b\left(X\right) \right]-  \mathbb{E}\left[b\left(X\right)m_g\left(X\right) \right)\right] }{\mathbb{E}\left[ 1- b(X) \right] - \mathbb{E}\left[ \left( 1-b\left(X\right) \right) m_b\left(X\right)  \right]+  \mathbb{E}\left[b\left(X\right)m_g\left(X\right) \right] }  \frac{\mathbb{E}\left[1-b(X)\right]}{ \mathbb{E}\left[b(X)\right]} .
 \end{equation*}

\end{lemma}%\bc{C'est moche non ?}

\begin{proof}

Consider a binary outcome $Y$. Under the assumptions of  Lemma~\ref{lemma:intrication_model}, there exist probabilities $b(x)$, $ m_g(x)$, and $m_b(x)$ such that 
\begin{align*}
     \mathbb{P}\left[ Y^{(a)} = 1 \mid X = x\right]  &= b(x) +  a\, \left( \left( 1-b\left(x\right) \right) m_b\left(x\right) -  b\left(x\right)m_g\left(x\right) \right).
\end{align*}

Using such a decomposition, one has

    \begin{align*}
        \tau_{\text{\tiny RD}}^P &=  \mathbb{E}\left[ b(X)+  \left( \left( 1-b\left(X\right) \right) m_b\left(X\right) -  b\left(x\right)m_g\left(X\right) \right)\right] -  \mathbb{E}\left[ b(X)\right]\\
        &=  \mathbb{E}\left[ \left( 1-b\left(X\right) \right) m_b\left(X\right)\right] -  \mathbb{E}\left[ b\left(X\right)m_g\left(X\right) \right], \\
        \tau_{\text{\tiny NNT}}^P &=  \frac{1}{\mathbb{E}\left[ \left( 1-b\left(X\right) \right) m_b\left(X\right)\right] -  \mathbb{E}\left[ b\left(X\right)m_g\left(X\right) \right] },\\
         \tau_{\text{\tiny RR}}^P &= \frac{ \mathbb{E}\left[ b(X)+  \left( \left( 1-b\left(X\right) \right) m_b\left(X\right) -  b\left(X\right)m_g\left(X\right) \right)\right] }{ \mathbb{E}\left[ b(X)\right]} \\
         &= 1  + \frac{ \mathbb{E}\left[  \left( 1-b\left(X\right) \right) m_b\left(X\right) \right] }{ \mathbb{E}\left[ b(X)\right]} - \frac{\mathbb{E}\left[ b(X) m_g\left(X\right)\right]}{\mathbb{E}\left[ b(X)\right]}, \\
         \tau_{\text{\tiny SR}}^P &= \frac{ 1- \mathbb{E}\left[ b(X)+  \left( \left( 1-b\left(X\right) \right) m_b\left(X\right) -  b\left(X\right)m_g\left(X\right) \right)\right] }{ 1- \mathbb{E}\left[ b(X)\right]} \\
         &= \frac{\mathbb{E}\left[ 1 - b(X) -  \left( \left( 1-b\left(X\right) \right) m_b\left(X\right) +  b\left(X\right)m_g\left(X\right) \right)\right] }{ \mathbb{E}\left[ 1- b(X)\right]} \\
         &= 1 - \frac{\mathbb{E}\left[ \left( 1-b\left(X\right) \right) m_b\left(X\right) \right]}{\mathbb{E}\left[ 1- b(X)\right]} + \frac{\mathbb{E}\left[ b\left(X\right)m_g\left(X\right) \right]}{\mathbb{E}\left[ 1- b(X)\right]}, \\
    \tau_{\text{\tiny OR}}^P  &= \frac{\mathbb{P}[Y^{(1)} = 1]}{\mathbb{P}[Y^{(1)} = 0]}\, \left(  \frac{\mathbb{P}[Y^{(0)} = 1]}{\mathbb{P}[Y^{(0)} = 0]}\right)^{-1} \\
    &= \frac{\mathbb{E}\left[ b(X)+  \left( \left( 1-b\left(X\right) \right) m_b\left(X\right) -  b\left(X\right)m_g\left(X\right) \right)\right] }{1- \mathbb{E}\left[ b(X)+  \left( \left( 1-b\left(X\right) \right) m_b\left(X\right) -  b\left(X\right)m_g\left(X\right) \right)\right] }  \left(  \frac{\mathbb{E}\left[b(X)\right]}{1- \mathbb{E}\left[b(X)\right]}\right)^{-1} \\
    &=     \frac{\mathbb{E}\left[ b(X)+  \left( \left( 1-b\left(X\right) \right) m_b\left(X\right) -  b\left(X\right)m_g\left(X\right) \right)\right] }{\mathbb{E}\left[ 1- b(X) -  \left( \left( 1-b\left(X\right) \right) m_b\left(X\right) +  b\left(X\right)m_g\left(X\right) \right)\right] }  \frac{\mathbb{E}\left[1-b(X)\right]}{ \mathbb{E}\left[b(X)\right]} \\
    &=     \frac{\mathbb{E}\left[ b(X) \right]+  \mathbb{E}\left[\left( \left( 1-b\left(X\right) \right) m_b\left(X\right) \right]-  \mathbb{E}\left[b\left(X\right)m_g\left(X\right) \right)\right] }{\mathbb{E}\left[ 1- b(X) \right] - \mathbb{E}\left[ \left( 1-b\left(X\right) \right) m_b\left(X\right)  \right]+  \mathbb{E}\left[b\left(X\right)m_g\left(X\right) \right] }  \frac{\mathbb{E}\left[1-b(X)\right]}{ \mathbb{E}\left[b(X)\right]} \\
    &= \left( 1  + \frac{ \mathbb{E}\left[  \left( 1-b\left(X\right) \right) m_b\left(X\right) \right] }{ \mathbb{E}\left[ b(X)\right]} - \frac{\mathbb{E}\left[ b(X) m_g\left(X\right)\right]}{\mathbb{E}\left[ b(X)\right]} \right) \\ 
    & \quad \quad \cdot \left( 1 - \frac{\mathbb{E}\left[ \left( 1-b\left(X\right) \right) m_b\left(X\right) \right]}{\mathbb{E}\left[ 1- b(X)\right]} + \frac{\mathbb{E}\left[ b\left(X\right)m_g\left(X\right) \right]}{\mathbb{E}\left[ 1- b(X)\right]}\right)^{-1}.
\end{align*}
\end{proof}

\subsection{Proofs of Section~\ref{sec:generalization}}

\subsubsection{Proof of Theorem~\ref{theorem:all-covariates}}
\label{subsubsec:proof_generalization_conditional_outcomes}

\begin{proof}
Let $\tau$ be a causal measure defined as 
\begin{align}
    \tau^P = f (\mathds{E}[Y^{(0)}], \mathds{E}[Y^{(1)}] ).
\end{align}
Let $P_{\text{\tiny S}}(X, Y^{(0)}, Y^{(1)})$ and $P_{\text{\tiny T}}(X, Y^{(0)}, Y^{(1)})$ satisfying Assumption~\ref{a:overlap} (overlap assumption) and Assumption~\ref{a:transportability-wide}. By Lemma~\ref{lem_generative_models}, on the source population, for all $x \in \operatorname{supp}(P_{\text{\tiny S}})$, we have
\begin{align}
    \mathds{E}_{\text{\tiny S}}[Y^{(0)} | X =x ] = b(x) \quad \textrm{and} \quad \mathds{E}_{\text{\tiny S}}[Y^{(1)} | X =x ] = g_{b(x)}^{-1} (m(x)), 
\end{align}
where $g_z : z' \mapsto f(z,z')$. According to Assumption~\ref{a:transportability-wide}, for all $x \in \operatorname{supp}(P_{\text{\tiny T}}) \, \cap \, \operatorname{supp}(P_{\text{\tiny S}}) = \operatorname{supp}(P_{\text{\tiny T}})$  (by Assumption~\ref{a:overlap}),
\begin{align}
    &  \mathds{E}_{\text{\tiny T}}[Y^{(0)} | X = x] = \mathds{E}_{\text{\tiny S}}[Y^{(0)} | X = x ] \\
    \textrm{and} ~~ &  \mathds{E}_{\text{\tiny T}}[Y^{(1)} | X = x] = \mathds{E}_{\text{\tiny S}}[Y^{(1)} | X = x ].
\end{align}
Thus, 
\begin{align}
    \mathds{E}_{\text{\tiny T}}[Y^{(0)} | X ] = b(X) \quad \textrm{and} \quad \mathds{E}_{\text{\tiny T}}[Y^{(1)} | X ] = g_{b(X)}^{-1} (m(X)). 
\end{align}
We are interested in estimating the average treatment effect on the target population, that is 
\begin{align}
    \tau^{P_\text{\tiny T}} = f (\mathds{E}_{\text{\tiny T}}[Y^{(0)}], \mathds{E}_{\text{\tiny T}}[Y^{(1)}] ).
\end{align}
According to Definitions~\ref{def:two-kind-covariates} and \ref{def:shidted-covariates}, we have 
\begin{align}
     \mathbb{E}_{\text{\tiny T}}\left[ Y^{(0)} \right] & = \mathbb{E}_{\text{\tiny T}}\left[  b(X) \right] \\
     &=  \mathbb{E}_{\text{\tiny T}}\left[   \mathbb{E}_{\text{\tiny T}}\left[  b(X) \mid X_{\textrm{Sh}} \right] \right] \\
     & = \mathbb{E}_{\text{\tiny T}}\left[   \mathbb{E}_{\text{\tiny S}}\left[  b(X) \mid X_{\textrm{Sh}} \right] \right] \\
     & = \mathbb{E}_{\text{\tiny T}}\left[   \mathbb{E}_{\text{\tiny S}}\left[  b(X) \mid X_{B \cap \textrm{Sh}} \right] \right], \label{proof_th1_eq1}
\end{align}
where the third line comes from Assumption~\ref{a:overlap} and the definition of $X_{Sh}$. Similarly, 
\begin{align}
     \mathbb{E}_{\text{\tiny T}}\left[  Y^{(1)} \right] &= \mathbb{E}_{\text{\tiny T}}\left[ g_{b(X)}^{-1} (m(X)) \right] \\
     &=  \mathbb{E}_{\text{\tiny T}}\left[   \mathbb{E}^{\text{\tiny T}}\left[  g_{b(X)}^{-1} (m(X)) \mid X_{\textrm{Sh}} \right] \right] \\
     & = \mathbb{E}_{\text{\tiny T}}\left[   \mathbb{E}_{\text{\tiny S}}\left[  g_{b(X)}^{-1} (m(X)) \mid X_{\textrm{Sh}} \right] \right] \\
     & = \mathbb{E}_{\text{\tiny T}}\left[   \mathbb{E}_{\text{\tiny S}}\left[  g_{b(X)}^{-1} (m(X)) \mid X_{(M \cup B) \cap \textrm{Sh}} \right] \right]. \label{proof_th1_eq2}
\end{align}
Consequently, one can generalize $\tau$ to the target population by using the formula
\begin{align}
    \tau^{P_\text{\tiny T}} = f \left(\mathbb{E}_{\text{\tiny T}}\left[   \mathbb{E}_{\text{\tiny S}}\left[  b(X) \mid X_{B \cap \textrm{Sh}} \right] \right], \mathbb{E}_{\text{\tiny T}}\left[   \mathbb{E}_{\text{\tiny S}}\left[  g_{b(X)}^{-1} (m(X)) \mid X_{(M \cup B) \cap \textrm{Sh}} \right] \right] \right).
\end{align}

\end{proof}

\subsubsection{Proof of Theorem~\ref{theorem:restricted-set-for-Y-continuous-RD}}
\label{sec:proof_theorem_generalizing_local_effects}

\begin{proof}
Consider the Risk Difference $\tau_{\text{\tiny RD}}$. Let   $P_{\text{\tiny S}}(X, Y^{(0)}, Y^{(1)})$ and $P_{\text{\tiny T}}(X, Y^{(0)}, Y^{(1)})$ satisfying Assumption~\ref{a:overlap} (overlap assumption) and Assumption~\ref{a:transportability}. 
Since $\tau_{\text{\tiny RD}}$ satisfies Assumption~\ref{ass:injection_def_domain}, Corollary~\ref{lemma:working-model-continuous-Y} can be applied on the source population, that is, for all $x \in \operatorname{supp}(P_{\text{\tiny S}})$, we have
\begin{align}
    \mathds{E}_{\text{\tiny S}}[Y^{(0)} | X =x ] = b(x) \quad \textrm{and} \quad \mathds{E}_{\text{\tiny S}}[Y^{(1)} | X =x ] = b(x) + m(x). 
\end{align}
Thus, for all $x \in \operatorname{supp}(P_{\text{\tiny S}})$,
\begin{align}
m(x) = \mathds{E}_{\text{\tiny S}}[Y^{(1)} - Y^{(0)} | X =x ].
\end{align}
According to Assumption~\ref{a:transportability}, for all $ x \in \operatorname{supp}(P_{\text{\tiny T}}) \cap \operatorname{supp}(P_{\text{\tiny S}}) = \operatorname{supp}(P_{\text{\tiny T}})$ (Assumption~\ref{a:overlap}),
\begin{align}
m(x) = \mathds{E}_{\text{\tiny T}}[Y^{(1)} - Y^{(0)} | X =x ].
\end{align}

%We start by proving that Assumption~\ref{a:transportability} is satisfied for $\tau_{\text{\tiny RD}}(X_{M \cap \textrm{Sh}})$. 
Since $\tau_{\text{\tiny RD}}$ is directly collapsible, we have
\begin{align}
\tau_{\text{\tiny RD}}^{P_{\text{\tiny T}}} & = \mathds{E}_{\text{\tiny T}}[m(X)] \\
& = \mathds{E}_{\text{\tiny T}} \left[ \mathds{E}_{\text{\tiny T}}\left[    m(X) \mid X_{M \cap \textrm{Sh}} \right] \right],
\end{align}
where
\begin{align*}
%\tau^{\text{\tiny T}}_{\text{\tiny RD}}(X_{M \cap \textrm{Sh}}) & = \mathbb{E}_{\text{\tiny T}}\left[ Y^{(1)} - Y^{(0)} \mid X_{M \cap \textrm{Sh}} \right] && \\
%& = 
\mathbb{E}_{\text{\tiny T}}\left[    m(X) \mid X_{M \cap \textrm{Sh}} \right]  %\text{Lemma~\ref{lemma:working-model-continuous-Y}} \\
& = \mathbb{E}_{\text{\tiny T}}\left[    m(X) \mid X_{\textrm{Sh}} \right] && \text{Definition~\ref{def:two-kind-covariates}} \\
& = \mathbb{E}_{\text{\tiny S}}\left[    m(X) \mid X_{\textrm{Sh}} \right] && \text{Definition~\ref{def:shidted-covariates}} \\
& = \mathbb{E}_{\text{\tiny S}}\left[    m(X) \mid X_{M \cap \textrm{Sh}} \right] && \text{Definition~\ref{def:two-kind-covariates}}\\
& = \tau^{P_\text{\tiny S}}_{\text{\tiny RD}}(X_{M \cap \textrm{Sh}}) .
 \end{align*}
 Consequently, 
\begin{align}
\tau_{\text{\tiny RD}}^{P_{\text{\tiny T}}}  
& = \mathds{E}_{\text{\tiny T}} \left[  \tau^{P_\text{\tiny S}}_{\text{\tiny RD}}(X_{M \cap \textrm{Sh}}) \right],
\end{align}
%Thus, Assumption~\ref{a:transportability} is verified for $\tau_{\text{\tiny RD}}$ with covariates $X_{M \cap \textrm{Sh}}$. Furthermore, by assumption in Theorem~\ref{theorem:restricted-set-for-Y-continuous-RD}, Assumption~\ref{a:overlap} is verified with  covariates $X_{M \cap \textrm{Sh}}$. Thus, by Proposition~\ref{prop:generalization-of-local-effects}, 
and $\tau_{\text{\tiny RD}}$ is generalizable with covariates $X_{M \cap \textrm{Sh}}$.

\end{proof}

\subsubsection{Proof of Theorem~\ref{thm_homogeneous_independence_generalization}}

The proof is straightforward by recalling that any collapsible causal measure satisfies Definition~\ref{def:indirect-collapsibility}.

\section{Comments on logistic regression}\label{appendix:usual-point-of-view}

A common practice in applied statistics is to adopt a logistic regression model (or any model encapsulating a function taking values in $\mathbb{R}$), for example assuming that the following logistic model holds:
\begin{equation}\label{eq:typical-model-used-binary-Y-main}
   \operatorname{log}\left( \frac{\mathbb{P}(Y^{(a)} = 1 \mid X)}{\mathbb{P}(Y^{(a)} = 0 \mid X) } \right) = \beta_0 + \langle \boldsymbol{\beta}, \boldsymbol{X} \rangle + A\,m,
\end{equation}
%\es{à relire}\bc{partiellement relu et modifié suite à la lecture}
where $\beta_0, \boldsymbol{\beta}$ and $m$ are the coefficients of a linear model (see for example \cite{Daniel2020MakingApple}). When the discriminative model from Equation~\ref{eq:typical-model-used-binary-Y-main} holds, some nice properties arise.
Notably, one can show that this implies constant conditional odds ratio $\tau_{\text{\tiny log-OR}} (x) = m$ and $\tau_{\text{\tiny OR}} (x) = e^{m}$. The derivations are detailed below:

\begin{align*}
    \tau_{\text{\tiny OR}}(X) &:= \frac{\mathbb{P}(Y^{(1)} = 1 \mid X)}{\mathbb{P}(Y^{(1)} = 0 \mid X) } \cdot \left( \frac{\mathbb{P}(Y^{(0)} = 1 \mid X)}{\mathbb{P}(Y^{(0)} = 0 \mid X) } \right)^{-1}\\
    & = e^{\beta_0 + \langle \boldsymbol{\beta}, \boldsymbol{X} \rangle + m} \cdot e^{-\beta_0 - \langle \boldsymbol{\beta}, \boldsymbol{X} \rangle} \\
    &= e^{m}.
\end{align*}

Beyond \eqref{eq:typical-model-used-binary-Y-main} it is possible to encapsulate non-parametric functions in the logit. Such decomposition is present in the literature \citep{Gao2021DINA} (and see Section~\ref{appendix:usual-point-of-view}, and in particular Lemma~\ref{lem:conditional-odds-ratio} for details).

\begin{lemma}[Logit discriminative model for a binary outcome]\label{lemma:generative-model-binary-Y}
 Considering a binary outcome $Y$, assume that 
 
 \begin{equation*}
     \forall x \in \mathds{X},\, \forall a \in \{0,1\},\quad 0 < p_a(x) < 1,\quad \text{where } p_a(x)  = \mathbb{P}(Y^{(a)} = 1 \mid X=x).
 \end{equation*}

 Then, there exist two functions $b, m:\mathcal{X} \to \mathbb{R}$ such that

\begin{equation*}
\operatorname{ln}\left( \frac{\mathbb{P}(Y^{(a)} = 1 \mid X)}{\mathbb{P}(Y^{(a)} = 0 \mid X) } \right) = b(X) + a\, m(X).
\end{equation*}
\end{lemma}

\begin{proof}
{\footnotesize {\color{Blue} Consider $a \in \{0,1\}$, and assume that their exists a function $p_a:\mathbb{R}^d \to ]0,1[$ such that,

\begin{equation*}
    \mathbb{P}(Y^{(a)} = 1 \mid X) = p_a(X).
\end{equation*}

Because $p_a$ takes values in  $]0,1[$ the odds can be considered, so that,

\begin{equation*}
    \operatorname{ln}\left(\frac{\mathbb{P}(Y^{(a)} = 1 \mid X)}{\mathbb{P}(Y^{(a)} = 0 \mid X) }\right) =   \operatorname{ln}\left(\frac{p_a(X)}{1-p_a(X)}\right).
\end{equation*}

Denoting, 

\begin{equation*}
    b(X) := \operatorname{ln}\left(\frac{p_0(X)}{1-p_0(X)}\right),
\end{equation*}

and

\begin{equation*}
   m(X) \ := \operatorname{ln}\left(\frac{p_1(X)}{1-p_1(X)}\right) - \operatorname{ln}\left(\frac{p_0(X)}{1-p_0(X)}\right) = \operatorname{ln}\left(\frac{p_1(X)}{1-p_1(X)}\, \frac{1-p_0(X)}{p_0(X)} \right),
\end{equation*}

one can write the log-odds as

\begin{equation*}
    \operatorname{ln}\left(\frac{\mathbb{P}(Y^{(a)} = 1 \mid X)}{\mathbb{P}(Y^{(a)} = 0 \mid X) }\right) =   b(X) + A\, m(X).
\end{equation*}

Note that another link function could have been chosen, which impacts how $b(x)$ and $m(x)$ are defined.}}
\end{proof}

\begin{lemma}[Conditional log odds ratio]\label{lem:conditional-odds-ratio}
Ensuring conditions of Lemma~\ref{lemma:generative-model-binary-Y} leads to,

\begin{equation*}
     \mathbb{E}\left[ \tau_{\text{\tiny log-OR}} (X) \right]:= \mathbb{E}\left[\operatorname{ln}\left( \frac{\mathbb{P}(Y^{(1)} = 1 \mid X)}{\mathbb{P}(Y^{(1)} = 0 \mid X) } \, \left( \frac{\mathbb{P}(Y^{(0)} = 1 \mid X)}{\mathbb{P}(Y^{(0)} = 0 \mid X) }  \right)^{-1} \right) \right]=  \mathbb{E}\left[m(X)\right].
\end{equation*}

\end{lemma}

This result is apparently satisfying, where $\mathbb{E}\left[ \tau_{\text{\tiny log-OR}} (X) \right]$ somehow only grasps the modification function. Still, note that due to non-collapsibility of the odds ratio, this \underline{does not imply} that $\tau_{\text{\tiny log-OR}} = \tau$ (i.e. $\tau_{\text{\tiny OR}}= e^{\tau}$) because $ \mathbb{E}\left[ \tau_{\text{\tiny log-OR}} (X) \right] \neq  \tau_{\text{\tiny log-OR}}$ (except if treatment effect is null or if the outcome does not depend on $X$, that is $b(X)$ and $m(X)$ are both scalars). As an intermediary conclusion, the working model from Lemma~\ref{lemma:generative-model-binary-Y} leads to complex expression of causal measures, except for $\mathbb{E}\left[ \tau_{\text{\tiny log-OR}} (X) \right]$, but with the default that this measure shows bad property of non-collapsibility.

For example, a working model such that $m(x) = m$ is a constant don't lead to any measures to be constant.

\begin{lemma}\label{lemma:expression-of-causal-quantities-under-generative-model-binary-outcome}
Ensuring conditions of Lemma~\ref{lemma:generative-model-binary-Y} leads to,
\begin{align}
     \tau_{\text{\tiny RD}} &= \mathbb{E}\left[ \frac{e^{b(X) + m(X)}}{1+e^{b(X) + m(X)}}  \right] - \mathbb{E}\left[ \frac{e^{b(X)}}{1+e^{b(X)}}  \right]  \\
       \tau_{\text{\tiny ERR}} &= \mathbb{E}\left[ \frac{e^{b(X) + m(X)}}{1+e^{b(X) + m(X)}}  \right] \left(\mathbb{E}\left[ \frac{e^{b(X)}}{1+e^{b(X)}}  \right]  \right)^{-1} -1  \\ 
      \tau_{\text{\tiny NNT}} &= \left( \mathbb{E}\left[ \frac{e^{b(X) + m(X)}}{1+e^{b(X) + m(X)}}  \right] - \mathbb{E}\left[ \frac{e^{b(X)}}{1+e^{b(X)}}  \right] \right)^{-1}\\   
     \tau_{\text{\tiny RR}} &= \mathbb{E}\left[ \frac{e^{b(X)+m(X)}}{1+e^{b(X)+m(X)}}\right] \left( \mathbb{E}\left[ \frac{e^{b(X)}}{1+e^{b(X)}}\right] \right)^{-1} \\
     \tau_{\text{\tiny SR}} &= \mathbb{E}\left[ \left( 1 + e^{b(X)+m(X)} \right)^{-1}  \right] \left(\mathbb{E}\left[ \left(1 + e^{b(X)}  \right)^{-1} \right]\right)^{-1} \\
     \tau_{\text{\tiny OR}} &= \frac{\mathbb{E}\left[ \frac{e^{b(X)+m(X)}}{1 +e^{b(X)+m(X)} }\right]}{\mathbb{E}\left[ \frac{1}{1+e^{b(X)+m(X)}}\right]} \frac{\mathbb{E}\left[ \frac{1}{1 + e^{b(X)} }\right]}{\mathbb{E}\left[ \frac{e^{b(X)}}{1+e^{b(X)}}\right]}. 
\end{align}
\end{lemma}

All expressions from Lemma~\ref{lemma:expression-of-causal-quantities-under-generative-model-binary-outcome} now involve both $b(.)$ and $m(x)$. All other metrics show complex relation between the two functions.

Finally, note that the logistic model is unable to easily describe accurately the Russian Roulette thought being simple. Note that
\begin{equation*}
   \operatorname{log}\left( \frac{\mathbb{P}(Y^{(a)} = 1 \mid X = x)}{\mathbb{P}(Y^{(a)} = 0 \mid X =x) } \right) = \operatorname{log}\left( \frac{b(x)}{1-b(x)}\right)+ A\, \operatorname{log}\left( \frac{\left(\frac{1}{6} + b(x) \right)}{1-\left(\frac{1}{6} + b(x))(1-b(x))\right)} \cdot \frac{1-b(x)}{b(x)} \right),
\end{equation*}
is the equivalent to Equation~\ref{eq:intuition-of-intrication}.

\section{More details about the Russian Roulette example}\label{appendix:more-details-on-the-intrication-model}

\textit{We provide more details on how the Russian Roulette is stated in \cite{CinelliGeneralizing2019}. Note that the first reference we have found of this problem is in \cite{Huitfeldt2019LessWrong}. This section is just meant to recall how the problem was initially introduced by \cite{Huitfeldt2019LessWrong}.}\\

Suppose the city of Los Angeles decides to run a randomized control trial.  Running the experiment, the mayor of Los Angeles discovers that “Russian Roulette” is harmful: among those assigned to play Russian Roulette, 17.5\% of the people died, as compared to only 1\% among those who were not assigned to play the game (people can die due to other causes during the trial, for example, prior poor health conditions). This example is a good toy example as the mechanism is well-known, with a chance of one over six to die when playing. Even if it seems counter-intuitive, we consider the treatment as being forced to play to the russian roulette (we consider the player plays only one time). 
We denote by $\Pi$ the population from Los Angeles. In that case, we can already note that the RR is $17.5$ and the ATE is $0.165$ (outcome being $Y$ equals to 1 if death before the end of the period). With this notation $\mathrm{E}[Y^{(0)} | pop = \Pi] = 0.01$ and $\mathrm{E}[Y^{(1)} | pop = \Pi] = 0.175$\\

After hearing the news about the Los Angeles experiment, the mayor of New York City (a dictator, and we propose to denote the population of New York City by $\Pi^*$) wonders what the overall mortality rate would be if the city forced everyone to play Russian Roulette. Currently, the practice of Russian Roulette is forbidden in New York, and its mortality rate is at 5\% (4\% higher than LA, being $\mathrm{E}[Y^{(0)} | pop = \Pi^*] = 0.05$). The mayor thus asks the city’s statistician to decide whether and how one could use the data from from Los Angeles to predict the mortality rate in New York, once the new policy is implemented. But in fact, knowing the mechanism of the russian roulette we can already compute the value of interest being $\mathrm{E}[Y^{(1)} | pop = \Pi^*]$. Results are presented in Table~\ref{tab:summmary_russian_roulette}. Here we used the fact that mortality is a consequence of two “independent” processes (the game of Russian Roulette and prior health conditions of the individual), and while the first factor remains unaltered across cities, the second intensifies by a known amount (5\% vs 1\%).  Moreover, we can safely assume that the two processes interact disjunctively, namely, that death occurs if and only if at least one of the two processes takes effect. We can also - within the two cities - compute the associated RR, ATE and survival ratio (SR). We can observe they are not the same, but only the survival ratio comparing how many people dies with treatment on how many people would have died without treatement, transport the \textit{mechanism} of the Russian Roulette (note that $\frac{5}{6} \sim 0.83$).

\begin{table}[!h]
\begin{center}
\begin{tabular}{l|l|l}
\hline
Population &  Los Angeles ($\Pi$) & New York city ($\Pi^*$)  \\ \hline \hline
$\mathrm{E}\left[Y^{(0)}\right]$ & 0.01 & 0.05 \\ \hline
$\mathrm{E}\left[Y^{(1)}\right]$ & $\frac{1}{6}0.99 + 0.01 = 0.175$ & $\frac{1}{6}0.95 + 0.05 = 0.208$ \\ \hline
RR & 17.5  & 4.16 \\ \hline
ATE & 0.165 & 0.158 \\ \hline
SR & 0.83 & 0.83 \\
\end{tabular}
\caption{Summary of the different values. Note that none of the transport equation is applied, everything is computed within each population taking into account a distinct mechanism between the two reasons to die. SR corresponds to the survival ratio.}
\label{tab:summmary_russian_roulette}
\end{center}
\end{table}

\section{Different points of view}\label{appendix:different-point-of-views}

\textit{This section gathers quotes from research papers or books. The aim is to illustrate how diverse opinions are.}

\paragraph*{General remarks about the choice of measure}
\begin{quote}
    Physicians, consumers, and third-party payers may
be more enthusiastic about long-term preventive treatments when benefits are stated as relative, rather than absolute, reductions in the risk of adverse events.
Medical-journal editors have said that reporting only relative reductions in risk is usually inadequate in scientific articles and have urged the news media to consider the importance of discussing both absolute and relative risks. For example, a story reporting that
in patients with myocardial infarction, a new drug
reduces the mortality rate at two years from 10 percent to 7 percent may help patients weigh both the
3 percent absolute and the 30 percent relative reduction in risk against the costs of the drug and its side
effects. -- \citep{moynihan2000coverage}
\end{quote}

\begin{quote}
    In general, giving only the absolute or only the relative benefits does not tell the full story; it is more
informative if both researchers and the media make
data available in both absolute and relative terms.  -- \citep{moynihan2000coverage}
\end{quote}

\begin{quote}
    The promotion of a measure often reflects personal preferences – those who are keen to promote the use of research in practice emphasize issues of interpretability of Risk Ratios and risk differences, those who are keen to ensure mathematical rules are always obeyed emphasize the limitations and inadequacies of the same measures. -- \citep{Deeks2022IssuesInSelection}
\end{quote}

\begin{quote}
    Failing to report NNT may influence the interpretation of study results. For example reporting RR alone may lead a reader to believe that a treatment effect is larger than it really is. -- \citep{Nuovo2002ReportingNNT}
\end{quote}

\begin{quote}
    As \textit{evidence-based practitioners}, we must decide which measure of association deserves our focus. Does it matter? The answer is yes. The same results, when presented in different ways, may lead to different treatment decisions. -- \citep{Cook2014UserGuide}
\end{quote}

\begin{quote}
    You must, however, distinguish between the RR and the RD. The reason is that the RR is generally far larger than the RD, and presentations of results in the form of RR (or RRR) can convey a misleading message. -- (focusing on binary outcome) \citep{Cook2014UserGuide}
\end{quote}

\begin{quote}
    Standard measures of effect, including the Risk Ratio, the odds ratio, and the risk difference, are associated with a number of well-described shortcomings, and no consensus exists about the conditions under which investigators should choose one effect measure over another. -- \citep{huitfeldt2018choice}
\end{quote}

\begin{quote}
    Additive treatment effect heterogeneity is also most informative for guiding public health policy that aims to maximize the benefit or minimize the harm of an exposure by targeting subgroups. The relative scale (Risk Ratios or odds ratios) can tend to overstate treatment benefits or harms. -- \citep{lesko2018considerations}
\end{quote}

\begin{quote}
   The way to express and measure risk may appear to be a pure technicality. In fact, it is a crucial element of the risk-benefit balance that underlies the dominant medical discourse on contraception. Its influence on the perception and communication of risk is decisive, especially among people without a solid statistical education, like most patients and doctors who prescribe the pill (mostly generalists and gynaecologists). The dispute over \textit{Non-rare thrombophilia} (NRT) screening sets an important difference between the absolute risk, the number of events occurring per time unit and the relative risk, which is the ratio between two absolute risks. Practically, whereas the relative risk may sound alarming, the absolute risk looks more reassuring. -- \citep{Bourgain2021Appraising}
\end{quote}

\begin{quote}
    We believe if an efficacy measure is
    \begin{itemize}
        \item well defined,
        \item understandable by human,
        \item desired by patients and clinicians,
        \item proven to be logic-respecting\footnote{see Definition~\ref{def:logic-respecting}.},
        \item readily implementable computationally,
    \end{itemize}

    them it is worthy of consideration. -- \citep{liu2022rejoinder}
\end{quote}

\paragraph*{The odds ratio as a complex measure to interpret}

\begin{quote}
    Odds ratios and parameters of multivariate models will often be useful in serving as or in constructing the estimates, but should not be treated as the end product of a statistical analysis of epidemiologic data or as summaries of effect in themselves. -- \citep{Greenland1987Interpretation}
\end{quote}

\begin{quote}
   The concept of the odds ratio is now well-established in epidemiology, largely because it serves as a link between results obtainable from follow-up studies and those obtainable from case-control studies. [$\dots$] This ubiquity, along with certain technical considerations, has led some authors to treat the odds ratio as perhaps a ``universal" measure of epidemiologic effect, in that they would estimate odds ratios in follow-up studies as well as case-control studies; others have expressed reservations about the utility of the odds ratio as something other than an estimate of an incidence ratio. I believe that such controversy as exists regarding the use of the odds ratio arises from its inherent disadvantages compared with the other measures for biological inference, and its inherent advantages for statistical inference.  -- \citep{Greenland1987Interpretation}
\end{quote}

\begin{quote}
   There is a problem with odds: unlike risks, they are difficult to understand.  -- \citep{davies1998can}
\end{quote}

\begin{quote}
    Another measure often used to summarise effects of treatment is the odds ratio. This is defined as the odds of an event in the active treatment group divided by the odds of an event in the control group. Though this measure has several statistical advantages and is used extensively in epidemiology, we will not pursue it here as it is not helpful in clinical decision making. -- \citep{Cook1995NNT}
\end{quote}

\begin{quote}
    In logit and other multiplicative intercept models (but not generally), OR also has the attractive feature of being invariant with respect to the values at which control variables are held constant. The disadvantage of OR is understanding what it means, and when OR is not the quantity of interest then its ‘advantages’ are not suficient to recommend its use. Some statisticians seem comfortable with OR as their ultimate quantity of interest, but this is not common. Even more unusual is to find anyone who feels more comfortable with OR than the other quantities defined above; we have found no author who claims to be more comfortable communicating with the general public using an odds ratio. -- \citep{king2002estimating}
\end{quote}

\begin{quote}
    The OR lacks any interpretation as an average. -- \citep{Cummings2009RelativeMeritsRRAndOR}
\end{quote}
\begin{quote}
    As is well established, the odds ratio is not a parameter of interest in public health research. -- \citep{Spiegelman2017Modeling}
\end{quote}

\begin{quote}
    Because of the exaggeration present, it is important to avoid representing ORs as RRs, and similarly, it is important to recognize that a reported OR rarely provides a good approximation of relative risks but rather simply provides a measure of correlation. -- \citep{George2020WhatsTR}
\end{quote}

\begin{quote}
    We agree with Liu et al. (2020) that (causal) odds ratios and hazard ratios are problematic as causal contrasts. The non-collapsibility of these parameters is a mathematical property which makes their interpretation awkward, and this is amplified for hazard by their conditioning on survival. Thus they are also unsuitable measures for transportability between different populations (Martinussen \& Vansteelandt, 2013). It is particularly concerning that meta-analyses pool odds ratios or hazard ratios from different studies each possibly using different variables for adjustment where the issue of non-collapsibility is typically ignored. -- \citep{Didelez2021collapsibility}
\end{quote}

\begin{quote}
    ORs are notoriously difficult to interpret. When people hear “odds” they think of “risks” and this leads to the common misinterpretation of the OR as a RR by scientists and the public, which is a serious concern. For example, an OR of 2 is not generally a doubling of risk (if the risk in the control group is 20\% and the OR is 2, then the risk in the treated group is 33.3\% not 40\%). In contrast, the RD and RR offer clearer interpretations. -- \citep{xiao2022IsORPortable}
\end{quote}

\begin{quote}
    The admitted mathematical niceties of the OR are not reason enough to accept such a confusing state of affairs. Of course, when the outcome is rare, the OR approximates the RR and is, therefore, approximately collapsible.-- \citep{xiao2022IsORPortable}
\end{quote}

\begin{quote}
    Because of the interpretability issues and lack of collapsibility, we urge researchers to avoid ORs when either the RD or RR is available. -- \citep{xiao2022IsORPortable} 
\end{quote}

\begin{quote}
    Odds ratios provoke similar discomfort—only 19\% of learners and 25\% of speakers at an annual meeting of the Canadian Society of Internal
Medicine (CSIM) understood odds ratios well enough to explain them to others. -- \citep{Lapointe2022FromMathToMeaning}
\end{quote}

\paragraph*{The OR is a better metric to use than RR}

\begin{quote}
    The results demonstrate the need to a) end the primary use of the RR in clinical trials and meta-analyses as its direct interpretation is not meaningful; b) replace the RR by the OR; and c) only use the post-intervention risk recalculated from the OR for any expected level of baseline risk in absolute terms for purposes of interpretation such as the number needed to treat. -- \citep{Doi2020callToChangePractice}
\end{quote}

\begin{quote}
    We can no longer accept the commonly argued for view that the relative risk is easier to understand. Once we realize that the RR depends more on prevalence than the exposure-outcome association, its interpretation becomes much more difficult to comprehend than the odds ratio. It is well known that, for common events, large values of the Risk Ratio are impossible and this should have rung the alarm bells much earlier regarding whether the RR is more a measure of prevalence than a measure of effect. However this was not the main focus of the derivation outlined previously and the latter was aimed at demonstrating why the OR is a true measure of effect against which the RR can be compared. -- \citep{Doi2020callToChangePractice}
\end{quote}

\begin{quote}
    Our response to this is that, although this is certainly a problem, there is an even bigger problem – \textit{the RR is not a portable measure of effect}. By "portable" we mean a numerical value that is not dependent on baseline risk and not transportability in causal inference. --- \citep{doi2022TimeToDoAway}
\end{quote}

\paragraph*{Relative versus absolute measures}
\begin{quote}
    In reviewing the different ways that benefit and harm can be expressed, we conclude that the RD is superior to the RR because it incorporates both the baseline risk and the magnitude of the risk reduction. -- \citep{Laupacis1988AnAssessmentOfClinically}
\end{quote}

\begin{quote}
    For clinical decision making, however, it is more meaningful to use the measure “number needed to treat.” This measure is calculated on the inverse of the absolute risk reduction. It has the advantage that it conveys both statistical and clinical significance to the doctor. Furthermore, it can be used to extrapolate published findings to a patient at an arbitrary specified baseline risk when the relative risk reduction associated with treatment is constant for all levels of risk. -- \citep{Cook1995NNT}
\end{quote}

\begin{quote}
    Medical journals need to be conscious that they will contribute to scaremongering newspaper headlines if they do not request authors to quantify Adverse Drug Reactions (ADR) into best estimates of absolute numbers. -- \citep{Mills1999PillScare}
\end{quote}

\begin{quote}
    As a relative measure of effect, the RR is most directly estimated by the multiplicative model when it fits the data. The risk difference is an absolute measure of effect, most directly estimated by the additive model when it fits the data. -- \citep{Spiegelman2017Modeling}
\end{quote}

\paragraph*{About portability or generalizability of causal effects}

\begin{quote}
    The numbers needed to treat method still presents a
problem when applying the results of a published
randomised trial in patients at one baseline risk to a
particular patient at a different risk. -- \citep{Cook1995NNT}
\end{quote}
\begin{quote}
    Some authors prefer odds ratios because they believe a constant (homogeneous) odds ratio may be more plausible than a constant Risk Ratio when outcomes are common. -- \citep{Cummings2009RelativeMeritsRRAndOR}
\end{quote} 

\begin{quote}
    All of this assumes a constant RR across risk groups; fortunately, a more or less constant RR is usually the case, and we suggest you make that assumption unless there is evidence that suggests it is incorrect. -- \citep{Cook2014UserGuide}
\end{quote}

\begin{quote}
    Although further and more formal quantitative work evaluating the relative degree of heterogeneity for Risk Ratio versus risk differences may be important, the previously mentioned considerations do seem to provide some indication that, for whatever reason, Risk Ratio modification is uncommon. -- \citep{Spiegelman2017Modeling}
\end{quote}

\begin{quote}
    It is commonly believed that the Risk Ratio is a more homogeneous effect measure than the risk difference, but recent methodological discussion has questioned the evidence for the conventional wisdom. -- \citep{huitfeldt2018choice}
\end{quote}

\begin{quote}
    In the real world of clinical medicine, doctors are usually given information about the effects of a drug on the Risk Ratio scale (the probability of the outcome if treated, divided by the probability of the outcome if untreated). With information on the Risk Ratio, a doctor may make a prediction for what will happen to the patient if treated, by multiplying the Risk Ratio and patient's risk if untreated (which is predicted informally based on observable markers for the patient's condition).  -- \citep{Huitfeldt2019LessWrong}
\end{quote}

\begin{quote}
    In this article we will show that the RR is not a measure of the magnitude of the intervention-outcome association alone because it as stronger relationship with prevalence and therefore is not generalizable beyond the baseline risk of the population in which it is computed. -- \citep{Doi2020callToChangePractice}
\end{quote}

\begin{quote}
    It is possible that no effect measure is “portable” in a meta-analysis. In cases where portability of the effect measure is challenging to satisfy, we suggest presenting the conditional effect based on the baseline risk using a bivariate generalized linear mixed model. The bivariate generalized linear mixed model can be used to account for correlation between the effect measure and baseline disease risk. Furthermore, in addition to the overall (or marginal) effect, we recommend that investigators also report the effects conditioning on the baseline risk. -- \citep{xiao2022IsORPortable}
\end{quote}

\begin{quote}
    Despite some concerns, the RR has been widely used because it is considered a measure with “portability” across varying outcome prevalence, especially when the outcome is rare. -- \citep{Suhail2022CallForChange}
\end{quote}

\section{Comments and answers to related articles}

As highlighted by the length of the references or even by Section~\ref{appendix:different-point-of-views}: the literature on the choice of causal measures is prolific. In this Section, we propose comments or answers to previous articles in order to show how our contributions either complete what was said or shed lights on a different apprehension of the problem.

\subsection{Comments of \cite{Cummings2009RelativeMeritsRRAndOR}}

\cite{Cummings2009RelativeMeritsRRAndOR} propose a review of how the OR and the RR differ. In particular, they review typical arguments for pro and cons, while providing examples. In this section, we want to comment how the entanglement model (Lemma~\ref{lemma:intrication_model}) allows to formalize many of their arguments and examples.

\begin{quote}
    \textit{Some authors prefer odds ratios because they believe a constant (homogeneous) odds ratio may be more plausible than a constant Risk Ratio when outcomes are common. Risk range from 0 to 1. Risk Ratios greater than 1 have an upper limit constrained by the risk when not exposed. For example the risk when not exposed is $0.5$, the Risk Ratio when exposed cannot exceed $2: 5\cdot 2 = 1$. In a population with an average Risk Ratio of $2$ for outcome $Y$ among those exposed to $X$, assuming that the risk for $Y$ if not exposed to $X$ varies from .1 to .9, the average Risk Ratio must be less than $2$ for those with risks greater than  $0.5$ when not exposed. Because the average Risk Ratio for the entire population is $2$, the average Risk Ratio must be more than $2$ for those with risks less than $.5$ when not exposed. Therefore, a Risk Ratio of $2$ cannot be constant (homogeneous) for all individuals in a population if risk when not exposed is sometimes greater than $.5$. More generally, if the average Risk Ratio is greater than $1$ in a population, the individual Risk Ratios cannot be constant (homogeneous) for all persons if any of them have risks when not exposed that exceed 1/average Risk Ratio. }
\end{quote}

The authors claim that if $\tau_{\text{\tiny RR}} > 1$, then

\begin{itemize}
    \item The RR has an upper limit linked to the risk of the unexposed ($p_0(x) = b(x)$),
    \item Or, the RR cannot be constant on every individuals if their risk is above a certain threshold being equal to 1/average Risk Ratio.
\end{itemize}

\begin{figure}[H]
    \begin{minipage}{.35\linewidth}
	\caption{\textbf{Illustration of the impossibility of having a constant $\tau_{\text{\tiny RR}}(x) > 1$ if allowing all ranges for baseline risks $p_0(x)$}: This plot illustrates \eqref{eq:cumming1} for several constant values of $\tau_{\text{\tiny RR}}(x)$ (from $1.2$ to $4$), showing how the baseline risk $p_0(x)$ implies different values of $m_b(x)$. If the baseline risk is too high, then there is no plausible $m_b(x)$ (the upper limit is highlighted with the dashed red line). The dark vertical dashed line illustrate the precise example of \cite{Cummings2009RelativeMeritsRRAndOR} with $\tau_{\text{\tiny RR}}(x)=2$.}
	\label{fig:cumming1}
    \end{minipage}%
    \hfill%
    \begin{minipage}{.62\linewidth}
     \includegraphics[width = 0.8\textwidth]{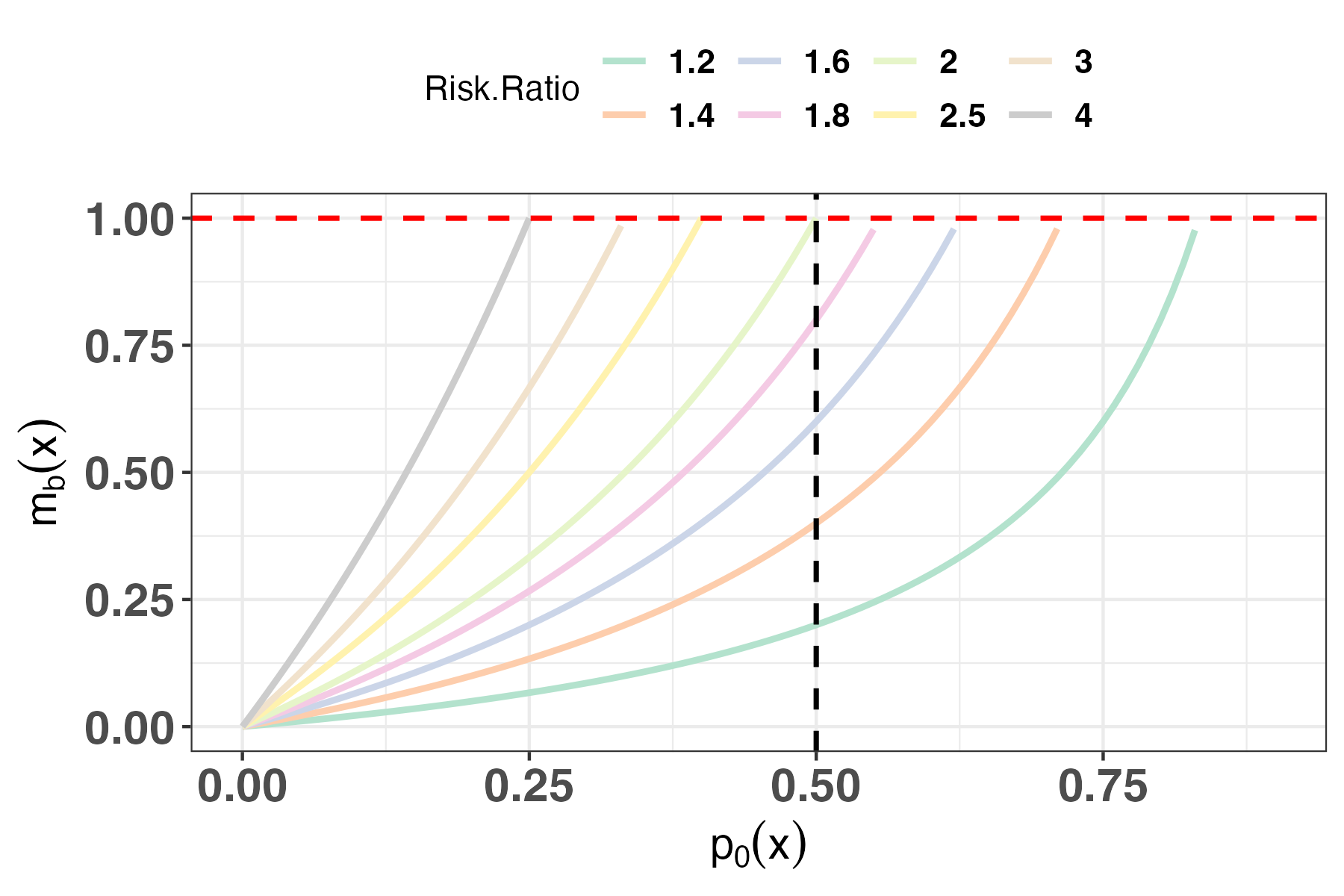}
    \end{minipage}
\end{figure}

The entanglement model perfectly describes such a situation, and we propose to illustrate why. As authors consider that $\tau_{\text{\tiny RR}} > 1$, then we use Lemma~\ref{lemma:intrication_model} with $\forall x,\,  m_g(x)=0$. More specifically, the authors mention that for $\tau_{\text{\tiny RR}} > 1$ (that we rather model as $\forall x,\,  m_g(x)=0$), it is not possible to have a constant RR on each subgroup. We recall that, 

\begin{align}\label{eq:cumming1}
 \forall x, \,   \tau_{\text{\tiny RR}}(x) =  1 + \frac{1-b(x)}{b(x)}m_b(x)
\end{align}
If $\tau_{\text{\tiny RR}}(x)$ is assumed constant, one can plot the probability $m_b(x)$ as a function of $b(x)$ and observe that indeed this quantity is bounded and/or that $m_b(x)$ can not exist for all baseline $b(x)$. We illustrate this equation on Figure~\ref{fig:cumming1}.

\begin{figure}[H]
    \begin{minipage}{.35\linewidth}
	\caption{\textbf{Illustration of the possibility to have a constant $\tau_{\text{\tiny SR}}(x) < 1$ when allowing all ranges for baseline risks $p_0(x)$}: This plot illustrates how several constant values of $\tau_{\text{\tiny SR}}(x)$ (from $0.2$ to $0.9$) is allowed for any baseline values $p(x)$. Note that this implies a constant $m_b(x)$.}
	\label{fig:cumming2}
    \end{minipage}%
    \hfill%
    \begin{minipage}{.62\linewidth}
     \includegraphics[width = 0.8\textwidth]{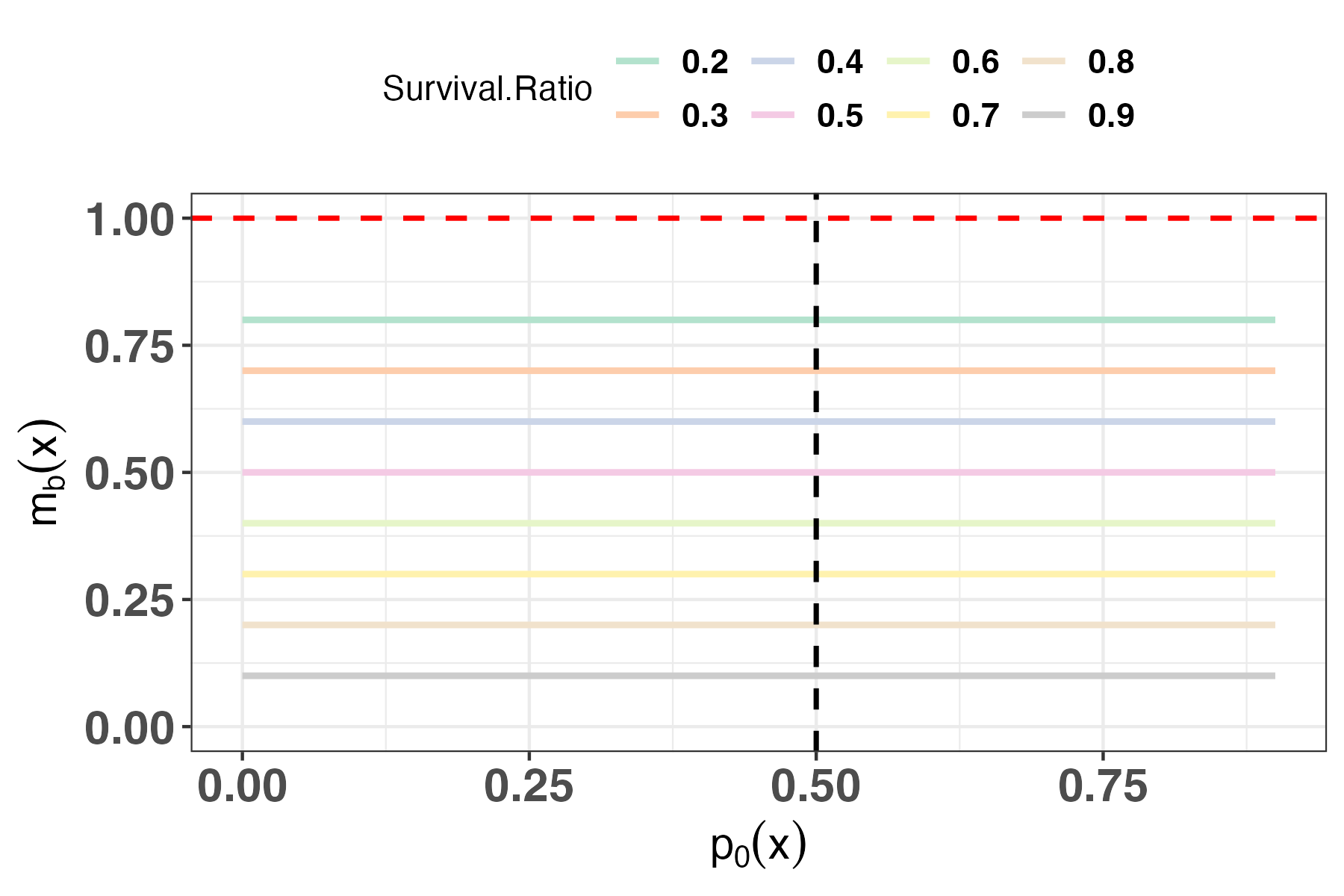}
    \end{minipage}
\end{figure}

We want to add that, as the treatment effect is assumed to increase the occurence of the event, then a better measure to use (at least if willing to maximise the chance to have a constant value for each individuals as claimed by the author) is the survival ratio. In particular, the Figure~\ref{fig:cumming1} can be adapted when considering a constant SR (see Figure~\ref{fig:cumming2}). One can observe that all ranges of the baseline risks are allowed.

Then, authors add the following comment.

\begin{quote}
   \textit{ Odds range from $0$ to infinity. Odds ratios greater than $1$ have no upper limit, regardless of the outcome odds for persons not exposed. If we multiply any unexposed outcome odds by an exposure odds ratio greater than 1 and convert the resulting odds when exposed to a risk, that risk will fall between $0$ and $1$. Thus, it is always hypothetically possible for an odds ratio to be constant for all individuals in a population.}
\end{quote}

We agree that it is always hypothetically possible for an odds ratio to be constant for all individuals (this corresponds to Lemma~\ref{lem:conditional-odds-ratio}, and $m(x)=m$ in the logistic working models). But note that this does not mean that the odds ratio at the individual level is then the same for the population level due to non-collapsibility.

\begin{quote}
    \textit{\textbf{Possibility of Constancy for Risk Ratios Less Than 1}. For both risk and odds, the lower limit is 0. For any level of risk or odds under no exposure, multiplication by a risk or odds ratio less than 1 will produce a risk or odds given exposure that is possible: 0 to 1 for risks and 0 to infinity for odds. Thus, a constant risk or odds ratio is possible for ratios less than 1. If the Risk Ratio comparing exposed persons with those not exposed is greater than 1, the ratio can be inverted to be less than 1 by comparing persons not exposed with those exposed. Therefore, a constant Risk Ratio less than 1 is hypothetically possible. This argument has been used to rebut the criticism of the Risk Ratio in the previous argument.}
\end{quote}

To us, this argument is a consequence of Lemma~\ref{lemma:monotonous-effect} accounting for the fact that a RR less than $1$ is comparable to $m_b(x) = 0$.

\subsection{Comment on Appendix 3 of \cite{huitfeldt2018choice}}

Many of our insights can be found in \cite{huitfeldt2018choice} (and in particular in their Appendix). 
What we want to highlight is that our notations and framework enable another view of the problem. First, we quote the authors.

\begin{quote}
    For illustration, we will consider an example concerning the effect of treatment with antibiotics ($A$), on mortality ($Y$). We will suppose that response to treatment is fully determined by bacterial susceptibility to that antibiotic ($X$). In the following, we will suppose that attribute $X$ has the same prevalence in populations s and t (for example because the two populations share the same bacterial gene pool) and that treatment with $A$ has no effect in the absence of $X$. Further, suppose that this attribute is independent of the baseline risk of the outcome (for example, old people at high risk of death may have the same strains of the bacteria as young people at low risk).
\end{quote}

Within the entanglement model, and denoting $X=0$ the absence of the mutation, this means that:
\begin{itemize}
    \item ``attribute $X$ has the same prevalence in populations s and t" which corresponds to Definition~\ref{def:shidted-covariates};
    \item``treatment with $A$ has no effect in the absence of $X$" $m_b(X=0) = m_g(X=0) = 0$,
    \item ``Further, suppose that this attribute is independent of the baseline risk of the outcome" Here, we think that this assumption could be easily transposed in our intrication model, clearly decomposing $X_B$ and $X_M$.
\end{itemize}

\subsection{Comment on the research work from Cinelli \& Pearl}

The way \cite{CinelliGeneralizing2019} deals with the problem is to encode the assumption of the problem with selection diagrams. In particular selection diagrams are an extension of DAGs with selection nodes, those nodes are used by the
analyst to indicate which local mechanisms are suspected to differ between two environments (in the Russian roulette example, the prevalence risk is suspected to differ between Los Angeles and New York, but not the mechanism).\\

A first difference to our work is that authors rather whant to predict in a target population $\mathbb{E}_{\text{\tiny T}}\left[Y^{(1)} \right]$ from $\mathbb{E}_{\text{\tiny T}}\left[Y^{(0)} \right]$ and $ \text{PS}_{01}$ and $ \text{PS}_{10}$ detailed below, while we focus on causal effects $\tau$. Another difference is that authors mostly reason marginally, while in our work we link subpopulations with larger populations relying on collapsibility.

Cinelli and Pearl introduce the following quantities:

\begin{equation*}
    \text{PS}_{01} := \mathbb{P}\left[ Y^{(1)} = 1 \mid Y^{(0)} = 0 \right],\quad \text{and}\quad   \text{PS}_{10} := \mathbb{P}\left[ Y^{(1)} = 0 \mid Y^{(0)} = 1 \right].
\end{equation*}

Those quantity corresponds to $\mathbb{E}\left[m_b(X)\right]$ and $\mathbb{E}\left[m_g(X)\right]$ defined in Lemma~\ref{lemma:intrication_model}. In their work, \cite{CinelliGeneralizing2019} assumes that $\mathbb{E}_\text{\tiny T}\left[m_b(X)\right] = \mathbb{E}_\text{\tiny S}\left[m_b(X)\right]$ and $\mathbb{E}_\text{\tiny T}\left[m_g(X)\right] = \mathbb{E}_\text{\tiny S}\left[m_g(X)\right]$. Therefore, their equation,

\begin{equation*}
    \mathbb{P}^{\Pi^*}\left[ Y^{(1)} = 1\right] = (1-\text{PS}_{10})  \mathbb{P}^{\Pi^*}\left[ Y^{(0)} = 1\right] + \text{PS}_{01}(1-\mathbb{P}^{\Pi^*}\left[ Y^{(0)} = 1\right]),
\end{equation*}

is completely equivalent to the entanglement model. Note that they do consider that $\mathbb{P}^{\Pi^*}\left[ Y^{(0)} = 1\right] $ ( which corresponds to $\mathbb{E}_{\text{\tiny T}}\left[ b(x) \right]$) varies when marginalized in another population. The entanglement model rather highlight the dependencies to covariates (i.e. chracteristics), while their equation rather models the fact that only the baseline risk is necessary to be known if 

\begin{equation*}
    Y^{(1)} \indep I \mid Y^{(0)},
\end{equation*}
where $I$ is the indicator of population's membership and if effect is monotonous (and they denote $Y^{(1)} \le Y^{(0)}$ or conversely depending on the direction assumed).

In our work, such assumption is equivalent with assuming monotonicity (either $m_b(x)=0$ or $m_g(x)=0$) and that all treatment effect modifiers are not shifted. Authors then propose to soften their assumptions deriving bounds on the target quantity $ \mathbb{P}^{\Pi^*}\left[ Y^{(1)} = 1\right] $. Our work rather keeps on targeting causal measure themselves, and  assume that we have access to the shifted covariates of $X_M$. We think this could be stated as,

\begin{equation*}
    Y^{(1)} \indep I \mid Y^{(0)}, X_M,
\end{equation*}
along with the monotonicity assumption.
Linking selection diagrams assumptions with results from Theorems~\ref{th_onlyRD_separates_baseline_tteffect} and  \ref{th_onlyRD_separates_baseline_tteffect_bounded_outcome} is an open work.

\section{Details about the simulations}
\label{appendix:additional_simulations}

\subsection{Comments on estimation}\label{appendix:comments-on-estimation}
In this paper, we have been focusing on identification rather than estimation. In this simulation, we illustrate the two approaches that can be taken when transforming identification formula (see Propositions~\ref{proposition:generalization-density} and \ref{prop:generalization-of-local-effects}) into estimation: Plug-in g-formula or Inverse Propensity Sampling Weighting (IPSW).
Existing consistency results of these approaches for the Risk Difference are reviewed in \cite{colnet2021causal}.
We assume that the data sampled from $P_{\text{\tiny S}}$ is a randomized trial $\mathcal{R}$ of size $n$ and the data sampled from $P_{\text{\tiny T}}$ is a cohort $\mathcal{T}$ of size $m$ which contains covariates information $X$ and possibly $Y^{(0)}$.

\subsubsection{Plug-in formula}

When considering \textit{generalization of the conditional outcome}, the plug-in g-formula consists in estimating the two surface responses $\mathbb{E}\left[ Y^{(a)} \mid X\right]$ using the RCT data from $P_{\text{\tiny T}}$. We denote by $\hat \mu_{a, n}(X)$ the estimates ($n$ is added to indicate that estimation is performed on the trial). Any approach can be proposed, for e.g. OLS or non-parametric learners. These models are then used on the target sample to estimate the averaged expected responses,

\begin{equation}\label{estimator-cond-outcome-g-formula}
    \hat{\mathbb{E}}_{\text{\tiny T}}\left[ Y^{(a)} \right] = \frac{1}{m} \sum_{i \in \mathcal{T}} \hat \mu_{a,n}(X),
\end{equation}
where $m$ denotes the target sample size. Doing so this estimate depends on the two sample sizes, $n$ and $m$. Finally, $ \hat{\mathbb{E}}_{\text{\tiny T}}\left[ Y^{(0)} \right]$ and $ \hat{\mathbb{E}}_{\text{\tiny T}}\left[ Y^{(1)} \right]$ are then used to estimate any causal measures on the target population: RD, RR, OR, and so on. Consistency of procedure \eqref{estimator-cond-outcome-g-formula} has been proven for any consistent estimator $\hat \mu_a$ of $\mathbb{E}\left[ Y^{(a)} \mid X\right]$ in \cite{colnet2022sensitivity}.\\

\textit{Generalizing local effects} using a plug-in formula suggests to estimate the local treatment effect (or CATE) $\hat \tau_n(x)$ using $\mathcal{S}$. This can be done using the previously introduced $\hat \mu_{a}(X)$ too (this is called T-learner), and then making a difference or a ratio of the two depending on the causal measure someone wants to generalize. Then, one has to estimate $\hat  g_m(X, P(X,Y^{(0)}))$ using $\mathcal{T}$, for exemple using a linear model (or any other model). Finally, one can obtain the target treatment effect with

\begin{equation}\label{estimtator-local-effect-g-formula}
    \hat \tau = \frac{1}{m}\sum_{i \in \mathcal{T}} \hat g_m(X_i, P(X_i,Y_i^{(0)})) \hat \tau_n(X_i),
\end{equation}

where $m$ denotes the target sample size. Note that \eqref{estimtator-local-effect-g-formula} relies on the estimation of $\tau(X)$ directly.
While the estimation of the conditional risk difference is well described in the literature \citep{wager2018estimation, nie2020quasioracle} (to name a few), estimation of conditional ratios is way less described. We have found only one recent work dealing with such questions \citep{yadlowsky2021estimation}. Consistency of such procedure for another metric than the Risk Difference is an open research question.

\subsubsection{Inverse Propensity Sampling Weighting (IPSW)}
IPSW uses the ratio of densities to re-weight individual observation in the trial. Denoting $r(X):=\frac{p_{\text{\tiny T}}(X)}{p_{\text{\tiny S}}(X)}$ the density ratio, one has first to learn this ratio $\hat r_{n,m}(X)$ using both data set $\mathcal{S}$ and $\mathcal{T}$. 
One can \textit{generalize conditional outcomes} doing:

\begin{equation*}
    \hat{\mathbb{E}}_{\text{\tiny T}}\left[ Y^{(a)} \right] = \frac{1}{n} \sum_{i \in \mathcal{S}} \hat{r}_{n,m}(X_i)A_iY_i.
\end{equation*}
Those estimates ($\hat{\mathbb{E}}_{\text{\tiny T}}\left[ Y^{(0)} \right]$ and $\hat{\mathbb{E}}_{\text{\tiny T}}\left[ Y^{(a1} \right]$) are then used to estimate any causal measures on the target population.

Now, considering \textit{generalizing local effects} using a re-weighing approach rather suggest to also estimate $\hat  g_m(X, P(X,Y^{(0)}))$ using $\mathcal{T}$ (for example using a linear model). Then, for e.g when considering the Risk Difference, this consists in doing

\begin{equation*}
\hat \tau_{\text{\tiny RD}} = \frac{1}{n} \sum_{i \in \mathcal{S}}\hat{r}(X_i) \left( A_iY_i - (1-A_i)Y_i\right),
\end{equation*}

or when considering the Risk Ratio, a procedure could be

\begin{equation*}
\operatorname{ln}\left(\hat  \tau_{\text{\tiny RR}} \right) =  \frac{1}{n} \sum_{i \in \mathcal{S}}\hat{r}(X_i) \left( \operatorname{ln}\left(A_iY_i\right) - \operatorname{ln}\left((1-A_i)Y_i\right) \right) \hat  w_m(X_i, P(X_i,Y_i^{(0)})).
\end{equation*}
 We use these weighting approaches for the simulation with a binary outcomes. As the purpose is not estimation, we propose a simulation with categorical covariates only, in particular to propose an estimation of $\hat r_{n,m}(X)$ as in \cite{colnet2022reweighting}. $\hat  w_m(X, P(X,Y^{(0)}))$ is estimating by computing the empirical mean of $\mathbb{E}\left[ Y^{(0)} \mid X\right]$ in each category.

\subsection{Continuous outcomes}\label{appendix:continuous}

\subsubsection{Data generative process}\label{appendix:continuous-generative-model}
We assume that the outcome is generated linearly from six covariates in the two populations
\begin{equation}
\label{eq:Ymodel-simulation-continuous}
    Y(a) = 0.05 X_1 +  0.04 X_2 + 2 X_3 + X_4 + 2 X_5 - 2 X_6 + a\cdot \left(1.5 X_1 + 2 X_2 + X_5 \right) +\epsilon \mbox{ with } \epsilon \sim \mathcal{N}(0,2).
\end{equation}

The two data samples are directly sampled from two different baseline distributions.

Covariates $X_1, X_2, X_3$ are generated from 

\begin{equation*}
    \mathcal{N}\left(\left[\begin{array}{l}
6 \\
5 \\
8
\end{array}\right],\left[\begin{array}{lll}
1 & 0  & 0.5 \\
0 & 1 & 0.2 \\
0.5 & 0.2 & 1 
\end{array}\right]\right)
\end{equation*}

in $P_\text{\tiny S}$, and in

\begin{equation*}
    \mathcal{N}\left(\left[\begin{array}{l}
15 \\
7 \\
10
\end{array}\right],\left[\begin{array}{lll}
1 & 0  & 0.5 \\
0 & 1 & 0.2 \\
0.5 & 0.2 & 1 
\end{array}\right]\right)
\end{equation*}

for $P_\text{\tiny T}$. $X_4$ is such that $X_4 \sim \mathcal{B}(1, 0.8)$ in $P_\text{\tiny S}$ and $X_4 \sim \mathcal{B}(1, 0.3)$ in $P_\text{\tiny T}$. Then, $X_5$ and $X_6$ are non-shifted covariates, where $X_5 \sim \mathcal{B}(1, 0.8)$ and $X_6 \sim \mathcal{N}(4, 1)$ in both populations.

Within the trial sample of size $n$ we generate the treatment according to a Bernoulli distribution with probability equals to $0.5$.

\paragraph*{Estimation}\label{appendix:continuous-estimation-steps}
For this simulation we applied a plug-in g-formula approach, using Ordinary Least Squares (OLS) to estimate $\hat \mu_{a,n}$ and $\hat  g_m(X, P(X,Y^{(0)}))$. $\hat \tau_n$ is estimated combining $\hat \mu_{a,n}$ as a difference or ratio or else (T-learner). More precisely, in this simulation the different steps when generalizing the conditional outcomes are the following : 
\begin{itemize}
    \item Fit an OLS estimator on the subset of treated individuals ($A=1$) in the trial sample to obtain $\hat \mu_{1,n}(X)$,
    \item Fit an OLS estimator on the subset of control individuals ($A=0$) in the trial sample to obtain $\hat \mu_{0,n}(X)$,
    \item Estimate the expected outcome if treated and control on the target population using the following formulae 
    $$\hat{\mathbb{E}}_{\text{\tiny T}}\left[ Y^{(a)} \right] = \frac{1}{m} \sum_{i \in \mathcal{T}} \hat \mu_{a,n}(X), $$
    \item Use the two previous quantities to estimate 
    \begin{itemize}
        \item The risk difference $\hat \tau_{\text{\tiny RD}} = \hat{\mathbb{E}}_{\text{\tiny T}}\left[ Y^{(1)} \right]-\hat{\mathbb{E}}_{\text{\tiny T}}\left[ Y^{(0)} \right]$,
        \item The Risk Ratio $\hat \tau_{\text{\tiny RR}} = \hat{\mathbb{E}}_{\text{\tiny T}}\left[ Y^{(1)} \right]/\hat{\mathbb{E}}_{\text{\tiny T}}\left[ Y^{(0)} \right]$,
        \item The excess Risk Ratio $\hat \tau_{\text{\tiny ERR}} = \left( \hat{\mathbb{E}}_{\text{\tiny T}}\left[ Y^{(1)} \right]-\hat{\mathbb{E}}_{\text{\tiny T}}\left[ Y^{(0)} \right]\right) /\hat{\mathbb{E}}_{\text{\tiny T}}\left[ Y^{(0)} \right]$.
    \end{itemize}
\end{itemize}

To generalize the local effects, we perform the following list of steps :

\begin{itemize}
    \item Rely on the first two steps performed to generalize the conditional outcomes, namely fit an OLS estimator on the trial sample to obtain $\hat \mu_{1,n}(X)$ and $\hat \mu_{0,n}(X)$,
    \item The risk difference is estimated with the following formula\footnote{Note that by linearity we retrieve that for the risk difference and for a continuous outcome, generalizing conditional outcomes or local effects is strictly equivalent.} $$ \hat \tau_{\text{\tiny RD}}  = \frac{1}{m} \sum_{i \in \mathcal{T}} \hat \mu_{1,n}(X) - \hat \mu_{0,n}(X),$$
    \item The Risk Ratio is obtained by
    \begin{itemize}
        \item Fitting an OLS estimator on the target sample to obtain $\hat \mu_{0,m}(X)$,
        \item To finally compute 
        $$\hat \tau_{\text{\tiny RR}} = \frac{1}{m} \sum_{i \in \mathcal{T}} \frac{\hat \mu_{1,n}(X_i)}{\hat \mu_{0,n}(X_i)} \underbrace{\frac{\hat \mu_{0,m}(X_i)}{\frac{1}{m} \sum_{j \in \mathcal{T}} \hat \mu_{0,m}(X_j)}}_\text{weights estimated on $\mathcal{T}$}.$$
    \end{itemize}
\end{itemize}

\begin{figure}[H]
    \centering
    \includegraphics[width=0.9\textwidth]{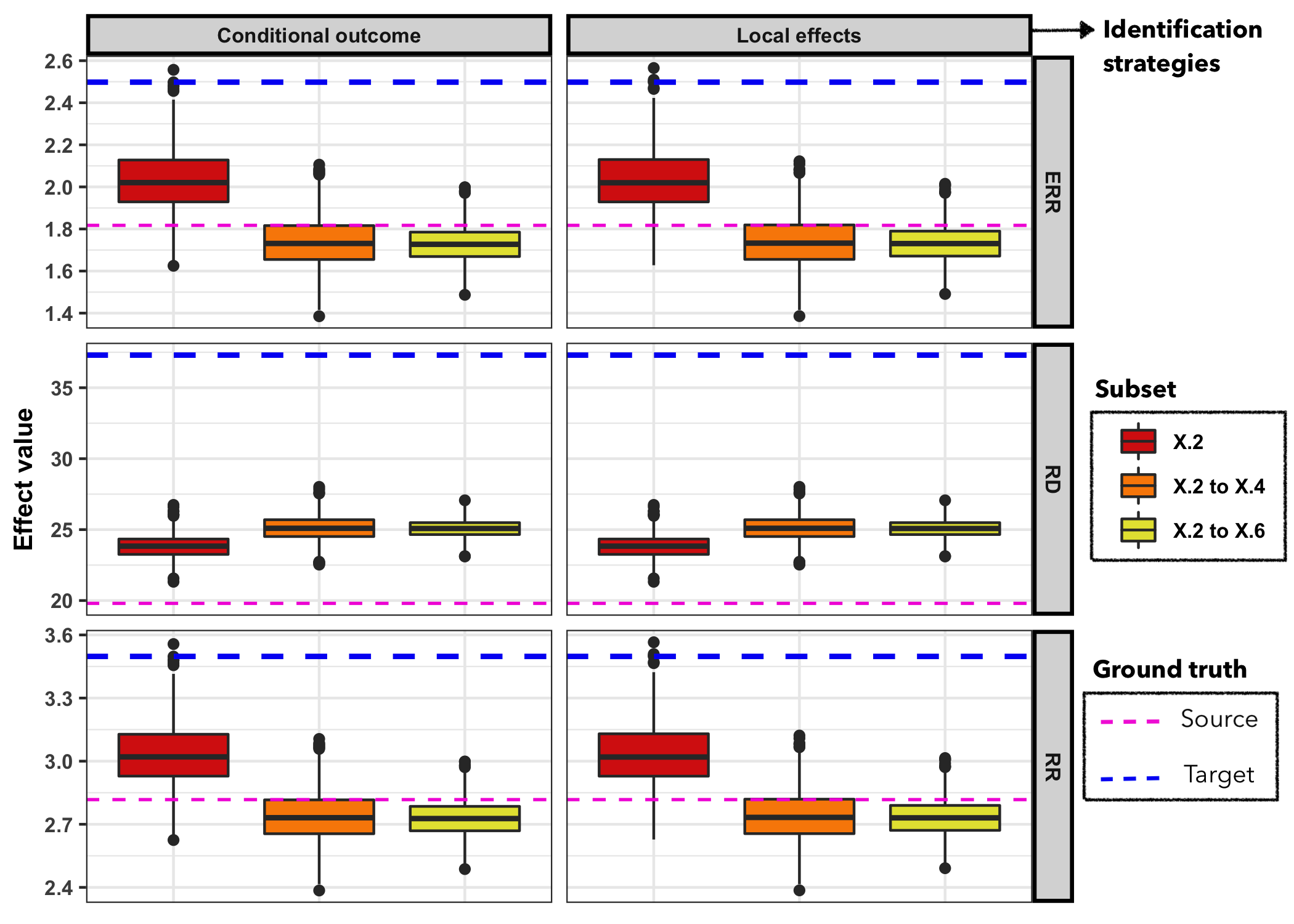}
    \caption{\textbf{Results of the simulations for a continuous outcomes without observing $X_1$}: where the generative model corresponds to \eqref{eq:simulation-continuous-generative-model}. Column 1 corresponds to generalizing conditional outcome, column 2 corresponds to generalizing local effect with the proper collapsibility weights. For these two approaches we use different covariates set, with $X_2$, $X_{2 \dots 4}$, and $X_{2 \dots 6}$. According to Theorem~\ref{theorem:all-covariates} and \ref{theorem:restricted-set-for-Y-continuous-RD}, the target treatment effect can not be identified when a shifted treatment effect modifier is unobserved. Simulations are performed following the exact same procedure than Figure~\ref{fig:simulations-continuous-Y}, with $1000$ repetitions, a source sample size of $500$ and target sample size of $1,000$.}
    \label{fig:simulations-continuous-Y-missing-X1}
\end{figure}

\paragraph*{What if a shifted treatment effect modifier is missing?} This situation leads to a biased estimate \citep{nguyen2018sensitivitybis, colnet2022sensitivity}. To illustrate such situation we replicated simulations presented in Figure~\ref{fig:simulations-continuous-Y} but without covariate $X_1$. Results are presented on Figure~\ref{fig:simulations-continuous-Y-missing-X1}.

\paragraph*{What if mispecification occurs?} This situation leads to a biased estimate. To illustrate such situation we introduced a different generative model for the outcome such that eq.~\ref{eq:Ymodel-simulation-continuous} becomes

\begin{equation}
\label{eq:Ymodel-simulation-continuous-mispe}
    Y(a) = 0.05 X_1^2 +  0.04 X_2 + 2 X_3 + X_4 + 2 X_5 - 2 X_6 + a\cdot \left(1.5 X_1^2 + 2 X_2 + X_5 \right) +\epsilon \mbox{ with } \epsilon \sim \mathcal{N}(0,2).
\end{equation}

\begin{figure}[H]
    \centering
    \includegraphics[width=0.9\textwidth]{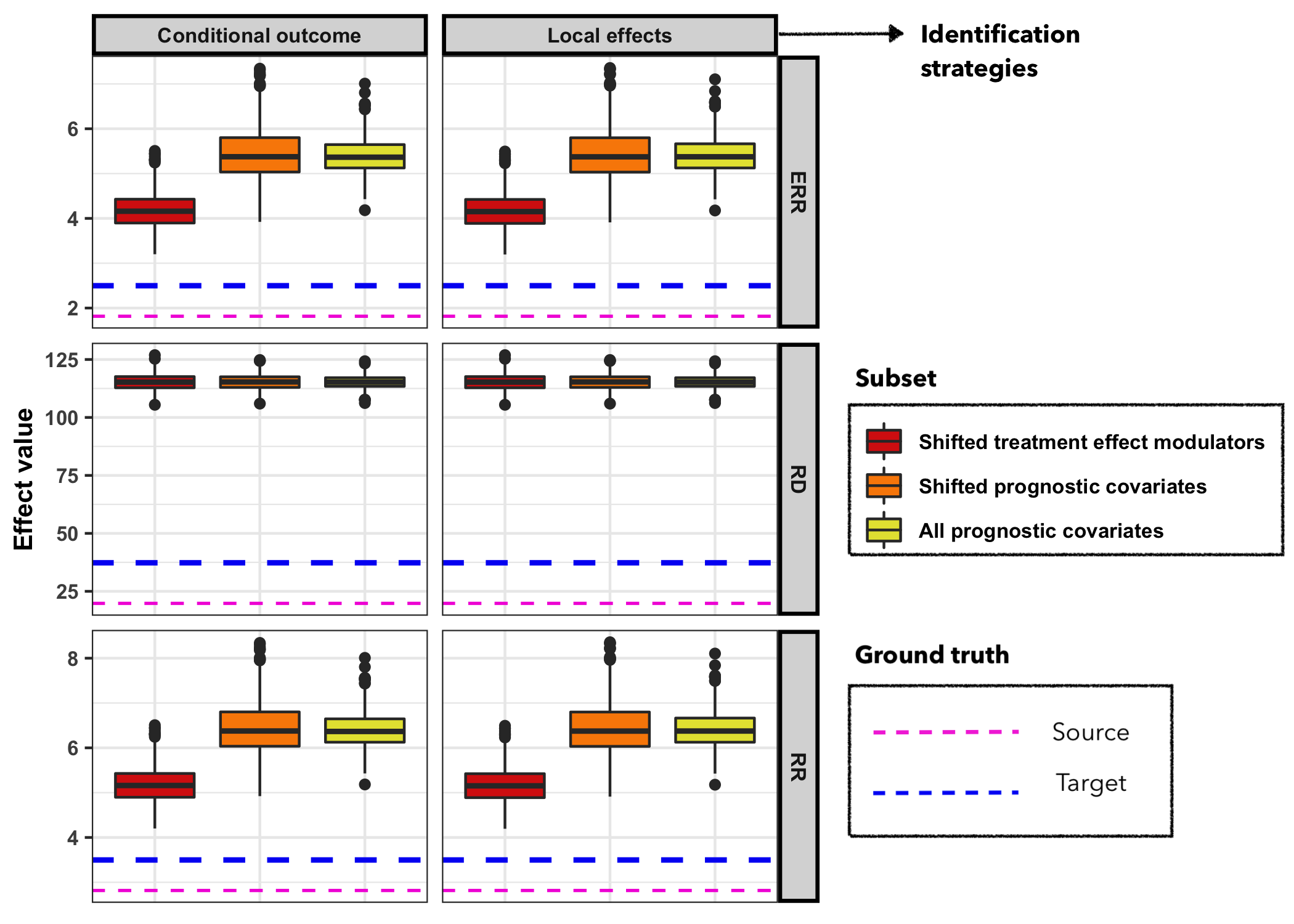}
    \caption{\textbf{Results of the simulations for a continuous outcomes with mispecification}: where the generative model corresponds to \eqref{eq:Ymodel-simulation-continuous-mispe} rather than \eqref{eq:Ymodel-simulation-continuous}. Column 1 corresponds to generalizing conditional outcome, column 2 corresponds to generalizing local effect with the proper collapsibility weights. For these two approaches we use different covariates set, with $X_2$, $X_{2 \dots 4}$, and $X_{2 \dots 6}$. According to Theorem~\ref{theorem:all-covariates} and \ref{theorem:restricted-set-for-Y-continuous-RD}, the target treatment effect can not be identified when a shifted treatment effect modifier is unobserved. Simulations are performed following the exact same procedure than Figure~\ref{fig:simulations-continuous-Y}, with $1000$ repetitions, a source sample size of $500$ and target sample size of $1,000$. Estimation is performed with plug-in g-formula modeling all responses with an OLS approach as detailed in Section~\ref{appendix:continuous-estimation-steps}.}
    \label{fig:simulations-continuous-Y-mispe}
\end{figure}

Then, and using a simple OLS estimator without squared terms such as described in Section~\ref{appendix:continuous-estimation-steps} and presented in Figure~\ref{fig:simulations-continuous-Y}, leads to a biased estimate in all situations (generalizing local effects or conditional outcomes, and with any of the covariates subsets). Results are presented on Figure~\ref{fig:simulations-continuous-Y-mispe}.

\subsection{Binary outcomes}\label{appendix:simulation-binary}

\subsubsection{Data generative process}\label{appendix:simulation-binary-data-generative-model}
For this simulation the covariates are categorical to ease the estimation strategy, and as the purpose of this work is not on estimation. The data generative model is build on top of \eqref{eq:intuition-of-intrication}, and adapted to give,
\begin{equation*}
    \mathbb{P}\left[ Y^{(a)} = 1 \mid X = x \right] = b(X_1, X_2, X_3) + a\,\left(1-b\left(X_1, X_2, X_3\right)\right)\,m_b(X_2, X_3),
\end{equation*}
where $X_1=\texttt{lifestyle}$, $X_2=\texttt{stress}$, and $X_3=\texttt{gender}$.

Each of the three covariates are sampled following a Bernoulli distribution. 
In $P_\text{\tiny S}$, one has $X_1 \sim \mathcal{B}(1, 0.4)$, $X_2 \sim \mathcal{B}(1, 0.8)$, and $X_3 \sim \mathcal{B}(1, 0.5)$. In $P_\text{\tiny T}$, one has $X_1 \sim \mathcal{B}(1, 0.6)$, $X_2 \sim \mathcal{B}(1, 0.2)$, and $X_3 \sim \mathcal{B}(1, 0.5)$.

The outcome is defined such as,

  \begin{equation*}
    b(X) =  \operatorname{ifelse}(X_1 = 1, 0.2, 0.05) \cdot  \operatorname{ifelse}(X_2  = 1, 2, 1) \cdot  \operatorname{ifelse}(X_3 = 1, 0.5, 1), 
  \end{equation*}

where $ \operatorname{ifelse}$ corresponds to the function with the same name in \texttt{R}. And,

  \begin{equation*}
    m_b(X) = \operatorname{ifelse}(X_2 = 1, 1/4, \operatorname{ifelse}(X_3 = 1, 1/10, 1/6)).
  \end{equation*}

Within the trial sample of size $n$ we generate the treatment according to a Bernoulli distribution with probability equals to $0.5$. 

\subsubsection{Estimation}\label{appendix:simulation-binary-estimation-steps}

First we estimate $\mu_{a}(.)$ on the trial sample. As covariates are categorical this corresponds to computing average values of $Y$ in each bin, namely

\begin{equation*}
   \forall x \in \mathcal{X},\;\; \hat \mu_{1,n} (x):=  \frac{\sum_{i \in \mathcal{S}; X_i = x} Y_i A_i}{\sum_{i \in \mathcal{S}} \mathbbm{1}_{A_i = 1} \mathbbm{1}_{X_i = x}}\;\; \text{ and,    }\,\hat \mu_{0,n} (x):=  \frac{\sum_{i \in \mathcal{S}; X_i = x} Y_i (1-A_i)}{\sum_{i \in \mathcal{S}} \mathbbm{1}_{A_i = 0} \mathbbm{1}_{X_i = x}}.
\end{equation*}

This allows to estimate $ \hat{\mathbb{E}}_{\text{\tiny T}}\left[ Y^{(0)} \right]$ and $ \hat{\mathbb{E}}_{\text{\tiny T}}\left[ Y^{(1)} \right] $ with 

\begin{equation*}
    \hat{\mathbb{E}}_{\text{\tiny T}}\left[ Y^{(1)} \right] = \frac{\sum_{i \in \mathcal{T}} \hat \mu_{1,n}(X_i)}{m}  \, \text{ and,    }\,  \hat{\mathbb{E}}_{\text{\tiny T}}\left[ Y^{(0)} \right] = \frac{\sum_{i \in \mathcal{T}} \hat \mu_{0,n}(X_i)}{m} .
\end{equation*}

Then, these two quantities are used to estimate :
\begin{itemize}
    \item The risk difference $\hat \tau_{\text{\tiny RD}} = \hat{\mathbb{E}}_{\text{\tiny T}}\left[ Y^{(1)} \right] - \hat{\mathbb{E}}_{\text{\tiny T}}\left[ Y^{(0)} \right] $,

    \item The number needed to treat $\hat \tau_{\text{\tiny NNT}} = \tau_{\text{\tiny RD}}^{-1}$,

    \item The Risk Ratio $\hat \tau_{\text{\tiny RR}} = \hat{\mathbb{E}}_{\text{\tiny T}}\left[ Y^{(1)} \right] / \hat{\mathbb{E}}_{\text{\tiny T}}\left[ Y^{(0)} \right] $,

     \item The survival ratio $\hat \tau_{\text{\tiny SR}} = \left( 1-\hat{\mathbb{E}}_{\text{\tiny T}}\left[ Y^{(1)} \right]\right) / \left( 1- \hat{\mathbb{E}}_{\text{\tiny T}}\left[ Y^{(0)} \right] \right) $,

\item The odds ratio $\hat \tau_{\text{\tiny OR}} =  \left( \hat{\mathbb{E}}_{\text{\tiny T}}\left[ Y^{(1)} \right] /  \left( 1- \hat{\mathbb{E}}_{\text{\tiny T}}\left[ Y^{(1)} \right] \right)\right)  \cdot \left( \hat{\mathbb{E}}_{\text{\tiny T}}\left[ Y^{(0)} \right] /  \left( 1- \hat{\mathbb{E}}_{\text{\tiny T}}\left[ Y^{(0)} \right] \right)\right)^{-1} $.

\end{itemize}

This procedure corresponds to the generalization of  the outcome.

On the other hand, for each categories in the trial sample, each causal measure $\tau$ is estimated on each strata to obtain $\tau(x)$, using the fitted outcome models $\hat \mu_{1,n}(.)$ and $\hat \mu_{0,n}(.)$. This corresponds to local effects. For example for the Risk Ratio, one has :

\begin{equation*}
    \forall x \in \mathcal{X},\, \hat \tau_{\text{\tiny RR},n}(x) :=  \frac{\hat  \mu_{1,n}(X_i)}{\hat \mu_{0,n}(X_i)}.
\end{equation*}

Then collapsibility weights are estimated as followed :

\begin{enumerate}
    \item Estimate the outcome model on the target population 

$$ \hat \mu_{0,m} (x):=  \frac{\sum_{i \in \mathcal{T}; X_i = x} Y_i }{\sum_{i \in \mathcal{S}} \mathbbm{1}_{X_i = x}}.$$

\item So that 

$$\hat w_{\text{\tiny T},m}(x) :=  \frac{\hat \mu_{0,m} (x)}{\frac{\sum_{i \in \mathcal{T}}Y_i}{m}}. $$
    
\end{enumerate}

Finally and for each causal measure $\tau$, the target effect is obtained computing :

\begin{equation*}
  \hat \tau_{\text{\tiny T},n,m} :=  \frac{1}{m}\sum_{i \in \mathcal{T}} \hat w_{\text{\tiny T},m}(X_i)  \hat \tau_n(X_i).
\end{equation*}

\newpage

\begin{landscape}

\bgroup
\def\arraystretch{2}%  1 is the default, change whatever you need

 \begin{table}
 \begin{center}
 {\scriptsize
\begin{tabular}{|l|l|l|l|}
\hline

 \multicolumn{1}{|c|}{\textbf{Name}} &  \multicolumn{1}{|c|}{\textbf{Outcome type}} & \multicolumn{1}{c|}{\textbf{Definition}}  & \multicolumn{1}{c|}{\textbf{\begin{tabular}[c]{@{}c@{}}Invariant\\ to encoding\end{tabular}}}  \\ \hline \hline
 Risk Difference  (RD)       & Continuous           &     $\tau_{\text{\tiny RD}} := \mathbb{E}\left[Y^{(1)}\right] - \mathbb{E}\left[Y^{(0)}\right]$                              &  Not applicable               \\ \hline
Risk Ratio (RR)            & Continuous                 &       $\tau_{\text{\tiny RR}} := \mathbb{E}\left[Y^{(1)}\right] / \mathbb{E}\left[Y^{(0)}\right]$                                                                                        &   Not applicable                                                                                                                   \\ \hline
 Excess Risk Ratio (ERR)       & Continuous               &    $\tau_{\text{\tiny ERR}}  := \tau_{\text{\tiny RD}}/\mathbb{E}\left[Y^{(0)}\right] = \tau_{\text{\tiny RR}} - 1 $                                                                               &        Not applicable                                                                                                                 \\ \hline \hline  \hline 
Risk Difference (RD)       & Binary                 &               $\tau_{\text{\tiny RD}} := \mathbb{P}\left[Y^{(1)} = 1\right] - \mathbb{P}\left[Y^{(0)} = 1\right]$                                                                 & Multiplied by $-1$                                        \\ \hline
 Number Needed to Treat (NNT)    & Binary         &  $\tau_{\text{\tiny RD}} := 1 / \left( \mathbb{P}\left[Y^{(1)} = 1\right] - \mathbb{P}\left[Y^{(0)} = 1\right] \right)$                 &  Multiplied by $-1$                                                                                                                                             \\ \hline
Risk Ratio (RR)        & Binary              &  $\tau_{\text{\tiny RR}} := \mathbb{P}\left[Y^{(1)} = 1\right] / \mathbb{P}\left[Y^{(0)} = 1\right]$                     &   $= \tau_{\text{\tiny SR}}$                                                                                             \\ \hline
Survival Ratio (SR)      & Binary              &   $\tau_{\text{\tiny SR}} := \mathbb{P}\left[Y^{(1)} = 0\right] / \mathbb{P}\left[Y^{(0)} = 0\right]$                 &      $= \tau_{\text{\tiny RR}}$                                                                                                                                     \\ \hline
Excess Risk Ratio (ERR)    & Binary                 &    $\tau_{\text{\tiny ERR}}  := \tau_{\text{\tiny RD}}/\mathbb{P}\left[Y^{(0)} = 1\right] = \tau_{\text{\tiny RR}} - 1 $                                                                          &        $= \tau_{\text{\tiny SR}} - 1$                                                                                                                 \\ \hline 
Relative Susceptibility (RS)      & Binary             &    $\tau_{\text{\tiny RS}}  := \tau_{\text{\tiny RD}}/\mathbb{P}\left[Y^{(0)} = 0\right] = 1 - \tau_{\text{\tiny SR}}$                                                     &        $= 1-\tau_{\text{\tiny RR}}$                                                                                                                      \\ \hline 
Odds Ratio (OR)  & Binary   &$\tau_{\text{\tiny OR}}  := \frac{\mathbb{P}[Y^{(1)} = 1]}{\mathbb{P}[Y^{(1)} = 0]}\, \left(  \frac{\mathbb{P}[Y^{(0)} = 1]}{\mathbb{P}[Y^{(0)} = 0]}\right)^{-1}  = \tau_{\text{\tiny RR}} \cdot \tau_{\text{\tiny SR}}^{-1}$           &  Reciprocal                                                                                                                                  \\ \hline
Log Odds Ratio (log-OR)     & Binary           & $\tau_{\text{\tiny log-OR}}  := \operatorname{log}\left( \frac{\mathbb{P}[Y^{(1)} = 1]}{\mathbb{P}[Y^{(1)} = 0]}\right) - \operatorname{log}\left( \frac{\mathbb{P}[Y^{(0)} = 1]}{\mathbb{P}[Y^{(0)} = 0]}\right)$                   &  Multiplied by $-1$                                                                                                                                                                                                                                      \\ \hline
\end{tabular}}
\caption{\textbf{Typical causal measures reported in clinical practice}: The upper part of the Table mentions the three typical measures found when the outcome is ordinal or continuous, and the lower part mentions measures for binary outcomes. For each measure we provide the explicit formulae, and invariance to encoding (also called symetry in the literature).}
\label{tab:list-measures-with-all-properties}
\end{center}
\end{table}

\egroup

\newpage

\begin{table}[h!]
\centering
\begin{tabular}{l cc ccc}
\toprule
\multirow{1}{*}{Measures}
& \multicolumn{2}{c}{Intrinsic properties} & \multicolumn{3}{c}{Generalization properties} \\
\cmidrule(lr){2-3} \cmidrule(lr){4-6}
& Collapsible  & Favorable CATE setting &  Under \Cref{a:transportability-wide}, using  & Under \Cref{a:transportability} & Under \Cref{a:transportability}\\
&   & &  generalization via potential outcomes & generalization via local effects & generalization via local effects \\
&   & &  generalization via potential outcomes & using the target baseline & without the target baseline \\\midrule
Risk Difference & \checkmark \checkmark & Additive effect & \checkmark & \checkmark & \checkmark \\
Number needed to treat & \checkmark & ? & \checkmark & \xmark & \xmark \\
Risk Ratio & \checkmark \checkmark & Multiplicative beneficial effect & \checkmark & \checkmark & \xmark \\
Survival Ratio  & \checkmark \checkmark & Multiplicative detrimental effect & \checkmark & \checkmark & \xmark \\
Odds Ratio & \xmark & ? & \checkmark & \xmark & \xmark \\
Log Odds Ratio  & \xmark & ? & \checkmark & \xmark & \xmark \\
\bottomrule
\end{tabular} 
\caption{Properties of causal measures presented in \Cref{tab:list-measures-with-all-properties}. Excess Risk Ratio and Relative Susceptibility are not presented are they equal to the Risk Ratio and Survival Ratio respectively, up to a linear transformation. In the first column, double $\checkmark $ stands for collapsibility, a single checkmark for measures that are logic-respecting but not collapsible and $\xmark$ corresponds to non logic-respecting measures. Generalization via potential outcomes correspond to \Cref{proposition:generalization-density} and generalization via local effects to \Cref{prop:generalization-of-local-effects}.}
\label{tab:measures_properties}
\end{table}

\end{landscape}
\end{document}